\documentclass[12pt,twoside,a4paper]{article}
\usepackage{epsfig}
\usepackage{a4}
\usepackage{graphics}
\usepackage{subfigure}
\usepackage{amsmath}
\usepackage[dvips]{color}

\newcommand{\bmath}[1]{\mbox{\boldmath ${#1}$}}

\newcommand{\half}{\mbox{${\textstyle \frac{1}{2}}$}}           
\newcommand{\fourth}{\mbox{${\textstyle \frac{1}{4}}$}}         
\newcommand{\fmn}[2]{\mbox{${\textstyle \frac{#1}{#2}}$}}
\newcommand{\dd}{\textrm{d}}
\newcommand{\bfg}[1]{{\mbox{\boldmath $#1$}}}
\newcommand{\boldtau}{\mbox{\boldmath $\tau$}}
\newcommand{\boldpi}{\mbox{\boldmath $\pi$}}
\newcommand{\shalf}{\mbox{$\sqrt{\textstyle \frac{1}{2}}$}}     
\newcommand{\ket}[1]{\left| #1\right>}              
\newcommand{\Att}{\mbox{$A_{TT}$}}
\newcommand{\ww}{\mbox{$\,$}}
\def\sigup{\sigma^{\uparrow\uparrow}}
\def\sigdw{\sigma^{\uparrow\downarrow}}
\begin{document}
\pagestyle{empty}
\baselineskip 4ex%
\begin{titlepage}
\begin{center}
\textbf{\Large \textcolor{blue} {Proposal}} \\
for the \\
\textbf{\Large \textcolor{blue} {SPIN PHYSICS FROM COSY TO FAIR}}\\[2ex]
\textbf{\large A proposed programme for polarisation experiments
in the COSY ring which could open the way to a polarised
antiproton facility at FAIR}
\\[3ex]
J\"ulich, August 2005
\end{center}
\vfill
\begin{figure}[hbt]
\begin{center}
  \psfig{figure=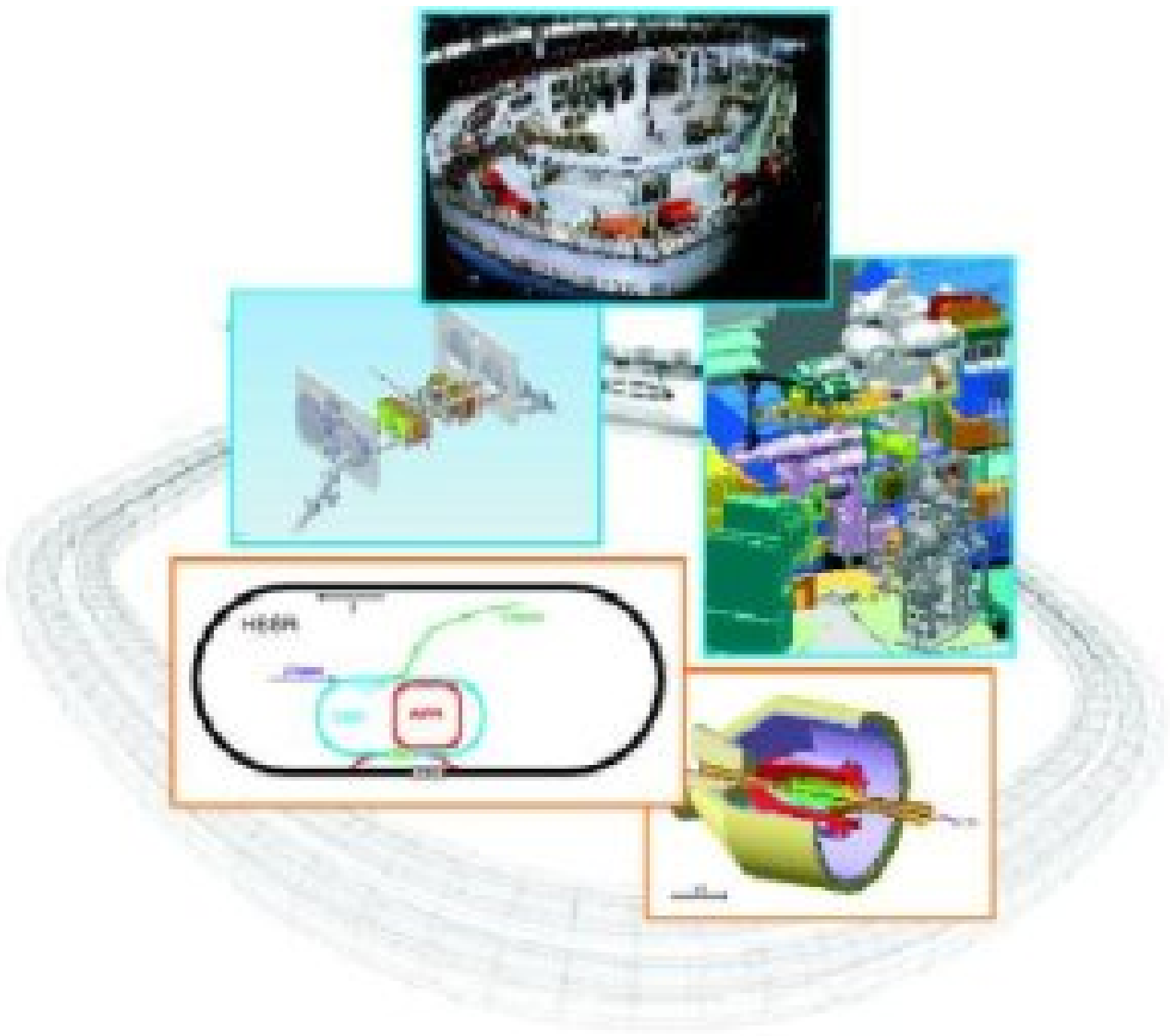,width=\textwidth}
\end{center}
\end{figure}
\end{titlepage}
\cleardoublepage
%
%
\vspace*{-2cm}

\baselineskip 3ex
\begin{center}
A.~Kacharava$^{1,a}$, F.~Rathmann$^{2,b}$, and
C.~Wilkin$^{3,c}$\\[1ex]
for the ANKE Collaboration:\\[1ex]
\end{center}

{\footnotesize
\begin{center}
S.~Barsov$^{4}$, %
V.G.~Baryshevsky$^{5}$, %
M.~B\"uscher$^{2}$, %
M.~Capiluppi$^{6}$, %
J.~Carbonell$^{7}$, %
G.~Ciullo$^{6}$, %
D.~Chiladze$^{2}$, %
M.~Contalbrigo$^{6}$, %
P.F.~Dalpiaz$^{6}$, %
S.~Dymov$^{8}$, %
A.~Dzyuba$^{2}$, 
R.~Engels$^{2}$, %
P.D.~Eversheim$^{9}$, %
A.~Garishvili$^{10}$, %
A.~Gasparyan$^{11}$, %
R.~Gebel$^{2}$, %
V.~Glagolev$^{12}$, %
K.~Grigoriev$^{4}$, %
A.~Gussen$^{11}$, %
D.~Gussev$^{8}$, %
J.~Haidenbauer$^{2}$, %
C.~Hanhart$^{2}$, %
M.~Hartmann$^{2}$, %
V.~Hejny$^{2}$, %
P.~Jansen$^{13}$, %
I.~Keshelashvili$^{2}$, %
V.~Kleber$^{2}$, %
F.~Klehr$^{13}$, %
H.~Kleines$^{14}$, %
A.~Khoukaz$^{15}$, %
V.~Koptev$^{4}$, %
P.~Kravtsov$^{4}$, %
A.~Lehrach$^{2}$, %
P.~Lenisa$^{6}$, %
V.~Leontiev$^{2}$, %
B.~Lorentz$^{2}$, %
V.~Komarov$^{8}$, %
A.~Kulikov$^{8}$, %
V.~Kurbatov$^{8}$, %
I.~Lehmann$^{16}$,%
G.~Macharashvili$^{8}$, %
Y.~Maeda$^{2}$, 
S.~Martin$^{2}$, %
T.~Mersmann$^{15}$, %
I.~Meshkov$^{8}$, %
M.~Mielke$^{15}$, 
M.~Mikirtytchiants$^{4}$, %
S.~Mikirtytchiants$^{4}$, %
U.-G.~Mei{\ss}ner$^{2}$, %
S.~Merzliakov$^{8}$, %
A.~Mussgiller$^{2}$, %
M.~Nekipelov$^{2}$, %
N.N.~Nikolaev$^{2}$, %
M.~Nioradze$^{10}$, %
D.~Oellers$^{2}$, %
H.~Ohm$^{2}$, %
Z.~Oragvelidze$^{10}$, 
M.~Papenbrock$^{15}$, 
D.~Prasuhn$^{2}$, %
T.~Rausmann$^{15}$, %
M.~Rentmeester$^{17}$, %
J.~Sarkadi$^{2}$, %
R.~Schleichert$^{2}$, %
V.~Serdjuk$^{8}$, %
H.~Seyfarth$^{2}$, %
A.~Smirnov$^{8}$, %
M.~Stancari$^{6}$, %
M.~Statera$^{6}$, %
E.~Steffens$^{1}$, %
H.~Str\"oher$^{2}$, %
A.~Sydorin$^{8}$, %
M.~Tabidze$^{10}$, %
P.~Engblom-Th\"orngren$^{16}$,
S.~Trusov$^{8}$, %
Yu.~Uzikov$^{8}$, %
Yu.~Valdau$^{2}$, %
A.~Vassiliev$^{4}$, %
M.~Wang$^{6}$, %
S.~Yaschenko$^{1}$, %
and I.~Zychor$^{18}$\\
\end{center}
}
\vspace{1cm}

{\footnotesize
\noindent
$^{1}$Physikalisches Institut II, Universit\"at Erlangen--N\"urnberg, 91058 Erlangen, Germany\\
$^{2}$Institut f\"ur Kernphysik, Forschungszentrum J\"ulich, 52425 J\"ulich, Germany\\
$^{3}$Physics and Astronomy Department, UCL, London WC1E 6BT, U.K. \\
$^{4}$High Energy Physics Department, PNPI, 188350 Gatchina, Russia\\
$^{5}$Research Institute for Nuclear Problems, Belarusian State
University, Minsk 220050, Belarus\\
$^{6}$University of Ferrara and INFN, 44100 Ferrara, Italy\\
$^{7}$Laboratoire de Physique Subatomique et de Cosmologie, 38026
Grenoble Cedex, France \\
$^{8}$Joint Institute for Nuclear Research, DLNP, 141980 Dubna, Russia\\
$^{9}$Helmoltz Institut f\"ur Strahlen und Kernphysik, Universit\"at Bonn, 53115 Bonn, Germany\\
$^{10}$High Energy Physics Institute, Tbilisi State University, 0186 Tbilisi, Georgia\\
$^{11}$Institute for Theoretical and Experimental Physics, 117259 Moscow, Russia\\
$^{12}$Joint Institute for Nuclear Research, VBLHE, 141980 Dubna, Russia\\
$^{13}$Zentrallabor Technologie, Forschungszentrum J\"ulich, 52425 J\"ulich, Germany\\
$^{14}$Zentrallabor Elektronik, Forschungszentrum J\"ulich, 52425 J\"ulich, Germany\\
$^{15}$Institut f\"ur Kernphysik, Universit\"at M\"unster, 48149 M\"unster, Germany\\
$^{16}$Department of Radiation Sciences, Box 535, S--751 21,
Uppsala, Sweden\\
$^{17}$Institute for Theoretical Physics, Radboud University Nijmegen, Nijmegen, The Netherlands\\
$^{18}$The Andrzej Soltan Institute for Nuclear Studies, 05400, Swierk, Poland \\
}

\vfill\noindent
$^a$ Email: a.kacharava@fz-juelich.de\\
$^b$ Email: f.rathmann@fz-juelich.de\\
$^c$ Email: cw@hep.ucl.ac.uk

\baselineskip 3ex
\cleardoublepage
%
%
\setcounter{page}{3}
\pagestyle{plain}
\newpage\vspace{-2cm}
\section{Executive Summary}

\subsection{Introduction} It is the aim of the ANKE--COSY spin
collaboration to carry out a well directed programme of
internationally competitive experiments involving polarised beams
and targets, using the outstanding facilities available at the
storage ring. These activities, at the same time, are good
preparation for our participation in the PAX{\tt @}FAIR project.
This Executive Summary will present a short description of the
apparatus that can be used for this purpose at COSY and then
discuss some of the principal experiments that will be undertaken
within the scope of this collaboration. It concludes by outlining
the PAX proposal.

\subsection{Experimental Facilities}

COSY accelerates and stores unpolarised and polarised protons and
deuterons in a momentum range between 0.3{\ww}GeV/c and 3.7{\ww}GeV/c. To
provide high quality beams, there is an Electron Cooler at
injection and a Stochastic Cooling System from 1.5{\ww}GeV/c up to the
maximum momentum. Transversally polarised beams of protons are
available with intensities up to $1.2\times10^{10}$ (with multiple
injection and electron cooling and stacking) and polarisations of
more than 80\%. For deuterons an intensity of about $3\times
10^{10}$ was achieved in the February 2005 run with vector and
tensor polarisations of more than 70\% and 50\% respectively.

Fast particles can be measured in the ANKE magnetic spectrometer
installed at an internal beam position of COSY. Detection systems
for both positively and negatively charged particles include
plastic scintillator counters for TOF measurements, multi--wire
proportional chambers for tracking, and range telescopes for
particle identification. A combination of scintillation and
\v{C}erenkov counters, together with wire chambers, allow one to
identify negatively charged pions and kaons. The forward detector,
comprising scintillator hodoscopes, \v{C}erenkov counters, and
fast proportional chambers, is used to measure  particles with
high momenta, close to that of the circulating COSY beam. There is
also a detector that can be used as a spectrometer for
backward--emitted particles.

Although strip and cluster--jet targets have been standard for use
at ANKE, we are currently in the process of installing a polarised
internal target (PIT) system, consisting of an atomic beam source,
feeding a storage cell, and a Lamb--shift polarimeter. The cell,
which will increase tremendously the available luminosity, was
tested \textit{in situ} in February 2005 and the whole apparatus
will be ready for commissioning experiments in early 2006. The
design is such that the target and polarimeter can be moved in and
out of the beam position, depending upon the requirements of the
experiment.

One of the major advantages of doing experiments at a storage ring
is that very low energy particles emanating from the very thin
targets can be detected in silicon tracking telescopes placed in
the target chamber. These are used to help in the measurement of
elastic scattering, which is vital for luminosity and polarisation
calibrations and, for a long target, establishing the vertex
position. However, their most exciting use is for measuring the
angles and energies of low energy protons ($<10${\ww}MeV) that emerge
as so--called \textit{spectators} from the interactions of beam
protons with the neutrons in the deuterium target. This
information allows one to determine the proton--neutron
centre--of--mass energy with high accuracy and will permit the
study of a whole range of $pn$ elastic and inelastic reactions.
The development of very thick (5--20{\ww}mm) double--sided
micro--structured Si(Li) and very thin (69{\ww}$\mu$m) double--sided
Si detectors provides a very flexible system for the use of the
telescopes in particle identification and angle and energy
measurement which, in the case of protons, will be from about
2.5{\ww}MeV up to 40{\ww}MeV. Each spectator detector can have typically a
10\% geometrical acceptance with respect to a point target and,
depending upon the needs of an individual experiment, up to four
or six telescopes could be employed.

Although some information on the beam polarisation is available
from the source, the standard methodology for determining it will
be through the comparison with several reactions with known
analysing powers that can be measured simultaneously in ANKE. For
example, in the interaction of polarised deuterons with a hydrogen
target, the vector and tensor polarisations could each be
determined in three different ways at 1.17{\ww}GeV. At energies where
the calibration reactions are unavailable, it is possible to use
the \textit{polarisation export} technique where, say, the
deuteron beam polarisation is measured at 1.17{\ww}GeV, the energy is
ramped to the region of interest for the physics measurement,
before being reduced again to 1.17{\ww}GeV, where the beam
polarisation is remeasured to check that any depolarisation is
unimportant. We have shown that this method works very well for
both proton and deuteron beams. The polarisation of the target
cell will also be checked through known standards, some of which
we ourselves will establish with the polarised beams.

\subsection{Physics Programme}

With the equipment available, many reactions will necessarily be
detected simultaneously. However, in order to give a flavour of
our rich programme, we here describe a few of the most important
ones for which there is minimal ambiguity in the interpretation. A
more complete compilation, listing our order of priorities, will
be found in the main part of this document.

\subsubsection{Proton--Neutron Spin Physics} \label{PNSP}

The nucleon--nucleon interaction is fundamental to the whole of
nuclear physics and hence to the composition of matter as we know
it. Apart from its intrinsic importance, it is also a necessary
ingredient in the description of meson production and other
processes. The meticulous investigation of the nucleon--nucleon
interaction must be a communal activity across laboratories, with
no single facility providing the final breakthrough. However, the
mass of EDDA data on $pp$ scattering has reduced significantly the
ambiguities in the $I=1$ phase shifts up to 2.1{\ww}GeV. Nevertheless,
the lack of good neutron--proton spin--dependent data make the
$I=0$ phase shifts very uncertain above 800{\ww}MeV and there are even
major holes in the knowledge between 515 and 800{\ww}MeV. We propose
to add significantly to the elastic scattering data set by making
measurements of cross sections, analysing powers, and
spin--correlation coefficients near both the forward and backward
directions by using the deuteron as a source of quasi--free
neutrons. This substitute target has been shown at other
laboratories to work well, though theoretical input is
necessary to extract the $pn$ amplitudes reliably, especially at small
momentum transfers.

Small angle neutron--proton scattering, which is difficult to
study with a neutron beam, will be investigated up to 1.1{\ww}GeV per
nucleon using the beam of polarised deuterons interacting in the
polarised hydrogen target. One \underline{fast} \textit{spectator}
proton will be detected in the ANKE magnetic system and the struck
proton in the silicon telescopes. This is possible in an interval
dictated by the telescope system, corresponding to momentum
transfers $0.005<|t|<0.1${\ww}GeV/c$^2$.

By using a deuterium target and detecting now the \underline{slow}
\textit{spectator} proton in the telescopes, the beam energy range
could be extended up to $\approx 3${\ww}GeV though, to connect with
the $pp$ phase shifts, the range up to 2.1{\ww}GeV is the most
important. In this configuration the kinematic interval is fixed
mainly by the measurement of the fast proton in ANKE
($4^{\circ}<\theta_p^{lab}<11^{\circ}$). Though in both
configurations there are deuteron effects that suppress certain
amplitudes, the necessary corrections can be
largely handled, and the measurement of the slow proton leads to a
good vertex identification even in the long target provided by the
target cell. Since the projected counting rates are very high, it
will be reasonable to take data in 100{\ww}MeV steps.

It has already been shown that ANKE is an efficient tool for
measuring small angle charge exchange of polarised deuterons,
$\vec{d}p\to (pp)n$, where the final $pp$ pair is at such low
excitation ($< 3${\ww}MeV) that it is almost exclusively in an $S$
state. In this case the reaction provides a \textit{spin filter}
that selects an $np$ charge--exchange spin--flip from the
($^3\!S_1$, $^3\!D_1$) states of the deuteron to the $^1\!S_0$ of
the diproton. Measurements of the deuteron tensor analysing powers
then allow one to extract the magnitudes of the different
spin--spin $np\to pn$ amplitudes in the backward direction. The
same type of experiments carried out with a polarised target
determines the relative phases of these amplitudes. Though the
selectivity of the $^1\!S_0$ region is clear, experience at Saclay
shows that valuable information on the charge--exchange amplitudes
is contained also in the higher $pp$ excitation data.

We have already shown in practice that the charge--exchange
reaction can also be carried out in inverse kinematics, with both
protons from a deuterium target being detected in the spectator
counters. This would allow the energies up to 3{\ww}GeV to be used,
though over a rather smaller momentum--transfer interval.

It is important to stress that, with the apparatus available, the
studies of the small and large angle elastic neutron--proton
scattering will be carried out simultaneously at ANKE.

\subsubsection{Deriving the chiral three--body force from
pion production}

Chiral perturbation theory represents the best current hope for a
reliable and quantitative description of hadronic reactions at low
energies. One important step forward in our understanding of pion
reactions at low energies will be to establish that the same
short--range $NN\to NN\pi$ vertex contributes to both $p$--wave
pion production and to low energy three--nucleon scattering, where
the identical operator plays a crucial role. In the chiral
Lagrangian, at leading and next--to--leading order, all but one
term can be fixed from pion--nucleon scattering. The missing term
corresponds to an effective $NN\to NN\pi$ vertex, where the pion
is in a $p$--wave and both initial and final $NN$ pairs are in
relative $S$ waves.

To second order in the pion momentum, nine observables are
required to perform the full amplitude analysis in order to
extract in a model--independent way the effective coupling
constant. Of these, data from TRIUMF and CELSIUS yield seven at a
beam energy around 350{\ww}MeV. Experiments designed to provide the
necessary overconstraints will be carried out at ANKE through
measurements of the analysing powers and spin--correlation
parameters in the reactions $pp\to pp\pi^0$ and $pn\to pp\pi^-$.

It should be noted that the diproton detection in ANKE to isolate
the $^1\!S_0$ state is just the same as that needed also for the
$dp$ charge--exchange programme described earlier. The resolution
on the $pp$ excitation energy is estimated to be around $0.3${\ww}MeV
and that on the missing mass about 5.5{\ww}MeV (RMS) which, at these
low energies, will allow us to distinguish unambiguously the pion
production reaction from any background. The counting rates are
quite high and, even in the $pn\to pp\pi^-$ case, where the
spectator proton has to be detected, more than $10^3$ events per
hour could be accumulated over the full range of pion
centre--of--mass angles.

Our facility offers the exciting possibility of extracting
pion--production amplitudes in a model independent way and thus
determining a vital parameter for chiral perturbation theory.

\subsubsection{Strangeness production: The
$\Lambda$--$N$ scattering length}

Effective field theories provide the bridge between the hadronic
world and QCD. For systems with strangeness, there are still many
open questions and it is not even clear if the kaon is more
appropriately treated as heavy or light particle. To improve
further our understanding of the dynamics of systems containing
strangeness, better data are needed. The insights to be gained are
relevant, not only for few--body physics, but also for the
formation of hypernuclei, and might even be of significance for
the structure of neutron stars. The hyperon--nucleon scattering
lengths are obvious quantities of interest in this context.

The IKP theory group has developed a method to enable one to
deduce a scattering length directly from data on a production
reaction, such as $pp\to pK^+\Lambda$, in terms of an integral
over the invariant $\Lambda p$ mass ($m_X$) distribution. Using
this method it can be seen that the inclusive Saclay $pp\to K^+X$
data, which had a mass resolution of 4{\ww}MeV, allow the extraction
of a scattering length with an experimental uncertainty of only
0.2{\ww}fm. However, the actual value of the scattering length
obtained in this way is not meaningful, since it represents the
incoherent sum of the $^3\!S_1$ and the $^1\!S_0$ $\Lambda p$
final states with unknown relative weights. It is important to try
to separate them.

The $\Lambda N$ triplet final state could be isolated
unambiguously by measuring the unpolarised $K^+$ spectrum in the
forward direction and this weighted with the incident spin
correlation, obtained using a transversally polarised beam and
target. This will be achieved by using the
$\vec{p}\,\vec{d}\to p_{sp} K^+ X$  reaction, since the spectator
proton ($p_{sp}$) will provide a better determination of the
vertex in the long polarised target cell.

In fact, by measuring the $K^+$ production rates in the
near--threshold region in hydrogen and deuterium, and the
spin--correlation in the deuterium case, we will also determine in
a model--independent way the magnitudes of the three $S$--wave
spin--isospin amplitudes, two corresponding to the spin--triplet
final state and one to the singlet. It is also possible with a
deuterium target to measure spin--transfer coefficients from the
initial proton or neutron to the final $\Lambda$, and these will
fix the relative phases of the three amplitudes. A significant
$m_X$ variation in the singlet--triplet interference would point
at a strong spin dependence of the $\Lambda n$ final state
interactions, though much more theoretical work would be required
to extract quantitative differences in the scattering lengths from
such data.

\subsection{{$\mathcal PAX$}: \textbf{P}olarised
\textbf{A}ntiproton e\textbf{X}periments at FAIR}

The possibility of testing nucleon structure through double--spin
asymmetries in polarised proton--antiproton reactions at the High
Energy Storage Ring (HESR) at the future Facility for Antiproton and
Ion Research (FAIR) at GSI was suggested by the PAX collaboration in
2004. Since then, there has been much progress, both in understanding
the physics potential of such an experiment and in studying the
feasibility of efficiently producing polarised antiprotons.  The
physics programme of such a facility would extend to a new domain the
exceptionally fruitful studies of the nucleon structure performed in
unpolarised and polarised deep inelastic scattering (DIS), which have
been at the centre of high energy physics during the past four
decades.

A viable practical scheme using a dedicated low--energy antiproton
polariser ring (APR) has been developed and published, which
allows a polarisation of the stored antiprotons at HESR--FAIR of
$\simeq 0.3$ to be reached. The approach is based on solid QED
calculations of the spin transfer from electrons to antiprotons,
which were confirmed experimentally in the FILTEX experiment. The
method is routinely used at J--Lab for the electromagnetic form
factor separation.

The polarised antiproton--proton interactions at HESR will provide
unique access to a number of new fundamental physics observables,
which cannot be studied without transverse polarisation of protons
and antiprotons.

The transversity distribution is the last missing leading--twist
piece of the QCD description of the partonic structure of the
nucleon.  It describes the quark transverse polarisation inside a
transversely polarised proton. Unlike the more conventional
unpolarised quark distribution $q(x,Q^2)$ and the helicity
distribution $\Delta q(x,Q^2)$, the transversity $h^q_1(x,Q^2)$
can neither be accessed in inclusive deep--inelastic scattering of
leptons off nucleons nor can it be reconstructed from the
knowledge of $q(x,Q^2)$ and $\Delta q(x,Q^2)$.  It may contribute
to some single--spin observables, but always coupled to other
unknown functions.  The transversity distribution is directly
accessible uniquely \emph{via} the \emph{double transverse spin
asymmetry $\bf A_{TT}$} in the Drell--Yan production of lepton
pairs, which is expected to be in the range 0.3--0.4. With the
expected beam polarisation from the APR and the luminosity of
HESR, the PAX experiment would make a definitive observation of
$h^q_1(x,Q^2)$ of the proton for the valence quarks.  The
determination of $h^q_1(x,Q^2)$ will open new pathways to the QCD
interpretation of single--spin asymmetry (SSA) measurements.  In
conjunction with the data on SSA from the HERMES collaboration,
the PAX measurements of the SSA in Drell--Yan production on
polarised protons would for the first time provide a test of the
theoretical prediction of the reversal of the sign of the Sivers
function from semi--inclusive DIS to Drell--Yan production.

The origin of the unexpected $Q^2$--dependence of the ratio of the
magnetic and electric form factors of the proton, as observed at
J--Lab, could be clarified by a study of their relative phase in
the time--like region. This can be measured \emph{via} SSA in the
annihilation $ \bar{p} p^{\uparrow} \to e^+e^-$ on a transversely
polarised target. The first ever measurement of this phase at PAX
would also contribute to the understanding of the onset of the
pQCD asymptotics in the time--like region and would serve as a
stringent test of dispersion theory approaches to the relationship
between the space--like and time--like form factors. The
double--spin asymmetry would allow an independent $G_E-G_M$
separation and serve as a check of the Rosenbluth separation in
the time--like region.

Arguably, in $p\bar{p}$ elastic scattering the hard scattering
mechanism can be checked beyond the $|t| = \frac{1}{2}(s-4m_p^2)$
accessible in the crossed--symmetric $pp$ scattering, because in
the $p\bar{p}$ case the $u$--channel exchange contribution can
only originate from the strongly suppressed exotic dibaryon
exchange. Consequently, in the $p\bar{p}$ case the hard mechanisms
can be tested to transfers almost twice as large as in $pp$
scattering. Even unpolarised large angle $p\bar{p}$ scattering
data can shed light on the origin of the intriguing oscillations
around the $s^{-10}$ behavior of the $90^{\circ}$ scattering cross
section in the $pp$ channel and put stringent constraints on the
much disputed odd--charge conjugation Landshoff mechanism.

The PAX collaboration proposes an approach in two phases,
with the eventual goal of an asymmetric proton--antiproton
collider in which polarised protons with momenta of about
3.5{\ww}GeV/c collide with polarised antiprotons with momenta up to
15{\ww}GeV/c. These circulate in the HESR, which has already been
approved and will serve the PANDA experiment. The overall machine
setup of the HESR complex would consist of:\vspace{-3mm}

\begin{enumerate}
\item An Antiproton Polariser (APR) built inside the HESR area
  with the crucial task of polarising antiprotons at kinetic energies
  around $\approx 50${\ww}MeV ($p\approx 300${\ww}MeV/c), to be accelerated
  and injected into the other rings.
\item A second Cooler Synchrotron Ring (CSR, COSY--like) in which
protons or antiprotons could be stored with momenta up to
3.5{\ww}GeV/c.  This ring should have a straight section, where a PAX
detector could be installed, running parallel to the experimental
straight section of HESR.
\item By deflecting the HESR antiproton beam into the straight
section of the CSR, both collider and fixed--target modes become
feasible.\vspace{-2mm}
\end{enumerate}

In Phase~I a beam of unpolarised or polarised antiprotons with
momenta up to 3.5{\ww}GeV/c in the CSR, will collide with a polarised
hydrogen target in the PAX detector. This phase, which is
independent of the HESR performance, will allow the first
measurement of the time--like proton form factors in single and
double polarised $\bar{p}p$ interactions over a wide kinematical
range, from close to threshold up to $Q^2=8.5${\ww}GeV$^2$. Several
double--spin asymmetries in elastic
$\bar{p}^{\uparrow}p^{\uparrow}$ scattering could be determined.
By detecting back--scattered antiprotons one could also explore
hard scattering regions of large $t$.

Phase~II will allow the first ever direct measurement of the quark
transversity distribution $h_1$, by studying the double transverse
spin asymmetry $\Att$ in the Drell--Yan processes $p^{\uparrow}
\bar{p}^{\uparrow} \rightarrow e^+ e^- X$ as a function of Bjorken
$x$ and $Q^2$ (= $M^2$)
\[\Att \equiv
\frac{d\sigup-d\sigdw}{d\sigup+d\sigdw}\,=\,
\hat{a}_{TT}\frac{\sum_q e_q^2
h_1^q(x_1,M^2)h_1^{\overline{q}}(x_2,M^2)} {\sum_q e_q^2
q(x_1,M^2)\overline{q}(x_2,M^2)}\,,\]%
where $q=u,d,\ldots$,
$\overline{q}=\overline{u},\overline{d}\ldots$, and $M$ is the
invariant mass of the lepton pair. The parameter $\hat{a}_{TT}$,
which is of the order of one, is the calculable double--spin
asymmetry of the elementary QED process $q\overline{q}\rightarrow
e^+ e^-$.

Two possible scenarios, an asymmetric collider or a high
luminosity fixed target experiment, might be foreseen to perform
the measurement.
%
A beam of polarised antiprotons from 1.5{\ww}GeV/c up to 15{\ww}GeV/c
circulating in the HESR, collides with a beam of polarised protons
with momenta up to 3.5{\ww}GeV/c circulating in the CSR.  This
scenario, however, requires one to demonstrate that a suitable
luminosity is reachable.  Deflection of the HESR beam to the PAX
detector in the CSR is necessary. By a proper variation of the
energy of the two colliding beams, this setup would allow a
measurement of the transversity distribution $h_1$ in the valence
region of $x>0.05$, with corresponding $Q^2=4 \ldots 100$ $\rm
GeV^2$. With a luminosity of $2\times 10^{30}${\ww}cm$^{-2}$s$^{-1}$
about $1000$ events per day can be expected. Recent model
calculations show that in the collider mode, luminosities in
excess of $10^{30}${\ww}cm$^{-2}$s${^-1}$ could be reached. For the
transversity distribution $h_1$, such an experiment can be
considered as the analogue of polarised DIS for the determination
of the helicity structure function $g_1$, \emph{i.e.}\ of the
helicity distribution $\Delta q(x,Q^2)$. The kinematical coverage
in $(x,Q^2)$ will be similar to that of the HERMES experiment.

If the required luminosity in the collider mode is not achievable,
an experiment with a fixed polarised internal hydrogen target can
be undertaken. In this case, an upgrading of the momentum of the
polarised antiproton beam circulating in the HESR up to 22{\ww}GeV/c
is envisaged. This scenario also requires the deflection of the
HESR beam to the PAX detector in the CSR. This measurement will
explore the valence region of $x>0.2$, with corresponding values
of $Q^2=4 \ldots 16${\ww}GeV$^2$, yielding about 2000 events per day.

\subsection{Conclusions}

There are unique opportunities at ANKE to measure the spin
dependence of many polarised reactions, primarily in the
proton--neutron sector. This is through the combination of
magnetic analysis of fast particles with the detection of slow
particles in the silicon telescope array. The proton--neutron
programme has already been started at ANKE by using polarised
deuterons incident on an unpolarised target and the full programme
with a polarised target will be initiated in 2006. In general the
requisite equipment exists, or has already been financed, though
minor upgrades may of course be necessary.

Many reactions will be measured simultaneously, but we will first
concentrate on the nucleon--nucleon programme, where
counting rates are high, before passing to the pion
production and then to the more challenging of the
strangeness experiments.

The experience that the team will gain in undertaking polarisation
measurements will be put to good use in the developments for PAX
at FAIR, to which the group as a whole is committed. The storage
of polarised antiprotons at HESR, as proposed by PAX, will open
unique possibilities to test QCD in hitherto unexplored regions,
thereby extending into a new domain the exceptionally fruitful
studies of nucleon structure performed in unpolarised and
polarised deep inelastic scattering.  This will provide another
cornerstone to the contemporary QCD physics programme with
antiprotons at FAIR.
%
%
\newpage
\tableofcontents
%
%
\newpage
\section{Introduction}

For several years COSY has provided circulating beams of polarised
protons. Used together with a polarised hydrogen target, these
beams have been successfully exploited by the EDDA collaboration
to measure the cross section, the analysing power and
spin--correlation observables in proton--proton elastic scattering
over much of the COSY energy range~\cite{EDDA}. Vector and tensor
polarised deuteron beams are also available up to an energy of
about $T_d\approx 2.3${\ww}GeV. Using such capabilities to the full is
one of the stated priorities of the laboratory: \emph{For
spin--physics experiments, the FZJ proposes to increase the
intensity of the polarized beams up to the space charge
limit}~\cite{PHN}.

In 2005, a polarised internal hydrogen and deuterium storage--cell
target (PIT) has been installed at the ANKE spectrometer. A
Lamb--shift polarimeter (LSP) will allow the adjustment of the
transition units of the polarised atomic beam source (ABS) that
feeds the storage cell. Thus it is expected that late in 2006 the
whole system of polarised beam and polarised target will be fully
operational inside the COSY ring at the ANKE position, where fast
charged particles can be magnetically analysed and slow particles
measured using telescopes of silicon counters placed around the
target. We will therefore soon be in a position to carry out many
of the recommendations of the 1998 workshop on \emph{Intermediate
Energy Spin Physics}~\cite{IESP}.

Under these circumstances it is incumbent on us to make a global
presentation to the PAC of the spin programme that will exploit
these advanced facilities over the years to come. We will be
concerned only with experiments that could be carried out within
the confines of the COSY storage ring by detecting charged
particles. In this spirit, the investigation of charge--symmetry
breaking in the polarised $\vec{d}\,d\to\alpha\,\pi^0$ reaction is
described rather in the WASA proposal~\cite{WASA}. Furthermore, the
relation to external experiments, such as TOF, must await the
preparation of further plans by these collaborations. No requests
are made here for beam time for particular experiments; these will
only follow later in conjunction with more detailed and specific
proposals.

However, this is also a period of transition for experimental
hadronic physics in Germany, with the plans to construct the
Facility for Antiproton and Ion Research (FAIR) at GSI
Darmstadt~\cite{cdr-gsi}. Though the scale of this operation,
involving the building of a high energy antiproton storage ring
(HESR), is vastly bigger than that at COSY, there is a great
potential synergy in respect of the spin physics programmes at the
two laboratories and, for the future of the field, an orderly
transfer of physics interest between the two is highly desirable.
Target development, polarimetry, cooled beams, spectator proton
detection \emph{etc.}\ are all areas that are covered in the
technical aspects of \S\ref{PWFP} in the context of an eventual
knowledge transfer to FAIR.

It is expected that antiprotons can be polarised through the
spin--transfer from the polarised electrons of the atoms in a
polarised target to orbiting
antiprotons~\cite{rathmann,meyer,horowitz-meyer,rathmann2}. This
could be carried out at FAIR in a separate antiproton polariser
ring, using a polarised internal storage cell
target~\cite{steffens}. A Letter--of--Intent, describing some of
the important experiments that can be carried out with polarised
beams and targets at FAIR, was submitted in February 2005 to the
GSI QCD--PAC by the PAX collaboration~\cite{pax-loi}. This was
well received by this committee and the relevant recommendations
are~\cite{QCD-PAC}:\\[1ex]%
\baselineskip 3ex
\begin{it} ``The PAC considers it is essential for the FAIR project to
commit to polarized antiproton capability at this time and include
polarized transport and acceleration capability in the HESR, space
for installation of the APR and CSR and associated hardware, and
the APR in the core project. We request the PAX collaboration
to:\\ %
1) Commit to the construction and testing of the APR (IKP J\"ulich
appears to be the optimal location)\\ %
2) Explore all options to increase the luminosity to the target
value specified above\\ %
3) Prepare a more detailed physics proposal and detector design
for each of the proposed stages.''\\
\end{it}

\baselineskip 3ex%
Though the principles of the polarising technique have been well
established for low energy
protons~\cite{rathmann,meyer,horowitz-meyer}, an investigation with
medium energy protons should be undertaken at COSY. The lead--up
to FAIR physics is discussed in \S\ref{PWFP}. Before that, in
\S\ref{EF}, we describe in detail the technology that we will have
at our disposal in terms of polarised beams/targets and detectors
to carry out the COSY programme. It is clear that we must
ourselves be able to measure the polarisation of the beams and/or
targets and not rely exclusively on outside polarimetry. How this
will be done is the subject of \S\ref{BandTP}.

It is universally agreed that a detailed understanding of
nucleon--nucleon scattering, at least at a phenomenological level,
up to high energies is a ``good thing'', and that this involves a
systematic compilation of experiments at different laboratories.
Though EDDA has clarified significantly the spin dependence of the
proton--proton scattering amplitudes up to at least
2{\ww}GeV~\cite{EDDA}, the data base of spin--observables in
neutron--proton scattering is very incomplete above 800{\ww}MeV, so
that there are large uncertainties in the isoscalar phase shifts.
Section~\ref{PNSP} shows how many of these holes can be filled by
internal COSY experiments using a polarised deuteron beam or
deuterium target. In addition to elastic neutron--proton
scattering, important data should be obtained on the excitation of
the $\Delta(1232)$ isobar, either explicitly through a $\pi N$
final state, or indirectly through $d\,\pi^0$ production. The
measurement of up to triple--spin observables will be achieved. It
is, however, very important to stress that many of the experiments
involve high counting rates and will be carried out simultaneously
with others provided that the trigger conditions allow this.

COSY generally operates at energies above the pion--production
threshold and in such a domain the Faddeev equations are no longer
able to describe proton--deuteron elastic scattering in a
quantitative way, especially at large angles. This is due, in
part, to the virtual excitation and de-excitation of the
$\Delta(1232)$ isobar, which can help to share the large momentum
transfer between the two nucleons in the deuteron. Although
reactions such as this, or the analogous backward proton--deuteron
charge exchange cannot, at present, be interpreted in an
unambiguous way, it is hoped that spin observables will provide
extra clues on the underlying dynamics. The possibilities here
will be surveyed in \S\ref{CPDR}.

The spin--dependence of the production of non--strange mesons in
polarised nucleon--nucleon collisions is the subject of
\S\ref{NSMP}. It is shown that, even far from threshold,
production of neutral mesons $X^0=\pi^0,\,\eta,\,\omega,\,\dots$
through the $\vec{p}\,\vec{p}\to pp X^0$ reaction, with the two
final protons at low excitation energy, contains valuable new
information because of the spin--filter effect resulting from the
two protons being in the $^1S_0$ state. Of especial interest is
the production on neutrons \emph{via} $\vec{p}\,\vec{n}\to
pp\pi^-$ which, at low energies, could be a valuable check on
effective field theory in the large momentum transfer regime. Also
in this section we discuss what one can learn from coherent single
and double pion production with the formation of a deuteron or
$^3$He nucleus.

Two of the prime considerations in the construction of ANKE were
the possibility of installing a polarised target and the ability
to detect positive kaons against a high background of other
particles. \S\ref{PSMB} shows that these two characteristics can
be combined in a unique way to advance studies of $K^+\Lambda$
production in nucleon--nucleon collisions. Near threshold, the
cross sections on protons and neutrons as well as the
spin--correlation and --transfer parameters can all be efficiently
measured and this allows one to isolate the spin--singlet and
triplet $\Lambda N$ final states in a model--independent way. Some
of these opportunities persist in the forward direction at higher
energies and open the door to a quantitative investigation of the
$\Lambda N$ scattering lengths in well--defined spin states.

Having stressed some of the strengths of ANKE that we shall
exploit, we must also recognise its limitations. For example, the
restricted phase space offered for exclusive reactions will not
allow a meaningful exploration of the spin dependence of rare
reactions, such as the production of the hypothesised exotic
$\Theta^+(1540)$ baryon. With its much larger acceptance, such
systematic studies might be better carried out at the TOF detector,
where evidence for the existence of the state has already been
presented~\cite{COSY-TOF0}.

The ambitious programme outlined in the following pages will take
of the order of four years to complete and a series of milestones
is suggested in the timetable presented in \S\ref{TS}. There are,
of course, many outstanding questions that cannot be answered at
the present time. Is it feasible to construct a partial Siberian
snake in COSY to rotate the polarisation axis of the beam using,
in part, the WASA solenoid~\cite{WASA}? Will it be feasible to
rotate the polarisation of the target? Finally, the future
relationship of COSY with FAIR is also uncertain, and this bears
directly on the importance of much of the work described here.
%
%
\section{Experimental Facilities}
\label{EF}
\subsection{Polarised beams at COSY}

The COoler SYnchrotron COSY~\cite{cosy} accelerates and stores
unpolarised and polarised protons and deuterons in a momentum
range between 0.3{\ww}GeV/c and 3.7{\ww}GeV/c. To provide high quality
beams, there is an Electron Cooler at injection and a Stochastic
Cooling System from 1.5{\ww}GeV/c up to the maximum momentum.
Vertically polarised proton beams of different momenta, with
polarisations of more than 80\%, are delivered to internal and
external experimental areas. An rf dipole has been installed to
induce artificial depolarising resonances.

Deuteron beams with different combinations of vector and tensor
polarisation became available in 2003. The first simultaneous
measurement of vector and tensor polarisation of the stored
deuteron beam using the ANKE spectrometer is described in
\S\ref{DBP}. The achieved intensities for polarised proton beams
are $5\times10^{9}$ (single injection with electron cooling) and
$1.2\times10^{10}$ with multiple injection with electron cooling
and stacking. For the polarised deuterons with single injection
an intensity of about $3\times10^{10}$ was achieved during the
February 2005 runs.

Increasing the phase space density by electron cooling at
injection and conserving the beam emittance during internal
experiments at high momenta through stochastic cooling are the two
of the outstanding characteristics of COSY.

\subsection{ANKE magnetic spectrometer}

\begin{figure}[h]
\centering \resizebox{\textwidth}{!}
{\includegraphics*[0mm,65mm][210mm,230mm]{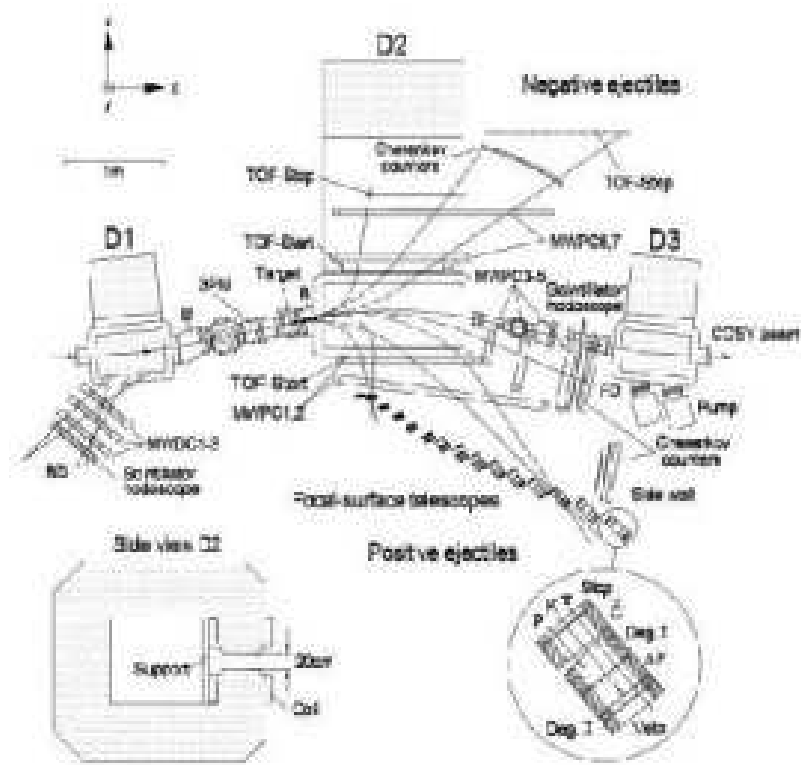}}
\caption{\footnotesize
Schematic drawing of ANKE Spectrometer. Typical trajectories of
ejectilies are indicated with emission angle of $0^{\circ}$ or
$\pm10^{\circ}$. } \label{fig:ANKE_full}
\end{figure}
It is proposed that the experimental programme outlined in this
document will be performed using the ANKE spectrometer, which is
described in detail in ref~\cite{anke}. The layout of this
facility, which is installed at an internal beam position of COSY,
is shown in Fig.~\ref{fig:ANKE_full}. The main components of the
spectrometer are: a magnetic system, an internal target and four
detection systems --- positive and negative side detectors,
forward and backward detectors. The ANKE magnetic system comprises
a dipole magnet D1, which deflects the circulating COSY beam
through an angle $\alpha$, a large spectrometer dipole magnet D2
to perform the momentum analysis (beam deflection $-2\alpha$), and
a third dipole magnet D3, identical to D1, to deflect the beam
through $\alpha$ back to the nominal orbit.

Strip and cluster--jet targets have been used for many years at
ANKE. The Polarised Internal Target (PIT), which was installed for
tests at the ANKE position in July 2005, will be described in
\S\ref{PIT}.

Detection systems for both positively and negatively charged
particles include plastic scintillator counters for TOF
measurements, multi--wire proportional chambers (MWPC) for
tracking, and range telescopes for particle identification. A
combination of scintillation and \v{C}erenkov counters, together
with wire chambers, allow one to identify negatively charged pions
and kaons. The forward detector (FD), comprising scintillator
hodoscopes, \v{C}erenkov counters, and fast proportional chambers,
is used to measure  particles with high--momenta, close to that of
the circulating COSY beam. A backward detector (BD), composed of
hodoscopes and multi--wire drift chambers, together with the D1
magnet, can be used as a spectrometer for backward--emitted
particles.

The silicon strip counters that are placed close to the target for
vertex reconstruction and detection of low--energy spectator
protons will be discussed separately in \S\ref{SST}.
%
%
\subsection{Silicon tracking telescopes for the detection of spectator
protons}\label{SST}

Modular Silicon Tracking Telescopes have been developed based on
double--sided silicon strip detectors~\cite{IEEE2001}. Serving in
general for
\begin{itemize}
\item low energy spectator proton detection/tracking and \item
vertex reconstruction into the ANKE target region,
\end{itemize}
they are optimised for the identification and measurement of low
energy protons, determining their four--momenta. They allow one to
use polarised deuterium gas as a polarised neutron target and to
study \emph{e.g.}\ reactions of the type $\vec{p}\,\vec{n}
\rightarrow p n X$ or $\vec{p}\,\vec{n} \rightarrow d X$.

\begin{figure}[h]
\begin{center}
\epsfig{file=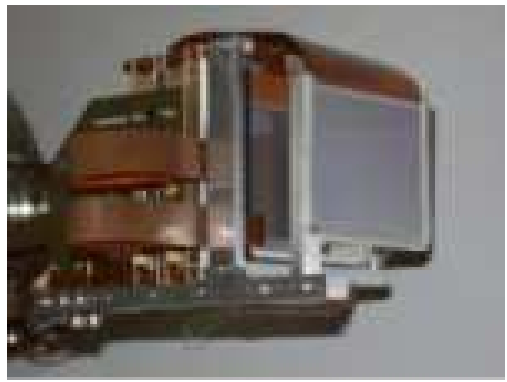,height=5cm} \caption{Telescope
arrangement of double--sided silicon strip detectors: 69$\,\mu$m
thin, $51 \times 66\,$mm$^2$ active area as first layer and
5100$\,\mu$m thick, $64 \times 64\,$mm$^2$ active area Si(Li)
detector as second layer. Protons with kinetic energies in the
range $2.5 < T_p < 35\,$MeV will be tracked and identified with
such an arrangement.} \label{FigTelescope}
\end{center}
\end{figure}

The telescopes are installed as close as 2$\,$cm from the COSY
beam inside the ultra high vacuum of the accelerator. Their basic
features are $\Delta E/E$ proton--deuteron identification from 2.5
to 40$\,$MeV and particle tracking over a wide dynamic range,
either $2.5\,$MeV spectator protons or minimum--ionising
particles. The recent development of very thick (5--20$\,$mm)
double--sided micro structured Si(Li) and very thin (69$\,\mu$m)
double--sided Si--detectors provides the modular use of the
telescopes for particle identification over a wide range of
energies. Fig.~\ref{FigTelescope} shows a telescope arrangement
with a thin and a thick detector.

\subsubsection{The design concept}

The basic design concept of a telescope is to combine particle
identification and tracking over a wide energy range. The tracking
of particles is accomplished through the use of double--sided
silicon strip detectors. The minimum energy of a proton to be
tracked is fixed by the thickness of the innermost layer. It will
be detected when it passes through the inner layer and be stopped
in the second. The maximum energy of protons that can be
identified is given by the range within the telescope and
therefore by the total thickness of all detection layers.
Measuring the energy losses in the individual layers of the
telescope permits the identification of stopped particles by the
$\Delta E/E$ method. Hence by tracking and subsequently measuring
precisely their energy, the telescopes determine the four--momenta
for stopping particles.

To measure the momentum of a particle from the track information
in the ANKE detection systems, the vertex of the reaction must be
known and this is a non-trivial task for an extended storage cell
target of up to 40$\,$cm length. Only by having additional
track(s) in the Silicon Tracking Telescopes close to the ANKE
target region inside the COSY vacuum, can the vertex be
reconstructed accurately.

Depending upon the requirements of the individual experiment, four
to six telescopes can be equipped with different sets of silicon
detectors and be positioned around the target region to serve for
several purposes:

\begin{itemize}
\item Spectator Detector: Low energy protons will be identified
and tracked in the range $2.5<T_p<35\,$MeV. Each telescope covers
about 10\% of the geometrical acceptance.
\item Vertex Detector:  One track in the Silicon Tracking
Telescopes defines the vertex in two coordinates (along and
perpendicular to the beam) with a precision of about 1$\,$mm. The
third coordinate can only be fixed using the spatial resolution of
the ANKE detection system, which gives about 10$\,$mm. Only two
tracks from the same reaction inside the telescopes allows a full
3--D vertex reconstruction with a precision to about 1$\,$mm. In
such a case reactions on the walls of the storage cell can be
easily identified.
\item Polarimeter: Two protons in the telescopes from the $pp$
elastic or quasi--elastic scattering allows one to analyse the
polarisation along the storage cell in parallel with the main
experiment.
\end{itemize}

\subsubsection{The detector performance}

Two telescopes have been assembled to check the performance of the
chosen detectors. For this purpose three types of double--sided
position sensitive detectors are arranged as silicon
tracking--telescopes.
\begin{itemize}
\item The inner layer is 69$\,\mu$m thick, has an active area of
$51 \times 66\,$mm$^2$, and an effective pitch of about
$400\,\mu$m. Its thickness sets the detection threshold for
protons in coincidence with the second layer at about 2.5$\,$MeV.
\item The second layer consists of a 300(500)$\,\mu$m thick
detector with an active area of $51 \times 66\,$mm$^2$ and a pitch
of $\approx 400\,\mu$m. It stops protons of kinetic energies up to
6.3(8)$\,$MeV.
\item The last layer is a 5500(10000)$\,\mu$m thick double--sided
Si(Li) detector with a pitch of 666$\,\mu$m and an active area of
$64 \times 64$mm$^2$~\cite{Protic}. It stops protons with energies
up to 40$\,$MeV and therefore covers most of the dynamic range of
the telescope.
\end{itemize}

The recent development of very thick ($>10\,$mm) double--sided
micro structured Si(Li)~\cite{Protic} and very thin (69$\,\mu$m)
double--sided Si--detectors enables the use of the telescopes over
a wide range of particle energies.

\begin{figure}[h]
\begin{center}
\epsfig{file=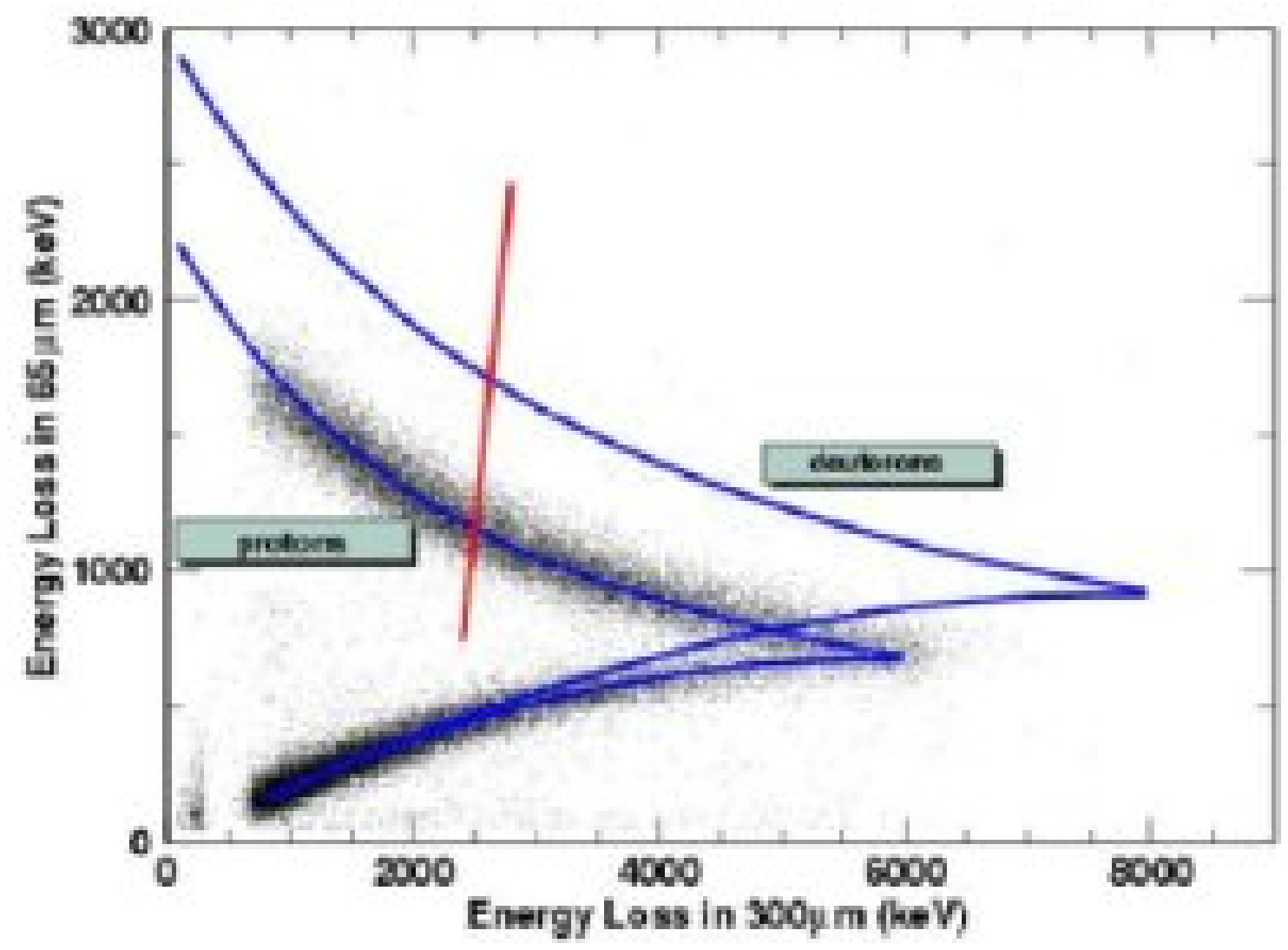,height=4.8cm}
\epsfig{file=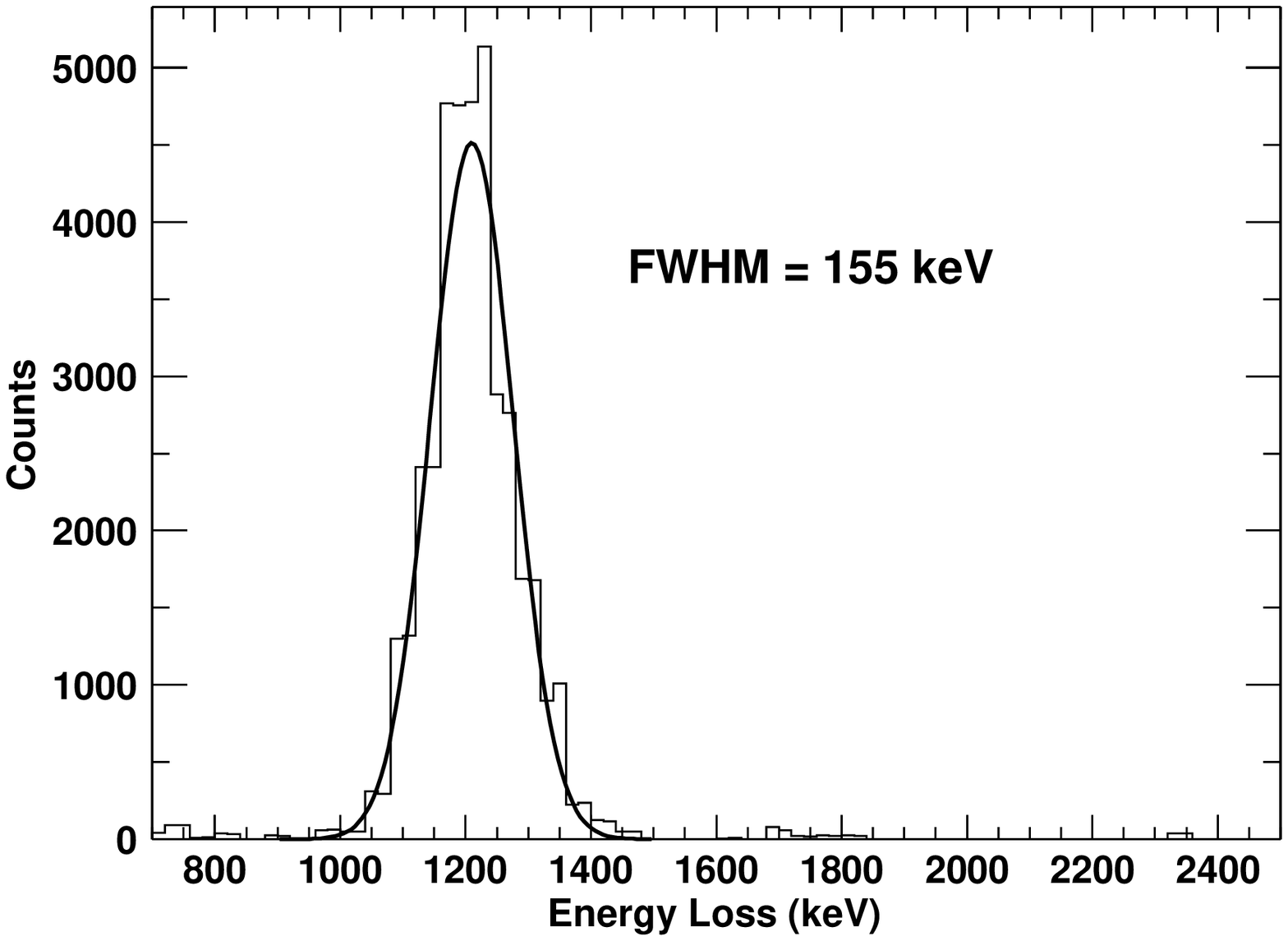,height=4.8cm} \caption{ The
energy loss in a 60$\,\mu$m \emph{vs} that in a 300$\,\mu$m thick
detector. Deuterons are not seen here because the detectors were
placed in the backward hemisphere of the target. The right figure
shows the energy resolution along the indicated slice
perpendicular to the proton band. } \label{FigdEE60vs300}
\end{center}
\end{figure}

\begin{figure}
\begin{center}
\epsfig{file=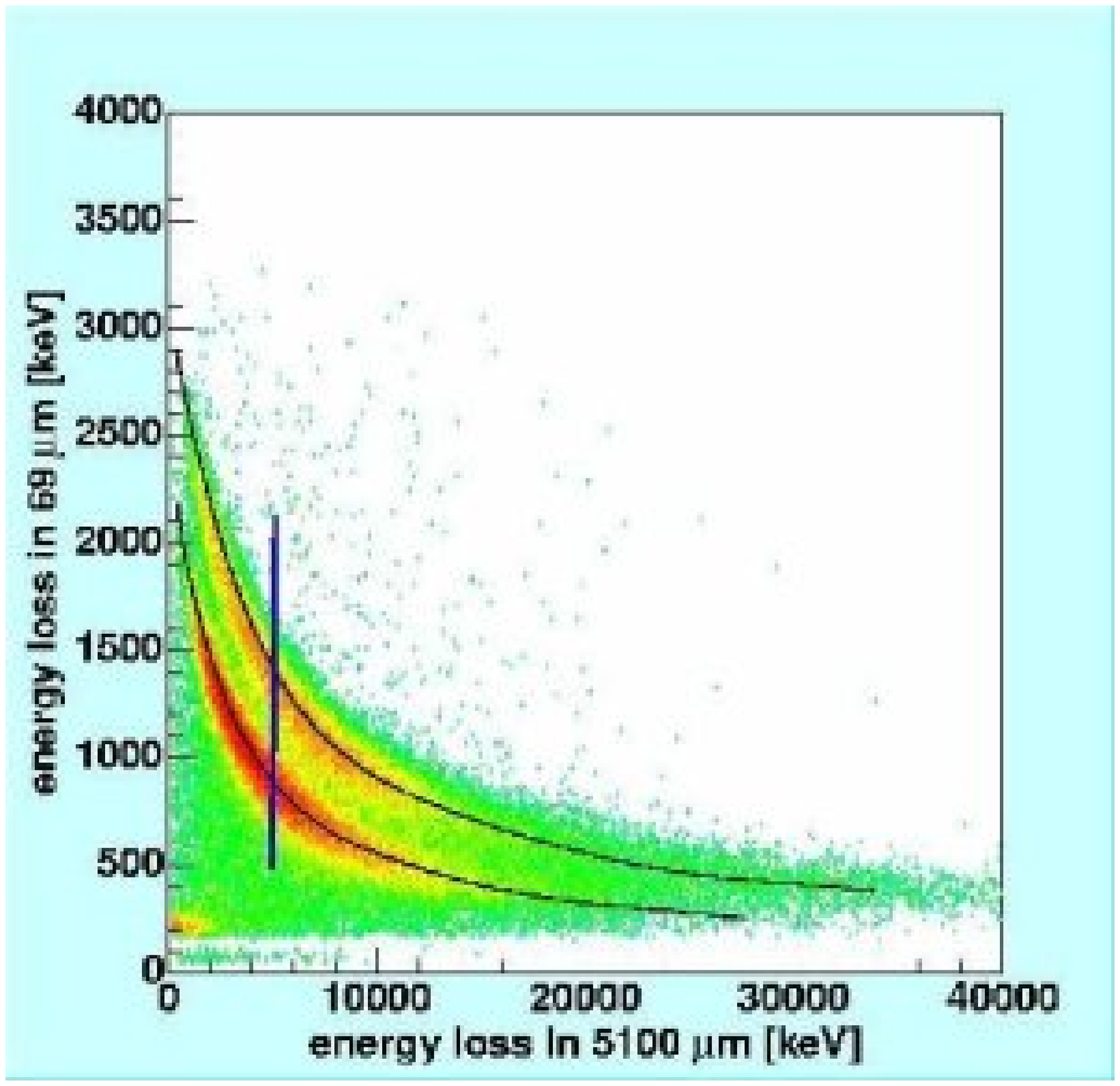,height=5cm}
\epsfig{file=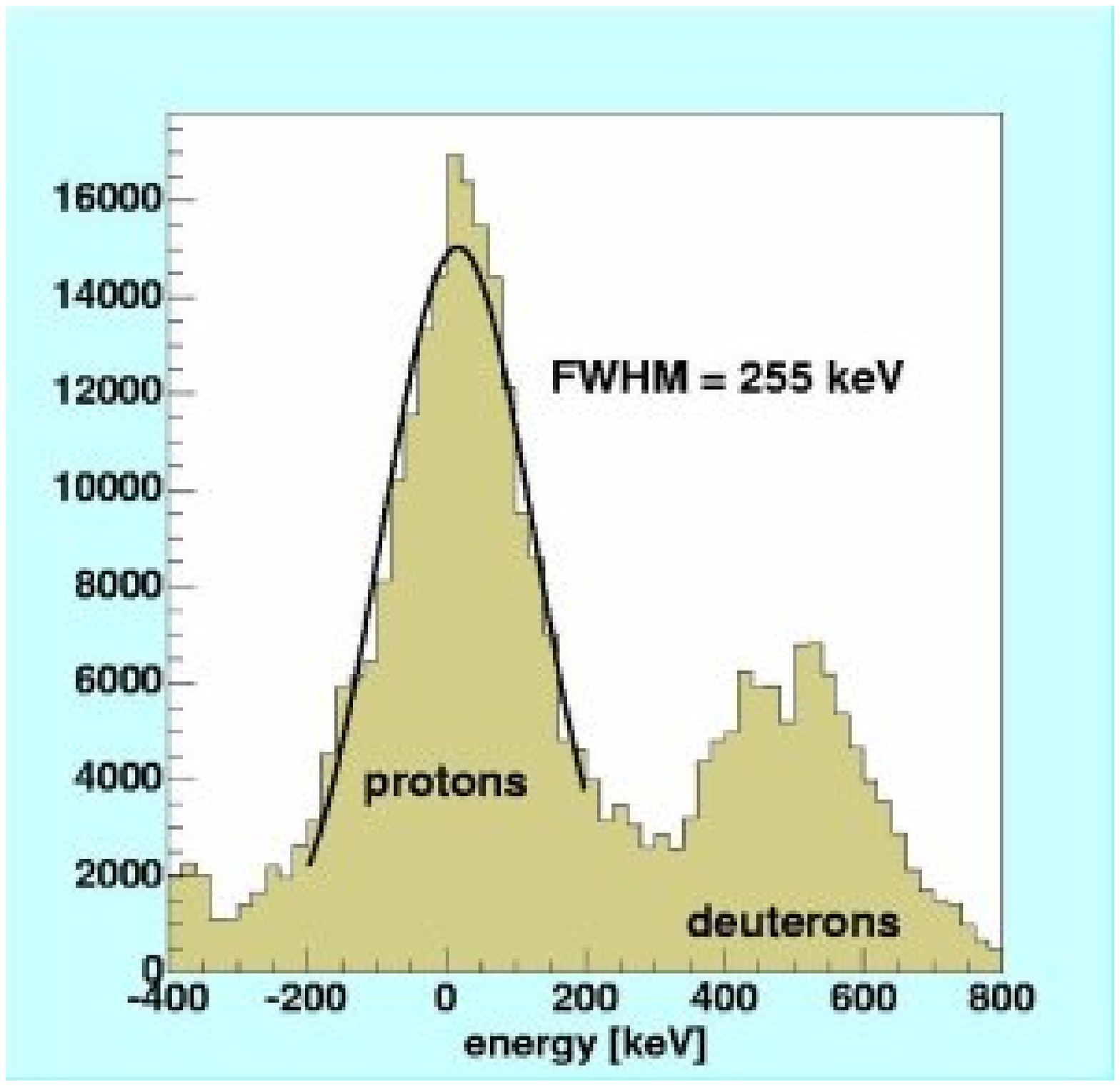,height=5cm} \caption{ The energy loss
in a 69$\,\mu$m \emph{vs} that in a 5100$\,\mu$m thick detector.
The right figure shows the energy resolution along the indicated
slice perpendicular to the proton band. } \label{FigdEE69vs5100}
\end{center}
\end{figure}

The $\Delta E$/E performance of the detection system is
demonstrated in Figs.~\ref{FigdEE60vs300} and
\ref{FigdEE69vs5100}. In addition to the experimental data the
SRIM estimations~\cite{SRIM} for the energy losses of protons and
deuterons are drawn. With a careful calibration of the system they
coincide to about $<3\,$\%.

The layout of these modular, self--triggering silicon tracking
telescopes provides
\begin{itemize}
\item $\Delta$E/E proton identification from 2.5 up to
40(50)$\,$MeV with an energy resolution of 150--250$\,$keV (FWHM).
The telescope structure of 69/300/500/5000$\,\mu$m thick
double--sided Si--strip detectors, read out by high dynamic range
chips~\cite{IDEAS}, allows $\Delta$E/E particle identification
over this wide dynamic range.
\item Particle tracking over a wide range of energies, either
2.5$\,$MeV spectator protons or minimum--ionising particles. The
angular resolution varies from 1$^\circ$--6$^\circ$ (FWHM). It is
on the one hand limited by the angular straggling within the
detectors and is therefore influenced by the track inclination. On
the other hand (\emph{e.g.}\ for minimum--ionising particles) it
is limited by the strip pitch of about 400--700$\,\mu$m and the
distances between the detectors. A typical vertex resolution for
two low energy protons in the telescopes is on the order of
$\approx 1\,$mm.
\item Self--triggering capabilities. The telescopes identify a
particle passage within 100$\,$ns and provides the possibility to
set fast timing coincidences with other detector components of the
ANKE spectrometer.
\item High rate capability. This becomes especially important for
the polarimetry studies because, for this application, two
telescopes have to be placed in the forward hemisphere. The
fast--timing option of the amplifier chips allows one to suppress
significantly accidentals.
\end{itemize}

\subsubsection{The in-vacuum electronics}

To combine a high dynamic range for the energy measurements with
the requirement of self--triggering electronics the VA32TA2 chip
has been developed~\cite{IDEAS}. The VA32TA2 houses 32
preamplifiers and 32 slow shaper amplifiers together with 32
corresponding fast shaper amplifiers and discriminators to get
fast timing and trigger signals. The slow shapers provide charge
integration with a peaking time of 2$\,\mu$s. The peak amplitude
is sampled by applying a hold signal, supplied externally with the
appropriate timing. The read--out is done over an up to 10$\,$MHz
multiplexed analogue output.

\begin{figure}[h]
\begin{center}
\epsfig{file=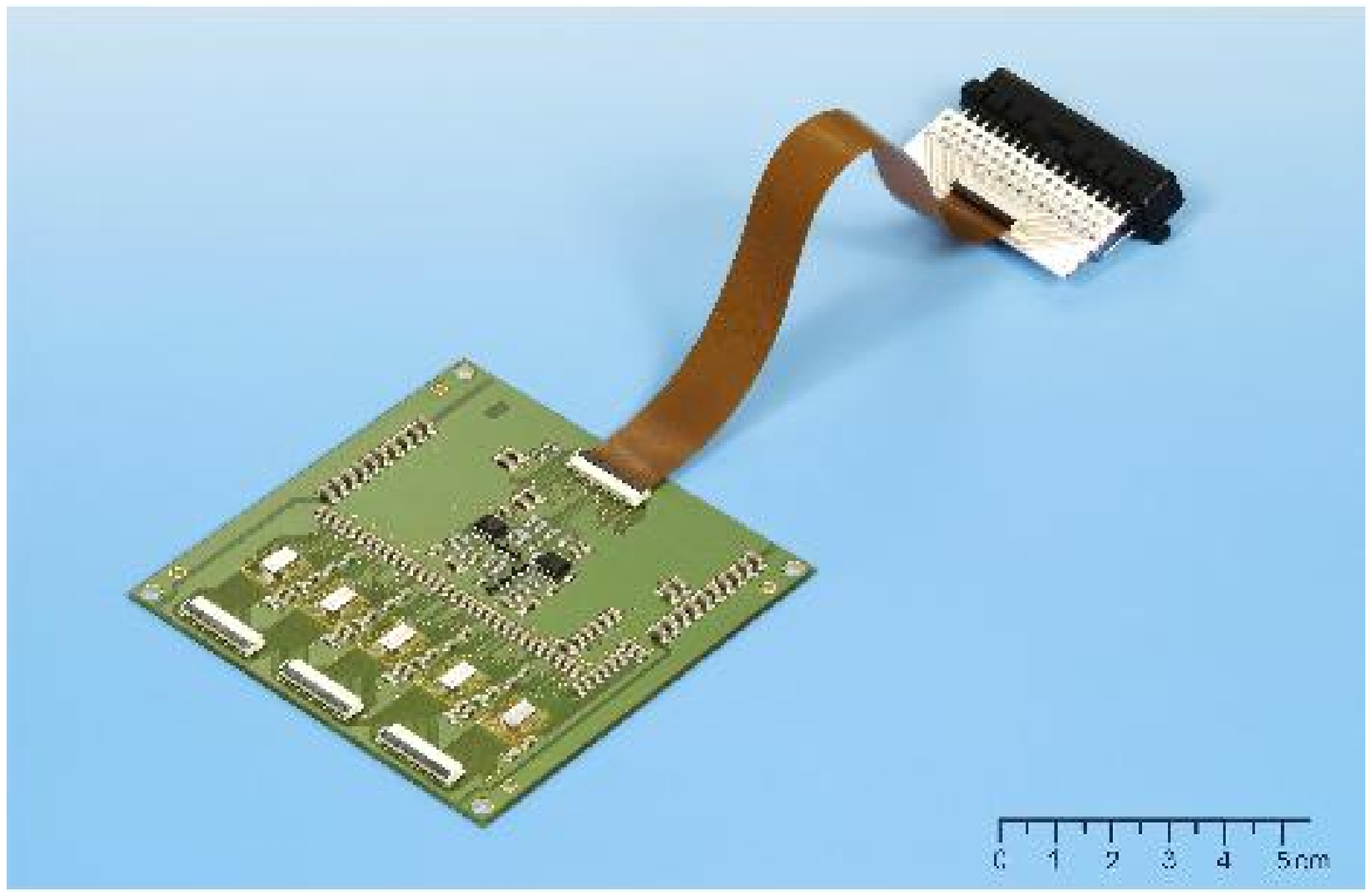,height=4.5cm}
\epsfig{file=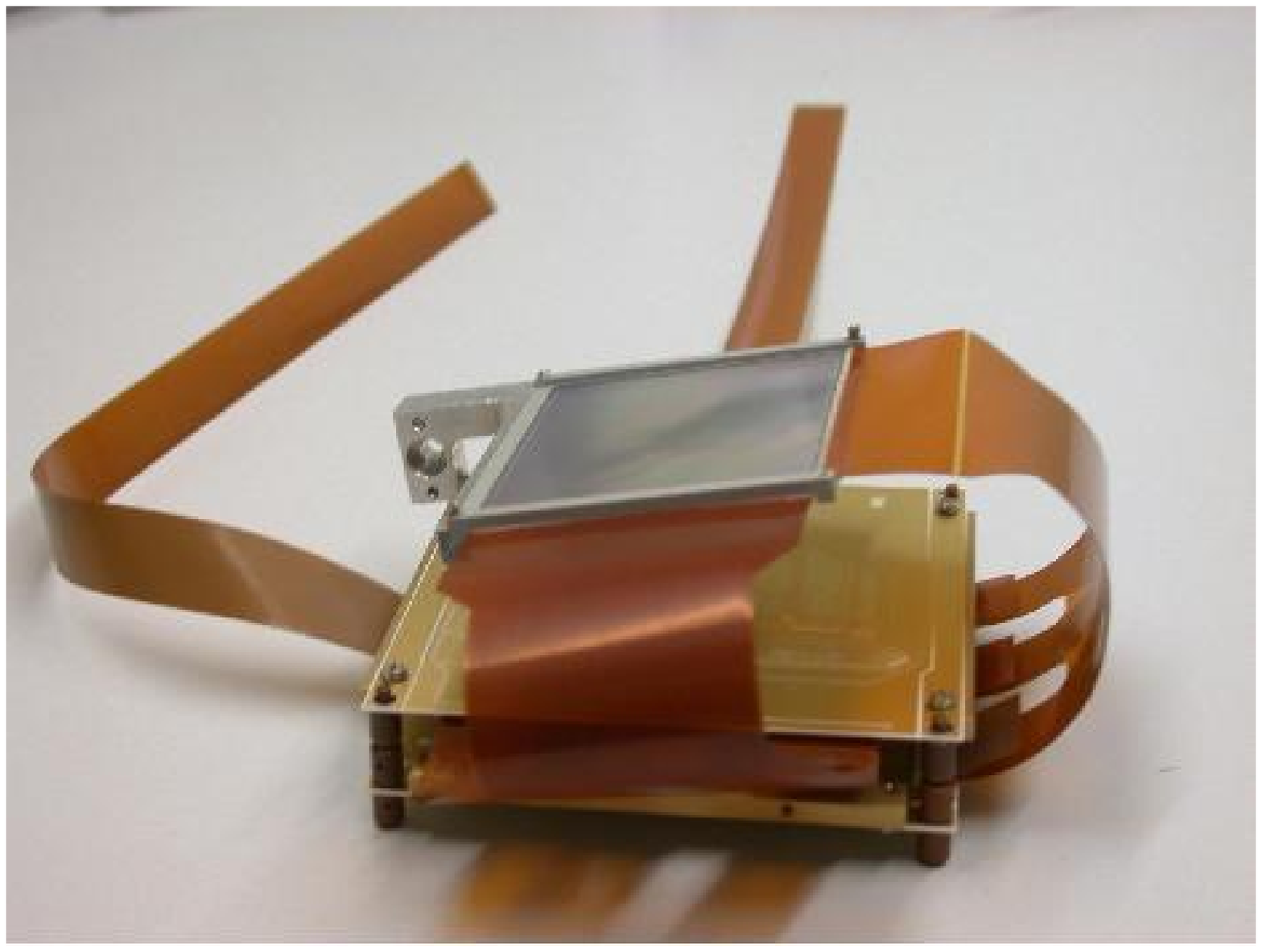,height=4.5cm} \caption{ The left
photo shows the chip board (here a G10 prototype) with 5 chips.
Three input connectors couple it to one side of a detector,
whereas one output connector interfaces to the vacuum
feed--through. The right photo shows the assembly of one
300$\,\mu$m detector with its two ceramic boards. }
\label{FigAssembly}
\end{center}
\end{figure}

The in--vacuum assembly of the chips is based on the use of
$90\times 90\,$mm$^2$ double--sided Al$_2$O$_3$ ceramic boards
(Fig.~\ref{FigAssembly}, left). Five chips are glued and bonded
onto one ceramic board. This correspond to a maximum number of 160
read--out channels where 151 are actually fed to the connectors.
Two of these boards serve to read out the front and backside of a
double--sided detector (Fig.~\ref{FigAssembly}, right).

\subsubsection{The read--out system}

\begin{figure}[h]
\begin{center}
\epsfig{file=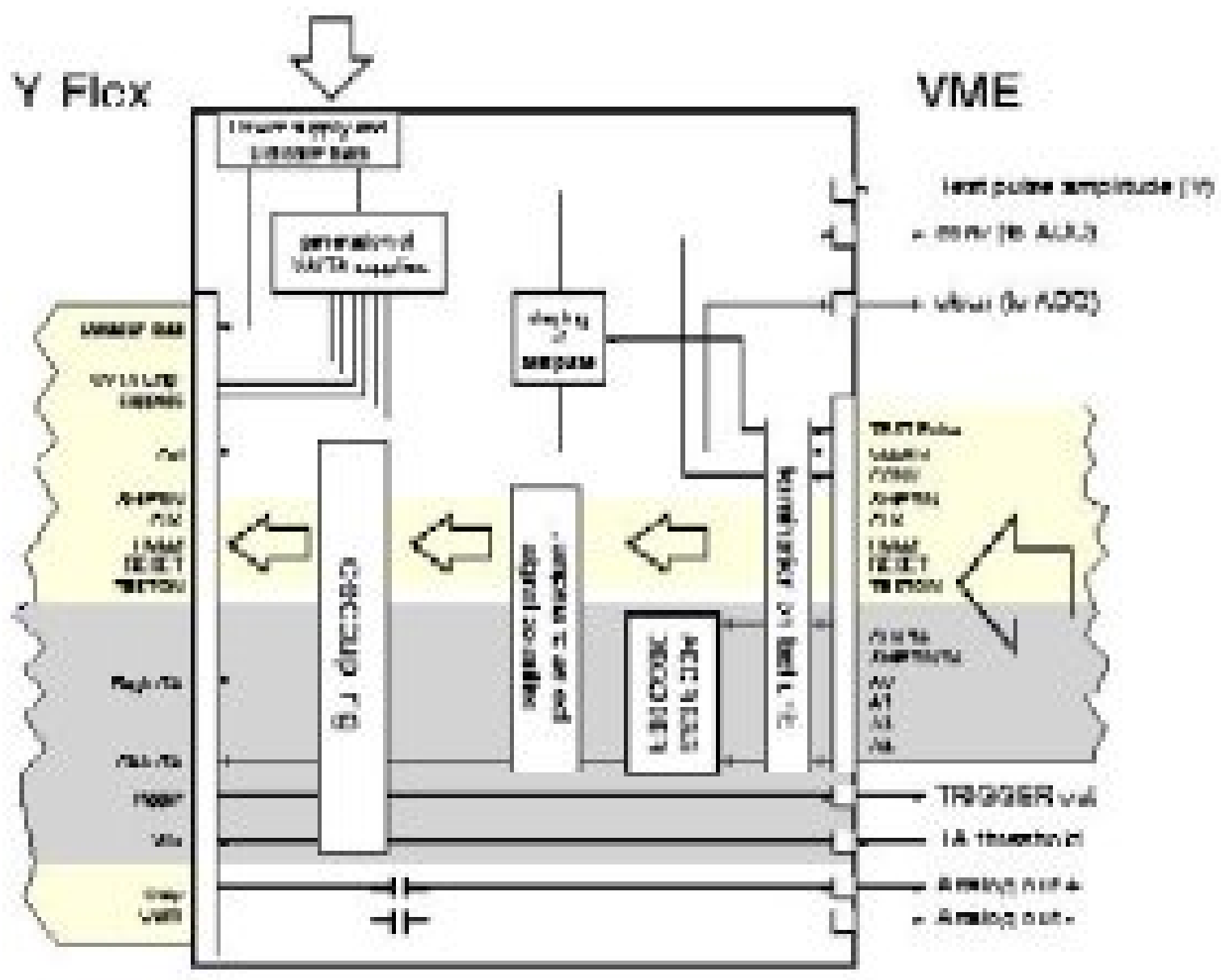,height=5cm}%
\epsfig{file=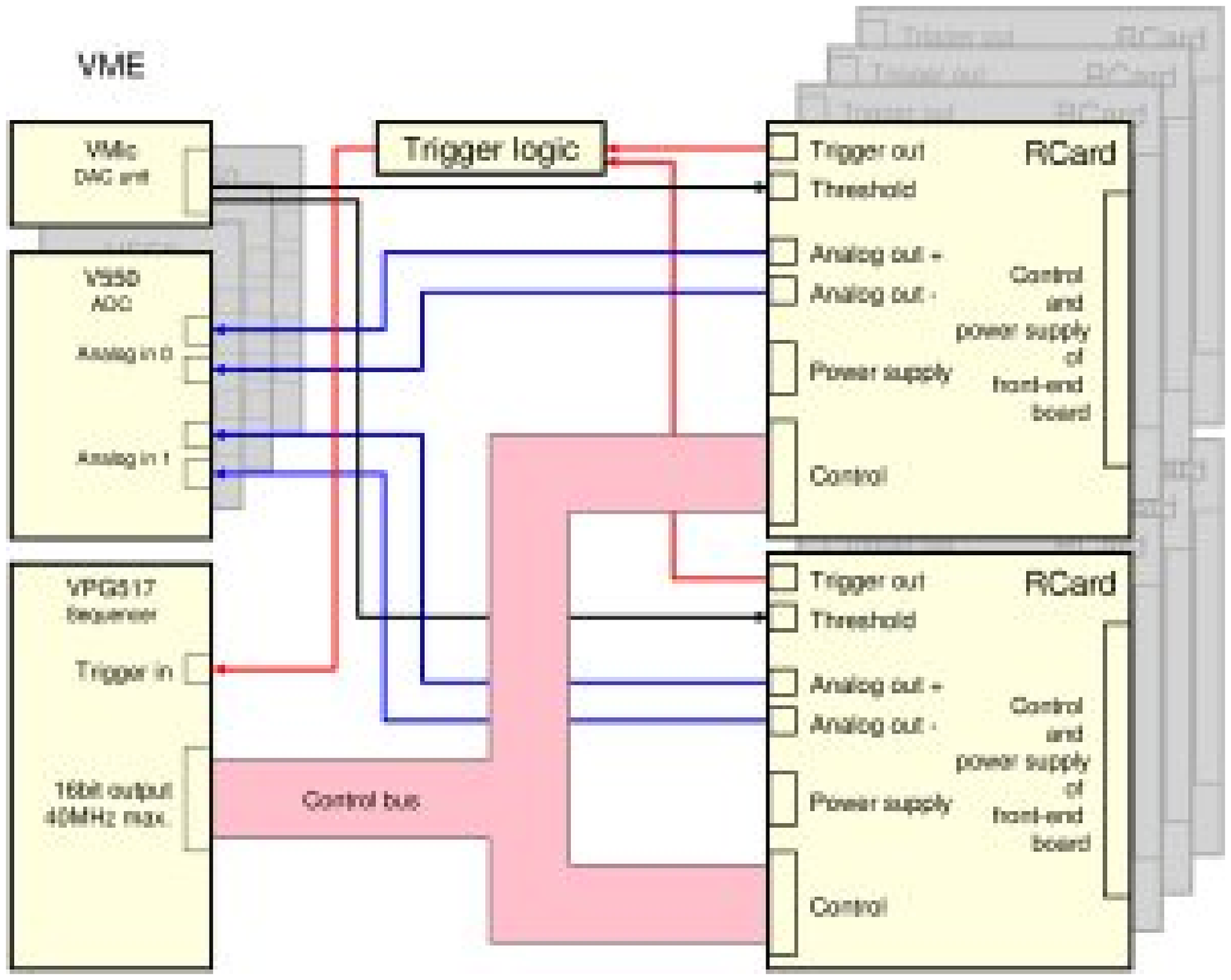,height=5cm}%
\caption{ Left: The RCard block scheme. The Y--Flex part coming
from the detector is electrically fully decoupled from the VME
control part. Right: The block diagram of the VME read--out
scheme. Two RCards are foreseen to read one double--sided silicon
strip detector. } \label{FigBlock}
\end{center}
\end{figure}


Fig.~\ref{FigBlock} shows the block scheme of the interface
electronics between VME and the in--vacuum ceramic boards, the
so--called RCard. Its main purpose is to decouple all bias and
control lines of the chips on the ceramics that are operated at
detector biases up to 1.5$\,$kV. One card is needed for each side
of a detector.

On the VME side the board provides a flat cable connector for all
digital control signals of the board and the front--end
electronics. Up to 16 RCards can be connected and addressed on a
single common bus. All necessary control signals to read out the
amplifier chips and to set the trigger pattern of the addressed
RCard are provided over this flat--cable connection. Since the
timing of the hold--signal for one VA32TA2 read--out chain is
crucial for good performance, an adjustable delay is provided for
this signal on each RCard. Two voltage inputs are provided which
allow one to control the discriminator thresholds and the
calibration pulse amplitude. The block scheme of the complete
setup with all VME components is shown in Fig.~\ref{FigBlock}
(right).

External 16-bit DACs are used for the generation of the VA32TA2
trigger thresholds on the RCards; Each threshold can be controlled
individually. The ADCs have 10$\,$bit resolution and are
especially designed for the read--out of multiplexed analogue
signals from silicon--strip detectors.

\subsubsection{The target cell arrangement}

\begin{figure}[h]
\begin{center}
\epsfig{file=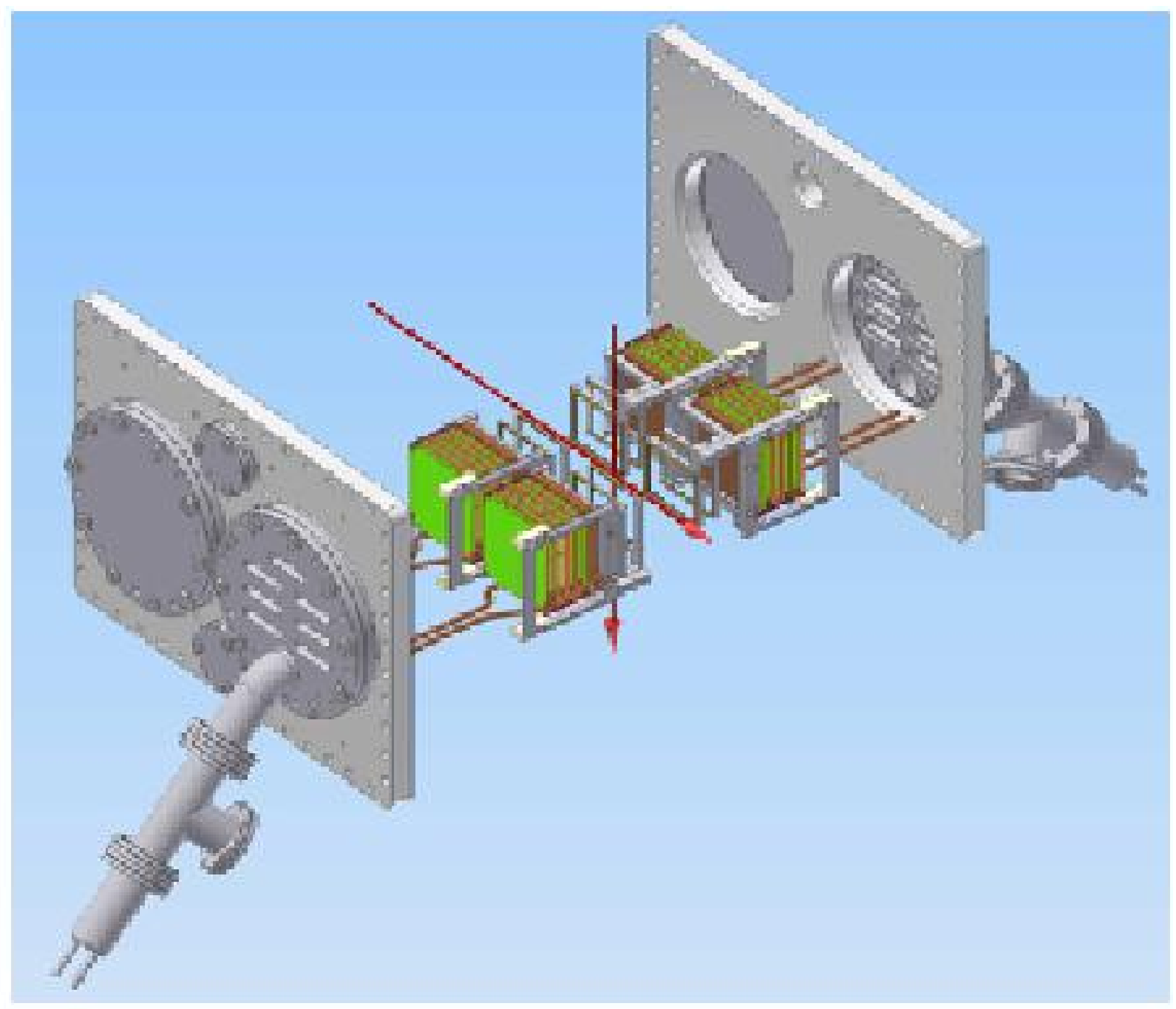,height=7cm} \caption{One of
the possible arrangements with telescopes being placed side--by--side
to cover a longer target cell.} \label{FigABS}
\end{center}
\end{figure}

The telescope systems are very flexible and their arrangement will
depend upon the particular requirements of the experiment being
carried out. For a point target one will generally try to cover a
large part of the solid angle whereas for a target cell, one needs
to cover its length, as illustrated in Fig.~\ref{FigABS}.

%
\subsection{Polarised internal target}
\label{PIT} The polarised internal target system consists of an
atomic beam source feeding a storage cell and a Lamb--shift
polarimeter.  The status of the different components is here
discussed.

The polarised internal hydrogen and deuterium storage--cell target
(PIT)~\cite{rathmann-pit} is already installed at the ANKE
spectrometer for the first test measurements without the COSY
beam. Three weeks have been foreseen by the COSY infrastructure
group for the preparatory work and this should be carried out
during the available maintenance weeks. Once these preparations
have been completed, it will be possible for future exchanges of
the PIT and the cluster target to take place within one of the
maintenance weeks.

Following these three weeks of preparatory work, one week of beam
time has been granted by the COSY PAC for PIT commissioning at
ANKE. The Lamb--shift polarimeter (LSP)~\cite{engels} will be used
as a tool to adjust the transition units of the polarised atomic
beam source (ABS)~\cite{max} that feed the storage cell. An
additional week of COSY--deuteron beam has been allocated for the
PIT commissioning phase to establish that one can measure the
nuclear polarisation of the target as well as its density through
elastic $d\vec{p}$ scattering. These measurements will also serve
to calibrate the LSP. After these five weeks of installation,
commissioning, and initial research in 2005, the PIT and the LSP
can be moved to their off--beam positions, outside the COSY ring,
depending on the ANKE and COSY experimental schedule.

The results achieved during the first phase of research with the
PIT will form the experimental basis for the future programme of
single and double--polarisation measurements at ANKE.

\subsubsection{Status of the PIT and LSP development}

\begin{figure}[hbt]
\begin{center}
  \psfig{figure=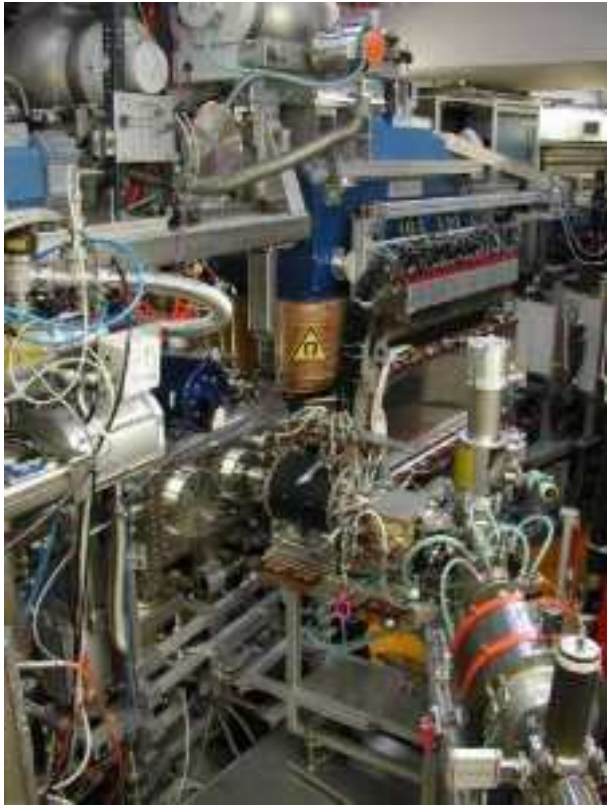,width=8cm}
\end{center}
\caption{\label{fig:abs-cosy}
  Polarised Internal Target at its in--beam position at ANKE.}
\end{figure}

\begin{figure}[hbt]
\begin{center}
  \psfig{figure=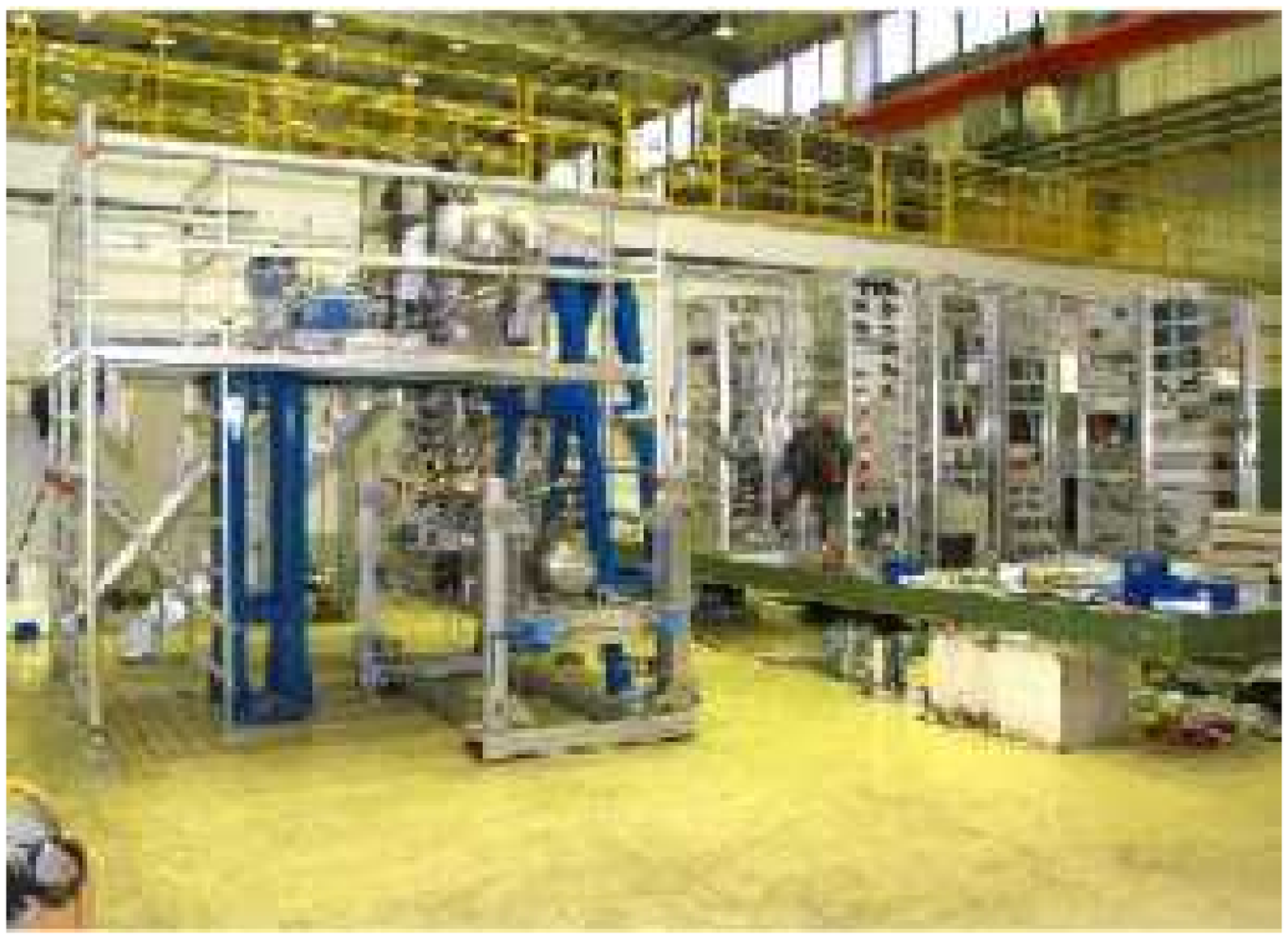,width=10cm}
\end{center}
\caption{\label{fig:abs-lkw} Polarised Internal Target
installation at the off--beam position. The blue pillars on the
left that support the bridge on which the ABS is mounted, exactly
mimic the D1 and D2 magnets of ANKE.}
\end{figure}

The new large--volume target chamber, the differential pumping
system in the ANKE section, two additional beam--position monitors
in front of the storage--cell and between the ANKE dipole magnets
D2 and D3, and additional vertical beam steerers, have already
been installed in order to facilitate the use of the PIT. In
October 2004 the ABS and the LSP were transferred from the
laboratory to their off--beam positions in the COSY hall outside
the tunnel and are ready for installation. The ABS is presently
mounted on a new bridge, designed to support it at the in--beam
position above the ANKE--target chamber. The LSP has been be
placed on a separate support, designed taking into account the
spatial boundary conditions in the target area and the movement of
both the D2 dipole magnet and the target chamber. All the supply
units for the ABS and the LSP, as well as the slow--control
system, are mounted on a common transport platform. An additional
vacuum chamber, of dimensions identical to those of the
ANKE--target chamber, has been produced and this allows the
necessary preparatory tests in the off--beam position. A very
limited number of crane movements is thus required to transfer the
complete setup to the ANKE position. Adjacent to the ANKE target
place, an elevated support has been created that will carry the
supply platform. Figure~\ref{fig:abs-cosy} shows the setup in the
in--beam position, whereas the off--beam configuration is
illustrated in Fig.~\ref{fig:abs-lkw}.

\begin{figure}[!hbt]
\begin{center}
  \psfig{figure=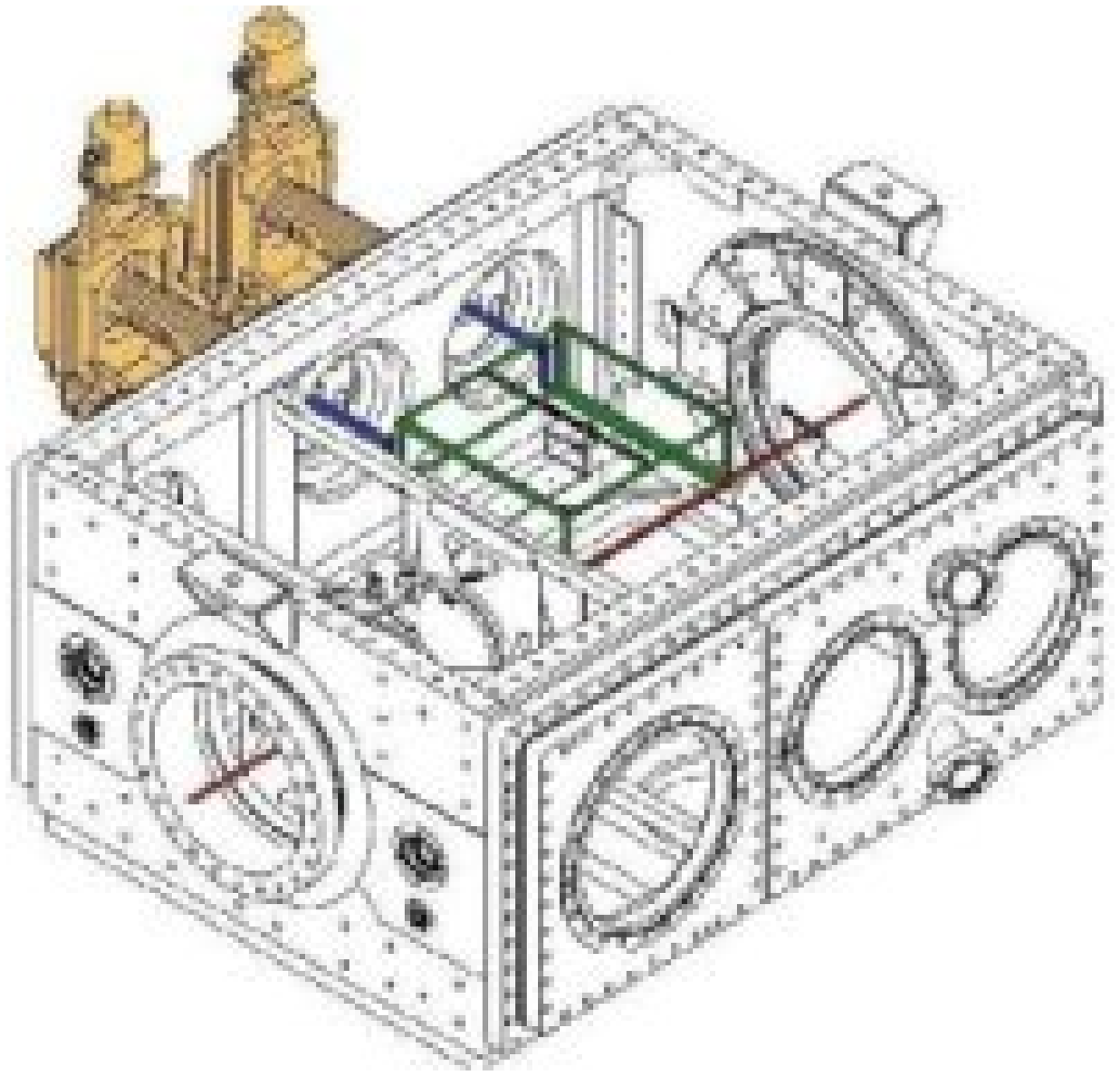,width=7cm}
\hspace{0.2cm}
  \psfig{figure=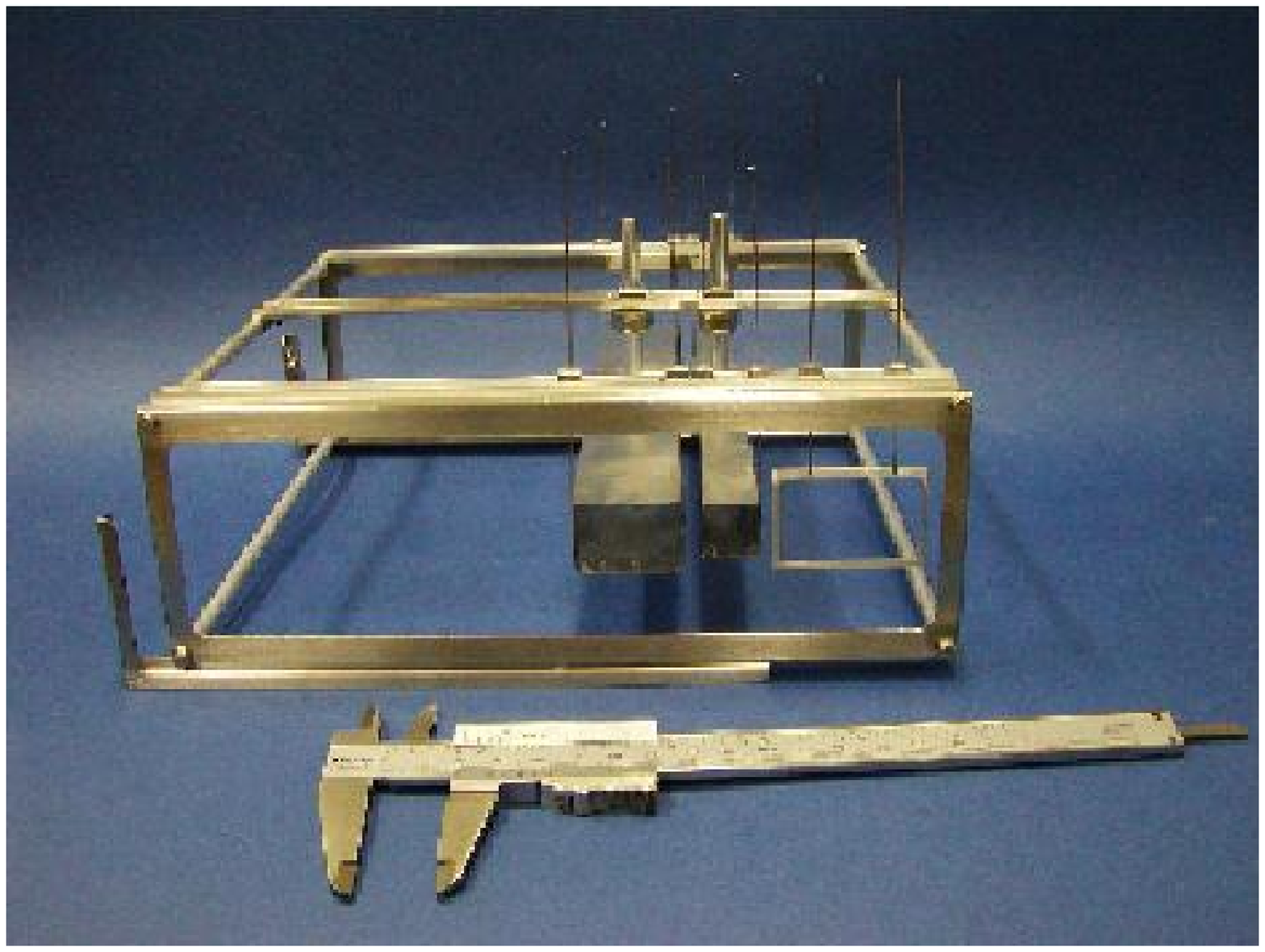,width=7cm}
\end{center}
\caption{\label{fig:cell-setup}Left: Setup of the movable support
for the storage cell inside the new target chamber at ANKE. Right:
Support frame, indicated in green on the left panel, with two
installed storage cells and a single aperture,. The cross section
of the cells are $30\times20${\ww}mm$^2$ and $15\times15${\ww}mm$^2$, the
ones of the aperture are $40\times25${\ww}mm$^2$. The unpolarised gas
supply to the two cells is attached from the top in the cell
centre.}
\end{figure}

The measurements to study the COSY--beam properties at the ANKE
target, \emph{i.e.}\ at the storage--cell position, and determine
the lateral dimensions for an optimised storage--cell have been
started with the setup shown in Fig~\ref{fig:cell-setup}.

The external positioning control, which is part of the PIT
slow--control system, allows one to centre different diaphragms
and prototype cells onto the COSY beam axis and to move them
step-wise for beam cross section and lifetime studies. According
to our preliminary results, a cell tube of about 15{\ww}mm diameter
(and 350{\ww}mm length) can be installed at the beam. However, due to
the fact that these measurements had to be done without dedicated
COSY beam optimisation and, in view of the strong dependence of
the target density upon the lateral extension of the cell tube,
further measurements are needed.

These studies have been continued at the beginning of 2005 during
one week of beam time allocated for that purpose. During these
measurements, the calibrated supply system for unpolarised gases
was utilised to feed the storage cell for investigations of
\emph{e.g.}\ beam-heating effects and for measurements of the
pressure distribution in the section in and around the ANKE target
chamber. It was possible to inject, store and accelerate to
2.4{\ww}GeV/c about $10^{10}$ polarised deuterons in the presence off
the large cell, shown on the right panel of
Fig.~\ref{fig:cell-setup}. This amounts to about 70\% of the
number of deuterons that could be stored at injection energy
(45{\ww}MeV). These test were carried out with a flux of about
$10^{-3}${\ww}mbar{\ww}l/s, leading to a target density in the large cell
of $9.4\times 10^{13}${\ww}cm$^{-2}$. It was also possible to take the
first data from a storage cell target at ANKE in this mode.

%
%
\clearpage
\section{Beam and Target Polarimetry}
\label{BandTP}
\subsection{Deuteron beam polarimetry}
\label{DBP} The polarised $H^{-}$ or $D^{-}$ ion beam delivered by
the source~\cite{ionsource}, is pre-accelerated in the cyclotron
JULIC and injected by charge exchange into the COSY ring. The
acceleration of vertically polarised protons and deuterons at COSY
is discussed in detail for example in ref.~\cite{Lehrach}.
Although beam polarisations in the ring can be established at
certain energies by using the EDDA polarimeter~\cite{EDDA}, in
order to ensure that all the conditions of the actual measurement
are met, it is preferable to be able to measure oneself the beam
polarisation during any experiment. We here describe briefly the
first test measurements~\cite{PROPOSAL} that were carried out at
ANKE using a polarised deuteron beam ($p_{d}=2.4${\ww}GeV/c) and an
unpolarised hydrogen cluster target to show how such polarimetry
can be carried out in practice.

The scheme for the polarised deuteron beam consisted of eight
different polarisation states, including one unpolarised mixture
and seven combinations of vector and tensor polarisations. The
states and the nominal values of polarisations ($P_z$ and
$P_{zz}$) and intensities are shown in Table~\ref{polmodes}. For
each injection into COSY, the polarised ion source was switched to
a different polarisation state. The duration of a cycle was
sufficiently long (200$\,$s) to ensure stable conditions for the
injection of the next state.  After the seventh state, the source
was reset to the zeroth mode and the pattern repeated. The ANKE
data acquisition system received status bits from the source,
latched during injection, that ensured the correct identification
of the polarisation states during the experiment.
\begin{table}[h]
\centering
\includegraphics[width=\textwidth]{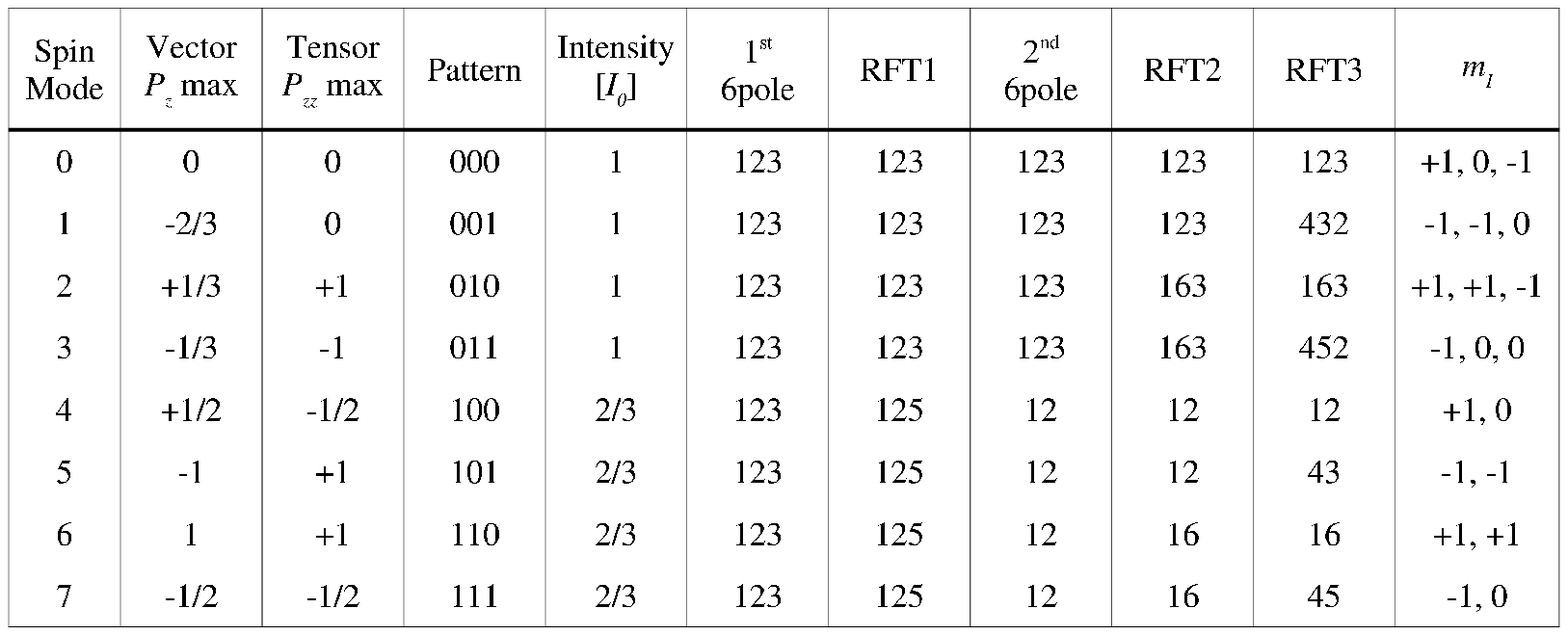}
\caption{\footnotesize Modes of the polarised deuteron ion source. The
        intensity modulations between the different modes constitute a
        compromise in order to achieve higher polarisations. $I_{0}$ refers to
        the maximum number of deuterons delivered by the source and stored in
        COSY.}
\label{polmodes}
\end{table}

\begin{figure}[htb]
\begin{center}
\centerline{\epsfxsize=2.4in\rotatebox{-90}{\epsfbox{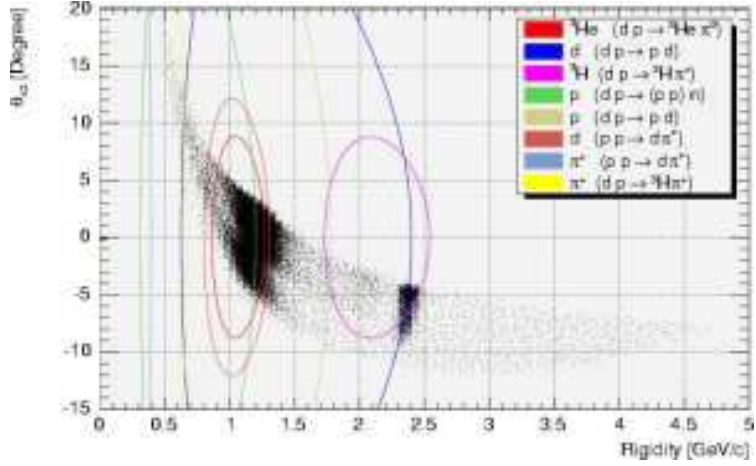}}}
    \caption{ANKE experimental acceptance for different reactions
            in $dp$ collisions at $T_d=1170${\ww}MeV.\label{accep}}
\end{center}
\end{figure}

Fig.~\ref{accep} shows the ANKE experimental acceptances for
singly charged particles for different reactions as functions of
the laboratory production angle and magnetic rigidity, together
with loci for the kinematics of different allowed processes. The
$dp$ elastic scattering reaction has a significant acceptance for
$4^{\circ}<\theta_\textrm{lab}^d<10^{\circ}$. The observables
$A_{y}$, $A_{yy}$, and $A_{xx}$ of this reaction were carefully
measured at Argonne~\cite{Igo} and SATURNE~\cite{Arvieux} for
$T_d=1198${\ww}MeV.

\begin{figure}[htb]
\centering
\includegraphics[width=\textwidth]{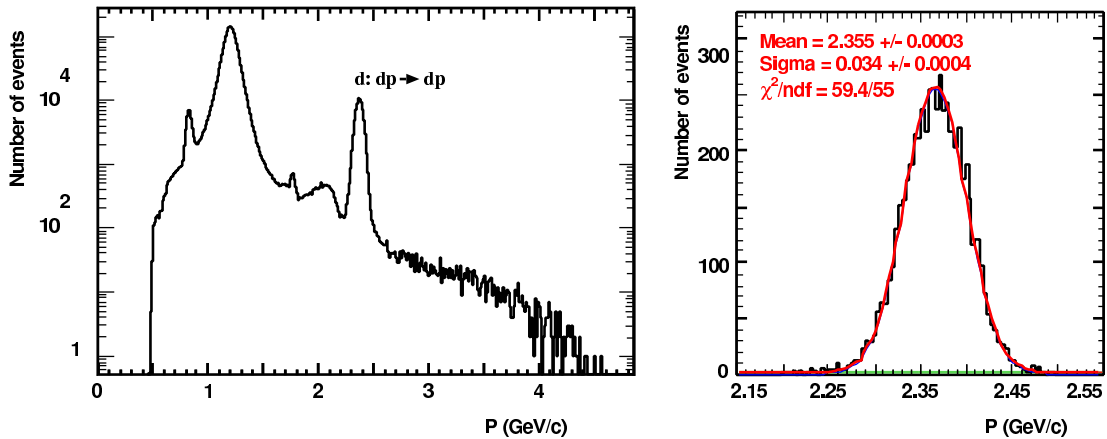}
\caption{\footnotesize Left: Single--track momentum spectrum for
the $dp$ data at 2.40{\ww}GeV/c. Right: Fit result of the elastic peak
region with the sum of a Gaussian and linear function.}
\label{fig:p_distrubition}
\end{figure}

The elastic peak region in the momentum spectrum of the single
track events (left panel of Fig.~\ref{fig:p_distrubition}) was
fitted with the sum of a Gaussian and linear function, and events
selected within 3$\sigma$ of the mean. An example of such a fit is
shown the right panel.

The $dp\to\,^3\textrm{He}\,\pi^0$ reaction can be investigated
using simply the $^3\textrm{He}$ information. The high momentum
branch of $^3\textrm{He}$ particles was isolated well in off--line
analysis by applying two--dimensional cuts in $\Delta E$
\emph{versus} momentum and $\Delta t$ \emph{versus} momentum for
individual layers of the forward hodoscope. The tensor analysing
power of this reaction has been measured at $0^{\circ}$ as a
function of beam energy at Saclay~\cite{Kerboul}.

The quasi--free $np\to d\pi^0$ can also be clearly identified by
detecting the two final charged particles in the $dp\to
p_{sp}d\pi^0$ reaction, where $p_{sp}$ is a spectator proton which
has about half the beam momentum. Though, by isospin, the
differential cross section should be half of that of $pp\to
d\pi^+$, all the analysing powers should be equal for $\pi^+$ and
$\pi^0$ production.

The charge--exchange process was selected from the missing--mass
with respect to the observed proton pairs (see \S\ref{ece}) and
time--difference information. The spectra for all spin modes
reveal a well defined peak at M$_{miss}$ equal to the neutron mass
to within 1\%. The background was less than $2\%$ and stable, so
that the charge--exchange process could be reliably identified.

Using the $\vec{d}p\to\,dp$, $\vec{d}p \to\,(2p)n$, $\vec{n}p\to
d\pi^0$, and $\vec dp\to\,^3\textrm{He}\,\pi^0$ reactions, which
all have large and well known analysing powers,
a simultaneous calibration of the vector and tensor components of
the polarised deuteron beam at COSY became possible for the first
time. In all cases the beam polarisation was consistent with being
proportional to the ideal values nominally supplied by the source.
The results are therefore summarised in Table~\ref{pol} in terms of
vector and tensor proportionality parameters $\alpha_{z}$ and
$\alpha_{zz}$.

\begin{table}[h]
\renewcommand{\arraystretch}{1.5}
\begin{center}
\begin{tabular}{|c|c|c|c|}
\hline
Reaction&Facility&$\alpha_{z}$&$\alpha_{zz}$\\
\hline%
$\vec{d}p\to dp$&EDDA&$0.74\pm0.02$&$0.59\pm0.05$\\
\hline%
$\vec{d}p\to dp$&ANKE&$0.73\pm0.02$&$0.49\pm0.02$\\
\hline%
$\vec{n}p\to d\pi^0$&ANKE&$0.70\pm0.03$&---\\
\hline%
$\vec{d}p\to\,^3\textrm{He}\,\pi^0$&ANKE&---&$0.58\pm0.05$\\
\hline%
$\vec{d}p\to (pp)n$&ANKE&---&$0.48\pm0.05$\\
\hline
\end{tabular}
\caption{Values of vector and tensor polarisation parameters. The
errors quoted are only statistical.} \label{pol}
\end{center}
\end{table}

The average of the ANKE measurements is
$\alpha_{z}^\textrm{ANKE}=0.72\pm0.02$ and
$\alpha_{zz}^\textrm{ANKE}=0.52\pm0.03$, which are compatible with
EDDA results~\cite{EPAC} measured prior to the ANKE run but at
lower beam energy and intensity.

\subsection{Polarisation export technique}
\label{PBT}

The absolute value of the beam polarisation is clearly needed in
any measurement with polarised projectiles. This is usually
determined from the scattering asymmetry in a suitable nuclear
reaction for which the analysing power is already known.
Calibration standards of the type discussed in \S\ref{DBP} are few
and only exist at discrete energies. It is therefore of great
practical interest to be able extend their application to
arbitrary energies where standards are not yet available. Now, if
care is taken to avoid depolarising resonances in the machine, the
beam polarisation should in general be conserved during the
process of ramping the beam energy up or down~\cite{pollock}. Such
tests measurements have been carried out at COSY for both
polarised proton and deuteron beams.

Results for proton beam polarimetry are described in
ref.~\cite{AY}. The absence of azimuthal symmetry of the ANKE
spectrometer does not permit one to measure a vector analysing
power from the left--right count rate asymmetry. We therefore
determined $A_y$ by reversing the orientation of the polarisation
every two cycles. Careful monitoring of the relative luminosity
$L_{\uparrow} /L_{\downarrow}$ was achieved by detecting single
particles in the FD either at $\theta_\textrm{lab} < 1^{\circ}$ or
at $\phi=90^\circ\pm5^\circ$ and $\phi=270^\circ\pm5^\circ$, where
the rates are insensitive to the vertical beam polarisation.

The beam polarisation at $T_p=0.800${\ww}GeV was determined by
measuring $pd$ elastic scattering, where the scattering angles
were fixed by the energy deposit of the identified deuterons in
the silicon telescopes. It should be noted that there are good
$pd$--elastic analysing power data at 0.796{\ww}GeV~\cite{irom}.

Since the corresponding data are not available at 0.5{\ww}GeV, we
resorted to the polarisation--export technique~\cite{pollock} to
obtain a polarisation calibration. This was achieved by setting up
a cycle with a flat top at energy $T_p=0.8${\ww}GeV (I), followed by
deceleration to a flat top at 0.5{\ww}GeV (II), and subsequent
re-acceleration to the 0.8{\ww}GeV flat top (III). The measured beam
polarisations $P_I=0.564 \pm 0.003^{\mathrm{stat.}} \pm
0.004^{\mathrm{syst.}}$ and $P_{III} = 0.568 \pm
0.004^{\mathrm{stat.}} \pm 0.005^{\mathrm{syst.}}$ agree within
errors, and this shows that we have avoided significant
depolarisation while crossing of the resonances. The systematic
errors arise from the uncertainties in the relative luminosity.
The weighted average of $P_I$ and $P_{III}$ was used to export the
beam polarisation to flat top II and to determine the angular
distribution of the previously unknown analysing power of $pd$
elastic scattering at 0.5{\ww}GeV. A small angle--independent
correction of $-0.0024$ was applied in the export procedure to
account for the 4{\ww}MeV difference in beam energy, using the energy
dependence of $A_y$ between 500 and 800{\ww}MeV.

\begin{figure}[htb]
\centering \centerline{\epsfxsize=7cm\epsfbox{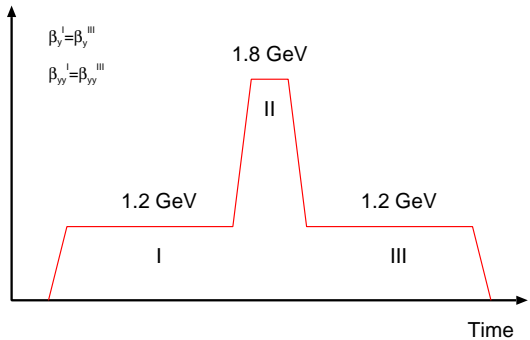}}
\caption{Schematic picture of the three different flat-top regions
used in a single cycle of the February 2005 run. The identity of
the deuteron polarisation in regions I and III means that the
1.2{\ww}GeV polarisation could be exported to 1.8{\ww}GeV.} \label{ramper}
\end{figure}

Beam time was allocated in February 2005 in order to measure the
polarised $p(\vec{d},pp)n$ reaction at three different beam
energies, \emph{viz.}\ $T_d=1.2${\ww}GeV (for polarimetry purposes),
1.6{\ww}GeV, and 1.8{\ww}GeV. In order to verify the polarisation export
technique with a circulating deuteron beam at COSY, the scheme
shown schematically in Fig.~\ref{ramper} was implemented.

The polarimetry was carried out using small angle $dp$ elastic
scattering, as described in \S\ref{DBP}. The preliminary results
of this test measurement are shown in table~\ref{d-ramp} in terms
of the non--normalised parameters $\beta$ where the analysing
power of the reaction has not been introduced. Given that, within
the small error bars, $\beta_{y/yy}^{I}=\beta_{y/yy}^{III}$ no
depolarisation has been observed. We can therefore conclude that
the beam polarisation at 1.8{\ww}GeV is the same as that at 1.2{\ww}GeV.
The export technique can therefore be used for both proton and
deuteron beams.

\begin{table}[h]
\renewcommand{\arraystretch}{1.5}
\begin{center}
\begin{tabular}{|l|l|}
\hline
$\beta_y^{I}=-0.213\pm0.005$&$\beta_y^{III}=-0.216\pm0.006$\\
\hline
$\beta_{yy}^{I}=-0.053\pm0.003$&$\beta_{yy}^{III}=-0.060\pm0.003$\\
\hline
\end{tabular}
\caption{Values of the non--normalised deuteron vector and tensor
polarisation parameters obtained before and after beam ramping
from 1.2{\ww}GeV to 1.8{\ww}GeV and back.} \label{d-ramp}
\end{center}
\end{table}

\subsection{Target polarimetry}
\label{TP}

\begin{figure}[hbt]
\begin{center}
 \psfig{figure=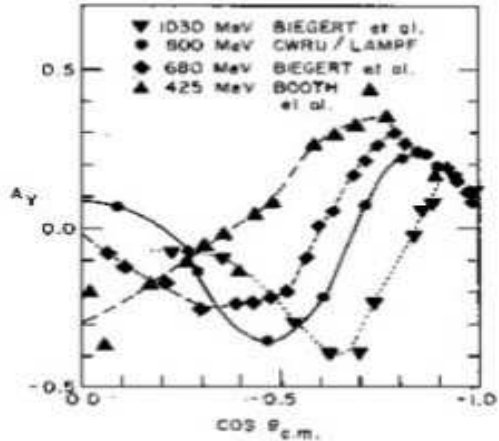,width=7cm,height=6.0cm}
\end{center}
\caption{\label{fig:winkelmann}Analysing power data for $d\vec{p}$
elastic scattering~\cite{winkelmann}. Note that the ANKE acceptance is
over the range $-0.77>\cos\theta>-0.98$.}
\end{figure}

Two weeks were granted by the COSY PAC for initial research with
the PIT and these are to be scheduled for autumn 2005. The main
goal is to accomplish a measurement of the target performance, in
particular the target polarisation and density. A suitable
reaction to measure the target density is $d\vec{p}$ elastic
scattering since, as shown in Fig.~\ref{fig:winkelmann}, the cross
section and analysing power have been well studied in the angular
range representing the ANKE acceptance~\cite{winkelmann}.

Once the target polarisation has been determined, the data sample
obtained can be used to derive the analysing power of the
charge--exchange reaction $d\vec{p}\to (pp)n$. However, experience
gained during the short test experiment with a polarised deuteron
beam and a hydrogen target~\cite{Chiladze} has shown that the
acceptance of ANKE is such that several reactions with
well--studied analysing powers will be recorded simultaneously.
Among these will be, for example, quasi--free $n\vec{p}\to
d\pi^0$, which has a fast \emph{spectator} proton, and
$d\vec{p}\to\,^3\textrm{He}\,\pi^0/\,^3\textrm{H}\,\pi^+$. There
will therefore be several reactions that can be used to provide a
calibration.

%
%
\subsection{Luminosity determination}

Though this whole document is biassed towards the determinations
of analysing powers and spin correlations \emph{etc.}, values of
differential cross sections are at least as important and for this
the luminosity has to be fixed. Inside a storage ring such as
COSY is it customary to do this by measuring in parallel a
reaction for which the cross section is known from other
experiments. This sometimes limits the energies at which
experiments can be reliably standardised. There are, however, two
other possibilities that we will exploit for normalisation
purposes.

When experiments are carried out using a target cell, the density
of the polarised gas target can also be inferred by comparison
with detector rates obtained with a calibrated flux of unpolarised
hydrogen gas admitted into the centre of the storage cell. This
method is described in detail in ref.~\cite{zapfe}. More
imaginatively, the energy loss of the beam due to electromagnetic
interactions in the target is a measure of the integrated
luminosity. The energy shift gives rise to a corresponding
frequency shift, which can be measured through the study of the
Schottky noise spectrum of the coasting beam. It is hoped that
this method, which is the subject of a detailed study at
COSY~\cite{Stein}, will be operational by the end of 2005.
%
%
\noindent
\section{Proton--Neutron Spin Physics}
\label{PNSP2}

The nucleon--nucleon interaction is fundamental to the whole of
nuclear physics and hence to the composition of matter as we know
it. Apart from its intrinsic importance, it is also a necessary
ingredient in the description of meson production processes.

In the case of proton--proton scattering, the data set of
differential and total cross sections and the various single and
multi--spin observables is very extensive. This allows one to
obtain reliable isospin $I=1$ phase shifts up to at least 800{\ww}MeV
at the click of a mouse~\cite{SAID,Nijmegen}. Furthermore, the
mass of new high quality EDDA data~\cite{EDDA} reduces
significantly the $I=1$ phase shift ambiguities up to
2.1{\ww}GeV~\cite{EDDA_rev}. This is, of course, only possible by
taking the new data in conjunction with the results of earlier
painstaking systematic work. The meticulous investigation of the
nucleon--nucleon interaction must therefore be a communal activity
across laboratories, with no single experiment providing the final
breakthrough.

Although the extra information required to fix the $I=1$
proton--proton amplitudes uniquely up to 2.1{\ww}GeV is limited, the
same cannot be said for the isoscalar $I=0$ case, since this would
require more good data on neutron--proton scattering. The
situation is broadly satisfactory up to around 515{\ww}MeV but the
only fairly complete data set above that is at the LAMPF energy of
around 800{\ww}MeV, though many of the measurements were carried out
at Saclay~\cite{SAID}.

The limited intensity, the large momentum bite, and the general
difficulty of working with neutral particles, makes one seek
alternatives to using neutron beams for the study of $np$
scattering. For many years the deuteron has been used as a
substitute for a free neutron target. The corrections required in
order to extract proton--neutron observables are generally quite
small and fairly well calculable at high energies because the
typical internucleon separation in the deuteron ($\approx 4\:$fm)
is large compared to the range of the projectile--nucleon force.
It is therefore plausible to assume that the projectile generally
interacts with either the target proton or neutron, with the other
nucleon being largely a \emph{spectator}, moving with the Fermi
momentum that it had before the collision. Nevertheless, the
nature of the corrections have to be studied carefully for each
individual reaction. For example, it has been shown that the spin
correlation and transfer parameters in $pp$ quasi--elastic
scattering in the 1.1 to 2.4{\ww}GeV range are very close to those
measured in free $pp$ collisions~\cite{Ball} and the Saclay group
find exactly the same reassurance for $pn$ quasi--elastic
scattering~\cite{Lesquen}. The investigation was, however, carried
out far away from the forward direction whereas other deuteron
corrections can be important at small angles, when it is not clear
which is the spectator and which the struck
nucleon~\cite{Aladashvili1}.

With the current and projected facilities positioned inside the
COSY ring, we expect to contribute to the elastic proton--neutron
data base in two distinct regions. By detecting a slow proton in
the silicon counters and a fast proton in ANKE, we will measure
elastic $pn$ scattering up to the maximum COSY proton beam energy
for laboratory angles of the fast proton with
$|\theta_p^{lab}|<12^{\circ}$ . Using a transversally polarised
beam and/or target, this will give access to the unpolarised cross
section, $\dd\sigma/\dd\Omega$, the proton and neutron analysing
powers, $A_y$, and the spin correlation parameters, $A_{yy}$ and
$A_{xx}$, as described in \S\ref{pn-elastic}.

In parallel with elastic scattering, measurements will also be
made of the cross section and spin dependence of the
$\vec{p}\,\vec{n}\to d\pi^0$ reaction up to 3{\ww}GeV by detecting the
spectator proton in the silicon counters and the deuteron in ANKE.
This reaction~\cite{pid}, which is the prototype of all
pion--production processes, can be measured near both the forward
and backward cm directions, provided that
$|\theta_d^{lab}|<8^{\circ}$. This is discussed further in
connection with other non--strange mesons in \S\ref{npdpi}.

The large angle (\emph{i.e.}\ charge exchange) region in $np$
elastic scattering is currently being investigated at ANKE by
studying the charge exchange reaction of a tensor polarised
deuteron beam on an unpolarised target~\cite{PROPOSAL,Chiladze}.
It has been shown~\cite{Bugg-Wilkin} that for low $pp$ excitation
energies such experiments are very sensitive to the spin--spin
terms in the $np$ charge--exchange amplitude. The deuteron tensor
analysing powers are then essentially equivalent to the
spin--transfer parameters in $\vec{n}p\to\vec{p}n$. Using both
vector and tensor polarised deuterons incident on a polarised
hydrogen target, it is possible to investigate additionally both
the spin--correlation parameters and triple--spin parameters, such
as $A(0s';sn)$. More details of this proposal are given in the
\S\ref{np-ce}, where it is seen that one of the biggest drawbacks
of this approach is that it is limited by the maximum COSY
deuteron energy of $T_d\leq 2.3${\ww}GeV, which means that the neutron
flux dies out beyond 1.15{\ww}GeV.

The same $\vec{p}\,\vec{d}\to ppn$ reaction can, however, be
studied in inverse kinematics up to the maximum COSY proton energy
of nearly 3{\ww}GeV by using a polarised proton beam incident on a
polarised deuterium target. The two protons from the reaction then
have low energies and both can be very efficiently measured in the
silicon telescopes, which cover a significant fraction of the
angular domain. It is shown in \S\ref{pn-ce} that the resolution
expected in the $pp$ excitation energy $Q_{pp}$ is even better
than that obtainable with a deuteron beam but the price that one
has to pay is that very small momentum transfers are not covered
for low values of $Q_{pp}$. It should be noted that the magnetic
spectrometer is \emph{not} used when obtaining such data. As a
consequence this experiment can be run in parallel with the
small--angle $\vec{p}\,\vec{n}$ elastic scattering described in
\S\ref{pn-elastic}. In fact, provided that one triggers on at
least one low energy proton, relevant data will be accumulated
whenever a deuterium target is in position.

At energies well below the pion production threshold, one can
model the $NN$ interaction in terms of a purely elastic two--body
problem, where the only role played by the mesons is as mediators
of the nuclear force~\cite{Machleidt}. However, at 1{\ww}GeV about
40\% of the total $np$ cross section corresponds to pion
production, mainly involving the excitation of the $\Delta$
isobar. This can be either implicit, as in the $pn\to d\pi^0$
reaction to be discussed in \S\ref{npdpi}, or explicit, as in
the $pn\to \Delta^0p\to pp\pi^-$ reaction. Even below the pion
production threshold, such processes give rise to dispersive
forces that affect elastic $NN$ scattering~\cite{Machleidt},
though they are sometimes modelled in terms of effective heavy
meson exchange. In quark language, the $N\to\Delta$ transition
just involves the spin flip of one of the constituent quarks
without changing its orbital angular momentum. Any description of
the $NN$ interaction above the pion threshold should, at the very
least, consider the coupled channels of $NN\rightleftharpoons
N\Delta$~\cite{JH}, for which experimental information is required
on the spin dependence of the transition amplitudes.

In addition to detecting the quasi--elastic charge exchange
$p(\vec{d},2p)n$ reaction~\cite{Ellegaard1,Sams95a}, the SPESIV
spectrometer allowed the extraction of the strength and analysing
power of $\Delta(1232)$ production, $p(\vec{d},2p)\Delta^0$, from
the missing mass in the reaction~\cite{Ellegaard2,Sams}. These
investigations of the spin--flip excitation of the $\Delta$ will
be extended at ANKE with a much bigger $pp$ phase space than at
SPESIV, using in addition a polarised hydrogen target. However, an
even greater improvement is offered through the use of the
polarised deuterium target, which would allow the studies to be
pursued all the way up to $T_p\approx 3${\ww}GeV. As described in
\S\ref{pn-ice}, the larger missing masses thus accessible would
permit also the study of the spin excitation of higher nucleon
isobars.

Just as for quasi--elastic charge exchange, the excitation of the
$\Delta^0$ can be investigated using just the information gathered
from the silicon counters. However, further information can be
extracted if one measures in ANKE a $\pi^-$ or a proton that comes
from the decay of the $\Delta^0$, \emph{viz} $\vec{p}\,\vec{d}\to
p\,p\,\Delta^0 \to p_s\,p_s\,p_f\,\pi^-_f$, where the slow ($s$)
and fast ($f$) subscripts indicate where the particles would be
detected. In the case of the $\Delta^0$, the angular distribution
of the decay proton or pion in the $\Delta$ rest frame would
determine the alignment of the $\Delta$. In this way we would be
measuring some triple--spin observables, which has never been done
before for $\Delta$ excitation. The decay pion/proton would also
facilitate the separation of the contribution of the
$\Delta(1232)$ from those of the other $\pi^-p$ resonances.

Small angle elastic proton--deuteron scattering is sensitive to
the $I=0$ exchange amplitude, \emph{i.e.}\ the sum of the $pp$ and
$pn$ amplitudes. Measurements here will therefore provide a
qualitatively different check on the phase shifts by removing
single pion exchange from the data set. Both polarised and
unpolarised data can be taken by detecting the recoil deuteron in
the silicon telescopes but, provided that the forward proton does
not emerge at too large an angle, the reaction is more clearly
identified by measuring also the proton in ANKE.

The physics arguments and the practical implementation of these
various programmes, which are listed in
Table~\ref{pn-experiments}, are reviewed in greater depth in the
subsequent subsections.\clearpage

\begin{table}[h]
\caption{Summary of experiments within the proton--neutron
programme. Note that the maximum proton beam energy is almost
3$\:$GeV.} \label{pn-experiments}
\begin{center}
\begin{footnotesize}
\begin{tabular}[c]{|l|l|l|l|}
\hline &&&\\
 Reaction& Primary detectors&Observables&Kinematic ranges \\
&&&\\
\hline%
&&&\\
$\vec{d}\,\vec{p}\to p_sp(n)$&
ANKE&$\vec{n}\,\vec{p}$
elastic scattering&$0.005<|t|<0.1${\ww}(GeV/c)$^2$\\
&Si telescopes&$\frac{\dd\sigma}{\dd\Omega},\:A_y,\:C_{nn},\:C_{ss}$&\\
&&&\\
\hline%
&&&\\
$\vec{p}\,\vec{d}\to p_sp(n)$&Si telescopes,
ANKE&$\vec{p}\,\vec{n}$
elastic scattering&$4^{\circ}<\theta_p^{lab}<11^{\circ}$\\
&&$\frac{\dd\sigma}{\dd\Omega},\:A_y,\:C_{nn},\:C_{ss}$&\\
&&&\\
\hline%
&&&\\
$\vec{d}\,\vec{p}\to pp(n)$&ANKE&$\vec{n}\,\vec{p}$
charge--exchange&$T_d<2.3\:$GeV \\
&&scattering&($T_p<1.15\:$GeV)\\
&&$\frac{\dd\sigma}{\dd\Omega},\:A_y,\:C_{nn},\:C_{ss}$&$\theta_p^{lab}<7^{\circ}$\\
&&&\\
\hline%
&&&\\
$\vec{p}\,\vec{d}\to pp(n)$&Si telescopes&$\vec{p}\,\vec{n}$
charge--exchange&$1.0<T_p<2.8${\ww}GeV\\
&&scattering&$|t|<0.25\:$(GeV/c)$^2$\\
&&$\frac{\dd\sigma}{\dd\Omega},\:A_y,\:C_{nn},\:C_{ss}$&\\
&&&\\
\hline%
&&&\\
$\vec{p}\,\vec{d}\to pp(\Delta^0)$&Si telescopes&$\vec{p}\,\vec{n}\to
\Delta^0p$&$0.01<|t|<0.25\:$(GeV/c)$^2$\\
&&$\frac{\dd\sigma}{\dd\Omega},\:A_y,\:C_{nn},\:C_{ss}$&\\
&&&\\
\hline%
&&&\\
$\vec{p}\,\vec{d}\to ppp\pi^-$&Si telescopes, ANKE&
$\vec{p}\,\vec{n}\to \Delta^0p$&$\theta_{p}^{lab}<12^{\circ}$\\
&&$\frac{\dd\sigma}{\dd\Omega},\:A_y,\:C_{nn},\:C_{ss}$&$|t|>0.01\:$(GeV/c)$^2$\\
&&$A(0s;nn)$&\\
&&&\\
\hline &&&\\
$\vec{d}\,\vec{p}\to dp$&ANKE&
$\frac{\dd\sigma}{\dd\Omega},\:A_y,\:C_{nn},\:C_{ss}$
&$4^{\circ}<\theta_d^{lab}<11^{\circ}$\\
&Si telescopes, ANKE&&$T_d<2.3${\ww}GeV\\
&&&\\
\hline &&&\\
&&&$0.5<T_p<2.8\:$GeV\\
$\vec{p}\,\vec{d}\to pd$&Si
telescopes&$\frac{\dd\sigma}{\dd\Omega},\:A_y,\:C_{nn},\:C_{ss}$
&$0.06<|t|<0.46\:$(GeV/c)$^2$\\
&Si telescopes, ANKE&&$|\theta_{p}^{lab}|<12^{\circ}$\\
&&&\\
\hline%
\end{tabular}
\end{footnotesize}
\end{center}
\end{table}

\clearpage
%
%
\subsection{Proton--neutron small angle elastic scattering}
\label{pn-elastic}%

\begin{figure}[hbt]
\begin{center}
\subfigure[All events]{\epsfig{file=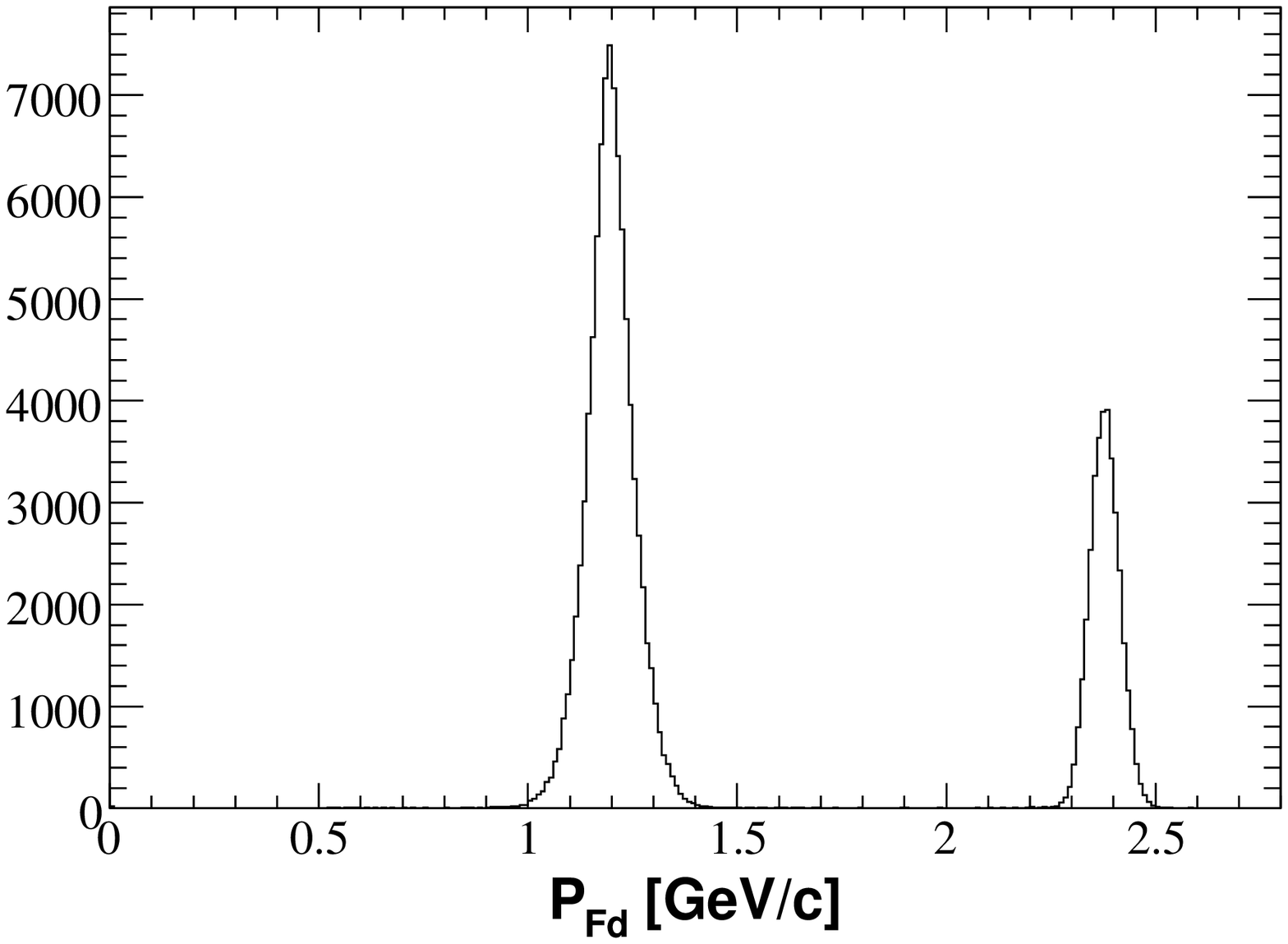,height=5cm}}%
\subfigure[Events that have a momentum of within $\pm 0.3${\ww}GeV/c
of half the beam momentum and survive the $\theta_{sp}$ \emph{vs}
$\theta_{fd}$ cut to enhance the $pp$ quasi--elastic reaction. ]
{\epsfig{file=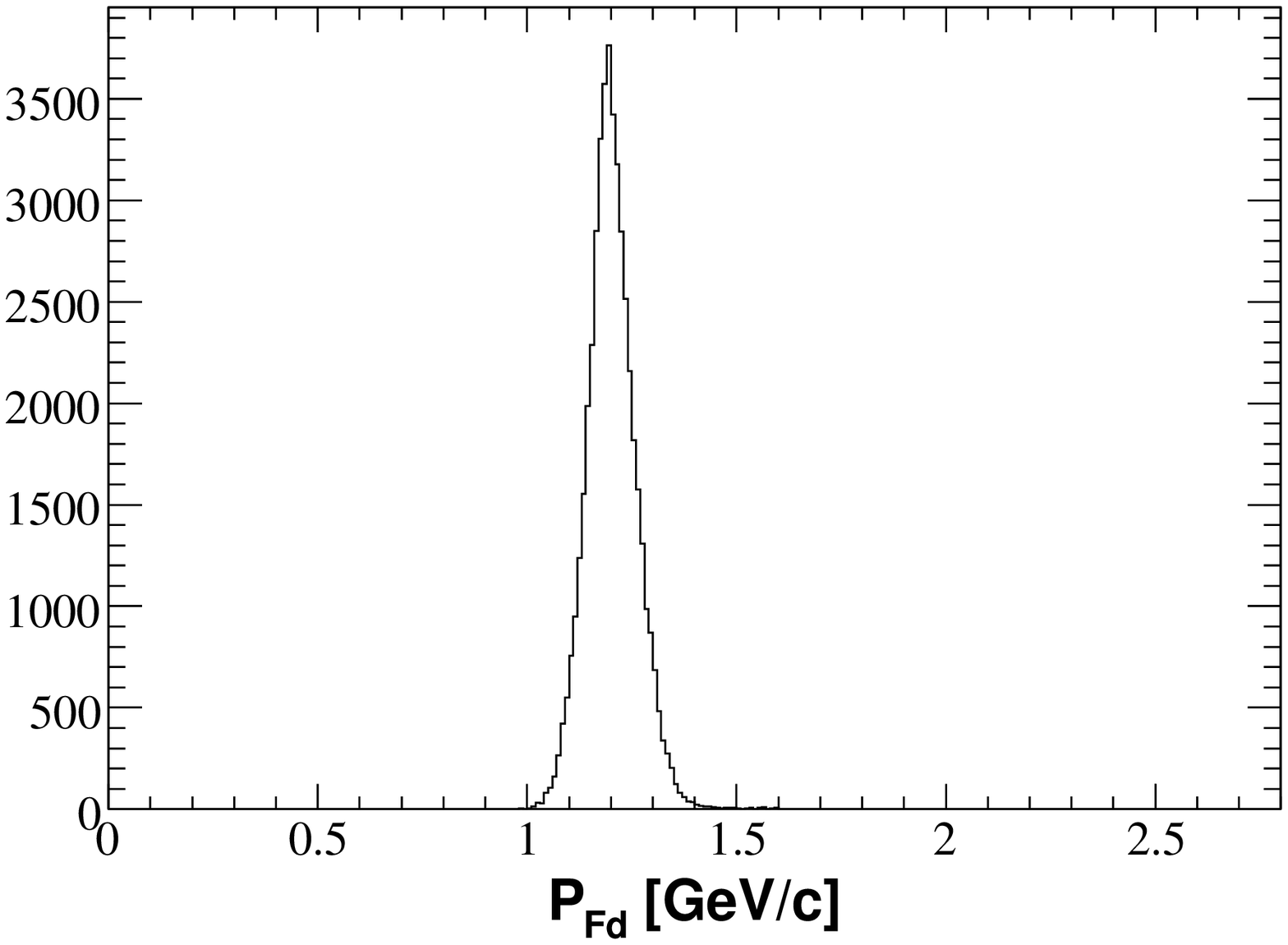,height=5cm}}%
\caption{Momentum distribution of forward-going particles for the
Forward--Spectator coincidence trigger.\label{FdMom}}
\end{center}
\end{figure}

To illustrate how proton--neutron elastic scattering can be
studied in the small angle region through the combination of the
silicon telescopes and the ANKE magnetic analysis, consider the
case of a deuteron beam. In Fig.~\ref{fig:p_distrubition} we
showed the momentum distribution of charged particles arising from
the interaction of 2.4{\ww}GeV/c deuterons with a hydrogen target on a
logarithmic scale. This yields only two significant peaks. The
first around 2.4{\ww}GeV/c corresponds to small angle $dp$ elastic
scattering whereas the second, close to half the beam momentum,
arises from deuteron break--up induced by small angle $pp$ and $np$ scattering.
A detailed investigation of the break--up results benefits from
information from the silicon telescopes described in \S\ref{SST}.
The subset of events of Fig.~\ref{fig:p_distrubition} where a slow
proton was detected in coincidence in the telescope is
presented in Fig.~\ref{FdMom}a on a linear scale. The elastic $dp$
peak is easily eliminated by requiring that the fast particle has
a momentum between $0.9$ and $1.5${\ww}GeV/c.
These data have as yet been the subject only of a very preliminary
analysis~\cite{Mussgiller} and the large corrections arising from
final--state--interactions have still to be fully implemented.
%
%
%
%

At large momentum transfers, where one ``knows'' which particles have
taken part in the collision, the cross section is basically the sum of
that on the proton and neutron separately with the other particle
being a \emph{spectator}. We first discuss the data in this limiting
(classical) picture and return later to the small $q$ region, where quantum
mechanical interferences between scattering by the proton and neutron
play crucial roles.

\begin{figure}
\begin{center}
\epsfig{file=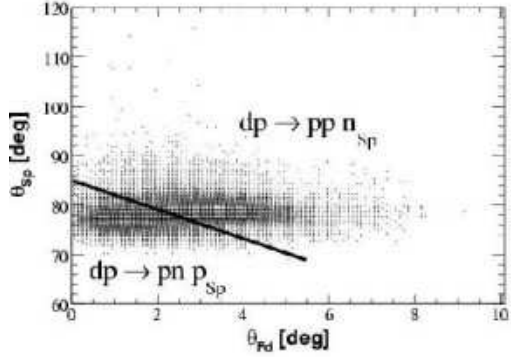,height=5cm}
\caption{Reconstructed value of the $\theta_{sp}$ angle of the
proton in the spectator detector \emph{vs} the $\theta_{fd}$ of
the proton in the forward detector. The solid line indicates the
cut chosen to separate the $pn$ and $pp$ quasi--elastic channels.
\label{ThetaVsTheta_Cut}}
\end{center}
\end{figure}

The separation of $pp$ from $pn$ quasi--elastic scattering in the
classical picture requires us to study the
correlation of the polar angles in the forward detector
($\theta_{fd}$) and the spectator counters ($\theta_{sp}$) shown
in Fig.~\ref{ThetaVsTheta_Cut}. Now for elastic $pp$ scattering at
a beam energy $T_p$ these two angles are related by
\begin{equation}
\label{angles}%
\tan(90^{\circ}-\theta_{sp}) =\left(1+\frac{T_p}{2M_p}\right)
\tan\theta_{fd}\:.
\end{equation}
Though this relation is shifted slightly by the deuteron binding
energy, and smeared significantly by the deuteron Fermi momentum,
after taking the counter geometry into account it suggests that
the majority of events to the right of the solid line in
Fig.~\ref{ThetaVsTheta_Cut} corresponds to quasi--elastic $pp$
scattering whereas those to the left arise dominantly from $np$.
Events that survive both the momentum and the $pp$ polar angle
cuts are illustrated in Fig.~\ref{FdMom}b.
%
%
%
%

\begin{figure}[!hbt]
\begin{center}
\subfigure[Events surviving the momentum cut]
{\epsfig{file=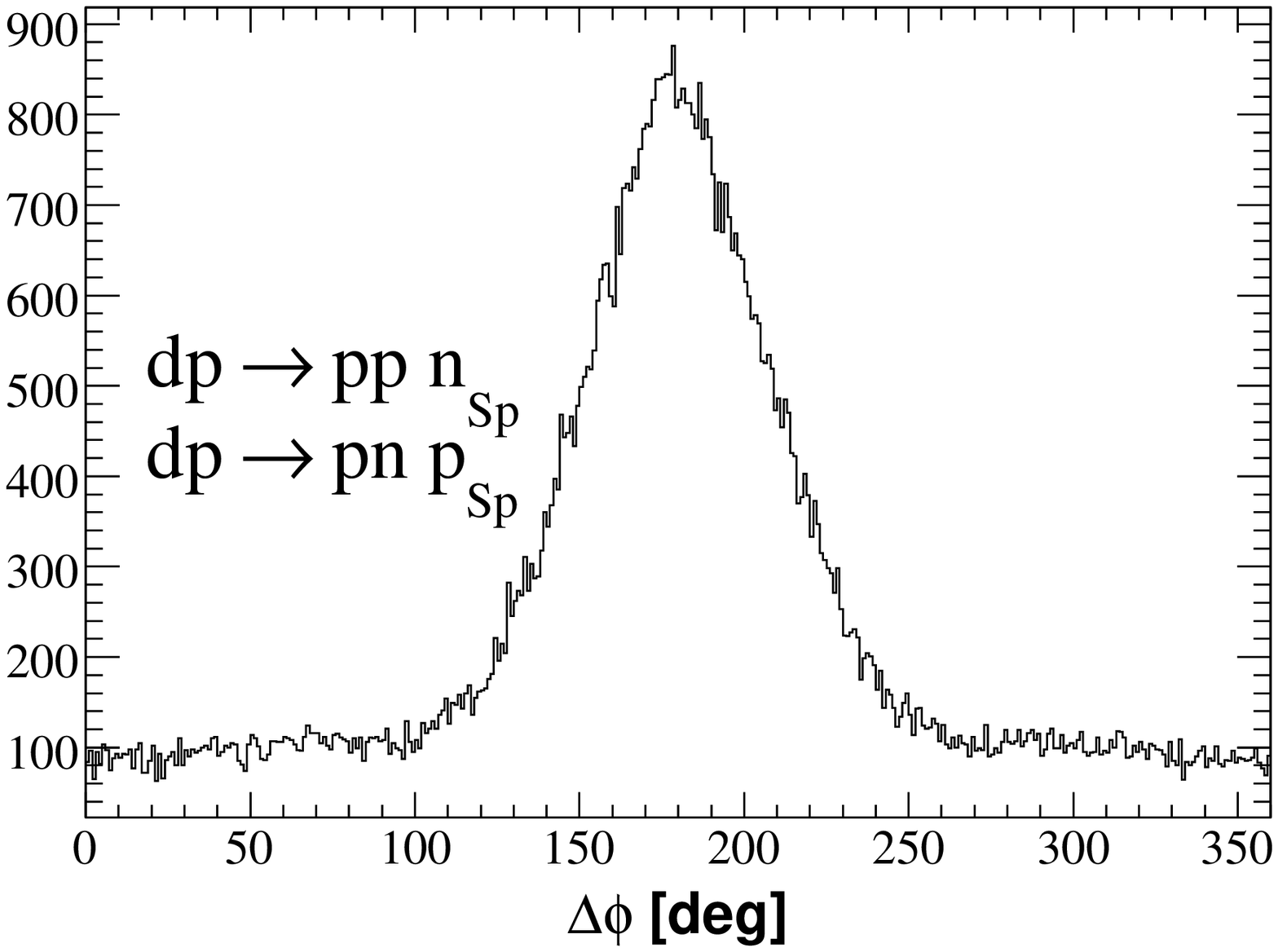,height=5cm}
}%
\subfigure[Events surviving both the momentum and $\theta_{sp}$
\emph{vs} $\theta_{fd}$ cut.]
{\epsfig{file=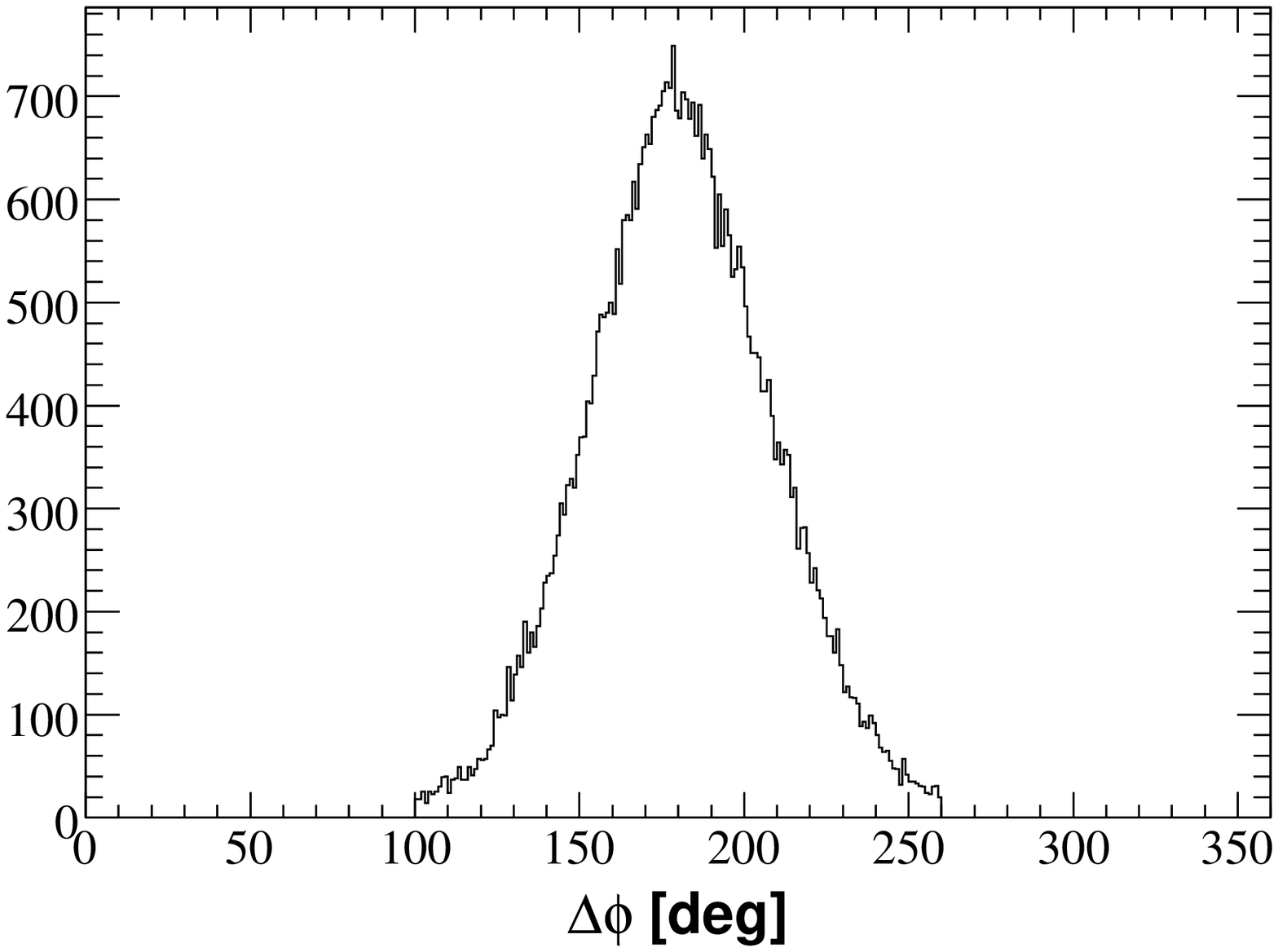,height=5cm}}%
\caption{Difference in $\phi$ of the forward--going proton and the
proton in the spectator detector. The peak around $180^{\circ}$
originates from the $pp$ quasi--elastic channel and the flat part
reflects mainly the $pn$ elastic channel.\label{DeltaPhi_Cut}}
\end{center}
\end{figure}

The corresponding azimuthal angles should also be correlated since
for elastic $pp$ scattering one has
\begin{equation}
\label{azimuths}%
\phi_{sp} = \phi_{fd}-180^{\circ}\:.
\end{equation}
The azimuthal correlation is illustrated in
Fig.~\ref{DeltaPhi_Cut}a for events where the only selection is
that coming from the momentum cut. Since for events where the
proton is the spectator there should be essentially no azimuthal
correlation, the $pp$ quasi--elastic peak sits on a relatively
flat background. This background is almost completely suppressed
in Fig.~\ref{DeltaPhi_Cut}b by the imposition of the polar angle
cut shown in Fig.~\ref{ThetaVsTheta_Cut}.
%
%
%
%
%
\begin{figure}[!hbt]
\begin{center}
\epsfig{file=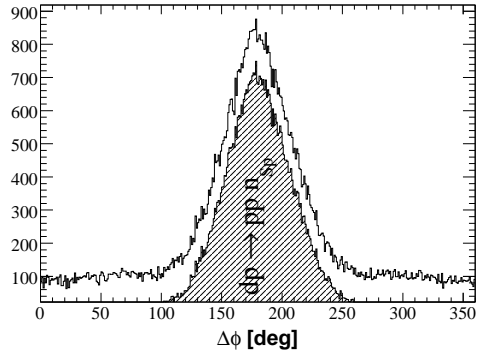,height=5cm}
\caption{Difference in $\phi$ of the protons in the forward and
spectator detectors after applying just the momentum cut and after
applying the momentum and $\theta_{sp}$ \emph{vs} $\theta_{fd}$
cuts. The shaded peak represents mainly the $pp$ quasi--elastic
events that have been selected by both cuts.
\label{DeltaPhi_Cut+DeltaPhi_Cut2}}
\end{center}
\end{figure}

\clearpage

The comparison of the $\phi$--correlation spectrum with and
without the polar angular cut is presented in
Fig.~\ref{DeltaPhi_Cut+DeltaPhi_Cut2}. This shows a very clean
peak which dominantly contains quasi--free $pp\to pp$ events,
though it would take a Monte Carlo simulation to try to estimate the
neutron contamination. Since the kinematics of each event have
been fully identified, one simple consistency test in this classical
picture would be to
investigate the angular correlations between the slow proton and
fast neutron to see if one obtains the same classification of
events. This analysis has demonstrated that the
silicon telescopes can function well in coincidence with the ANKE
magnetic system and that clean data can be obtained in this way.

However, in reality, at low momentum transfers it is not possible
even in principle to separate completely the $pp$ from the $pn$
interactions in $dp$ collisions. A naive identification of the
slower particle in the deuteron rest frame with the
\emph{spectator} quickly leads to
inconsistencies~\cite{Aladashvili1}. For small values of $t$ there
are coherent effects associated with the addition of the $pp$ and
$pn$ amplitudes. Furthermore, much of the transition strength is
actually soaked up by the elastic deuteron--proton channel. Such
effects are not essentially different in nature from those studied
extensively in low momentum transfer deuteron--proton charge
exchange~\cite{Bugg-Wilkin,Carbonell} and, provided that the $pp$
amplitudes are known, the corrections in the present case depend
primarily on the low energy $pn$ final state interaction. Such
corrections will be introduced into the analysis of future data
taken with the more advanced telescope system with a larger solid
angle coverage.

\begin{figure}[htb]
\begin{center}
\centerline{\epsfxsize=3in\rotatebox{-90}{\epsfbox{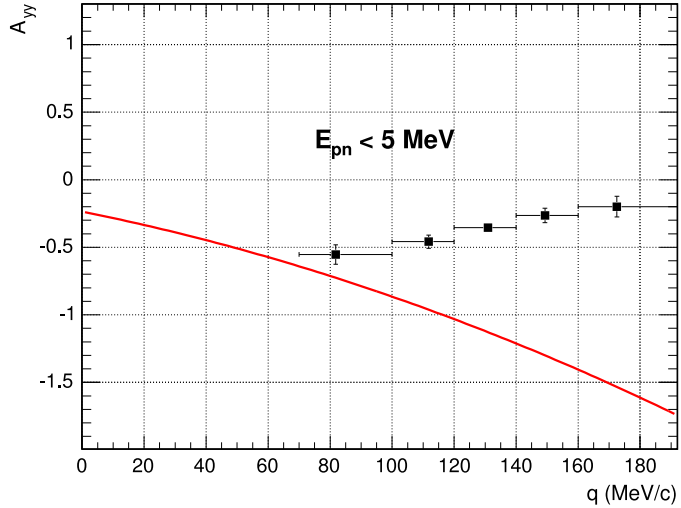}}}
    \caption{$A_{yy}$ of the $\vec{d}p\to (pn)p$ reaction at
    $T_d=1170\,$MeV for events where $E_{pn}<5\,$MeV in bins of
    momentum transfer $q$. Also shown is the interpolation of the
    analysing power of the charge--exchange data of Fig.~\ref{Ann}
    where account has been taken of the dilution caused by the finite
    $\phi$ acceptance. Both curve and points are subject to a common
    overall normalisation uncertainty of 4\% arising from the beam
    polarisation.
 \label{Ayy_pn}}
\end{center}
\end{figure}

The classical picture fails most spectacularly when both the momentum
transfer and the excitation energy in the final $pn$ system is
small. In the quasi--free regime there can be no significant
dependence of the $\vec{d}p\to (pn)p$ counting rate on the tensor
polarisation of the deuteron beam and any such signal would reflect
the presence of \emph{two} nucleons in the deuteron beam. Preliminary
values of $A_{yy}$ for the $\vec{d}p\to (pn)p$ reaction with
$E_{pn}\leq 5\,$MeV are shown in Fig.~\ref{Ayy_pn}. The
signal is large and negative, though this decreases in strength as the
cut on $E_{pn}$ is relaxed while the vector analysing power increases
in this limit.

In the figure we show also a parameterisation of the $p(\vec{d},2p)n$
tensor analysing power $A_{yy}$ of Fig.~\ref{Ann}, where account has
been taken of the signal dilution due to the finite azimuthal
acceptance in the $(pn)$ case. At the smallest momentum transfer
($q\approx 80\,$MeV/c), $A_{yy}(\vec{d}p\to \{pn\}p)
\approx A_{yy}(\vec{d}p\to \{pp\}n)$, though the
values diverge as $q$ is increased. This is not an accident! If we
neglect the deuteron $D$--state then in impulse approximation at $q=0$
the only allowed transition in the $\vec{d}p\to (pn)p$ reaction
is $^3\!S_1\to\, ^1\!S_0$. This has a $(\Delta S,\,\Delta I,\,\Delta
I_z)=(1,\,1,\,0)$ character and is just
the isobaric analogue of the deuteron charge--exchange reaction
discussed in \S\ref{ece}. Furthermore, the $(0,\,0,\,0)$ transitions,
driven by the large isoscalar spin--non--flip $NN$ amplitudes, vanish
like $q^4$ at small $q$. This is because they correspond to final $^3\!D_1$ or
higher $S$--waves that are orthogonal to the deuteron wave
function. The only possible source of dilution of the $A_{yy}$ signal to
order $q^2$ arises therefore from the $^3\!S_1\to\, ^3\!P_{0,1,2}$
transitions, which also involve an isospin flip. The final transition
to this order is $^3\!S_1\to\, ^1\!P_{1}$, which is isoscalar.

This picture
will, of course, have to be modified somewhat to take into account
effects arising from the deuteron $D$--state. However, the basic
suppression of the scalar--isoscalar amplitude at small $q$ remains and this does explain
qualitatively our findings that $A_{yy}(\vec{d}p\to \{pn\}p)$ looks like
a diluted charge--exchange signal and that $A_y$, which should vanish
for the $^1\!S_0$ state~\cite{Bugg-Wilkin}, increases for larger $q$
and $E_{pn}$ through the excitation of the $^3\!P_{0,1,2}$ system.

%
%
\subsection{Proton--neutron elastic charge exchange}
\label{ece}
\addtocounter{figure}{1}

\hspace{-0.7cm}
\begin{minipage}[t]{7.2cm}
\hspace{0.5mm} \baselineskip 3ex The ANKE collaboration is making
measurements of the $\vec{d}p\to ppn$ reaction with the aim of
extracting spin--dependent $np$ charge--exchange
amplitudes~\cite{PROPOSAL,Chiladze}.

\hspace{0.5cm} Now the most complete investigation of deuteron
charge exchange in the COSY energy regime was carried out at
Saclay at 1.6 and 2.0{\ww}GeV~\cite{Ellegaard1,Sams95a} and the
results for the hydrogen and deuterium targets are shown in
Fig.~\arabic{figure}. Because of uncertainties in the acceptance
of the SPESIV spectrometer used in the experiment to detect the
pairs of protons, the overall cross section normalisation is
arbitrary, though the relative strength between deuterium and
hydrogen targets away from the forward direction is $0.68\pm0.04$.
At $q\approx 0$, it is reduced by a further factor of about $2/3$
due to the Pauli blocking the final $nn$ system.

\begin{center}
\rule{2cm}{0.3mm}\end{center}
\baselineskip 3ex Fig.~\arabic{figure}: Cross section and
Cartesian and spherical tensor analysing powers of the
$(\vec{d},2p)$ reaction on hydrogen and deuterium at
$T_p=1.6${\ww}GeV~\cite{Sams95a}. The broken curves represent
plane--wave predictions whereas the solid ones include eikonal
corrections~\cite{Glauber}. The overall cross section
normalisation was chosen to agree with theory at a momentum
transfer of $q=0.7${\ww}fm$^{-1}$.
\end{minipage}
\hfill
\begin{minipage}[t]{7.3cm}
\vspace*{-0.7cm}

\noindent
\input epsf
\begin{center}
\mbox{\epsfxsize=7cm \epsfbox{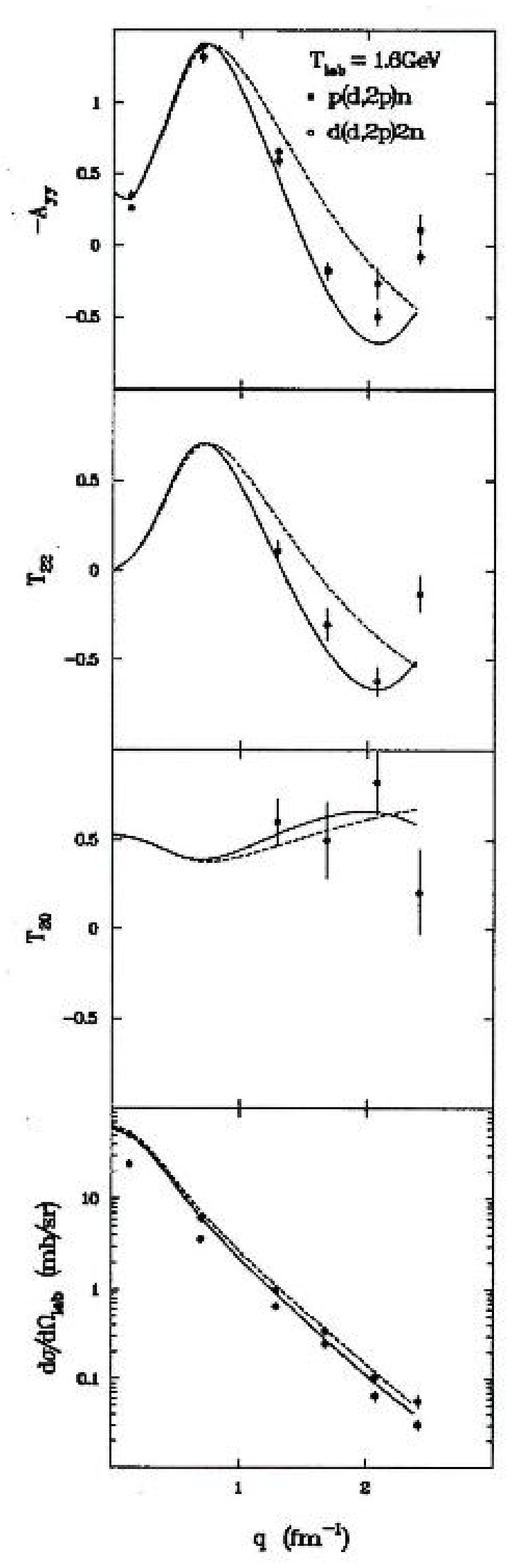}}
\end{center}
\end{minipage}
\baselineskip 3ex
\newpage
To show the basic sensitivity of these measurements, consider
neutron--proton charge--exchange amplitudes in the cm system:
\begin{equation}
\label{fpn} f_{np}=\alpha +i\gamma
(\bfg\sigma_{n}+\bfg\sigma_{p}){\bf n} +\beta (\bfg\sigma_{n}
\cdot {\bf n})(\bfg\sigma_{p} \cdot {\bf n})+ \delta
(\bfg\sigma_{n} \cdot {\bf m})(\bfg\sigma_{p} \cdot {\bf m})
 +\varepsilon (\bfg\sigma_{n} \cdot {\bf l})(\bfg\sigma_{p} \cdot {\bf l}),
\end{equation}
where $\bfg\sigma_{n} $ and $\bfg\sigma_{p}$ are the Pauli
matrices for neutron and proton, respectively. The orthogonal unit
vectors are defined in terms of the initial (${\bf k}$) and final
(${\bf k}'$) momenta as
\[
\mathbf{
 n=\frac{\mathbf{k}\times\mathbf{k'}}{|\mathbf{k}\times\mathbf{k'}|},~~
 m=\frac{\mathbf{k'}-\mathbf{k}}{|\mathbf{k'}-\mathbf{k}|},~~
 l=\frac{\mathbf{k'}+\mathbf{k}}{|\mathbf{k'}+\mathbf{k}|}\,.}
\]
The amplitudes are normalised such that the elementary $np\to pn$
differential cross section has the form
\begin{equation}
\label{secpn} \left(\frac{d\sigma}{d q^2}\right)_{\!\!np\to
pn}=I_{np}= |\alpha|^{2}+|\beta|^{2}+2|\gamma|^{2}
+|\delta|^{2}+|\varepsilon|^{2}.
\end{equation}

For low excitation energy $E_{pp}< 3${\ww}MeV of the final $pp$ pair,
and at low momentum transfer, the charge exchange reaction $dp\to
(pp)n$ mainly excites the $^1S_0$ state of the final $pp$ system,
which involves a spin--flip from $np$ triplet to $pp$ singlet. The
process therefore provides a \emph{spin--filter}. In
single--scattering approximation, the resulting amplitude depends
only upon the spin--dependent parts of $f_{np}$, \emph{i.e.}\
$\beta$, $\delta$ and $\varepsilon$. Bugg and
Wilkin~\cite{Bugg-Wilkin} have shown that, under these conditions,
there are two form factor describing the transition from the
deuteron to the $^1S_0$ $pp$ state:
\begin{eqnarray}
\nonumber
S^{+}(k, \half q)&=&\langle\psi_{k}^{(-)}|j_{0}(\half
qr)|u\rangle +\sqrt{2}\langle\psi_{k}^{(-)}|j_{2}(\half
qr)|w\rangle\,,\\
S^{-}(k, \half q)&=&\langle\psi_{k}^{(-)}|j_{0}(\half qr)|u\rangle
-\langle\psi_{k}^{(-)}|j_{2}(\half qr)|w\rangle/\sqrt{2},
\end{eqnarray}
where $u(r)$ and $w(r)$ are the $S$ and $D$ components of the
deuteron wave function and $\psi^{(-)}_k(r)$ is the $pp$
$(^1\!S_0)$ scattering wave function. Here $k$ is the $pp$
relative momentum, corresponding to an excitation energy
$E_{pp}=k^2/M$, where $M$ is the proton mass. Denoting the ratio
of the transition form factors by $R=S^{+}(k,\half
q)/S^{-}(k,\half q)$ and the sum of squared amplitudes
\begin{equation}
\label{idce} I=|\beta|^{2}+|\gamma|^{2}+|\varepsilon
|^{2}+|\delta|^{2}R^{2},
\end{equation}
the differential cross section, tensor analysing powers, and
transverse spin--spin correlation parameters of the $dp\to
(pp)_{^1S_0}n$ reaction take the forms~\cite{Bugg-Wilkin, barbaro}
\begin{eqnarray}
\nonumber
\frac{d^{4}\sigma}{dq^{2}d^{3}k}&=&\fmn{1}{3}I\left\{S^{-}(k,\half
q) \right\}^2,\\
\nonumber I\,
T_{20}&=&\fmn{1}{\sqrt{2}}\{|\gamma|^{2}+|\beta|^{2}+
|\delta|^{2}R^{2}-2|\varepsilon|^{2}\}\\
\nonumber I\,
T_{22}&=&\fmn{\sqrt{3}}{2}\{|\gamma|^{2}+|\beta|^{2}-
|\delta|^{2}R^{2}\}\\
\nonumber
I\, C_{x,x}&=&-2\textrm{Re}(\beta^* \varepsilon)\\
I\,C_{y,y}&=&-2\textrm{Re}(\varepsilon^*\delta)R. \label{observables}
\end{eqnarray}

After rotating these formulae for the $T_{2i}$ to the beam
direction~\cite{Ohlsen}, they were used by the Saclay
group~\cite{Sams95a} to interpret their data at 1.6{\ww}GeV. The only
significant correction comes from the multiple scatterings whose
effects increase steadily with momentum
transfer~\cite{Bugg-Wilkin}.

However, it is important to stress that data at larger $pp$
excitation energies also contain valuable information on the $pn$
charge--exchange amplitudes~\cite{Aladashvili2}, but for this a
detector of much larger acceptance than SPESIV is required. This
was provided at low beam energies by the EMRIC device, which
determined the cross section, tensor, and vector analysing powers
at 200 and 350{\ww}MeV~\cite{Kox}. These agreed well with impulse
approximation estimations~\cite{Carbonell} and provided the basis
for the design of the POLDER polarimeter~\cite{POLDER}. This has
been used very successfully in the determination of the
polarisation of the recoil deuteron in elastic electron--deuteron
scattering at J--Lab, which allows the separation of the deuteron
form factors~\cite{JLab}.

By design, the above $(\vec{d},2p)$ experiments were carried out
at energies where the $np\to pn$ amplitudes are relatively well
known~\cite{SAID}. The aim of COSY proposal 125~\cite{PROPOSAL} is
to carry out such measurements at energies where the $np$ data
base is far less complete, in particular above the Los Alamos
energy of 800{\ww}MeV per nucleon. Furthermore, by using polarised
beam and targets, one can gain access also to spin--correlation
parameters, which contain valuable relative phase information.
Such experiments can be carried out using a polarised deuteron
beam, as is currently being employed~\cite{PROPOSAL}, or a
polarised deuterium target and we now compare the merits of the
two approaches.
%
%
\subsubsection{Charge--exchange with a polarised deuteron beam}
\label{np-ce}

An initial measurement of the deuteron--induced charge--exchange
reaction was carried out at the ANKE spectrometer using a
polarised deuteron beam at $p_d=2400${\ww}MeV/c
($T_d=1170${\ww}MeV)~\cite{PROPOSAL}. Two fast protons, emitted in a
narrow forward cone with momenta around half that of the deuteron
beam, were detected by the Forward Detector (FD) system of the
ANKE set--up (see fig~\ref{accep}).

\begin{figure}[htb]
\begin{center}
\centerline{\epsfxsize=15cm{\epsfbox{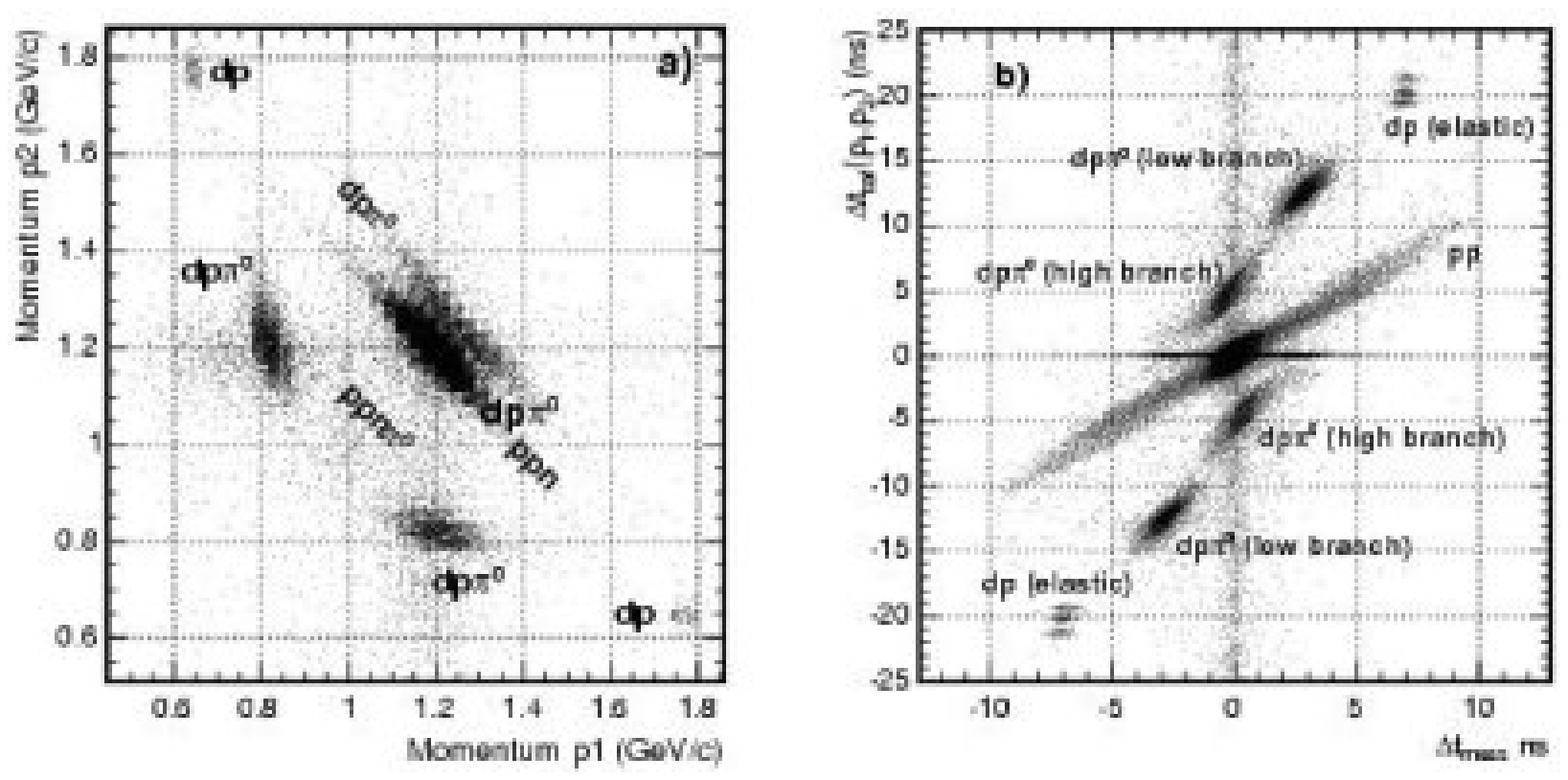}}}
\end{center}
\caption{\label{identification} a: Correlation of the momenta of
two charged tracks in ANKE resulting from the interaction of
1.17{\ww}GeV deuterons in a hydrogen target. b: Correlation of the
time differences $\Delta t_{meas}$ and $\Delta t_{tof}$.}
\end{figure}

The first step in processing the $dp$ charge--exchange breakup
data is to choose two--track events using the MWPC information.
The momentum vectors were determined with the help of the magnetic
field map of the spectrometer, assuming a point--like source
placed in the centre of a beam--target interaction region.
Fig.~\ref{identification}a displays double--particle events on a
scatter plot of particle momenta $p_1$ \emph{versus} $p_2$. The
smallness of the FD solid angle acceptance leads to a kinematic
correlation for events with two or three particles in the final
state (Fermi motion spreads slightly the correlation for the
corresponding quasi--free processes). No assumption on the masses
of the particles is required for such a correlation. The break--up
events clearly manifest themselves among several processes
recorded in the spectrometer. For the events with particles
hitting different counters in the hodoscope, the correlation of a
measured time difference $\Delta t_{meas}$ with a difference of
the time of flight $\Delta t_{tof}$ can be observed. The time of
flight from the target to the hodoscope is calculated from the
measured momentum assuming that the particle has the mass of the
proton mass. Real proton pair events are then located at the
diagonal of the scatter plot (Fig.~\ref{identification}b).

\begin{figure}[htb]
\begin{center}
\centerline{\epsfxsize=3.5in{\epsfbox{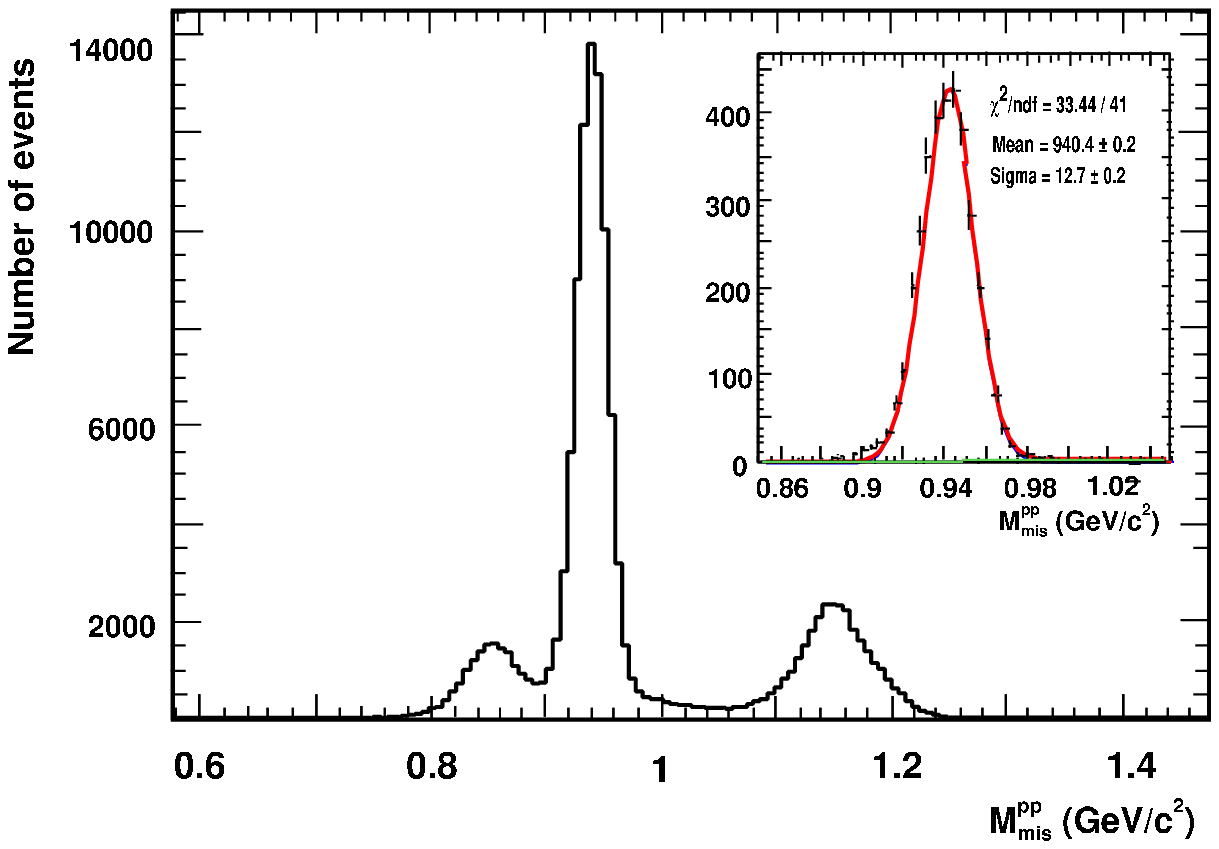}}}
\end{center}
\caption{\label{Mpp_distribution} Missing mass distribution of all observed proton pairs.
        The inset shows the distribution near the neutron mass for
        the pairs selected by the TOF. }
\end{figure}

The charge--exchange process was identified from the missing--mass
with respect to the observed proton pairs (see
Fig.~\ref{Mpp_distribution}) and time difference information. The
spectra for all spin modes reveal a well defined peak at
M$_{miss}$ equal to the neutron mass to within 1\%. The background
was less than $2\%$ and stable, so that the charge--exchange
process could be reliably identified.

\begin{figure}[!htb]
\begin{center}
\subfigure[Moduli of the two independent $np\to pn$ scattering
amplitudes at $t=0$.]
{\epsfig{file=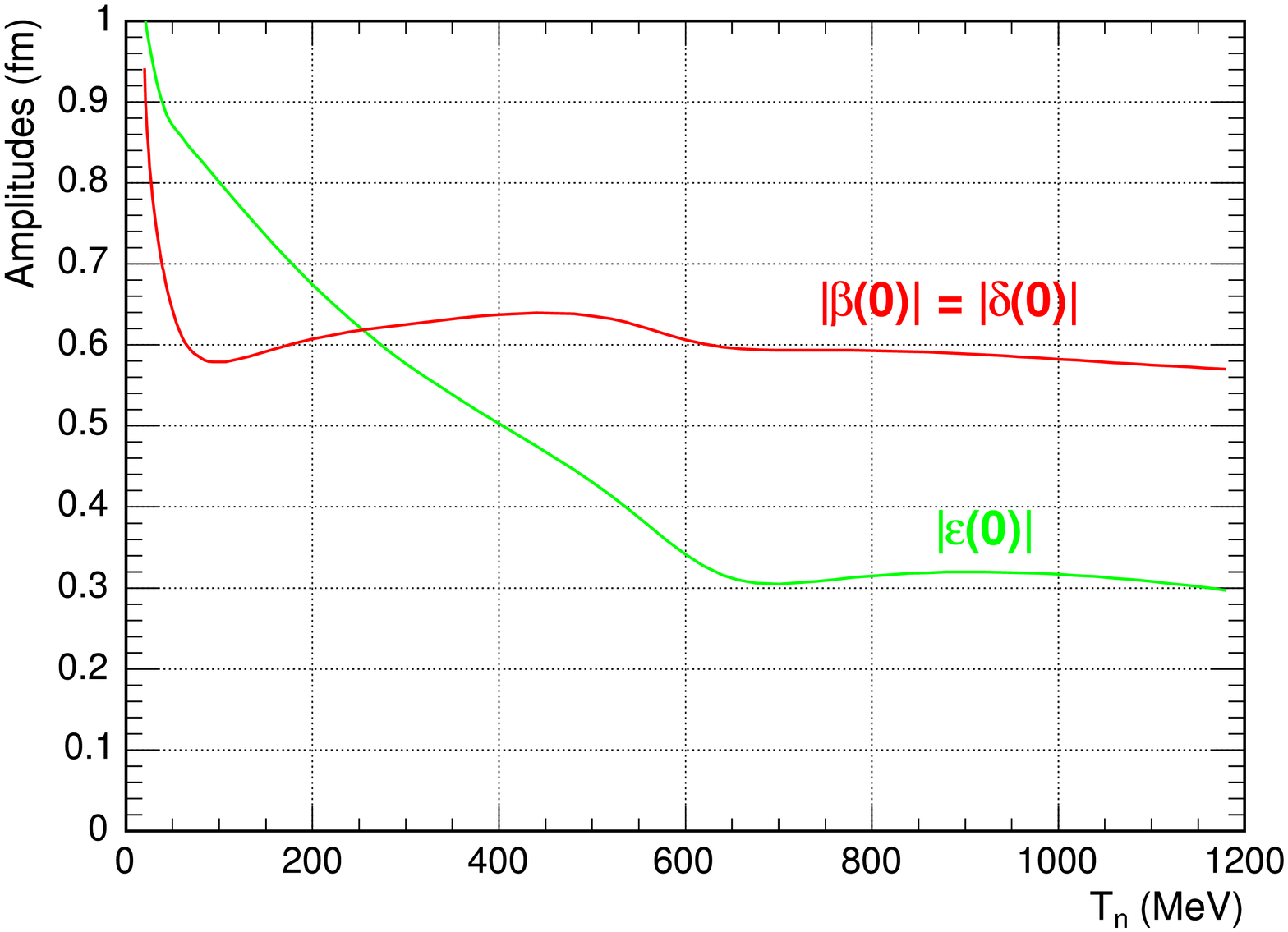,width=5cm,angle=-90} \label{zamps}}
\subfigure[$T_{20}$ for $\vec{d}p\to(pp)_{^1\!S_0}n$ in impulse
approximation compared to our value of $T_{20}=0.39\pm 0.04 $ at
$\frac{1}{2}T_d = 585${\ww}MeV.]
{\epsfig{file=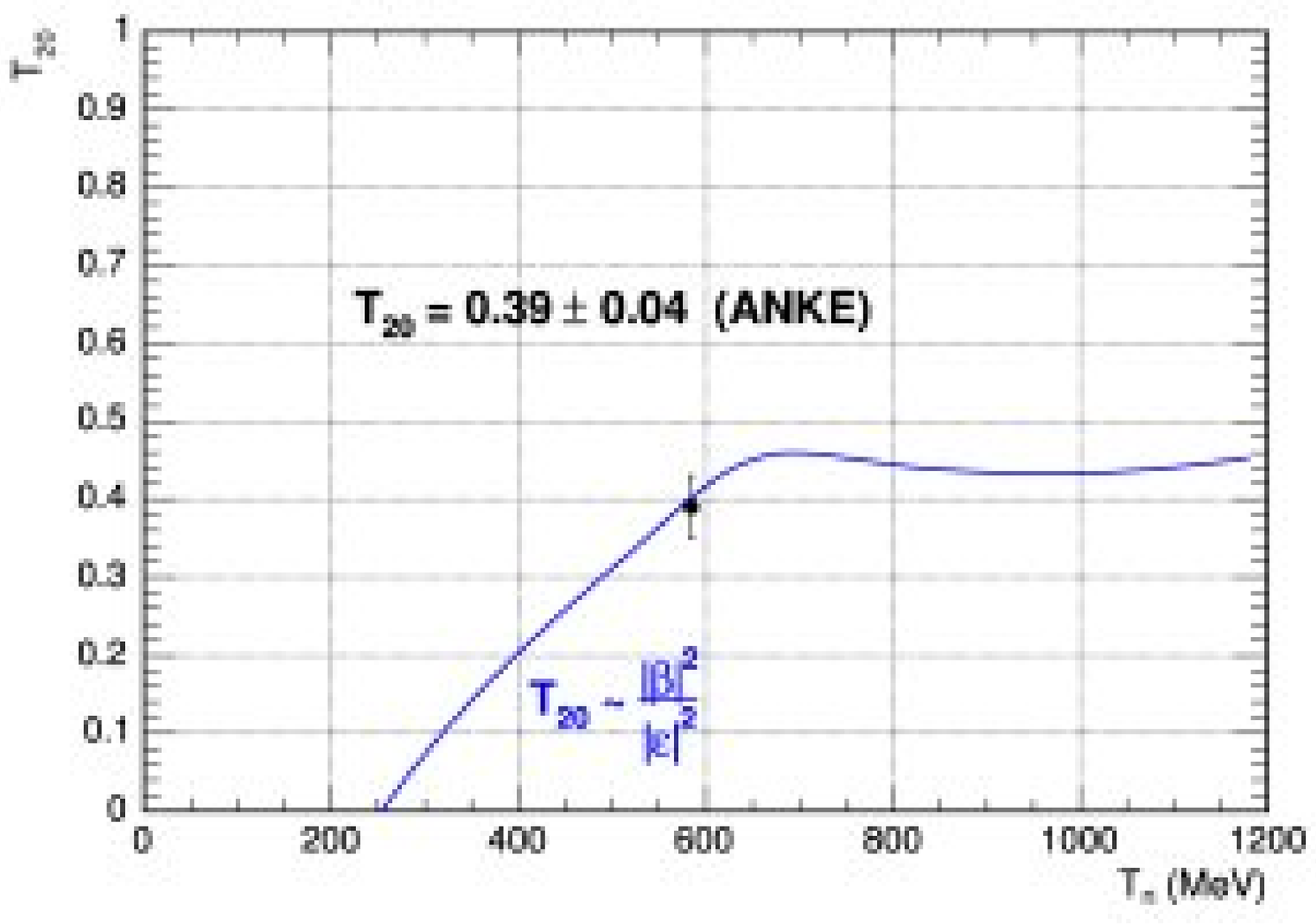,width=5cm,angle=-90} \label{zT20}}
\end{center}
\caption{Predictions for amplitudes and observables deduced using
the SAID program~\cite{SAID}.} \label{fig:as}
\end{figure}


\clearpage

Using the polarised deuteron charge--exchange (CE) break--up
reaction $p(\vec{d},2p)n$, where the final protons have an
excitation energy of less than 3{\ww}MeV and hence are in the
$^1\textrm{S}_0$ state, we can access the spin--dependent
amplitudes of the elementary $np$ elastic scattering using
Eq.~(\ref{observables}). For collinear kinematics we can directly
reconstruct the magnitude of the two spin amplitudes by measuring
the cross section and $T_{20}$ analysing power. The value of the
ratio $|\beta(0)|/|\varepsilon(0)|=1.86\pm0.15$, obtained from our
preliminary experiment, is shown in Fig.~\ref{fig:as}. Also shown
are predictions for the values of the moduli of the two forward
spin--flip amplitudes, as functions of energy~\cite{SAID}. Since
the SAID prediction of the ratio is
$1.79\pm0.27$~\cite{SAID,Arndt}, it is clear that our statistical
precision is already superior to that of the World data
base.

\begin{figure}[htb]
\begin{center}
\centerline{\epsfxsize=3in\rotatebox{-90}{\epsfbox{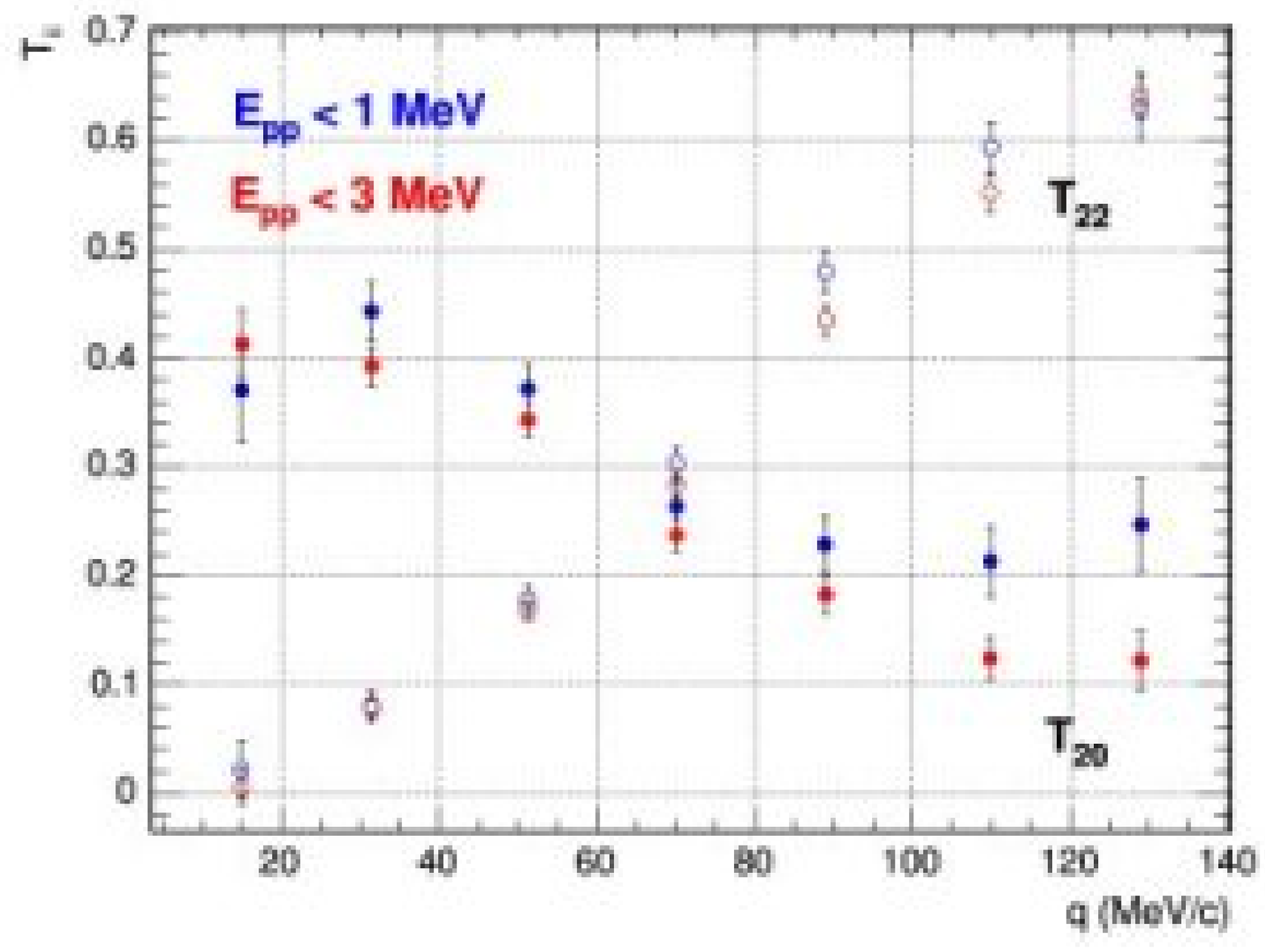}}}
    \caption{Spherical tensor analysing powers $T_{20}$ and $T_{22}$
    of the $p(\vec{d},2p)n$
    reaction at $T_d=1170\,$MeV in 20{\ww}MeV bins of momentum transfer
    $q$ with two different cuts on the excitation energy:
    $E_{pp}<1\,$MeV (blue) and $E_{pp}<3\,$MeV (red).   \label{Tnn}}
\end{center}
\end{figure}

\begin{figure}[htb]
\begin{center}
\centerline{\epsfxsize=3in\rotatebox{-90}{\epsfbox{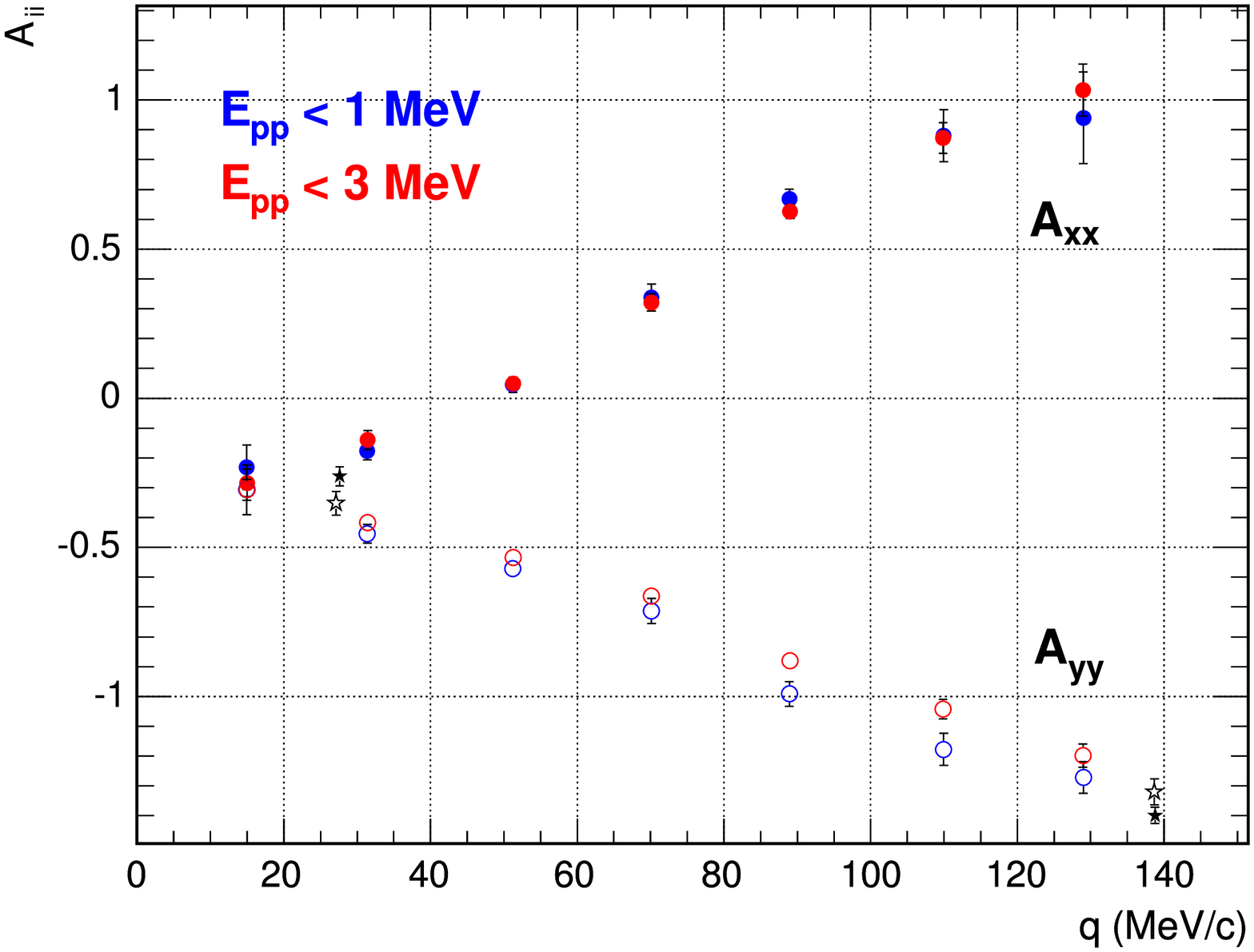}}}
    \caption{Cartesian tensor analysing powers $A_{yy}$ and $A_{xx}$
    of the $p(\vec{d},2p)n$
    reaction at $T_d=1170\,$MeV with cuts as in Fig.~\ref{Tnn}. Also
    shown by stars are the $A_{yy}$ measurements from SATURNE
at 1600$\,$MeV~\cite{Sams95a}.  \label{Ann}}
\end{center}
\end{figure}

Though the angular acceptance for the two fast protons in ANKE is
very limited, it is known that the tensor analysing powers should
change very fast with momentum transfers due to the
near--vanishing of the $\delta$ amplitude for $q\approx
m_{\pi}$~\cite{Bugg-Wilkin}. It is then possible to utilise the
dependence of the signal on the azimuthal angle to extract
separately values of $T_{20}$ and $T_{22}$, or the Cartesian
analysing powers $A_{yy}$ and $A_{xx}$, and preliminary estimates
are to be found in 20{\ww}MeV bins in momentum transfer in
Figs.~\ref{Tnn} and \ref{Ann} respectively. If the excitation
energy is not cut sufficiently, these tensor signals should be
diluted slightly at larger $q$ due to contamination from final
spin--triplet states~\cite{Bugg-Wilkin}. As seen from the figures,
the effects of changing the limit on $E_{pp}$ from 1$\,$MeV to
3$\,$MeV is only significant above about 100{\ww}MeV/c. The value
in the final bin is just compatible with the kinematic limit
$A_{xx}\leq 1$. Also illustrated in Fig.~\ref{Ann} are the
$A_{yy}$ results from SATURNE at the rather higher energy of
1600$\,$MeV~\cite{Sams95a}; at these small values of momentum
transfer this group did not have a clean separation of $A_{xx}$
and $A_{yy}$.

The variation with momentum transfer is generally as expected on the
basis of the Bugg--Wilkin model~\cite{Bugg-Wilkin} though the detailed
theoretical calculation~\cite{Carbonell}, as used to describe data at
lower energies~\cite{Kox}, has still to be implemented at our energy.

\clearpage
\subsubsection{Polarimetry reactions}
\label{polarimetry}

To illustrate the power of the spectrometer in identifying
two--body and quasi--two--body reactions, many of which are used
to calibrate the beam polarisation, we show some of the results
obtained in polarised deuteron--proton collisions at
$T_d=1.17${\ww}GeV~\cite{PROPOSAL,Chiladze}

Fig.~\ref{accep} showed the ANKE experimental acceptances for
singly charged particles for different reactions as functions of
the laboratory production angle and magnetic rigidity, together
with the loci representing the kinematics of different allowed
reactions.

\begin{figure}[htb]
\begin{center}
    \psfig{figure=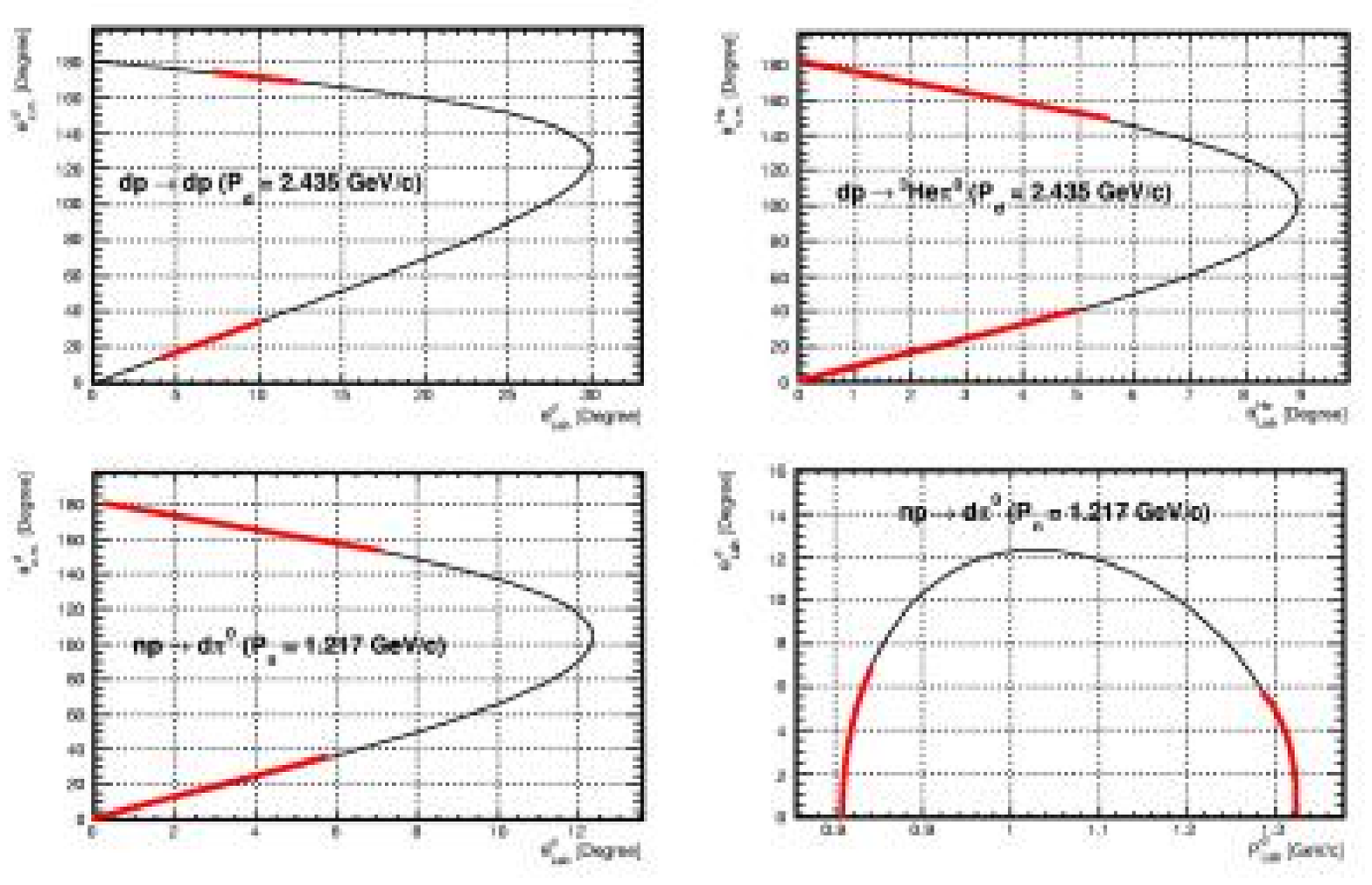,height=9cm,angle=0}
    \psfig{figure=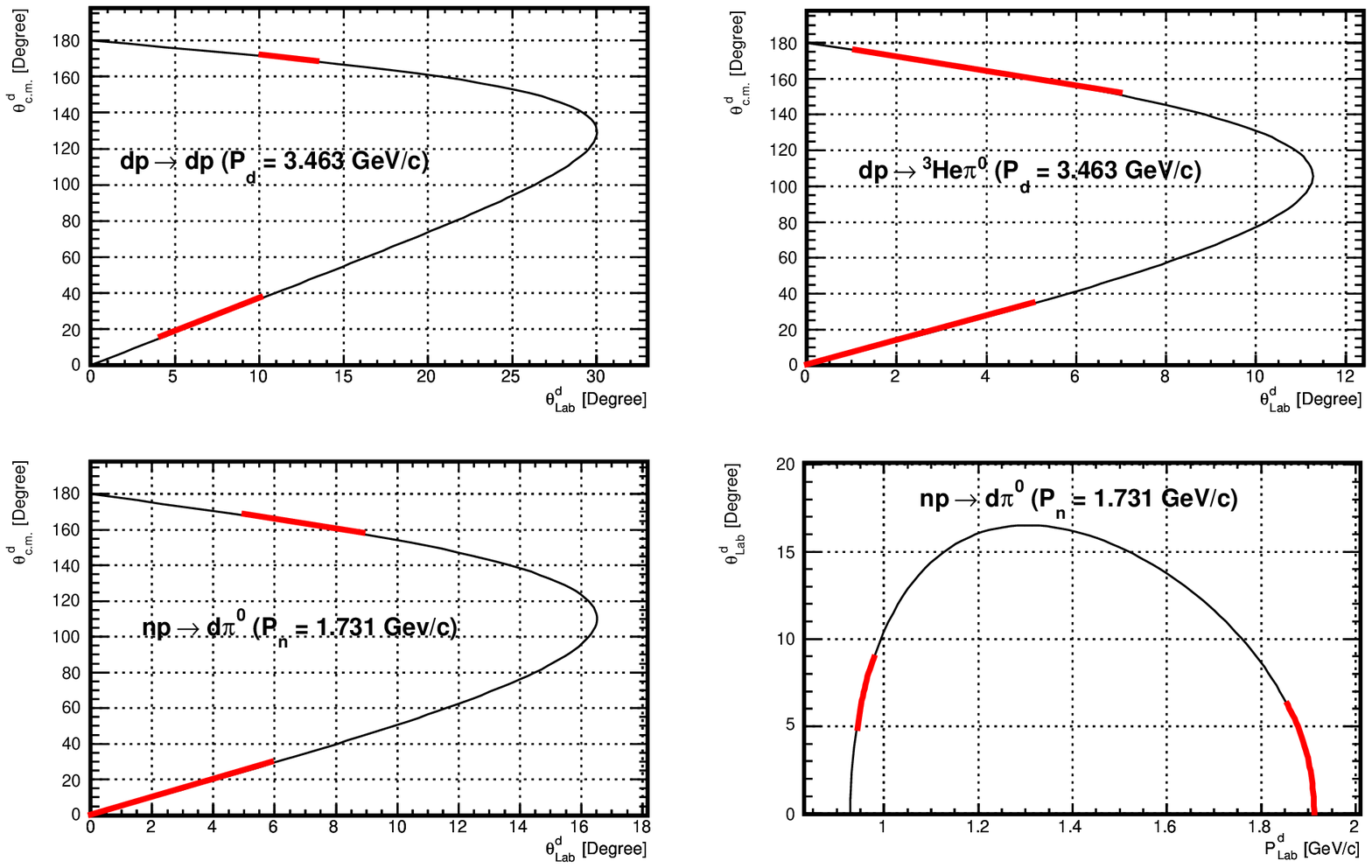,height=9cm,angle=0}
    \caption{Kinematic curves for different reactions at deuteron
    momenta of 2.435 and 3.463{\ww}GeV/c; the bold parts
    of the curves denote regions accessible in the current
    set--up.\label{kinem}}
\end{center}
\end{figure}

To facilitate the subsequent discussion, we show in
Fig.~\ref{kinem} kinematic curves relating the laboratory and cm
angles for three observed two--body reactions, \emph{viz}: $dp\to
dp$, $dp\to\,^3\textrm{He}\,\pi^0$, and $np\to d\pi^0$ at the
momentum of the 2004 run (2.435{\ww}GeV/c)~\cite{Chiladze} and
also at the higher COSY momentum (3.463{\ww}GeV/c). From these it
is seen that the $dp\to dp$ reaction has a significant acceptance
for $4^{\circ}<\theta_{lab}^d<10^{\circ}$, and that this depends
little on the beam momentum.

The quasi--free $np\to d\pi^0$ can be clearly identified in ANKE
through the detection of the two final charged particles in the
$dp\to p_{sp}d\pi^0$ reaction, where $p_{sp}$ is a spectator
proton which has essentially half the beam momentum,
$p_{sp}\approx 1.17${\ww}GeV/c~\cite{Chiladze}. In the
two--dimensional momentum spectrum of Fig.~\ref{identification}
are shown the bands arising from the high ($p_d\approx 1.3${\ww}GeV/c)
and low--momentum ($p_d\approx 0.8${\ww}GeV/c) branches, corresponding
to backward and forward production of the $\pi^0$ in the cm
system.

\begin{figure}[htb]
\begin{center}
    \psfig{figure=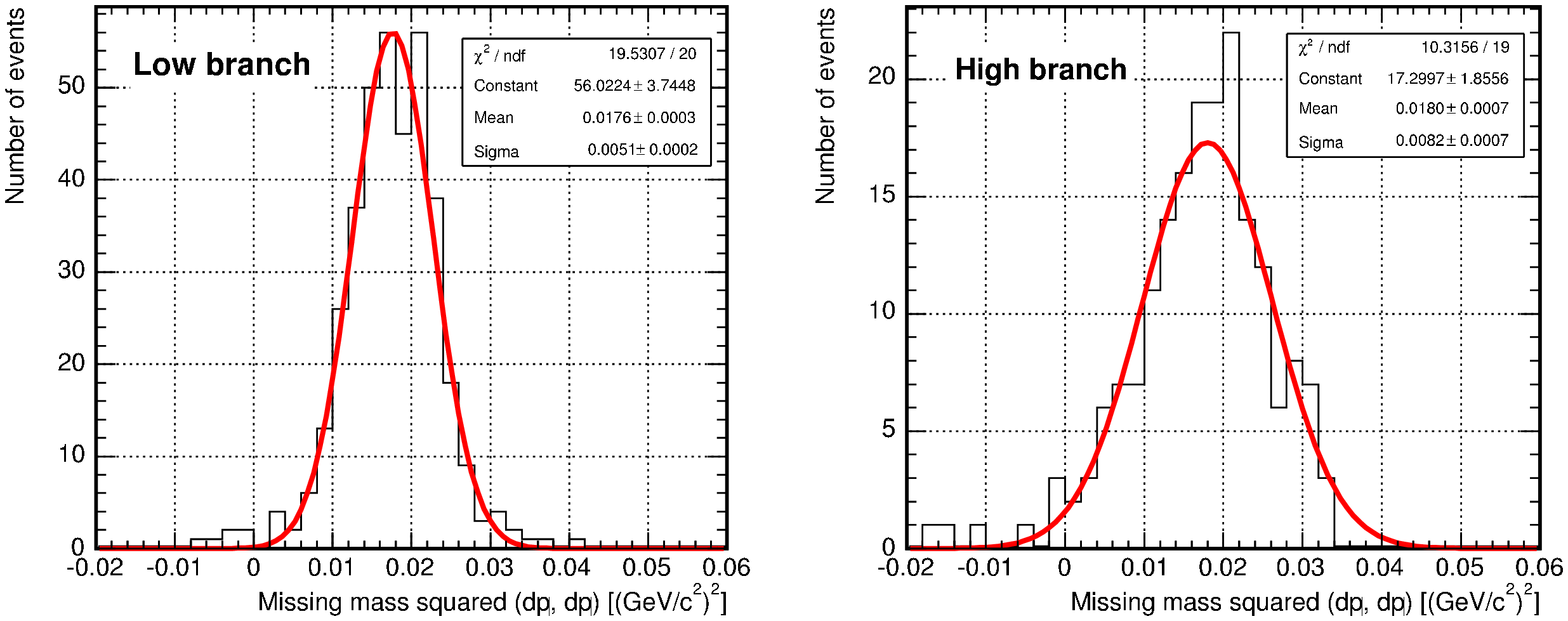,height=\textwidth,angle=-90}
    \caption{Missing masses from the $\vec{d}p\to p_{sp}dX$
    reaction showing clear $\pi^0$ peaks.\label{mx}}
\end{center}
\end{figure}

The two--dimensional spectrum in the differences of the times of
flight shown in Fig.~\ref{identification} proves that there is in
fact very little background for these events, and this is
supported by the missing masses for the two regions in
Fig.~\ref{mx}, which demonstrates well identified pion peaks,
though the single run presented here represents but a small part
of our overall statistics.
\clearpage

\vspace{-0.7cm}

\begin{figure}[htb]
\begin{center}
    \psfig{figure=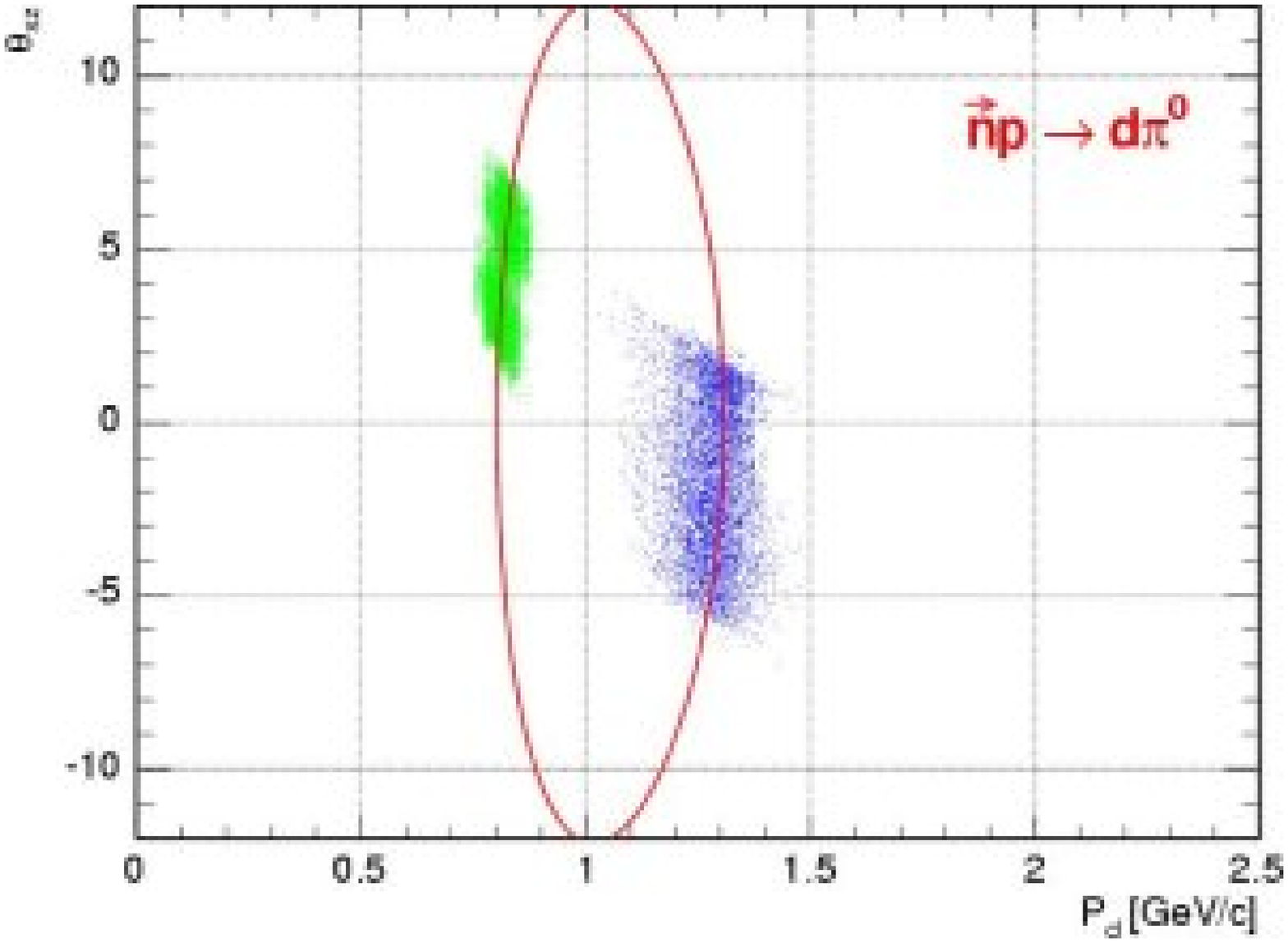,height=9cm,angle=-90}
    \caption{$\theta_d$ \emph{versus} $p_d$ scatter plot compared with
    the kinematic locus for the free $np\to d\pi^0$ reaction.\label{npdpi0}}
\end{center}
\end{figure}

The two--dimensional plot of data on the deuteron production angle
\emph{versus} momentum is shown in Fig.~\ref{npdpi0}, where it is
seen that results from both the high and low--momentum branches
are scattered around the kinematical curve corresponding to the
free $np\to d\pi^0$ reaction.
\begin{figure}[!htb]
\begin{center}
\subfigure[$^3$He
identification]{\epsfig{file=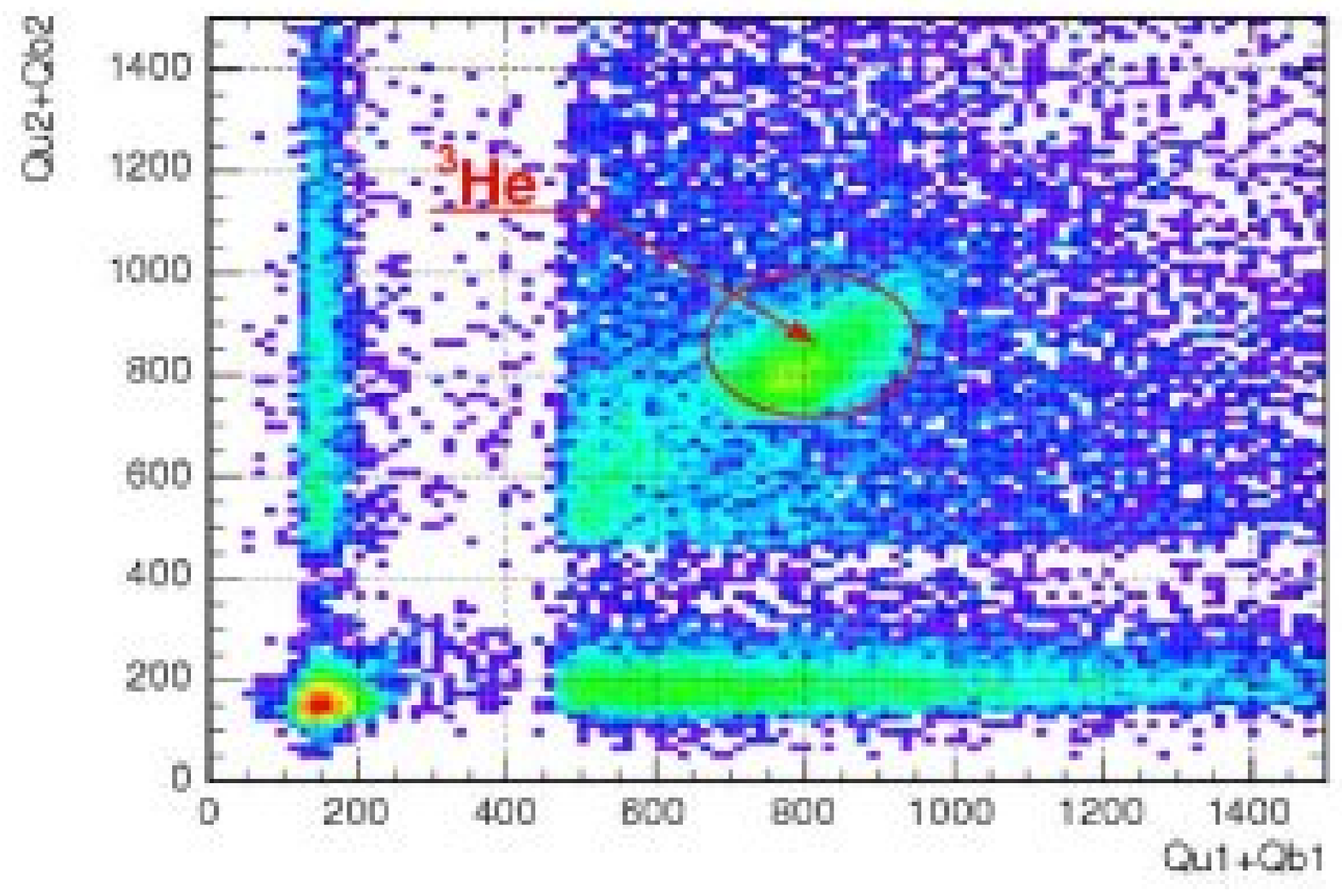,width=7cm}
\label{f:SP_pulser}} \subfigure[Missing mass squared for the
reaction $dp\to\,^3\textrm{He}\,X$. The result of the peak fit is
indicated.]{
    \epsfig{file=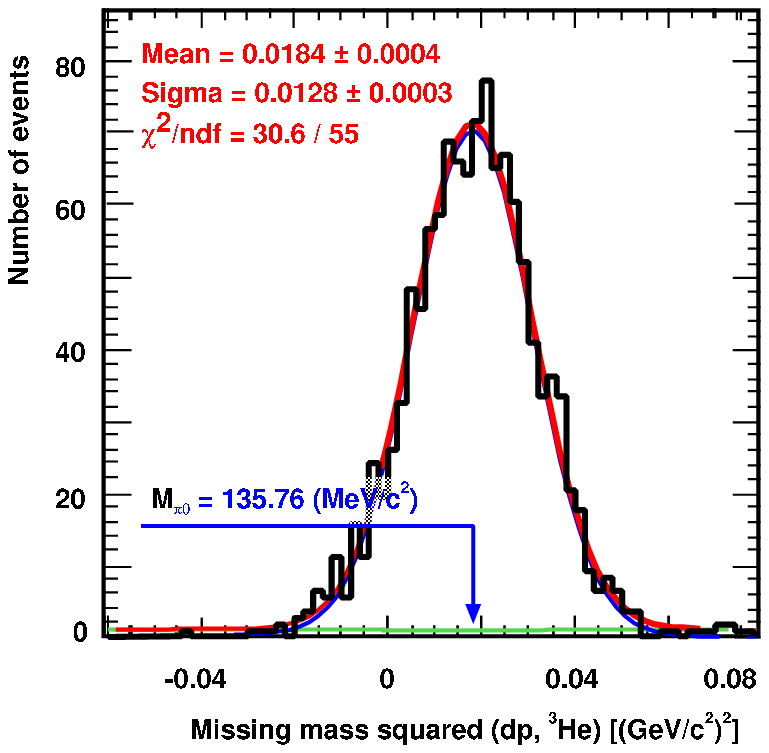,width=6cm} \label{f:SPalpha}}
\end{center}
\caption{\footnotesize
    Identification of the $dp\to\,^3\textrm{He}\,\pi^0$ reaction.}
\label{he3}
\end{figure}

In the $dp\to\,^3\textrm{He}\,\pi^0$ case, the $\pi^0$ is
recognised through the missing mass obtained from the
$^3\textrm{He}$ measurement. The result of this identification is
shown in Fig.~\ref{he3}. The high momentum branch of
$^3\textrm{He}$ particles was selected well in off--line analysis
by applying two--dimensional cuts in $\Delta E$ \emph{versus}
momentum and $\Delta t$ \emph{versus} momentum for individual
layers of the forward hodoscope. Though the peak in Fig.~\ref{he3}
is wide, this is not critical since, apart from the radiative
capture, there is no physical background over this region.

%
%
\subsubsection{Charge--exchange with a polarised deuterium target}
\label{pn-ce}%

\begin{figure}[ht!]
\begin{center}
\epsfig{file=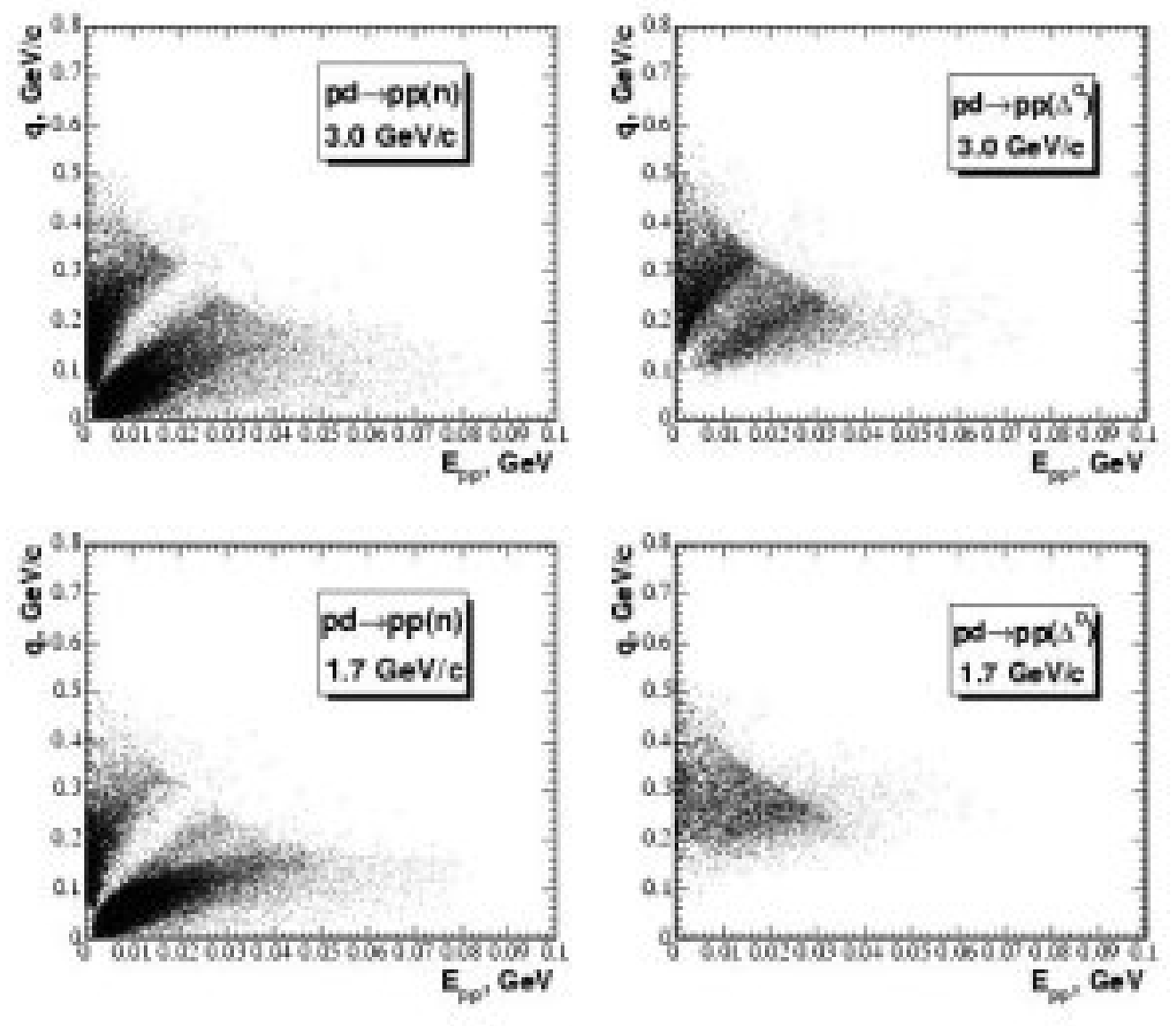, height=12cm} \caption{Simulations of
the acceptances for elastic and inelastic charge exchange on a
deuterium target where the two protons are detected in the
spectator counters.} \label{qepp}
\end{center}
\end{figure}

The advantages of studying charge exchange using just the
spectator counters in combination with a polarised deuterium
target have already been outlined. Preliminary Monte Carlo
simulations of the acceptances in momentum transfer \emph{versus}
$pp$ excitation energy are to be found in Fig.~\ref{qepp} for both
elastic and $\Delta$ production in charge exchange of protons with
momenta 1.7 and 3.0{\ww}GeV/c. Both protons are given Fermi momentum
distributions and the $pp$ \emph{fsi} is included, but no attempt
has yet been made to include a dynamic reaction mechanism.

\begin{figure}[ht!]
\begin{center}
\epsfig{file=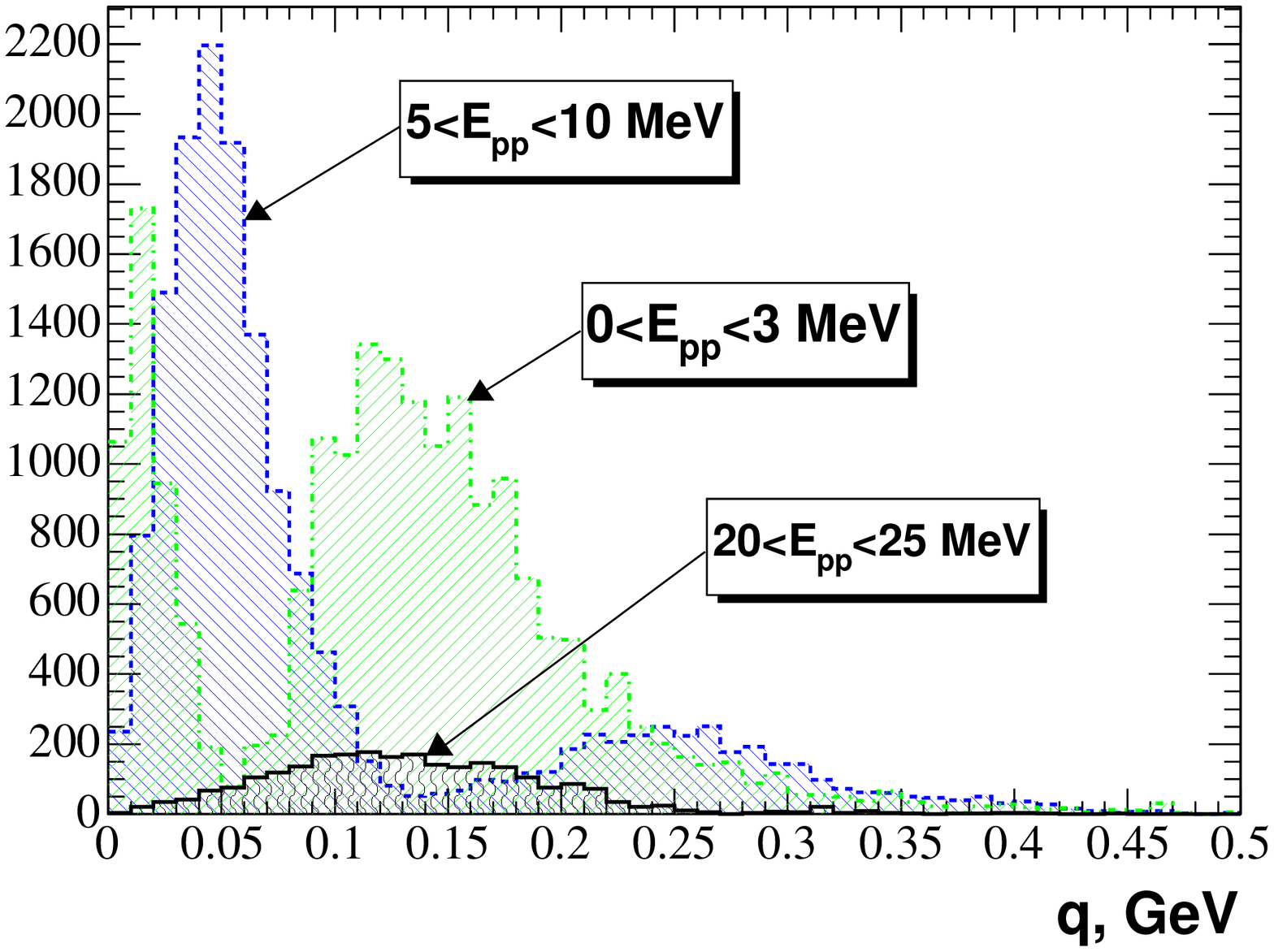, height=7cm} \caption{Projections of
the $pd\to ppn$ acceptance of Fig.~\protect\ref{qepp} at
3.0{\ww}GeV/c.} \label{slice}
\end{center}
\end{figure}

The details of the plots will depend critically upon the placing
of the silicon detectors and the values shown here are obtained
using the configuration described in \S\ref{SST}. Nevertheless,
several features are common to different settings. The wide valley
with no counts, running from the bottom left corner, indicates the
separation of the regions where the two protons go into the same
or different counters. If they go into the same counter at low
$E_{pp}$ a minimum momentum transfer is required in order to give
an energy sufficient to separate the signals (typically 1{\ww}MeV).
When the protons go into different telescopes, it is not possible
to get very small momentum transfers and so in the elastic case
there is a blind area when both $E_{pp}$ and $q$ are small. The
cut--off when these variables are both large is connected with the
maximum stopping power of the silicon detectors, here assumed to
be 60{\ww}MeV. There are still some events in the $S$--wave region of
$E_{pp}<3${\ww}MeV but, as shown in Fig.~\ref{slice}, it is not
possible to follow the $q$--dependence of these very far.

In the case of inelastic charge exchange, there is a minimum
longitudinal momentum transfer set by the kinematics of
$q_{min}\approx (M_{\Delta}^2-M_n^2)/2p_{lab}$, where $p_{lab}$ is
the laboratory beam momentum. Until one gets well above this lower
bound, at least one of the protons is likely to go to far forward
and miss the spectator counters. This problem gets worse at lower
momenta.
%
%
\subsection{Proton--neutron inelastic charge exchange}
\label{pn-ice}%
\addtocounter{figure}{1}

It was argued earlier that for energies somewhat above the pion
production threshold, it is hard to treat the $NN$ interaction in
isolation and that one must consider at least the coupling to the
$N\Delta$ channel~\cite{JH}. The number of amplitudes, the width
of the $\Delta$ and the associated difficulty of overlapping bands
in the Dalitz plots, and the weaker constraints provided by
unitarity, make this problem very challenging indeed, both
experimentally and theoretically. We believe, however, that COSY
can make significant contributions in this field. This is
important because serious problems arise when phenomenological
$NN\to N\Delta$ amplitudes are used as input for the estimation of
spin--observables in more complicated reactions. For example, the
inclusion of three--body forces arising from
$\Delta$--excitation~\cite{Nemoto,Deltuva} improve the agreement
between Faddeev calculations and the unpolarised differential
cross section for $dp$ elastic scattering at $T_p< 200${\ww}MeV.
However there is no corresponding systematic improvement for the
spin--observables measured in this process. A deficit in our
knowledge of the spin structure of the three--body forces was
pointed out recently in $pd\to (pp)n$ at 0.5{\ww}GeV~\cite{AY}.

The first amplitude analysis of $NN\to N\Delta$ was reported by
Wicklund \emph{et al.}~\cite{Wicklund}. The subsequent analysis of
Shypit \emph{et al.}~\cite{Shypit1988} used only data from their
own experiment but this was followed by an update~\cite{Bugg1989}
that included all the World data available at that time,
\emph{i.e}.
\begin{enumerate}
\item $\dd\sigma/\dd\Omega$, $A_{N0},A_{S0}$ and $A_{L0}$ near 576
and 792{\ww}MeV~\cite{Wicklund}%
\item $\dd\sigma/\dd\Omega$ and $A_{N0}$ at
800{\ww}MeV~\cite{Hancock1983},%
\item $A_{NN}$ and $A_{LL}$ at 643, 729 and 800
MeV{\ww}\cite{Bhatia1983},%
\item Five Wolfenstein parameters at 643, 729 and
800{\ww}MeV~\cite{Hollas1985,Riley1987},%
\item $\dd\sigma/\dd\Omega$, and $\sigma_{inel}$ over the complete
energy range~\cite{Shimizu1982}.
\end{enumerate}
It also included $\pi N$ partial waves $S_{31}$, $S_{11}$,
$P_{31}$, and $P_{11}$ using the OPE approximation for $NN\to
N(\pi N)$. The amplitude analysis of $pp\to d\pi^+$ and $pp\to
N\Delta$ from Ref.~\cite{Bugg1989} was used in
Ref.~\cite{Anisovich1992} as input to perform a re-analysis within
the $N/D$ method. The  aim was to distinguish between the pole or
cusp interpretation of the data near the $N\Delta$ threshold.

Let us discuss in more detail, for example, the Argonne
data~\cite{Wicklund}, which were obtained at proton beam energies
of 0.57, 0.81, and 1.01{\ww}GeV, and the first amplitude analysis of
these results. In Ref.~\cite{Mizutani1993} the data on the $NN\to
N\Delta$ reaction were investigated within the unitary model of
coupled $\pi NN-NN$ channels. This model, based on the
meson--exchange picture, describes in a unified manner the
following processes: $NN\to NN$, $NN\leftrightarrow \pi d$,
$NN\leftrightarrow \pi NN$, $\pi d\to \pi d$, and $\pi d \to \pi
NN$. The unpolarised differential cross section, production
asymmetry and spin correlation parameters were calculated for the
$NN\to N\Delta $ reaction. The global agreement with the available
Argonne data is reasonably satisfactory. The main problem concerns
the integrated asymmetry $A_y$, which is consistent with the data
only at 0.57{\ww}GeV. At higher energies, the model is unable to
reproduce the change from a broad positive maximum to the broad
positive minimum observed in $A_y$. The asymmetry problem at high
energy is directly related to the fact that none of the models
used is able to describe the helicity--$3/2$ spin--correlation
parameter $P_y\rho_{33}$.

A possible way to improve (partially) the agreement was pointed
out in Ref.~\cite{Sammarruca1994}, where the $NN\to N\Delta$
amplitude was extracted from the Bonn meson--exchange model of
elastic $NN$--scattering. Within such a model of the
$NN$--interaction, taking account of the coupling to the $N\Delta$
and $\Delta\Delta$ channels allows one to improve considerably the
phase shifts and inelasticity parameters of $NN$--scattering below
1{\ww}GeV~\cite{EDDA_rev}. At higher energies, only the $NN$ total and
integrated elastic cross sections are satisfactorily described
with Reggeised vector meson exchanges, whereas the spin
observables measured at 1--2.5{\ww}GeV are in strong disagreement with
the OBE model predictions. However, the Bonn model does not take
into account the correlated $2\pi$ and $\pi\rho$ exchanges, which
are included in the J\"ulich model~\cite{Janssen1994}.
\clearpage
\hspace{-0.7cm}
\begin{minipage}[t]{7.0cm}
\baselineskip 3ex\hspace{0.5cm}\vspace*{-5mm}

The polarised deuteron charge--exchange programme of the Saclay
group included some measurements of the excitation of
$\Delta(1232)$ through the $\vec{d}A\to pp\,A'$ reaction for
targets $A=p,\,d,\,^{12}$C, where the isobar was identified
\emph{via} the missing mass in the
reaction~\cite{Ellegaard2,Sams}. The near--forward differential
cross sections for hydrogen and deuterium targets are shown in
Fig.~\arabic{figure} at $T_d=2\:$GeV as functions of the
laboratory energy loss $\omega_{lab}$. To a good approximation,
the polarisation response is related to the analysing power by
$A_{yy}\approx -\sqrt{2}P/\rho_{20}$, where the beam polarisation
$\rho_{20}=0.61\pm0.01$.

\hspace{0.5cm} As mentioned in \S\ref{pn-ce}, quasi--elastic
charge exchange on deuterium was a factor of about $0.68$ smaller
than for hydrogen. However, in the pion--production region, the
quasi--free $dd\to pp \Delta^-p$ cross section should be bigger
than that for $dd\to pp \Delta^0n$ by an isospin factor
of three. The polarisation responses are essentially
indistinguishable and, when the deuterium data are divided by a
factor of $4\times 0.68$, the hydrogen and deuterium results
largely coincide. This agreement with the scaled cross sections is
similar
\end{minipage}
\hfill
\begin{minipage}[t]{7.0cm}
\vspace*{-0.7cm}

\noindent
\input epsf
\begin{center}
\mbox{\epsfxsize=7cm \epsfbox{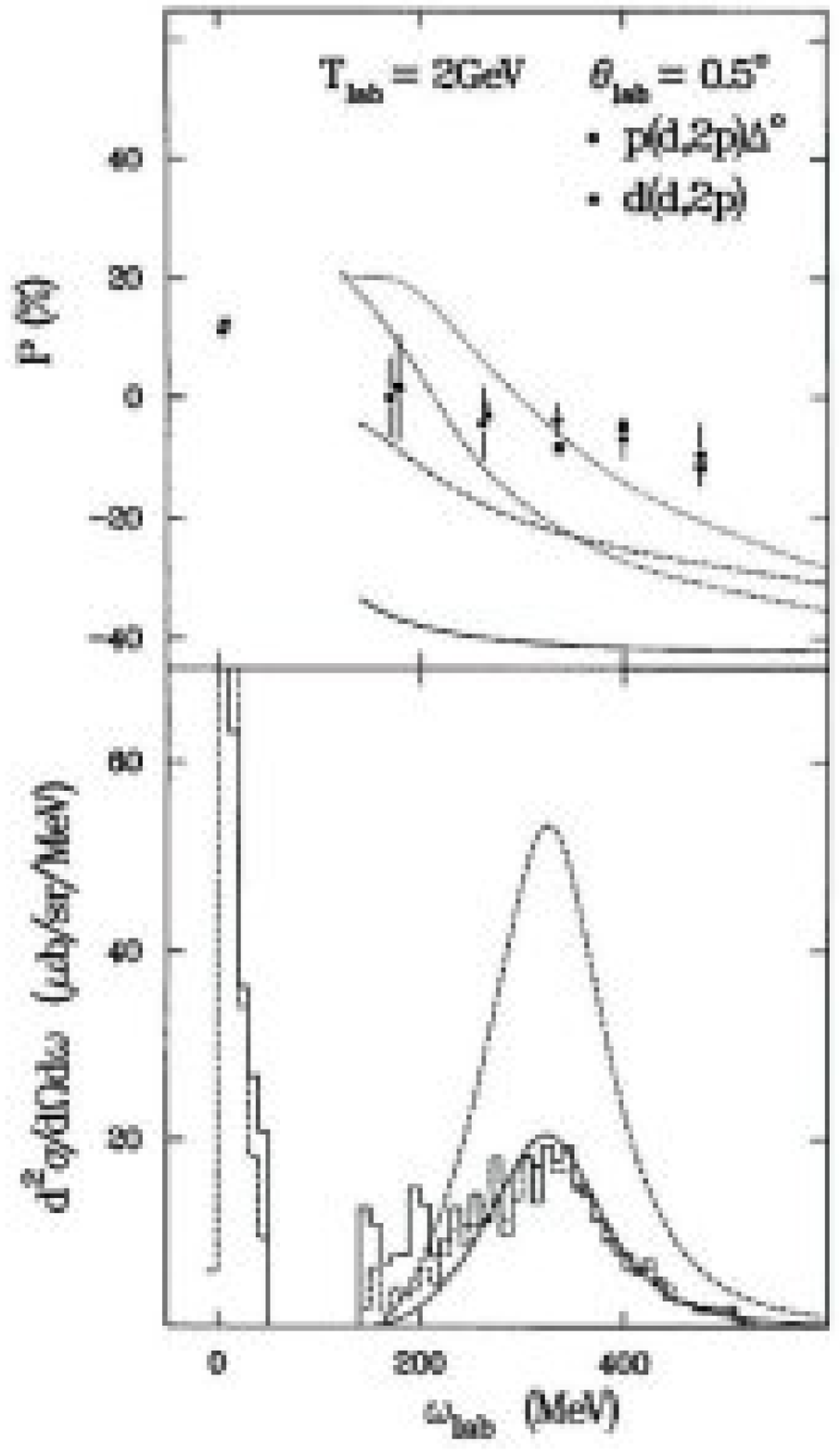}}
\end{center}\noindent%
\vspace{-8mm} \baselineskip 3ex

Fig.~\arabic{figure}: Polarisation
response and cross section for $p(\vec{d},2p)\Delta^0$ and
$d(\vec{d},2p)\Delta N$ at $T_d=2\:$GeV and
$\theta_{pp}^{lab}=0.5^{\circ}$. The broken histogram corresponds
to the scaled result for deuterium. The solid curves are models
with direct pion exchange whereas for the broken curves this is
modified using the \emph{poor man's
absorption model} cut off~\cite{Sams}.
\end{minipage}

\baselineskip 3ex%
\vspace{2mm}

\noindent
at the other angles measured and this proves the expected
dominance of the $I=\frac{3}{2}$ strength.

Unlike the $p(\vec{d},2p)n$ measurements at 1.6{\ww}GeV, discussed in
the previous subsection, no spin--rotator was used for $\Delta$
excitation at 2.0{\ww}GeV. On account of the small angular acceptance
of SPESIV, only one tensor analysing power combination could then
be measured and this principally determined $A_{yy}$.

Away from the forward direction, the \emph{poor man's absorption}
prescription to the one--pion exchange model for $pN\to\Delta N$
gives a plausible description of the cross section and analysing
power~\cite{Sams}. However, the density matrix elements for the
$\vec{p}\,p\to\Delta^{++}n$ reaction at the rather higher energy
of $T_d=2.2\:$GeV~\cite{Wicklund}, which depend upon interference
terms, are only qualitatively reproduced. Inelastic
charge--exchange experiments will be repeated at ANKE in the near
future with a polarised deuteron beam of a similar energy to
Saclay as a by--product of the approved $p(\vec{d},2p)n$
experiment~\cite{PROPOSAL,Chiladze}. A wider range of different
deuteron analysing powers will then be derived. Already in
Fig.~\ref{Mpp_distribution} we see some evidence of pion
production even at $T_d=1.17${\ww}GeV.

In the future it will be possible to go to much higher energies by
using the polarised deuterium internal target. Though there is
little chance of getting sufficient data in order to allow a
completely model--independent amplitude analysis,and the phase
space shown in \S\ref{pn-ce} is limited, we will measure the
$\vec{d}(\vec{p},2p)\Delta^0$ reaction with polarised beam and
target up to $T_p\approx 3\:$GeV, using just the spectator
telescopes in the manner described in the previous subsection.
Rank--two tensor observables, such as the $t_{20}$ and $t_{22}$
analysing powers and the transverse spin--correlation coefficient
of the proton with a vector polarised deuteron, are particularly
robust quantities that can pick out moduli of amplitudes rather
than the imaginary parts of interferences that are sensitive to
phase differences which can arise from the $NN$ or $N\Delta$
interactions.

However, the tensor polarisation of the outgoing $\Delta^0$ can
also be measured, as was demonstrated at Argonne for the
$\vec{p}\,p\to\Delta^{++}n$ reaction, by looking at the angular
distribution of the pion or proton from the $\Delta$ decay in the
$\Delta$ rest frame{\ww}\cite{Wicklund}. In our case, this would
require the detection in ANKE of the fast $\pi^-$ or proton from
the $\Delta^0$ disintegration in coincidence with the two slow
protons in the silicon counters. Now for the $NN\to \Delta N$ reaction
there are four independent amplitudes in the forward direction and it
can be shown~\cite{Yuri_new} that study of the
$\vec{d}\vec{p}\to (pp)\vec{\Delta}^0$ reaction will give access to
three combinations of these amplitudes.

In small momentum transfer reactions, such as this and $p(d,2p)n$,
it is to be expected that the major correction to a simple
quasi--free interpretation will come from multiple scatterings,
which can be handled in the eikonal approximation~\cite{Glauber}.
These generally give an overall damping and have less effect on
the spin observables until the momentum transfers are above around
$1\:$fm$^{-1}$~\cite{Sams95a,Sams}
%
%
\subsection{Small angle $\vec{p}\,\vec{d}$ or $\vec{d}\,\vec{p}$ elastic
scattering}
\label{pdsa}

\begin{figure}[ht!]
\begin{center}
\epsfig{file=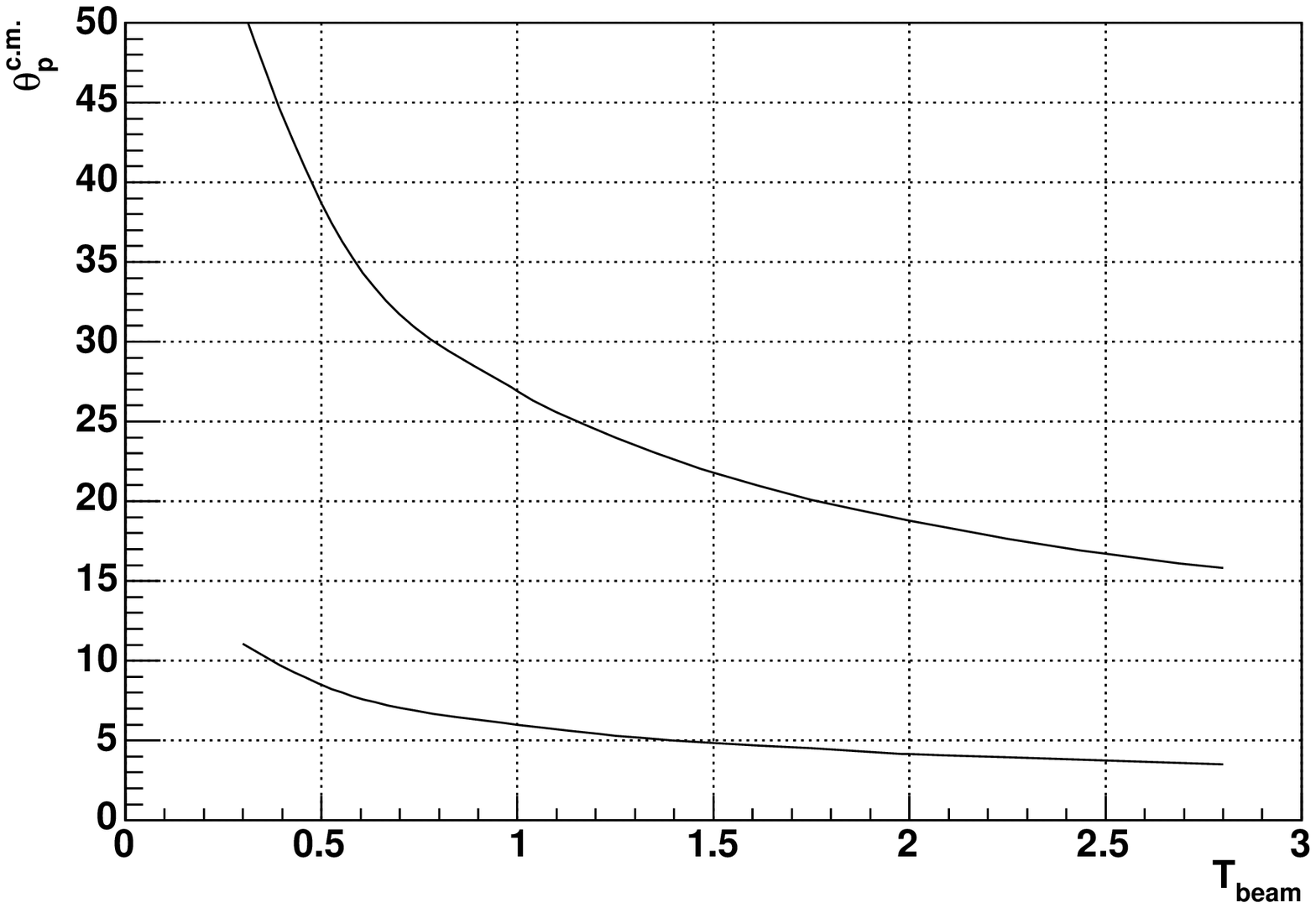, height=7cm} \caption{Predicted
upper and lower bounds on the proton cm angle in proton--deuteron
elastic scattering, with only the deuteron being detected in the
silicon telescopes.} \label{angle_l}
\end{center}
\end{figure}

Elastic proton--deuteron scattering at small angles can be
measured at ANKE with a deuteron beam or a deuterium target but
the kinematic ranges will differ. With a proton beam one could
detect a deuteron in the spectator counters and a proton in the FD
and, observed in this way, the reaction has already been used for
luminosity and beam polarisation determinations at ANKE~\cite{AY}.
However, the cross section is so big that merely measuring the
deuteron in the spectator counters is sufficient to identify
completely the process. Considering only recoil deuterons with
energies in the range $2.5<T_d<50${\ww}MeV leads to the plot of the
minimum and maximum proton cm angles shown in Fig.~\ref{angle_l}.
This corresponds essentially to a fixed range in momentum
transfer, $0.001<|t|<0.19${\ww}(GeV/c)$^2$, which we can cover up to
the maximum COSY energy of close to 3{\ww}GeV.

\vspace{-8mm}

Above about 1.5{\ww}GeV the fast proton would fall within the maximum
ANKE angular acceptance of $\theta_p^{lab}<11^{\circ}$ but even
then some of the small momentum transfer protons would be lost in
the forward cone of $\theta_p^{lab}<4^{\circ}$.

Small angle deuteron--proton elastic scattering has also been
measured at COSY by detecting just the deuteron in the ANKE
magnetic spectrometer. It could be used for luminosity and
polarisation purposes~\cite{Chiladze} since, as shown in
Fig.~\ref{fig:p_distrubition}, the proton does not need to be
measured in order to have a very clean signal for the reaction.
The acceptance for measuring the deuteron from the reaction at
$T_d=1170${\ww}MeV can be deduced from Fig.~\ref{accep} and, as shown
in Fig.~\ref{kinem}, this remains roughly the same in the
laboratory at different beam momenta;
$4^{\circ}<\theta_d^{lab}<10^{\circ}$.

Comparing the acceptances for $pd$ and $dp$ elastic scattering, it
is seen that smaller angles are covered with the proton beam. This
illustrates the power of the silicon telescope array when used at
the ANKE facility.

%
%
\newpage
\section{Proton--Deuteron Reactions at High Momentum Transfers}
\label{CPDR}%

The combination of the ANKE magnetic spectrometer and spectator
telescopes with polarised beams and a polarised deuterium gas cell
will lead to the detection in parallel of many non--mesonic
nuclear reactions in addition to those in the proton--neutron
programme outlined in \S\ref{PNSP2}. Some of these have a great
interest in their own right and will require dedicated beam time
if their potential is to be exploited to the full. This is
especially true of large momentum transfer proton--deuteron
elastic and charge exchange scattering, for which interesting
results have already been achieved at low $pp$ excitation energies
at ANKE~\cite{Komarov,AY}.

The nuclear three--body problem is fundamental, but the difficulty
that one faces with such studies at ANKE is that the main benefits
of COSY come when working above the pion--production threshold.
This is a region where, for a variety of reasons, the Faddeev
description of proton--deuteron elastic and inelastic reactions is
no longer appropriate. There is then no agreed calculational
scheme to model such data and there is much ambiguity in how to
interpret, for example, the energy dependence of the cross section
and tensor analysing power of $dp$ elastic scattering at
$180^{\circ}$~\cite{A-B}, though virtual $\Delta$ excitation is
certainly involved at some level~\cite{Craigie,Kolybasov}.
Nevertheless, it is hoped that the measurement of spin observables
will provide valuable clues to the dominant reaction mechanisms
involved. As an example of this, consider the $\vec{p}d \to
(pp)_{\!^1\!S_0}n$ reaction, where it has been shown that at
500{\ww}MeV the proton analysing power is almost maximal for
$\theta_n^{cm}\approx 167^{\circ}$~\cite{AY}. This indicates that,
out of the six possible spin amplitudes~\cite{Ohlsen}, only two
combinations are significant and that these, in addition to being
almost equal in magnitude, have just the right relative phase.
This feature disappears at 800{\ww}MeV, where $A_y$ is uniformly
small. It is therefore expected that measurements of spin
correlations and tensor analysing powers will provide further
insights into proton--deuteron large momentum transfer reactions.

%
%
\subsection{Large angle $\vec{p}\,\vec{d}$ or $\vec{d}\,\vec{p}$ elastic
scattering}
\label{pddp}

Electromagnetic probes are generally considered to be the most
clean instruments with which to study the structure of nuclei for
$r_{NN}<0.5${\ww}fm. Our present knowledge of the deuteron
structure at $r_{NN}<1${\ww}fm comes mainly from $ed$ elastic
scattering and photodisintegration $\gamma d\to pn$. However, the
shorter the distances probed in electromagnetic processes the
larger are the contributions from meson--exchange currents, which
are not purely electromagnetic in origin. The problem of
meson--exchange currents can only be solved within a consistent
theory of strong interactions, which is still absent at high but
finite values of $Q$ in the non-perturbative region of QCD.  The
interpretation of electromagnetic processes on the deuteron at
$Q>1${\ww}GeV/c therefore suffers from many
ambiguities~\cite{garson, gilmangross}.

Independent information on the dynamics of short-range $NN$
interaction and the short--range structure of nuclei can, in
principle, also be obtained from hadronic processes at high $Q$,
provided that the reaction mechanism can be well
identified\footnote{Some $NN$ models, which exhibit different
behaviour at $r_{NN}<1${\ww}fm, may be unitarily equivalent and
thus describe the same physics. Some examples, based on one pion
and rho exchanges in the $NN$ potential, were given by Desplanques
and Amghar~\cite{DesplanquesAmghar1, DesplanquesAmhar2}. In such a
case a model that requires fewer corrections due to the many body
forces or many body currents might be the preferred one.}.

Existing data on backward $pd$ elastic scattering at
0.5--2{\ww}GeV are in disagreement with the predictions of
one--nucleon--exchange (ONE) models based on the widely used Paris
or Reid soft core $NN$ potentials. These overestimate considerably
the unpolarised cross section for $T_p> 1.5${\ww}GeV. On the other
hand, the one--pion--exchange (OPE) model, driven by the $pp\to
d\pi^+$ subprocess~\cite{Craigie,Kolybasov}, is much less
sensitive to the high momentum components of the $NN$ wave
function and is in qualitative agreement for $T_p$ in the range
0.5--2.5{\ww}GeV. Furthermore, the deuteron analysing power in the
backward direction, $t_{20}(\theta_{cm}=180^\circ)$, agrees with
the ONE only at rather low energies $T_p<0.3${\ww}GeV. In contrast
to the ONE predictions, the experimental values of
$t_{20}(\theta_{cm}=180^\circ$) do not changes sign with
increasing beam energy~\cite{A-B}. Taken together, these
observations suggest that the ONE mechanism plays only a minor
role in this process.

Backward elastic scattering of 1.17{\ww}GeV (polarised) deuterons can
been clearly identified in the two--dimensional plot of the
momentum correlation of two charged particles shown in
Fig.~\ref{identification}a, where it is seen that there is
essentially no background~\cite{Chiladze}.

\begin{figure}[ht]
\centering
\includegraphics[height=10.0cm,angle=-90]{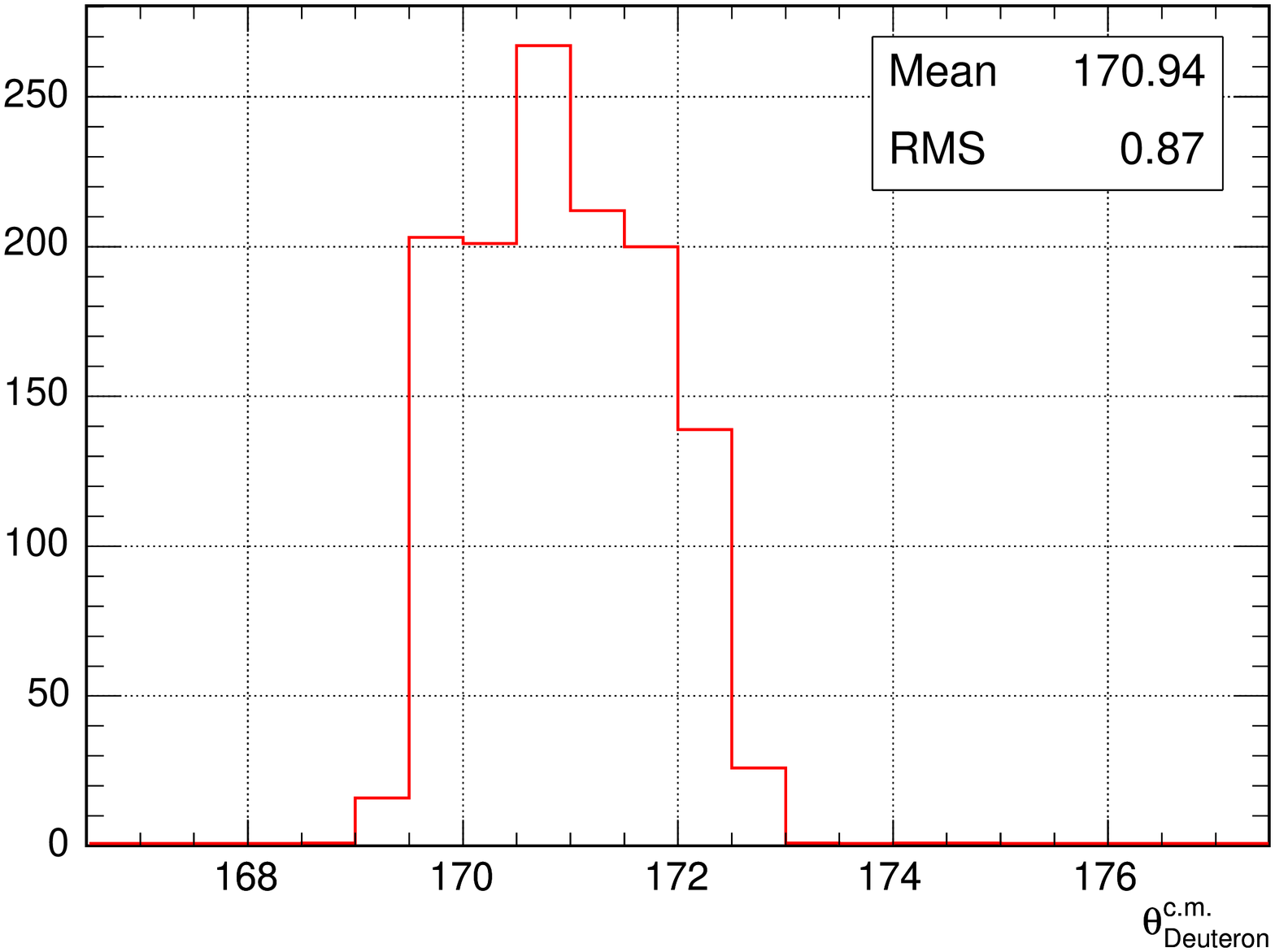}
\caption{Distribution in the cm deuteron scattering angle for
identified $dp\to pd$ events at 1.17{\ww}GeV.}\label{dpback}
\end{figure}

\begin{figure}[ht]
\centering \epsfig{file=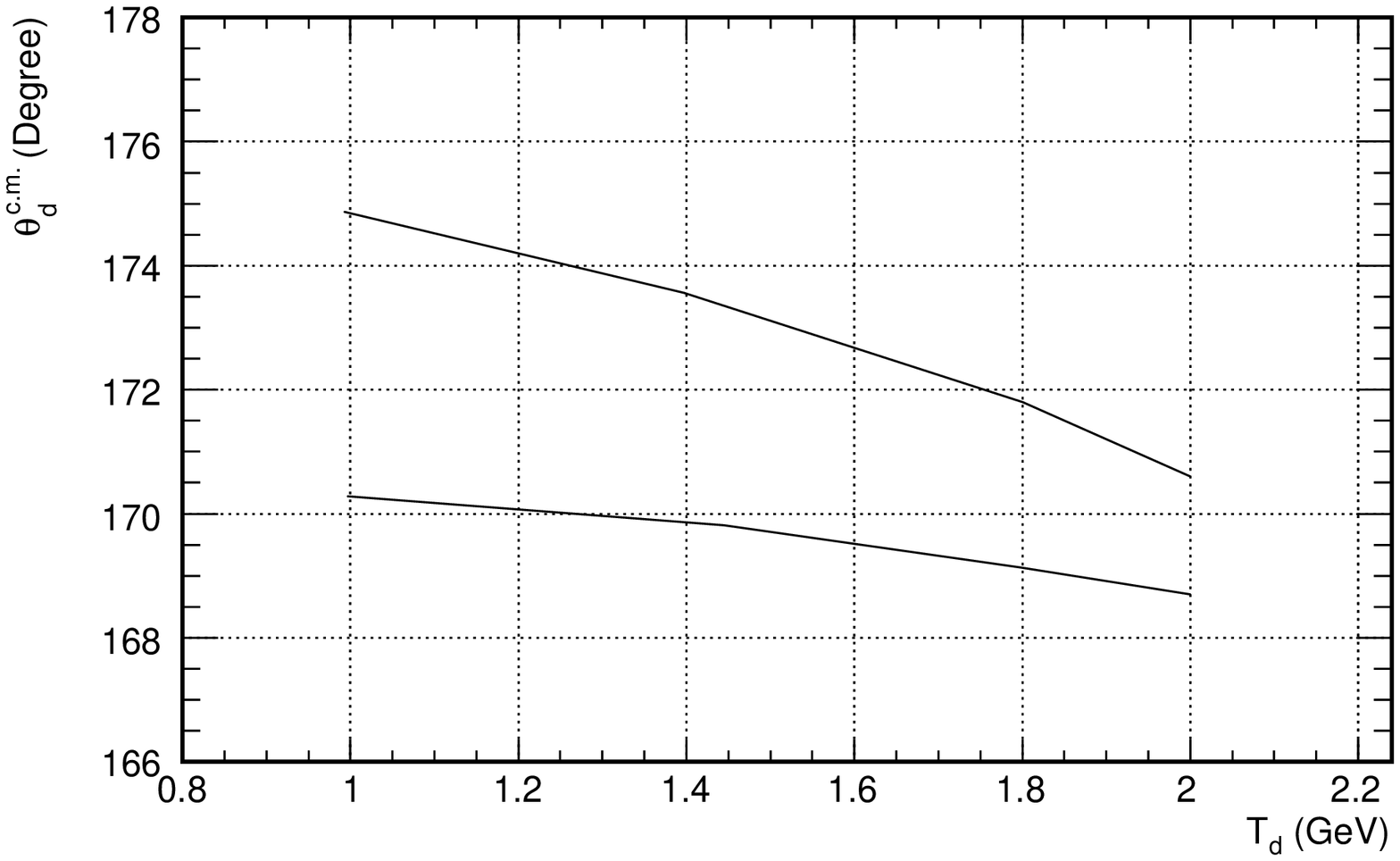, height=7cm}
\caption{Angular acceptance in the cm system for deuterons from
deuteron--proton elastic scattering as a function the deuteron
beam energy.}\label{ThCMvsE}
\end{figure}

The distribution of the number of events in deuteron cm angle
presented in Fig.~\ref{dpback} shows that the acceptance for the
reaction is approximately $169.5^{\circ} <\theta_d^{cm}
<172.5^{\circ}$ but, as shown in Fig.~\ref{ThCMvsE}, this range
decreases steadily with beam energy.

%
%
\subsection{$\vec{p}\,\vec{d}\to (pp)\,n$ at large momentum
transfers}
\label{pdppn}

To minimise the effects of the $\Delta$, which plays such a large
role in backward proton--deuteron elastic scattering, the ANKE
collaboration has a programme for measuring the reaction
\begin{equation}\label{breakup}%
pd \to (pp)_{^1\!S_0} n\,,
\end{equation}
where $(pp)_{\!^1\!S_0}$ denotes a proton pair with small
excitation energy. By taking $E_{pp} < 3${\ww}MeV, we can be fairly
certain that there is little contamination from higher $pp$
partial waves. In contrast to the small--angle charge exchange
discussed in \S\ref{ece}, the selection of fast diprotons in the
laboratory system corresponds to neutrons emerging with cm angles
close to 180$^{\circ}$ with respect to the incident proton. The
kinematics are then very similar to those of backward $pd\to dp$.
This reaction provides two new features compared to $pd$ elastic
scattering~\cite{imusmuz}:\\
\emph{i}) The contribution from three--body forces, arising from
the excitation of $\Delta$ and $N^*$ resonances in the
intermediate state, is suppressed by an isospin factor of three in
amplitude~\cite{yujetp}.\\
\emph{ii}) The \emph{uncoupled} $S$--wave dominates the internal
state of the diproton at $E_{pp} < 3${\ww}MeV. Due to the repulsive
nature of the $pp$ force at short distances, it is expected that
the $^1\!S_0$ diproton wave function should have a node at a
relative $pp$ momenta $q\approx 0.4${\ww}GeV/c~\cite{imusmuz}. This
should be easier to test than in $pd\to dp$, where minima are
filled in by quadrupole effects connected with the deuteron
$D$--state. For diproton production there should be regions in
energy that are dominated by different mechanisms and that can
test separately the ingredients of models~\cite{yujpg}.

\begin{figure}[ht!]
\begin{center}
\epsfig{file=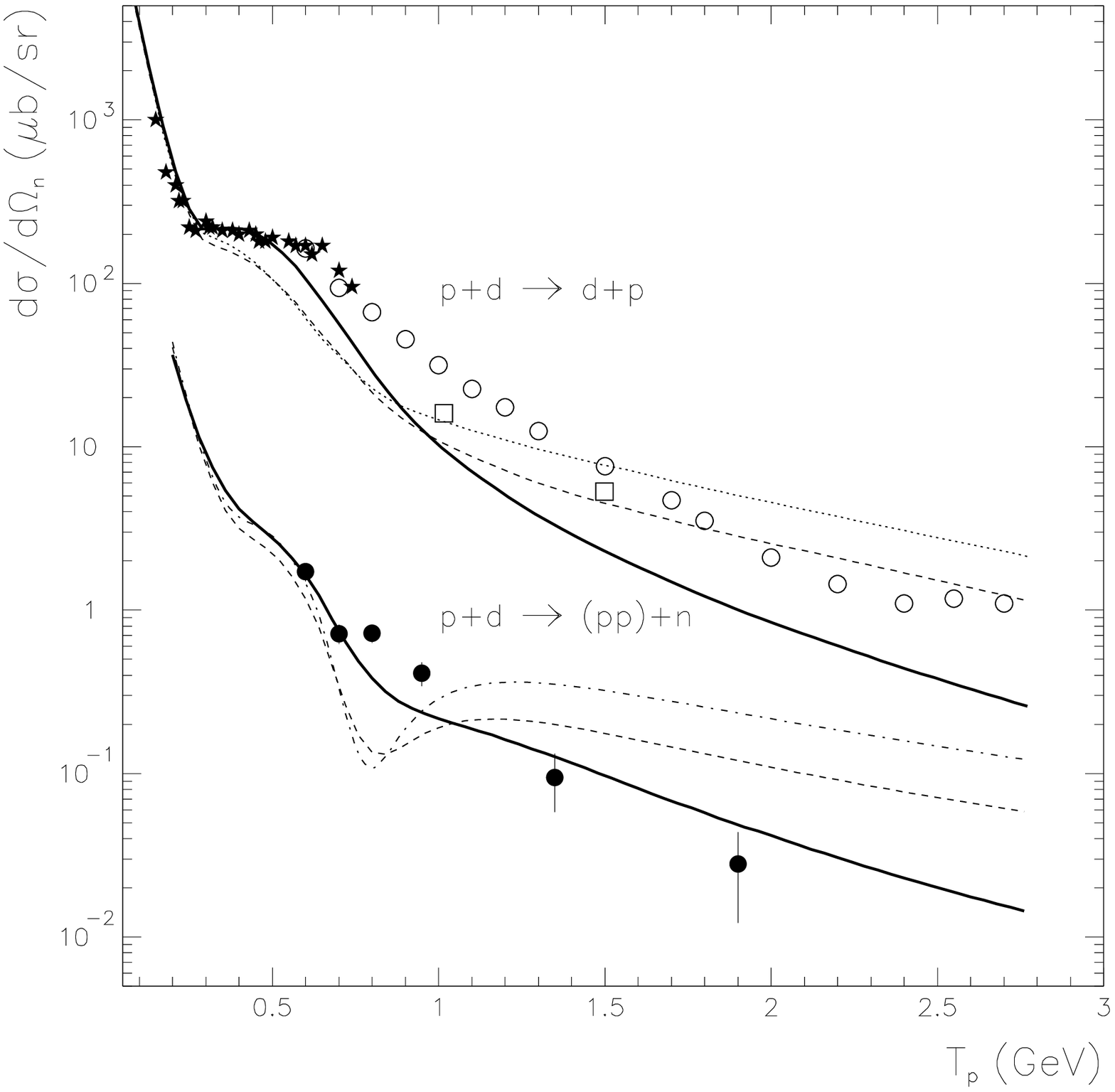, height=7cm} \caption{Comparison of the
differential cross section for backward elastic proton--deuteron
scattering with that for charge--exchange
break--up~\cite{Komarov}. The solid curves include contributions
from one--nucleon exchange, single and double scattering, with
$\Delta(1232)$ excitation~\cite{haidenbauer}.} \label{Komarov_1}
\end{center}
\end{figure}

\addtocounter{figure}{1}

The unpolarised cross section of the reaction was measured at COSY
at proton beam energies from $T_p=0.6$ to 1.9{\ww}GeV~\cite{Komarov}.
A reasonable agreement with these data is achieved in a model that
includes one--nucleon exchange, single scattering, and double $pN$
scattering with the excitation of a $\Delta(1232)$
isobar~\cite{haidenbauer}. This analysis takes into account
interactions in the initial and final states by employing modern
$NN$ potentials, \emph{e.g.}\ CD--Bonn~\cite{machleidt}. Older
potentials, such as the Paris~\cite{paris} and Reid Soft Core
(RSC)~\cite{rsc}, seem to overestimate the high--momentum
components of the $^1\!S_0$ wave function and this leads to a
strong disagreement with the data. Thus, within this model, one
has sensitivity to the $NN$ interaction that should be explored
further through measurements of the spin dependence of the
reaction.

The measured proton analysing power~\cite{AY} depends sensitively
upon interferences, but the deuteron tensor analysing power and
spin correlations are much more robust indicators of the reaction
mechanism. These have been predicted at
$180^{\circ}$~\cite{uziseyfarth} in the same model as that used
for the unpolarised cross sections~\cite{haidenbauer} and shown in
Fig.~\arabic{figure}. \vspace{-10mm}

\hspace{-0.7cm}
\begin{minipage}[t]{6cm}
\baselineskip 3ex
\vspace{6mm} Fig.~\arabic{figure}: %
Tensor analysing power $T_{20}$, and spin--spin correlation
parameters $C_{y,y}$, $C_{z,z}$ and $C_{xy,z}$ predicted for
different $NN$--potentials within a model that includes
one--nucleon exchange, single scattering, and $\Delta$
excitation~\protect\cite{uziseyfarth}. The curves correspond to
RSC (dotted)~\cite{rsc}, Paris (dashed)~\cite{paris}, CD--Bonn
(full)~\cite{Machleidt}.
\end{minipage}
\hfill\hspace{-7mm}
\begin{minipage}[t]{8.5cm}
\vspace*{-0.7cm}

\noindent
\input epsf
\begin{center}
\mbox{\epsfxsize=10cm \epsfbox{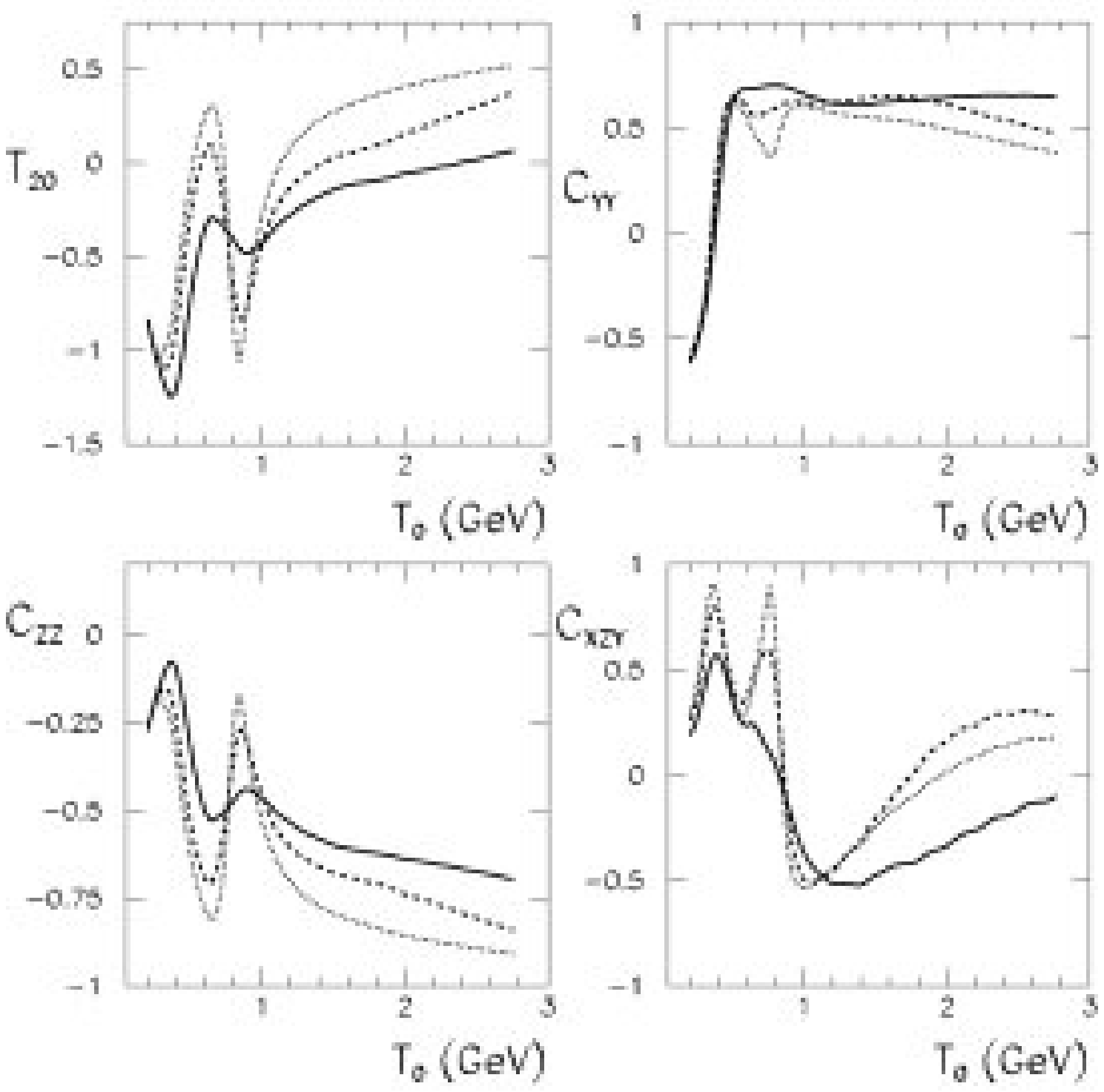}}
\end{center}
\vspace{3mm}
\end{minipage}

For energies $T_p> 1${\ww}GeV, the $\Delta$ contribution is expected
to die away and single--nucleon exchange might then dominate. In
this limit $T_{20}$ must change its sign at some energy, whose
value will depend upon the form of the $NN$--interaction. In
contrast, the existing data for $pd\to dp$ and $pd\to p(0^\circ)X$
show a $T_{20}$ that does not change sign, being large and
negative up to $T_p=4${\ww}GeV~\cite{ashgirey}. Another test for the
dominance of distorted one--nucleon--exchange is the equality
$A_y^p=A_y^d$, though this could be modified by interferences with
small amplitudes coming from other mechanisms.
%
%
\clearpage
\section{Non--strange Meson Production}
\label{NSMP}

\subsection{Deriving the chiral three--body force from pion \newline production}
\label{tbf}

\subsubsection{Motivation}

One of the major challenges in today's physics is to relate the
properties of few--nucleon systems and nuclei to the theory of strong
interactions, QCD. Over recent years there has been major theoretical
progress in establishing an effective field theory that, while having
a clear cut connection to QCD, allows one to study processes involving
strongly interacting particles within a well defined perturbative
scheme. It is chiral symmetry that provides the preconditions for the
construction of an effective field theory. It forces not only the mass
of the pion $m_\pi$, as the Goldstone boson of the chiral symmetry
breaking, to be low, but also the interactions to be weak, since the
pion needs to be free of interactions in the chiral limit for
vanishing momenta.

Following the pioneering works by Weinberg~\cite{sweinpi}, and
Gasser and Leutwyler~\cite{gasleu}, chiral perturbation theory is
now a well developed and powerful tool for investigations of the
$\pi\pi$~\cite{pipi}, $\pi N$~\cite{bernard197} as well as few
nucleon systems~\cite{evgeni}. Furthermore Weinberg also pointed
out how to calculate, in an equally controlled way, pion
scattering from nuclei as well as inelastic reactions on
nuclei~\cite{swein1}.

It was recently observed that there is one modification necessary
to the standard chiral perturbation theory when this is applied to
pion production in $NN$ collisions. The large scale introduced by
the initial momentum, given by $\sqrt{m_\pi M_N}$, has to be
considered explicitly~\cite{cohen,hankol,hankai}. Only then will
the chiral expansion converge, contrary to earlier claims based on
the assumption that all momenta are of order
$m_\pi$~\cite{dmit,ando}. Thus a proper expansion scheme for pion
production is now established and a complete calculation for the
reactions $NN\to NN\pi$ is currently under way.

However, we also need reliable few--nucleon wave functions, which
are based on the same chiral effective theory. Only in this way
can we guarantee that the transition operators are consistent with
the wave functions. Fortunately, these wave functions, or the
interactions necessary to generate them, do
exist~\cite{evgenynn,entem,evgenyspectr}. Furthermore, the
extension to few--nucleon systems has been
accomplished~\cite{evgeny3nf}, allowing processes on light nuclei
to be studied in the future.

\begin{figure}[ht!]
\begin{center}
\epsfig{file=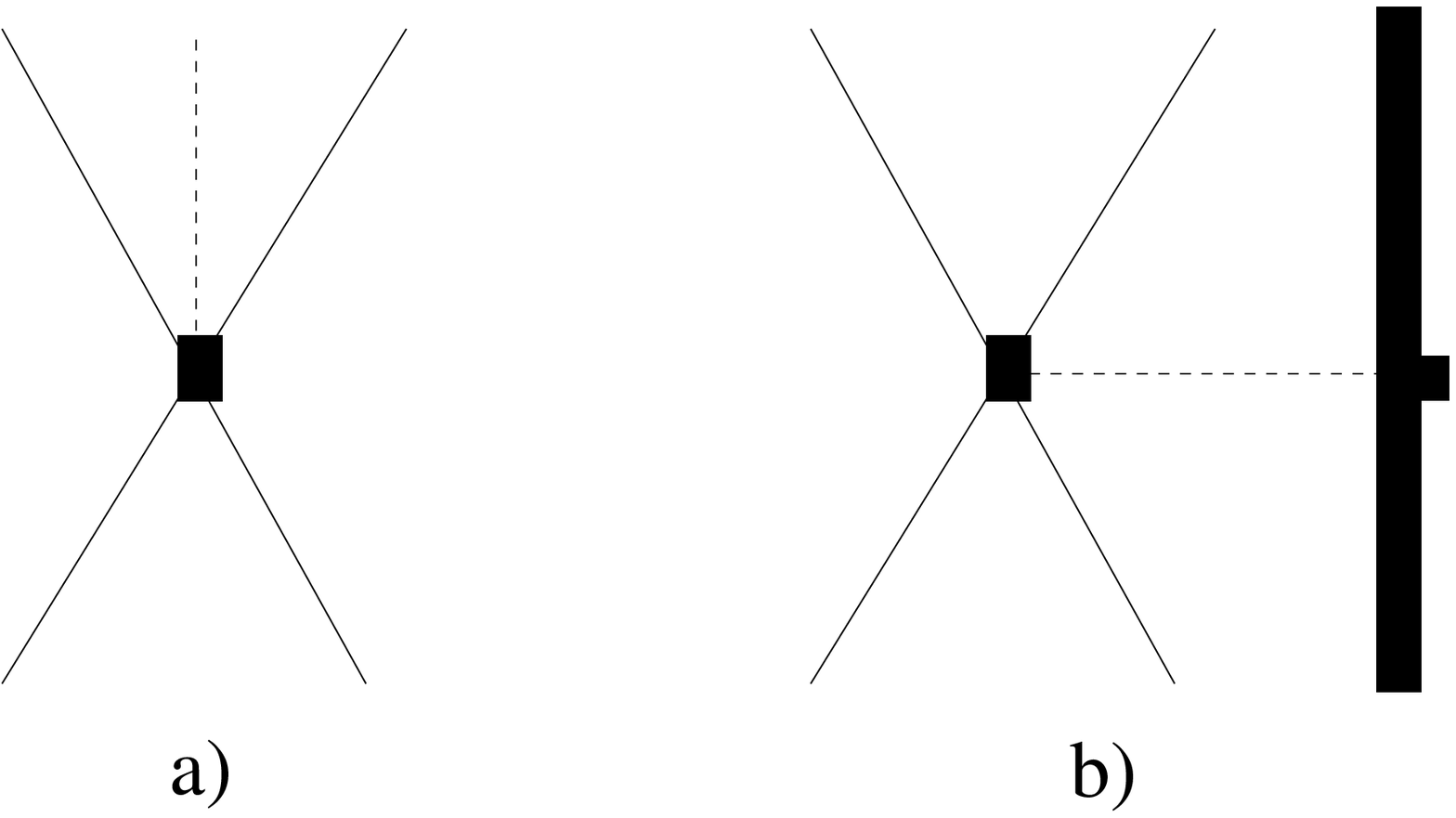, height=5cm} \caption{Illustration of
the role of the $4N \pi$ contact term in $NN\to NN\pi$ and three
nucleon scattering. Solid lines denote nucleons, dashed lines
pions.} \label{dill}
\end{center}
\end{figure}

One important step forward in our understanding of pion reactions
at low energies will be to establish that the same short--range
$NN\to NN\pi$ vertex contributes to both $p$--wave pion production
and to low energy three--nucleon scattering, where the identical
operator plays a crucial role~\cite{hankol,evgeny3nf}. The
connection of pion production operators to three--body forces is
illustrated in Fig.~\ref{dill}.

Once this consistency has been established, we will be well placed
to calculate also isospin--violating pion production in $NN$
collision. With the measurement of a non--zero forward--backward
asymmetry in $pn\to d\pi^0$ at TRIUMF~\cite{opper} and of a
non--zero total cross section in $dd\to
\alpha\pi^0$~\cite{stephen}, there is a chance to establish a
connection between static isospin violation, as manifested in the
proton--neutron mass difference, and dynamical isospin violation,
which may occur in isospin--violating $\pi N$ scattering. This is
possible because the latter appears to be the formally leading
operator contributing to the above mentioned
reactions~\cite{ddatheo}. A direct measurement of
isospin--violating $\pi N$ scattering is, of course, hindered by
the absence of neutral pion beams.

\subsubsection{Details}

The starting point for an analysis of pion production is an
appropriate Lagrangian density, constructed to be consistent with
the symmetries of the underlying more fundamental theory (in this
case QCD) and ordered according to a particular counting scheme.
At leading and next--to--leading order, all but one  term can be
fixed from $\pi N$ scattering data. The only new structure can be
expressed as
\begin{eqnarray}
  -\frac{d}{f_{\pi}}
        N^{\dagger}(\boldtau\cdot\vec{\sigma}\cdot\vec{\nabla}\boldpi)N\,
        N^{\dagger}N\:,
\end{eqnarray}
\noindent where $f_\pi$ denotes the pion decay constant in the
chiral limit. This term describes an effective $NN\to NN\pi$
vertex, where the outgoing pion is in a $p$--wave and both the
$NN$ initial and final state are in an $S$--wave. Thus, only two
transitions are possible: \\
$T\!=\!0\to T\!=\!1$, \emph{viz.}\
$^3\!S_1-\,^3\!D_1\to\,^1\!S_0p$, which can be
studied in $pn\to pp\pi^-$;\\
$T\!=\!1\to T\!=\!0$, \emph{viz.}\ $^1\!S_0\to\,^3\!S_1p$, which
can be studied in $pp\to pn\pi^+$.

In order for the counting scheme to work, we require that
$\delta=(f_\pi^2M_N)d = {\cal O}(1)$. As will be shown below, this
order of magnitude is indeed consistent with the existing data
from $pp\to pn\pi^+$.

So far in the literature, calculations have been carried out up to
N$^2$LO for $p$--wave pion production within chiral perturbation
theory~\cite{hankol}, where evidence was presented that the
approach is indeed convergent. However, to be sure of this, a
calculation up to one order higher is needed, and this project is
currently under way~\cite{CH0}.\\[1ex]

\begin{figure}[htb]
\vspace{6.2cm} \includegraphics{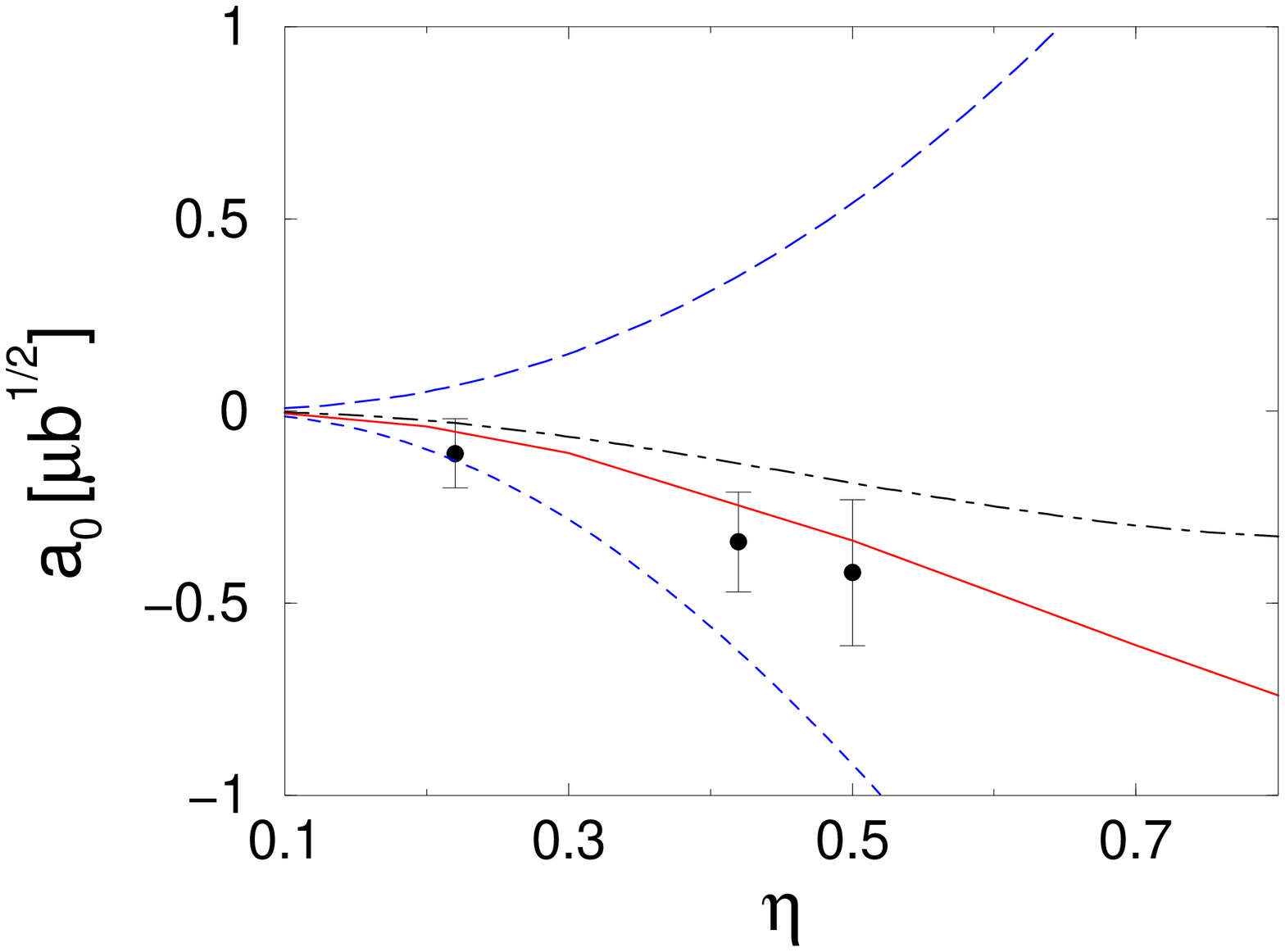} \caption{$a_0$ of  $pp \to
np\pi^+$ in chiral perturbation theory. The different lines
correspond to values of the parameter related to the
three--nucleon force: $\delta=1$ (long dashed line). $\delta=0$
(dot--dashed line), $\delta=-0.2$ (solid line), and $\delta=-1$
(short dashed line). Data are  from Ref. \protect\cite{flammang}.
} \label{a0pn}
\end{figure}

It is important to extract the parameter $d$ from experiment.  So
far this was done only for the reaction $pp\to pn\pi^+$---the
corresponding data is given in Ref.~\cite{flammang}.  As was
argued above, only the amplitude corresponding to the transition
$^1\!S_0\to\,^3\!S_1p$ (called $a_0$) is influenced by the
corresponding contact interaction. The results of the chiral
perturbation theory calculations are shown in Fig.~\ref{a0pn} for
different values of the parameter $\delta$.  Thus we find that the
results for $a_0$ are indeed rather sensitive to the strength of
the contact interaction.  The authors of Ref.~\cite{huebner} claim
that, with $\delta=-0.2$, the contact term gives an important
contribution to $A_y$ in $Nd$ scattering at energies of a few
MeV\footnote{The calculation of Ref.~\cite{huebner} suffers from
numerical problems.}

It turns out that the contribution of pion rescattering is very
sensitive to the regulator employed in the convolution of the
production operator and the final state wave function. This
scheme--dependence needs to be compensated by a counter term,
namely the $NN\to NN\pi$ contact term of interest here. It is thus
very important, to prove the consistency of the whole scheme, to
show that the same operator strength also contributes to
observables in other channels.

In Ref.~\cite{report} it was shown that the differential cross
section and analysing power for the reaction $pn\to pp\pi^-$ for
low $pp$ excitation energies is sensitive to an interference of
the $s$--wave  $A_{11}$ pion--production amplitude ($^3\!P_0\to
\,^1\!S_0s$) and the $p$--wave amplitudes of $A_{01}$,
\emph{viz.}\ $^3\!S_1\to \,^1\!S_0p$ and $^3D_1\to \,^1\!S_0p$.
Obviously, the four--nucleon contact interaction contributes to
both. Thus, once a proper chiral perturbation theory calculation
is available for the $s$--wave pion production, the reaction
$pn\to pp\pi^-$ close to the production threshold might well be
the best reaction from which to extract the parameter $d$. The
initial state is an isoscalar and thus the $\Delta$--nucleon
intermediate state does not contribute before pion emission.
Secondly, the leading $p$--wave amplitude is the one of interest,
in contrast to $pp\to pn\pi^+$, where $p$--wave pion production is
completely dominated by the transition $^1\!D_2\to \,^3\!S_1p$
involving production through the $\Delta$.

\begin{figure}[tb]
\vspace{6.2cm} \includegraphics{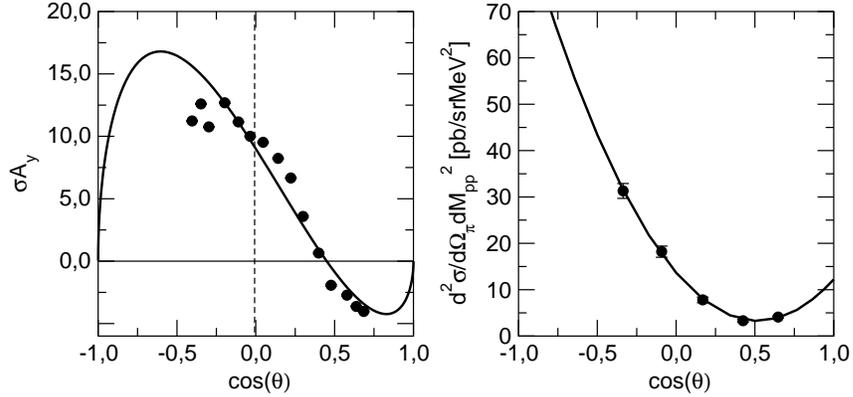} \caption{Analysing power and differential
cross section of the $np\to
  pp\pi^-$ reaction at T$_{Lab}$ = 353{\ww}MeV ($\eta = 0.65$) and $E_{pp}<3${\ww}MeV.
The experimental data are from Ref.~\protect{\cite{triumf1}} and
\protect{\cite{triumf2}}. The curves are polynomial fits up to
second order in the pion momenta.
 }
\label{pppimi}
\end{figure}

The goal of the proposed COSY measurement is to provide the
missing observables needed to extract the amplitudes for the
transitions $^3\!S_1-\,^3\!D_1\to \,^1\!S_0p$ unambiguously, where
the final $pp$ state is isolated by putting a tight cut on the
$pp$ excitation energy $E_{pp}$. The TRIUMF data, shown in
Fig.~\ref{pppimi}, are consistent with the assumption that at
$T_{lab}=353${\ww}MeV only terms up to quadratic in the outgoing pion
momentum need to be kept in the expressions. This implies that, in
addition to the amplitudes of interest, three additional ones have
to be considered. These, which all relate to the isospin--1
initial state, correspond to $^3\!P_0\to \,^1\!S_0s$,
$^3\!P_2\to\, ^1\!S_0d$, and $^3\!F_2\to \,^1\!S_0d$.

To extract the two $p$--wave amplitudes, nine independent
observables are required, of which the TRIUMF data provides five
(here each angular structure is counted as an individual
observable). Now, a recent measurement at Uppsala found a sizable
pion $d$--wave contribution, even quite close to the production
threshold~\cite{Jozef}, and these data give the values of another
two observables. At least two further measurements are required
and it would be preferable to measure more in order to eliminate
discrete ambiguities and improve the statistical and systematic
precision, especially when deuterium targets are employed. In
addition to repeating some of the earlier measurements, we now
consider the following possibilities for
finding new observables at ANKE:\\
1: $A_y (\vec{p}p\to pp\pi^0)$,\\
2: $A_{yy}(\vec{p}\vec{n}\to pp\pi^-)$,\\
3: $A_{y}(\vec{p}n\to pp\pi^-)$,\\
4: $A_{yy}(\vec{p}\vec{p}\to pp\pi^0)$.

\subsubsection{Experimental considerations}
\begin{figure}[hbt]
\input epsf
\begin{center}
\epsfig{file=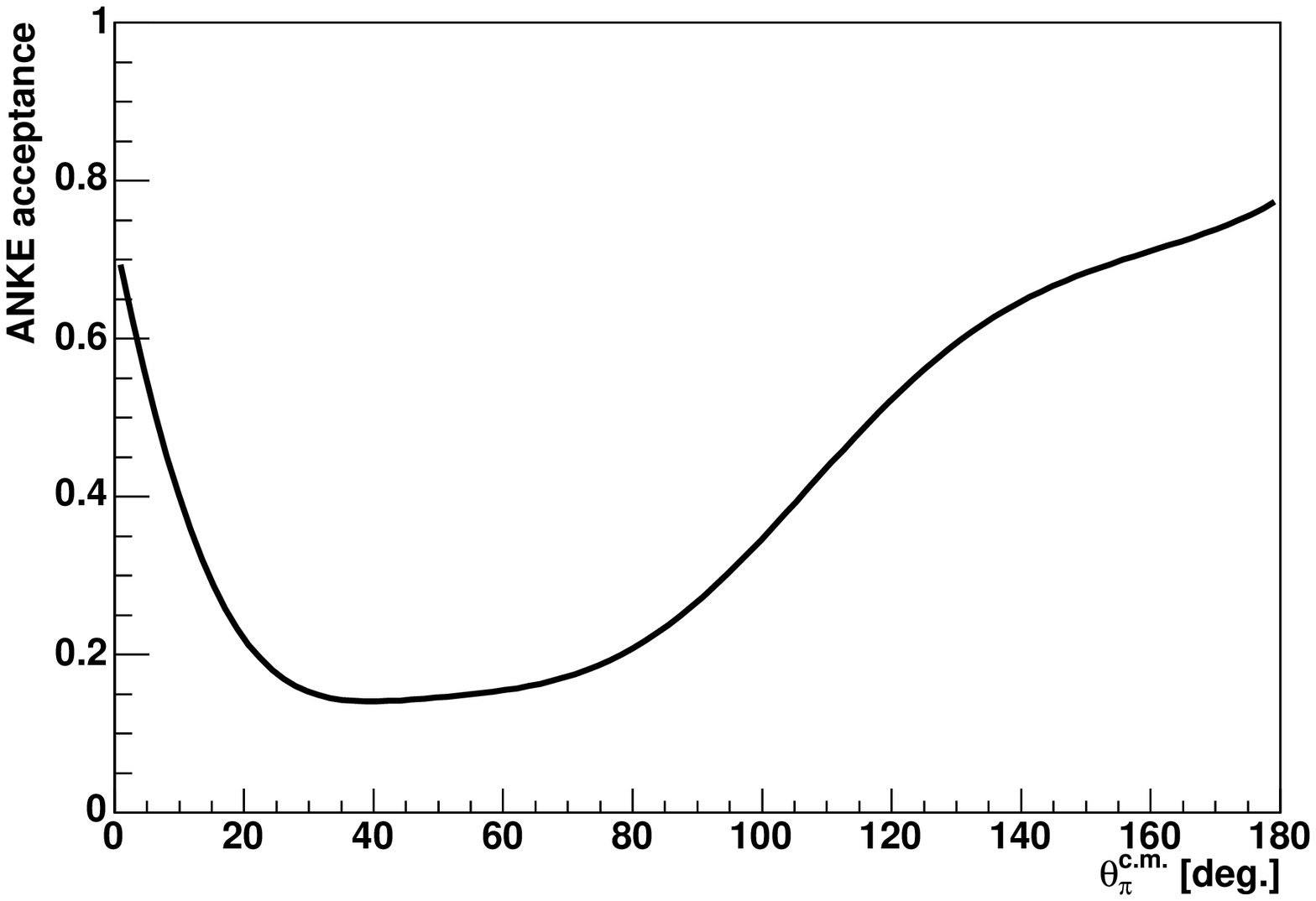, height=5cm} \caption{Predicted ANKE
acceptance as a function of the $\pi^0$ polar
angle.}\label{ThetaPiAcc}
\end{center}
\end{figure}

Simulations of the $pp \to (pp) \pi^0$ reaction were undertaken at
a beam energy of $T_p = 353${\ww}MeV, which was the same as in the
TRIUMF experiment~\cite{triumf1,triumf2}. At this energy, proton
pairs with small excitation energy ($E_{pp}<3${\ww}MeV) can be
registered in the ANKE positive side detector system. These
protons typically have momenta around 400{\ww}MeV/c and for these a
2$\%$ momentum and a $0.4^{\circ}$ angular resolution were assumed
in the simulation. From the acceptance of ANKE as function of the
$\pi^0$ polar angle shown in Fig.~\ref{ThetaPiAcc}, it is seen
that there are no blind spots in the angular distribution.

The missing mass resolution is expected to be about 5.5{\ww}MeV (RMS)
and this will allow one to distinguish unambiguously the pion
production reaction from any background.

\begin{figure}[hbt]
\input epsf
\begin{center}
\epsfig{file=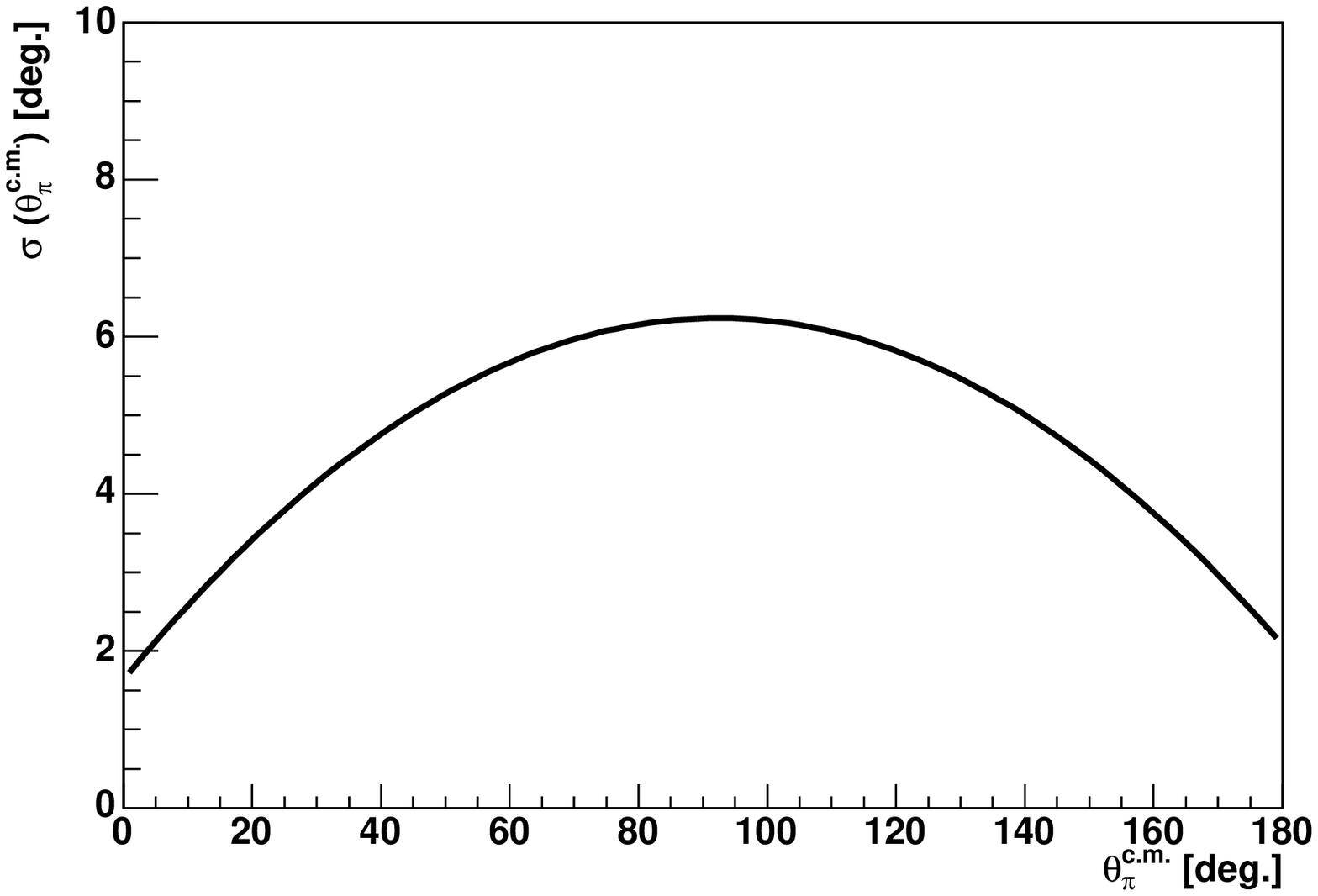, height=5cm} \caption{Predicted
resolution in the pion cm polar angle.}\label{ThetaPiRes}
\end{center}
\end{figure}

In Fig.~\ref{ThetaPiRes} the resolution pion cm polar angle is
shown as a function of this angle. Given that so few partial waves
are expected and that both the cross section and analysing power
vary smoothly with $\theta_{\pi}^{cm}$, this resolution is quite
sufficient for the purpose.

\begin{figure}[hbt]
\input epsf
\begin{center}
\epsfig{file=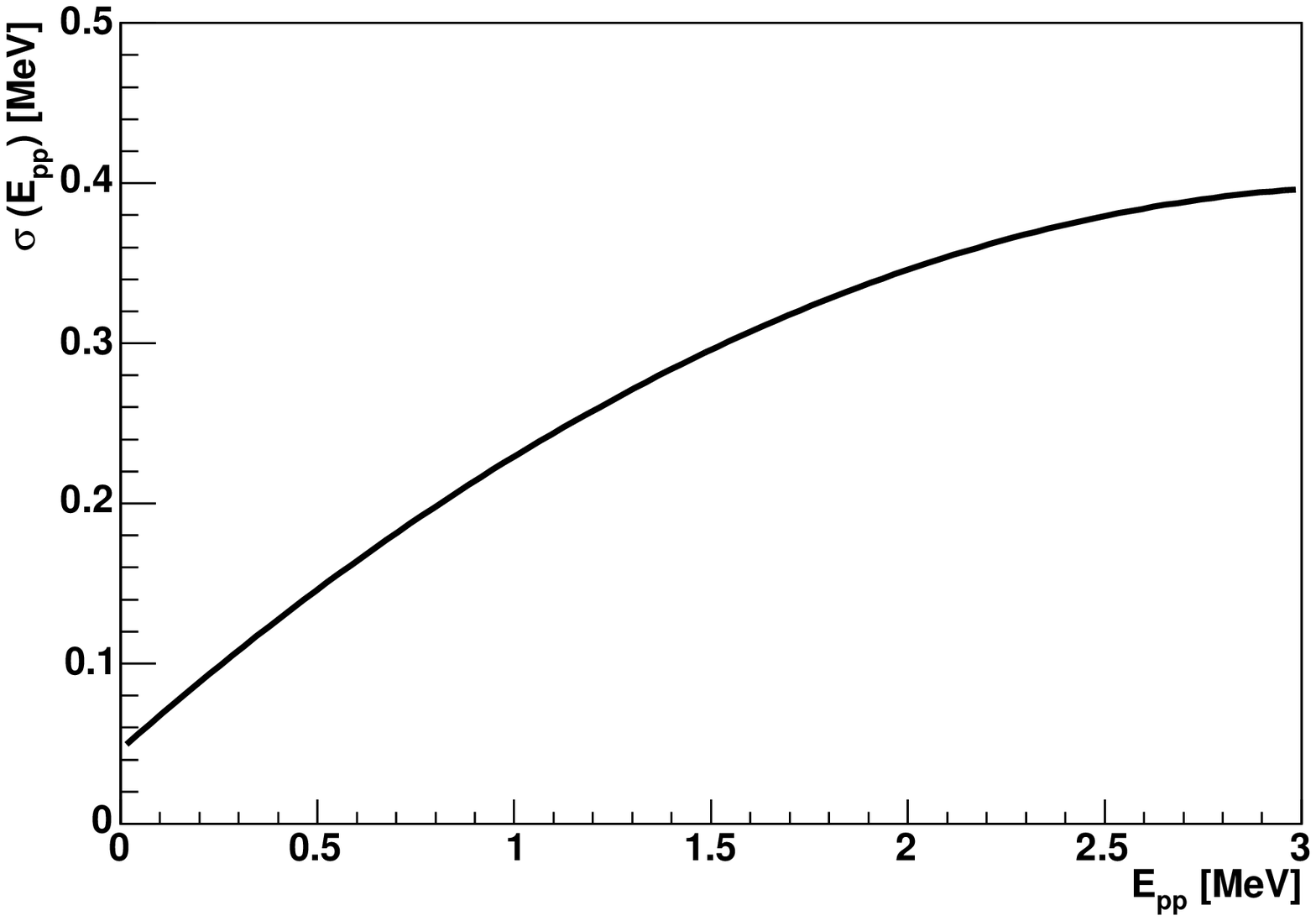, height=5cm} \caption{Predicted resolution
in the $pp$ excitation energy.}\label{EppRes}
\end{center}
\end{figure}

In order to ensure that the final proton pair is in the $^1\!S_0$
state, it is important to put a tight cut on their excitation
energy, to be below say 3{\ww}MeV. As shown in Fig.~\ref{EppRes}, the
resolution in excitation energy is better than $0.4${\ww}MeV at 3{\ww}MeV,
though this is more vital for the cross section than the
asymmetries.


The experimental conditions within the magnetic spectrometer for
measuring the quasi--free $pn \to (pp) \pi^-$ reaction on a
deuterium target are rather similar to those appertaining to the
$\pi^0$ production. The $\pi^-$ does not need to be detected but
the overall rate is reduced because of the acceptance of the
silicon counters for the spectator proton, as described in
\S\ref{SST}. It must also be recognised that the Fermi momentum of
the struck neutron spreads the cm energy in the initial
proton--neutron system over 10's of MeV so that some binning of
the results will be required in the analysis.

\begin{figure}[ht!]
\begin{center}
\epsfig{file=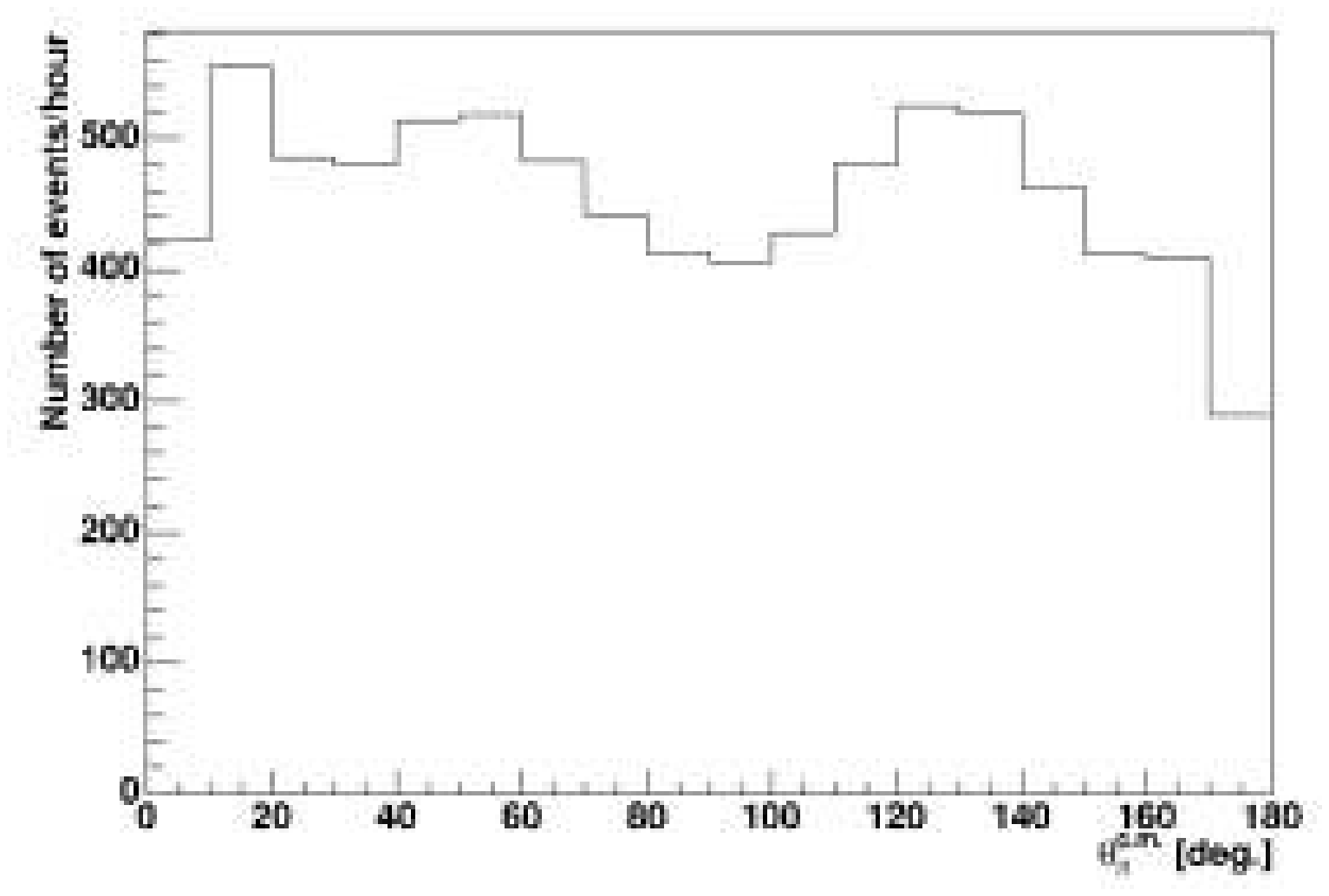, height=6cm} \caption{Predicted
counting rates for the $pn\to pp\pi^-$ reaction at 345{\ww}MeV for a
luminosity of $10^{30}$. These must be multiplied by the
acceptance of the spectator counters.} \label{CR3}
\end{center}
\end{figure}

It is therefore clear that both $\pi^0$ and $\pi^-$ production in
the 350{\ww}MeV region can be well studied in ANKE over the full range
of pion angles. The energy resolution is good enough allow one to
select final proton pairs with small excitation energy. The
counting rates in the case of the polarised target still have some
uncertainty but, if we assume a luminosity of $10^{30}$ together
with values of the $np\to pp\pi^-$ cross sections at
345{\ww}MeV~\cite{Daum}, we obtain the counting rates shown in
Fig.~\ref{CR3}.\\

These counting rates have not yet incorporated the reduction due
to the acceptance of the silicon telescopes. With the extended
system under discussion, this factor should be of the order of
20\%. From these estimates it is clear that statistical precision
should not pose a serious problem.

\begin{figure}[ht]
\centering
\includegraphics[height=10.0cm,angle=-90]{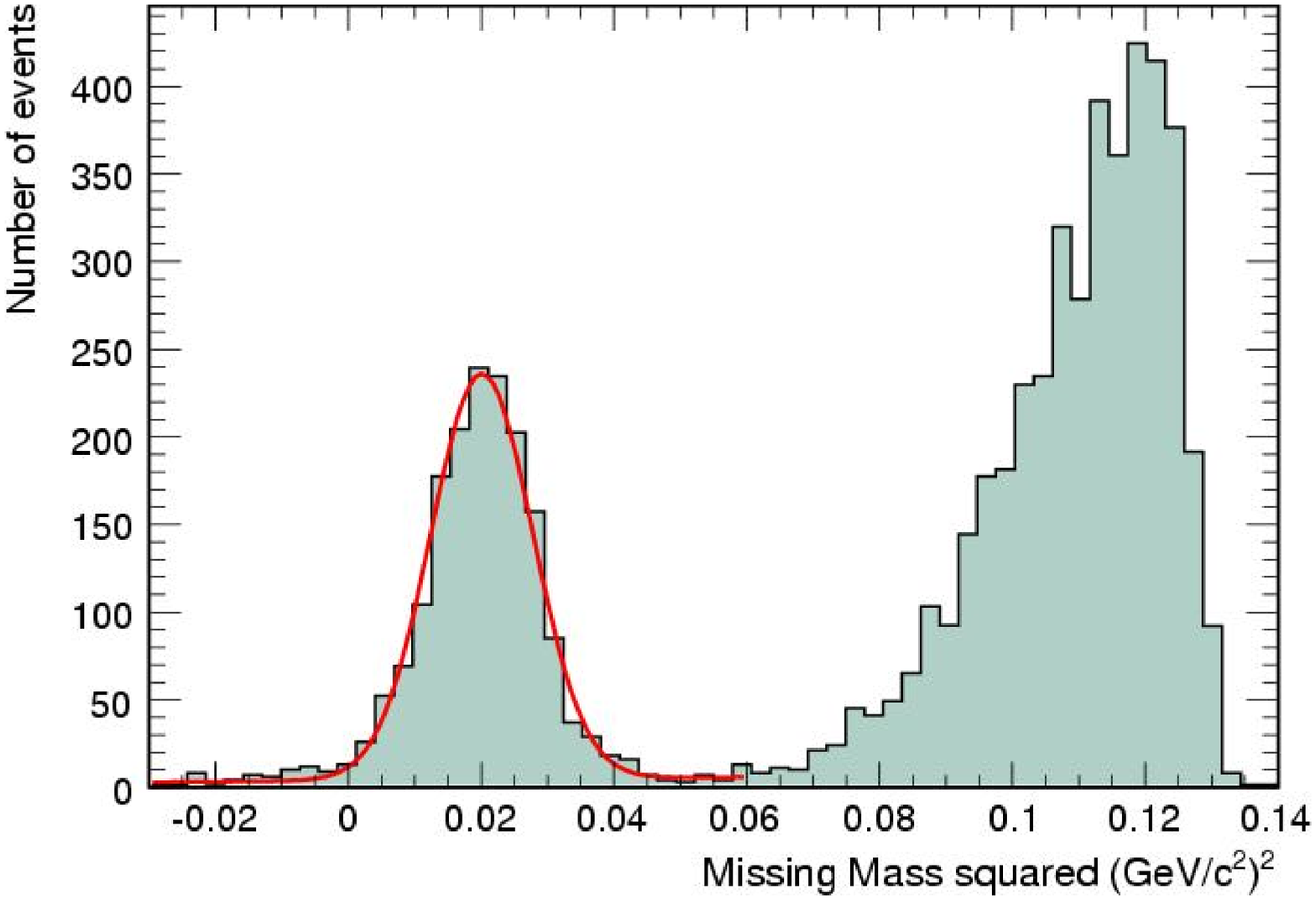}
\caption{Preliminary data on the $pp\to ppX^0$ reaction in the
forward direction at 800{\ww}MeV where the $pp$ excitation energy is
selected to be below 3{\ww}MeV. The tracks from both protons have here
been detected in the same counter.}\label{Kurbatov}
\end{figure}

Test data on the $pp\to pp\pi^0$ reaction were taken during the
same runs as for the $pd\to (pp)n$ reaction~\cite{Komarov}. The
preliminary analysis of the data at 800~MeV are shown in
Fig.~\ref{Kurbatov} with the same diproton selection
$E_{pp}<3${\ww}MeV~\cite{Kurbatov}. It should be stressed that in this
analysis only events where both tracks hit the same counter have
so far been retained, thus demonstrating that the missing mass is
sufficient for the clean identification of the $pp\to pp\pi^0$
reaction, despite the cross section being over two orders of
magnitude smaller than for $pp\to d\pi^+$. When the tracks are
observed in different counters, in addition to missing mass, we
have the information about ionisation losses and the difference in
time of flight. Such events will therefore be identified with even
greater reliability.
In addition to the well separated $\pi^0$ peak, there is clear
evidence for $(\pi\pi)^0$ production, for which the threshold is
at 0.073{\ww}(GeV/c$^2$)$^2$.

Since data corresponding to $pd\to pp X$
were also taken, we should be able to extract also the cross
sections for $pn\to (pp)\pi^-$.
Given that there are only two spin amplitudes for either $pp\to
pp\pi^0$ or $pn\to pp\pi^-$, provided that the final diproton is
constrained to be in the $^1\!S_0$ state, measurements of the
analysing powers and spin correlations are sufficient for a
complete amplitude analysis.
%
%
\subsection{$\vec{n}\,\vec{p}\to d\,X$}
\label{npdX}
%
%
\subsubsection{$\vec{n}\,\vec{p}\to d\,\pi^0$}
\label{npdpi}

It is doubtful whether a $pp\to d\pi^+$ reaction, taking place in
the PIT, could be identified purely by detecting the fast deuteron
but, as shown in Fig.~\ref{accep}, the $d\pi^+$ coincidence
generally falls outside the ANKE acceptance. However, the bands
corresponding to the quasi--free $\vec{n}p\to d\pi^0$ reaction at
585{\ww}MeV per nucleon are clearly seen in the polarised deuteron
test data of Fig.~\ref{npdpi0}. The ranges of cm angles covered,
which are illustrated in Fig.~\ref{kinem}, change only slowly with
beam energy. Not knowing the vertex in the PIT with such high
precision, the accuracy of the reconstruction will not be as good
as that shown in Fig.~\ref{npdpi0} but should still be sufficient
for measuring the reaction with polarised beam and target. At low
energies the very extensive data base on $pp\to
d\pi^+$~\cite{SAID} would allow useful checks on the systematics
of measurements with a polarised target but at high energies such
experiments would add new information to the World data set.

Now the unpolarised $pn\to d\pi^0$ reaction has also been studied
at around 556{\ww}MeV using a deuterium target at ANKE~\cite{Lehmann}.
The fast deuteron was measured in the Forward Detector and the
spectator proton in a prototype of the silicon telescopes
described in \S\ref{SST}.

\begin{figure}[htb!]
 \begin{center} \epsfig{file=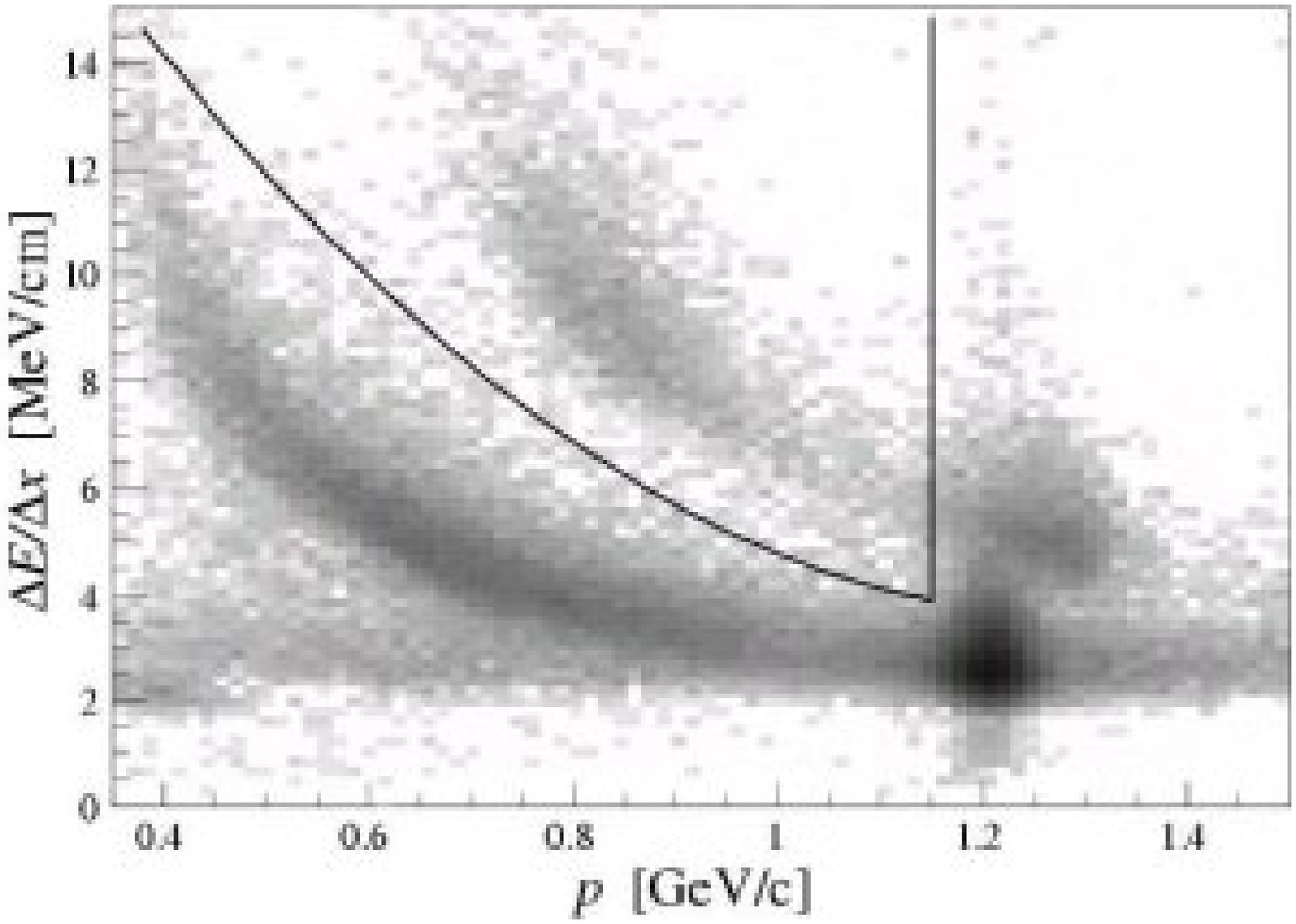,width=6cm,angle=270}
 \caption{The normalised energy loss per centimetre for particles in
 the first hodoscope layer of the ANKE forward detector {\em vs}
 their measured momentum. Clearly visible are the upper and lower
 bands originating from deuterons and protons respectively. Note that
 the entries are shown on a logarithmic scale of greyness so that the
 statistics for protons are orders of magnitude greater than for
 deuterons. The lines indicate the cuts applied to select deuterons
 shown in Fig.~\protect\ref{INTI}}
 \label{f:d_id}
 \end{center}
\end{figure}

Energy losses of particles in the first
plane of the scintillator hodoscope of ANKE are plotted
\textit{vs} their reconstructed momenta in Fig.~\ref{f:d_id}. The spectrum is
dominated by the proton peak around 1.17$\,$GeV/c corresponding to
small--angle deuteron break--up events. However, there are also
clear proton and deuteron bands and, by imposing a
momentum--dependent threshold between them, one can reduce the
proton contribution significantly.  Since the Landau tail from the
quasi--elastic protons cannot be suppressed very effectively at
high momenta, only the range below 1.15$\,$GeV/c was selected to
extract the $pn\to d\pi^0$ cross section, as indicated in
the figure.

\begin{figure}[ht!]
\begin{center}
\epsfig{file=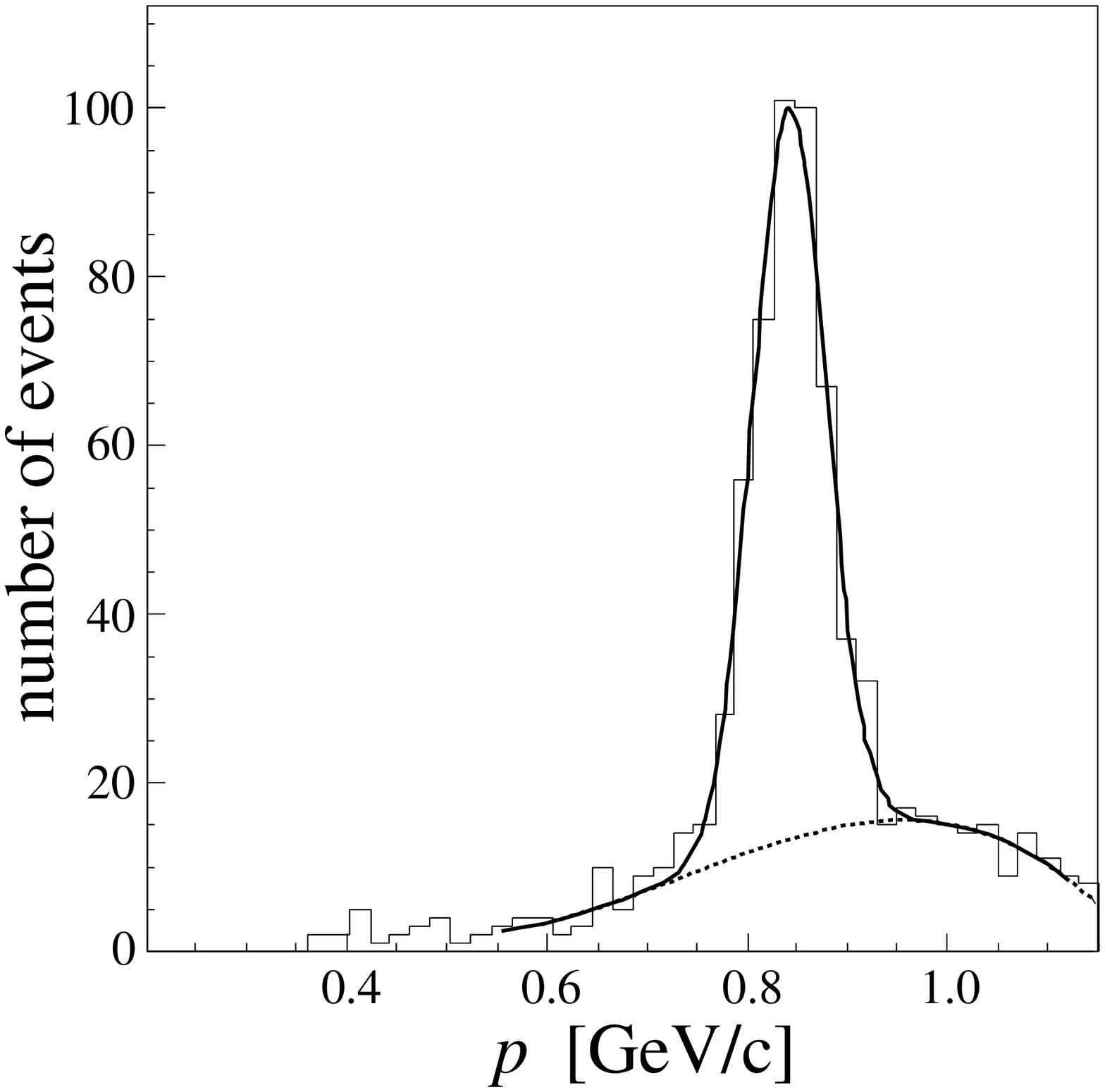, height=6cm} \caption{Deuteron momentum
distributions obtained using spectator protons with kinetic
energies in the range $2.6\leq T_\mathrm{sp} \leq 4.4\,$MeV
together with energy losses in the hodoscope for proton
suppression.} \label{INTI}
\end{center}
\end{figure}

The experimental momentum distribution is compared in
Fig.~\ref{INTI} to a Gaussian fit for the $d\pi^0$ events plus a
polynomial \emph{ansatz} for the background. Since the experiment
was carried out very close to the two--pion threshold, the
background must arise almost entirely from protons misidentified
as deuterons. This is the major problem of this type of
experiment.

A total cross section of $\sigma_\mathrm{tot}(pn\to
d\pi^0)=(1.62\pm 0.14)\,$mb was deduced from this experiment at an
effective mean beam energy of $T_\mathrm{beam} = 556\,$MeV. A
direct measurement of this cross section with a neutron beam at
this energy gave $\sigma_\mathrm{tot}(np\to d\pi^0)=(1.6\pm
0.27)\,$mb~\cite{Wil71}. This shows that, even with the modest
spectator counter system, we can be more than competitive with the
results obtained with neutron beams. Because the spectator proton
might give more information regarding the reaction vertex, this
approach looks promising for the measurement of
$\vec{p}\,\vec{n}\to d\pi^0$ with polarised beam and target.

%
%
\subsubsection{$\vec{n}\,\vec{p}\to d\,\pi\pi$}
\label{npd2pi}

Though the importance of the coupling of the $I=1$ $NN$ and
$N\Delta$ systems has been stressed earlier, the coupling to the
$\Delta\Delta$ channel should be significant for the $I=0$
observables~\cite{Garcilazo}. Experimental studies here
necessarily involve two--pion production.

The most prominent feature of two--pion production in nuclear
reactions is the so--called \emph{ABC} effect, which was first
detected as a sharp ($\Gamma\approx 50${\ww}MeV/c$^2$) $I=0$ $s$--wave
enhancement in the two--pion spectrum at a mass $m_{\pi\pi}\approx
310${\ww}MeV/c$^2$ in the reaction $pd\to\,^3$He$\,X^0$~\cite{ABC}.
Even sharper structure was seen in the $dd \to \,^4$He$\,X^0$
reaction, where the forward and backward (cm) peaks completely
dominate the spectrum~\cite{Banaigs2}. Since the positions and
widths of the peak tend to change with the kinematical conditions,
it was long realised that this was not a $\pi\pi$ resonance but
rather some dynamical effect. The cross section and deuteron
analysing powers in the $dd \to \,^4$He$\,X^0$ case~\cite{Ralf}
could all be well explained quantitatively in a model where there
was independent excitation of two $\Delta(1232)$
isobars~\cite{GFW2}. The \emph{ABC} structure then arises through
the $p$--wave nature of the $\Delta$ decay, where the pions tend
to be emitted forward or backward. When they come out together one
gets a low mass peak whereas if they emerge back--to--back there
is a rather broader enhancement at maximum missing mass, which is
also clearly seen in the data~\cite{Banaigs2}.

\begin{figure}[ht]
\begin{center}
\subfigure[Differential cross section at an \underline{average}
neutron momentum of 1.88{\ww}GeV/c and laboratory detection angle of
0$^{\circ}$~\protect\cite{Plouin}. The dashed curve represents
phase space and the solid one a double--$\Delta$
estimate~\protect\cite{Mosbacher}.]
{\epsfig{file=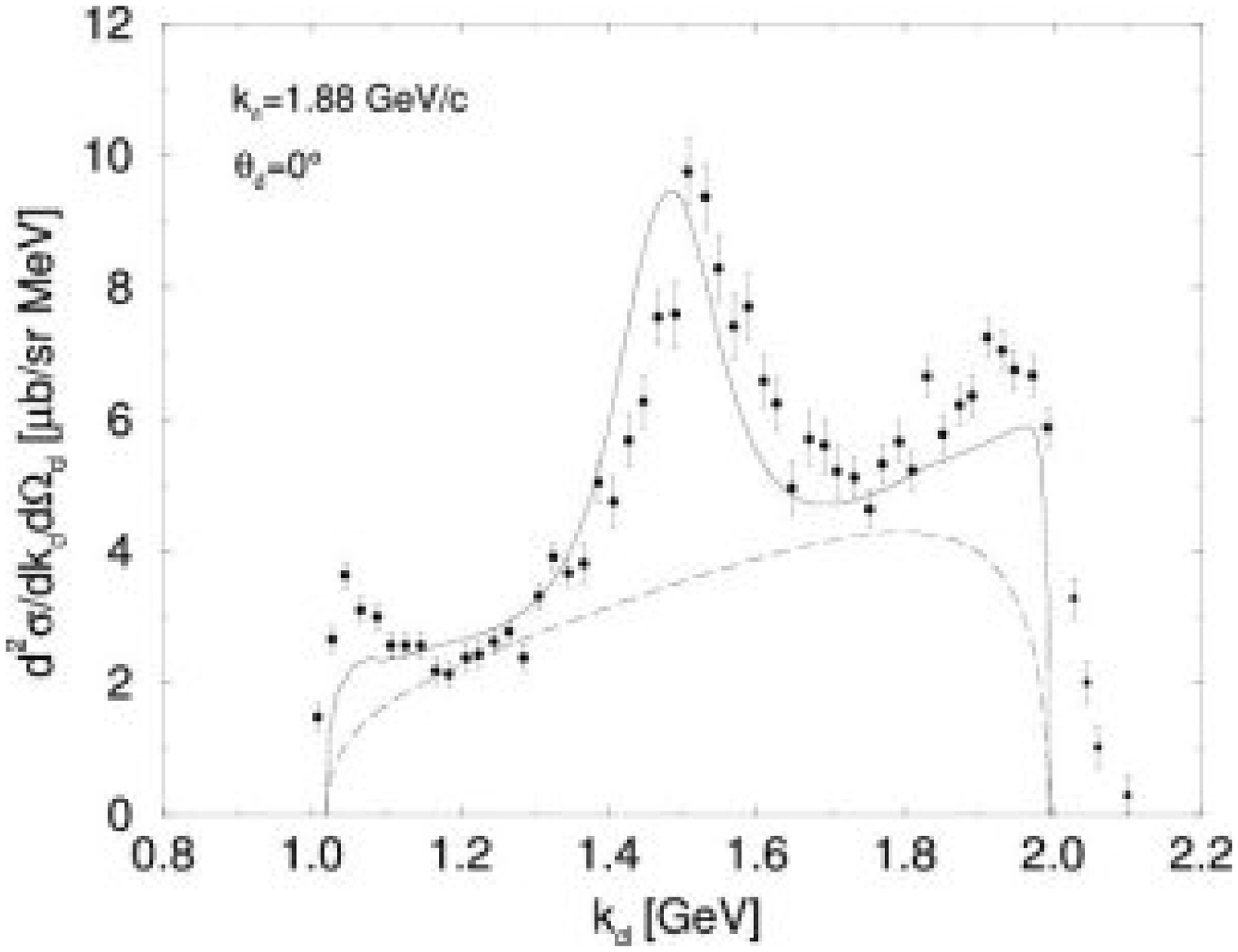,height=5cm} \label{g_3}}
  \subfigure[Bare double--$\Delta$ model]{
    \epsfig{file=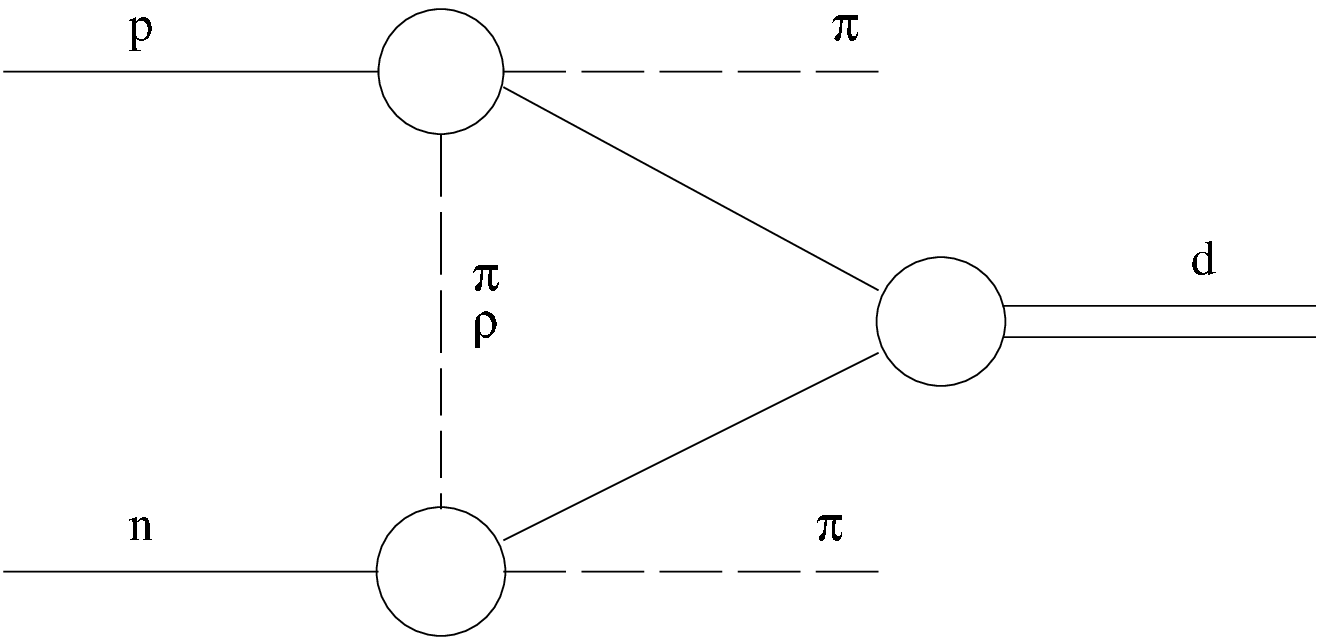,height=3cm} \label{g_6}}
  \caption{The ABC peaks and central bump in $np \to dX$.\label{NNABC}}
\end{center}
\end{figure}

Of most importance theoretically is the case of $np\to d\pi\pi$
since this is likely to be linked most directly to the
$\Delta\Delta$ degrees of freedom in the nuclear force. The
experimental data here are of much poorer quality, having been
produced through inclusive production on a deuterium
target~\cite{Vu_Hai} or with neutron beams~\cite{Plouin}. As a
result, though the low missing mass peaks of Fig.~\ref{NNABC}a are
clearly there, as is the central bump, they are a bit smeared out.

The bare double--$\Delta$ model of
Fig.~\ref{NNABC}b~\cite{Risser}, which has been refined to include
residual $\Delta\Delta$ and $\Delta N$
interactions~\cite{Mosbacher}, gives a fair description of the
experimental data at 1.88{\ww}GeV/c shown in Fig.~\ref{NNABC}a.
Deviations can be ascribed in part to the effects of smearing in
the incident neutron momentum. However, it should be noted that
this momentum corresponds to an average excitation energy of
$Q=511${\ww}MeV, which is quite close to $2(M_{\Delta}-M_N)\approx
586${\ww}MeV/c$^2$. On the other hand, at $Q=363${\ww}MeV, which is well
below the $\Delta\Delta$ threshold, the cross section is much
smaller and there is no sign of the \emph{ABC}
structure~\cite{Hollas}. The reaction at these lower energies may
be driven mainly by the Roper resonance that is excited on one of
the nucleons~\cite{Luis}.

\noindent

There are, as yet, no estimations available for the spin
dependence of double--pion production in either the
double--$\Delta$ or Roper models. For the excitation of a $0^+$
\emph{ABC} in the forward direction, there are two spin--dependent
amplitudes:
\begin{equation}
F(pn \to d\,ABC) = A\,(\bmath{\epsilon
}_{pn\parallel}
\bmath{\epsilon}_{d\parallel}^{\dagger} +
B\,(\bmath{\epsilon}_{pn\perp}\cdot
\bmath{\epsilon}_{d\perp}^{\dagger})\:, \label{FABC}
\end{equation}
where $\bmath{\epsilon}_{pn}$ represents the spin--1 combination
of the initial $pn$ system and parallel and perpendicular are with
respect to the beam direction. The magnitudes of the two
amplitudes can be separated by measuring the transverse spin
correlation~\cite{Wilkin1980}:
\begin{equation}
C_{NN}=\frac{|A|^2}{(|A|^2+2|B|^2)}\:,
\end{equation}
from which it can be seen that $C_{NN}$ can never be negative.

In the $\Delta\Delta$ model, the peak of the $pn\to d\,ABC$ cross
section comes in a region where there is relatively little
excitation energy with respect to $2M_{\Delta}$~\cite{Mosbacher}.
The main contributions are therefore likely to be dominated by
$L_{\Delta\Delta}=0$, in which case $S_{\Delta\Delta}=1$ or 3,
corresponding to initial angular momenta of $0$ and $2$. However
the transition from the $np$ to the $\Delta\Delta$ states depends
upon which meson exchanges dominate. The ratio $|B|/|A|$ will
therefore provide information on the spin dependence of the $NN\to
\Delta\Delta$ potential.

%
%
\subsection{The production of heavier non--strange mesons in $NN$
collisions}%
\label{heavier}
%
%
\subsubsection{Missing--mass experiments}
\begin{figure}[ht!]
\begin{center}
\epsfig{file=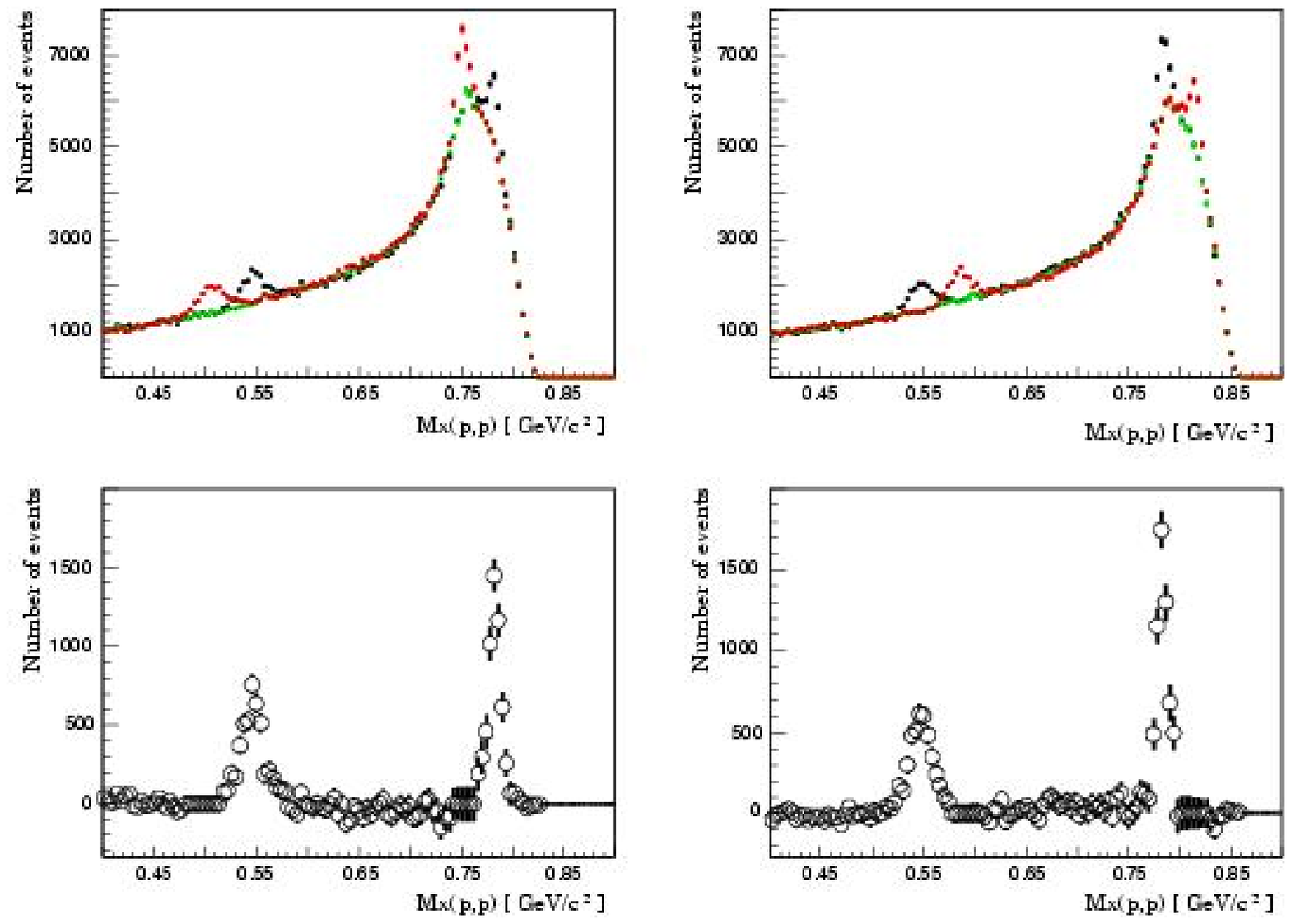, height=9cm} \caption{Missing--mass
spectra (black points) in proton--proton collisions at 2.85{\ww}GeV/c
(upper left panel) and 2.95{\ww}GeV/c (upper right panel). The
backgrounds (green) were estimated by using the data at the other
momentum, as described in the text. Once subtracted, these leave
the clear $\eta$ and $\omega$ peaks shown in the lower two
panels~\cite{SBarsov}.} \label{SBSB}
\end{center}
\end{figure}

The ANKE spectrometer is well suited to the measurement of
near--threshold production of mesons in $pp$ and, by using the
silicon telescopes, $pn$ collisions. Experiments are also possible
at higher energies though the phase--space coverage is then
restricted. As examples of the capabilities of the spectrometer,
we show in Fig.~\ref{SBSB} missing--mass results in the
$\eta/\omega$ region obtained at 2.85 and 2.95{\ww}GeV/c in $pp\to
ppX$, where one proton was measured in the forward detector and
the other in the positive side detector. These momenta correspond
to excess energies of $Q=60$ and $Q=92${\ww}MeV with respect to the
central mass of the $\omega$--meson and over 300{\ww}MeV for the
$\eta$~\cite{SBarsov}.

The $\omega$ missing--mass peaks at the two excess energies sit
close to the maxima in the multi--pion background and so a robust
treatment of this is necessary in order to extract the number of
$\omega$ counts. The shape of this background is largely
determined by the ANKE acceptance convoluted with a multi--pion
phase space. It varies little with beam momentum provided that the
curve is plotted with respect to the maximum missing mass allowed
at that beam energy. More quantitatively, it was shown in the
analysis of the SATURNE $pp\to pp\,\omega$ data~\cite{Hibou} that,
if the momenta of the final protons are kinematically transformed
from one beam momentum below the $\omega$ threshold to another
above, the background obtained in this way gives a very reliable
description of the multi--pion production under the $\omega$ peak.
Of course, when carrying out this procedure, the effects of the
relative luminosities have to be taken into account.

This method was subsequently used to extract signals for the
$pn\to d\,\omega$ reaction from quasi--free production on the
deuteron, where the spectator proton was detected in a prototype
silicon telescope~\cite{SBIL}.

The black points in Fig.~\ref{SBSB} show the total missing--mass
spectra at the two energies and, since below--threshold data were
not available, use had to be made of the fact that in the SATURNE
transformation only for the single--particle final states
$X=\eta,\,\omega$ does the peak change its position, as
illustrated by the red points. Taking the two data sets together,
there is sufficient information to remove the unwanted
single--meson peaks from the red background points to leave the
smooth green points. The subtraction of the green from the black
then resulted in the lower panels of Fig.~\ref{SBSB}, which shows
the $\eta$ and $\omega$ peaks with very little background. This
demonstrates that a model--independent approach to the treatment
of multi--pion background works quite well at ANKE, at least, in
the case of missing mass distributions obtained from the total
acceptance. For differential distributions in centre--of--mass
angle and excitation energy of the diproton, this method is more
sensitive to the details of the acceptance and this is currently
under investigation~\cite{SBarsov}.
%
%
\subsubsection{Exclusive measurements}

The decays of the $\eta$ and $\omega$ mesons lead to neutral
particles whose detection would require a spectrometer such as
WASA~\cite{WASA}. However, the $\phi$ meson has an almost 50\%
branching ratio to $K^+K^-$, which can be detected in the ANKE
system. Preliminary data on both $pp\to pp\,\phi$ and $pn\to
d\,\phi$ are available where kaon pairs were measured in
coincidence with fast protons and deuterons from hydrogen and
deuterium targets respectively~\cite{Yoshi}.

\begin{figure}[hbt]
\begin{center}
\subfigure[Data taken at 2.65{\ww}GeV ($Q=17${\ww}MeV)]
{\epsfig{file=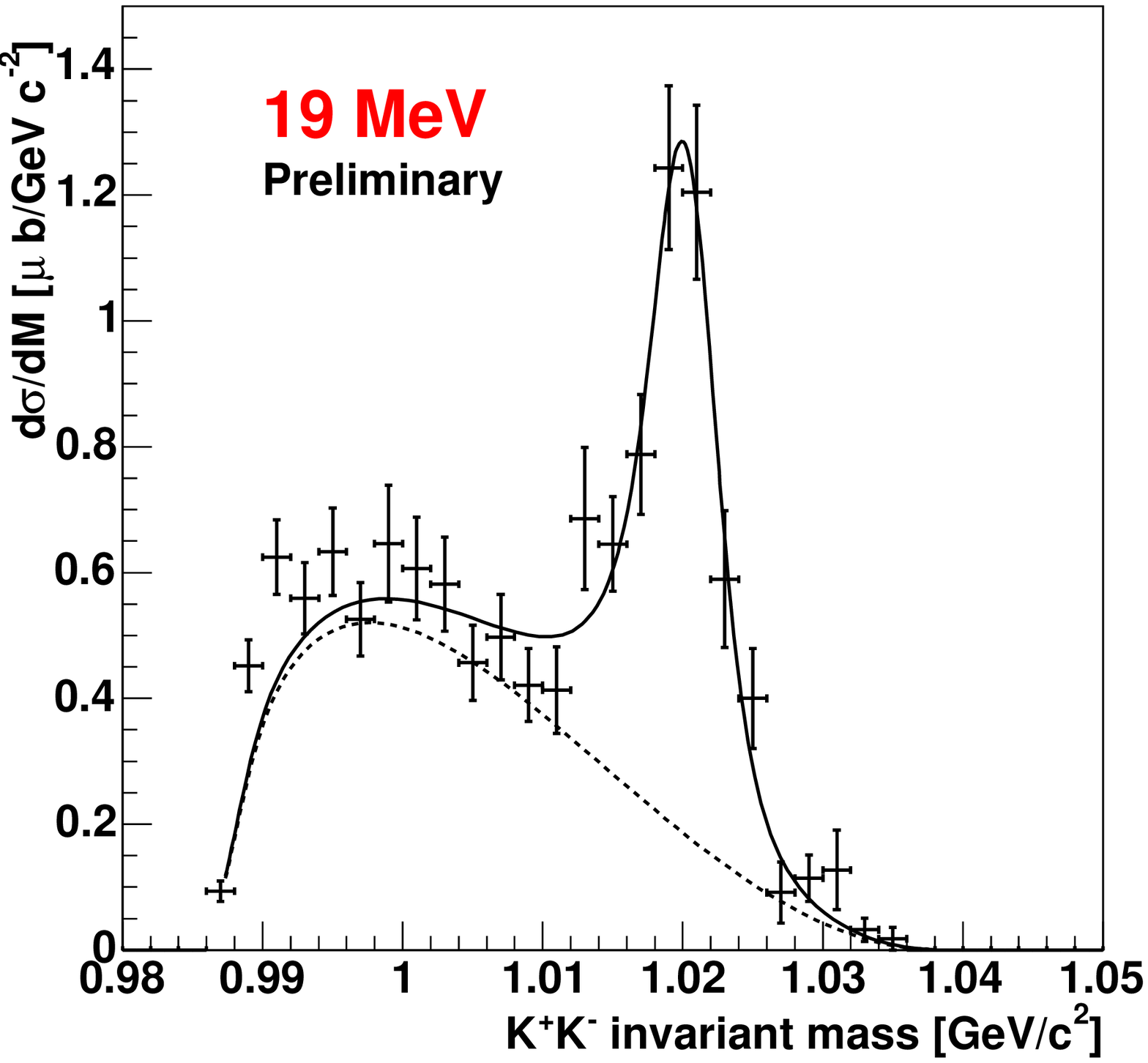,height=5.5cm}}%
\subfigure[Data taken at 2.83{\ww}GeV ($Q=76${\ww}MeV)]
{\epsfig{file=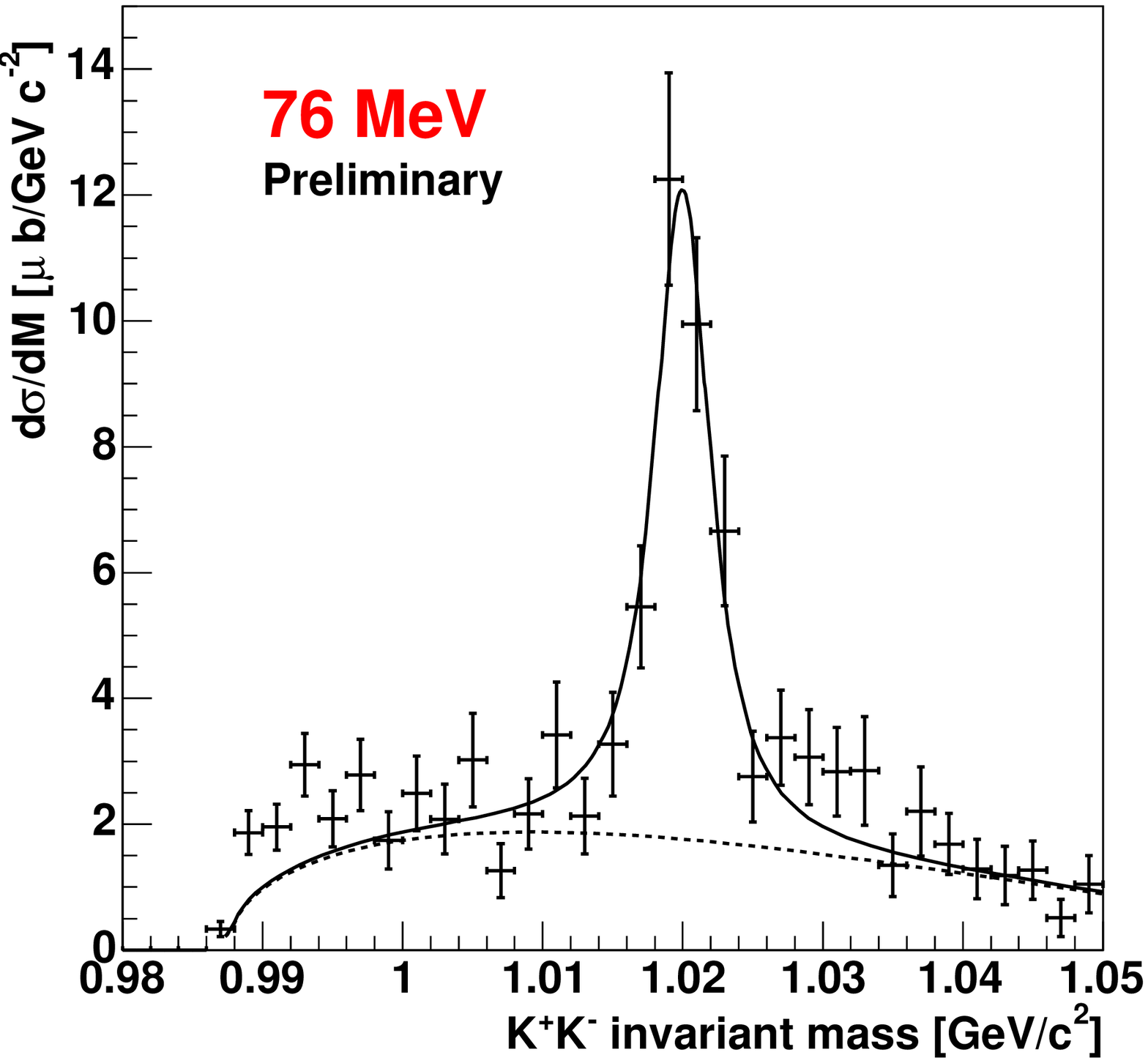,height=5.5cm}}%
\caption{$K^+K^-$ invariant mass spectra in the reaction $pp\to
ppK^{+}K^{-}$. The dotted curves show four--body phase--space
simulations of non--resonant $K^+K^-$ production, whereas the
solid ones include also contributions from $\phi$ production.
\label{Yoshipp}}
\end{center}
\end{figure}

From the $K^+K^-$ invariant mass spectra shown for the $pp$ case
in Fig.~\ref{Yoshipp} it is seen that the $\phi$ peak stands out
clearly from a background that might have its origins mainly in
the production of the much broader $a_0/f_0$ states.

\begin{figure}[hbt]
\begin{center}
\subfigure[$K^+K^-$ invariant mass distribution]
{\epsfig{file=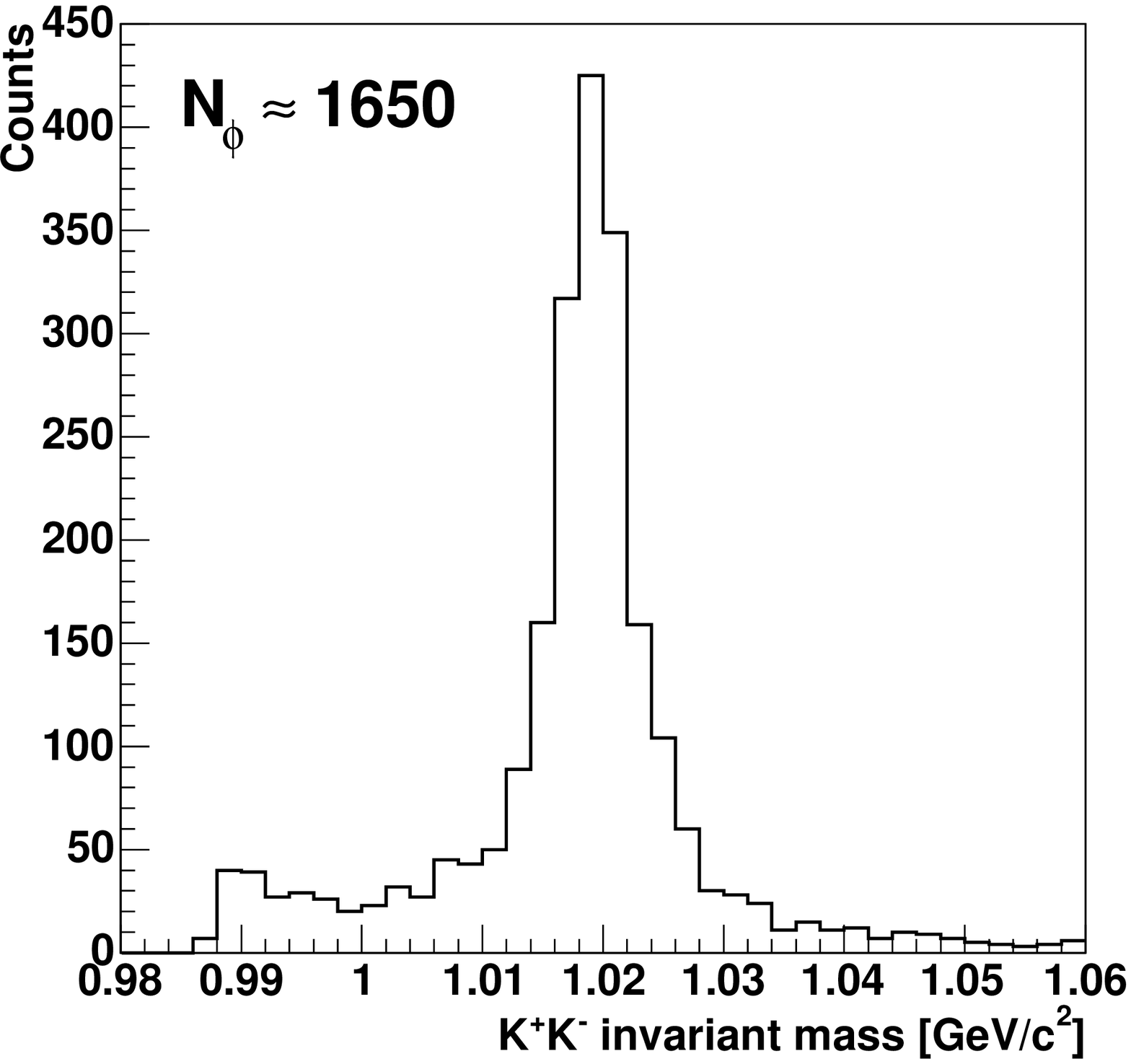,height=6cm}}%
\subfigure[Reconstructed proton spectator momentum distribution
compared to a simulation using the Bonn deuteron wave function.]
{\epsfig{file=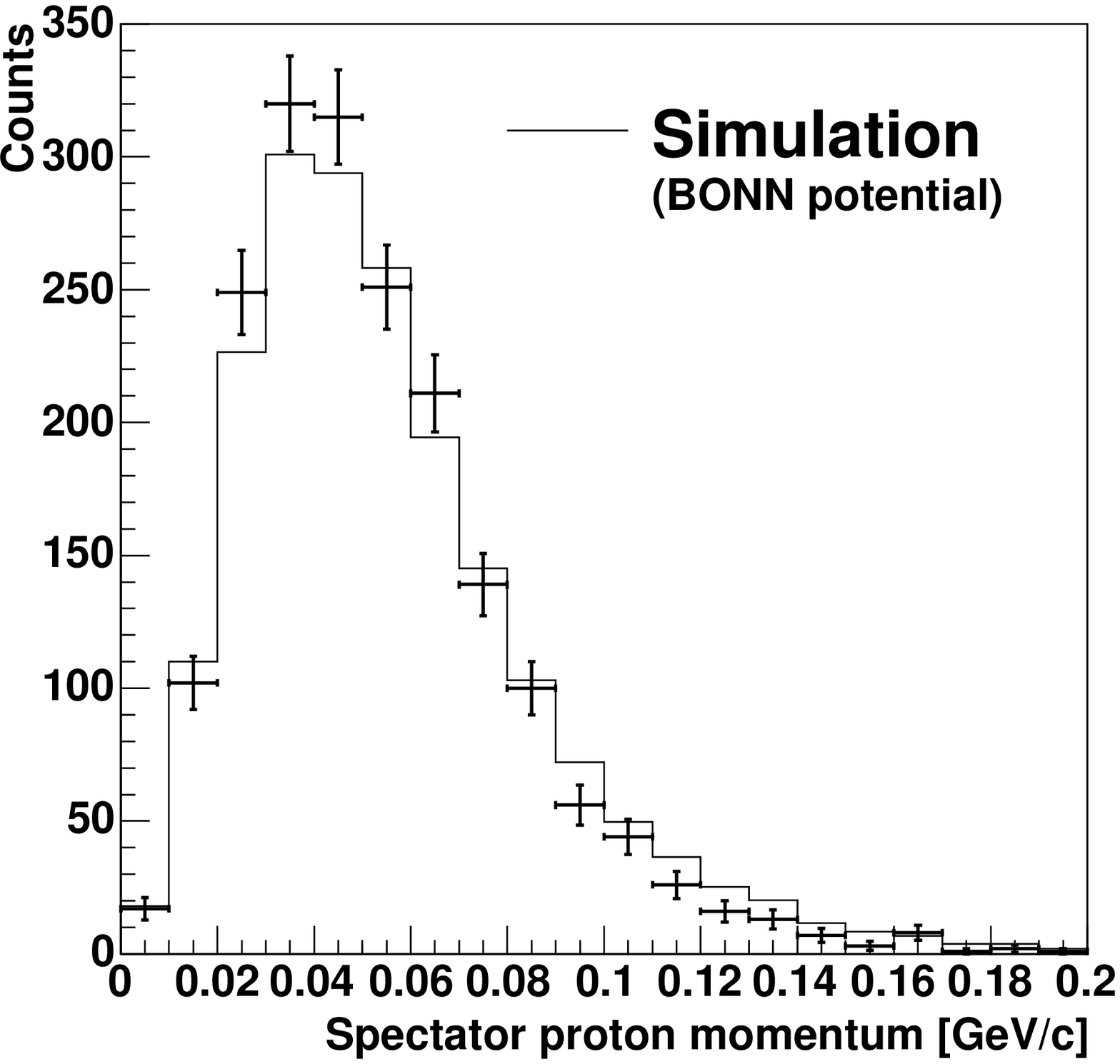,height=6cm}}%
\caption{Events corresponding to the $pd\to p_sd K^+K^-$ reaction
at a beam energy of 2.65{\ww}GeV.\label{Yoshipn}}
\end{center}
\end{figure}

The $pn\to d\phi$ reaction could be identified from $pd\to p_sd
K^+K^-$ data, where the momentum of the spectator proton ($p_s$)
was obtained by kinematically fitting the information from the
deuteron and kaon measurements. The spectator momentum
distribution shown in Fig.~\ref{Yoshipn}b agrees well with that
predicted from the deuteron wave function. The $K^+K^-$ invariant
mass spectrum of Fig.~\ref{Yoshipn}a shows a $\phi$ peak on a much
lower background than in the $pp$ case and this difference might
already contain information on the isospin dependence of $a_0/f_0$
production.

Though the beam energy was fixed at 2.65{\ww}GeV, the variation in the
magnitude and direction of $\bmath{p}_s$, allowed one to scan the
cross section in steps in excitation energy up to 80{\ww}MeV. This is
completely analogous to the CELSIUS extraction of the $pn\to
d\eta$ cross section where the $\eta$ decay into two photons was
used to provide the spectator momentum
reconstruction~\cite{Stina}.

\begin{figure}[hbt]
\begin{center}
\subfigure[Total cross section for $\phi$ production in $pp\to
pp\phi$ and $pn\to d\phi$ reactions as functions of the excess
energy. In addition to the ANKE measurements~\cite{Yoshi} the
DISTO result is also shown~\cite{DISTO}. The curves show the
behaviour expected from phase space.]
{\epsfig{file=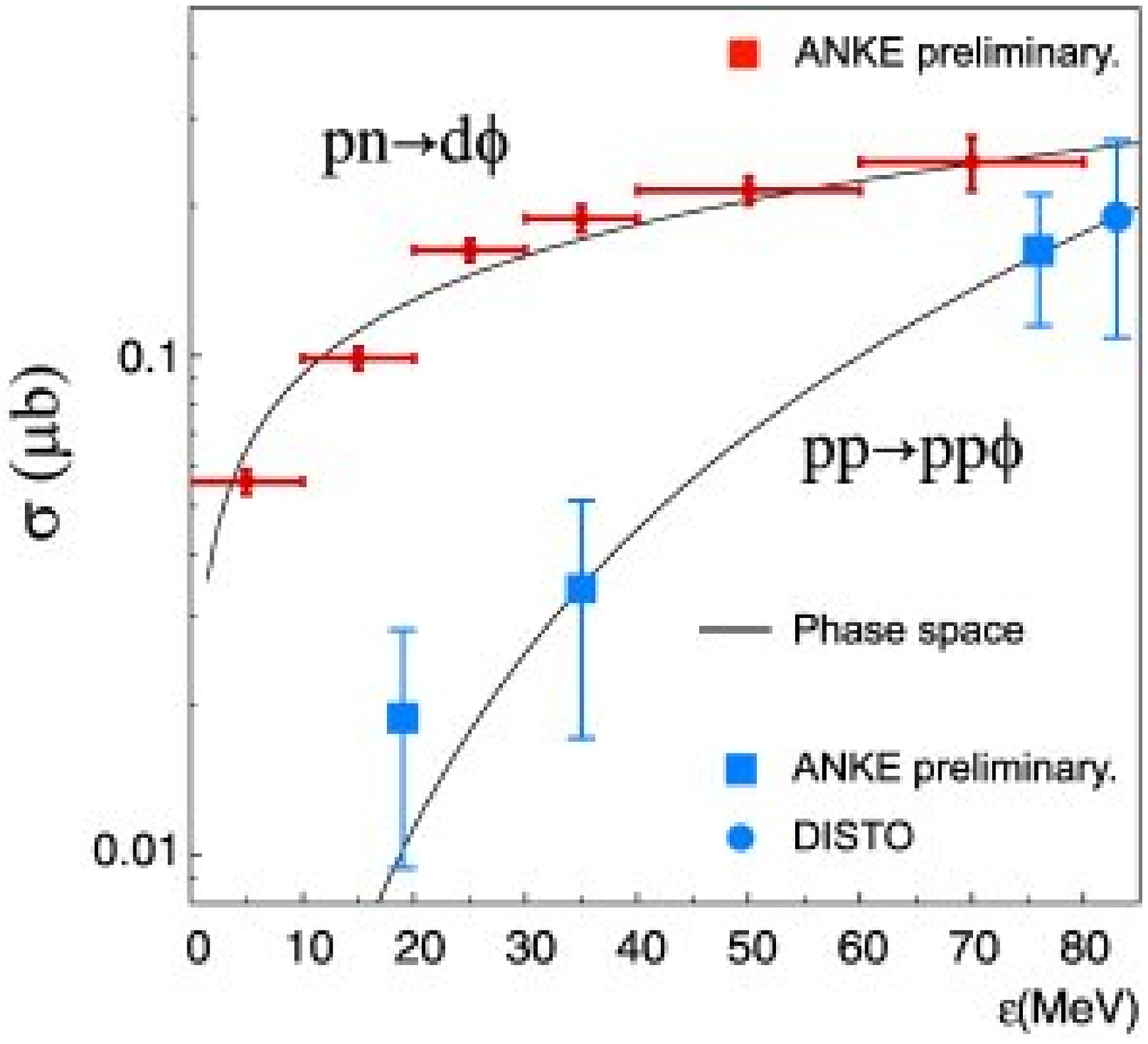,height=6cm}}%
\subfigure[The ratios of the total cross sections for $\phi$ and
$\omega$-meson production in $pp$ and $pn$ collisions. The ratio
is normalised using $R_{OZI}=4.2\times 10^{-3}$, which follows
from the naive OZI rule.]
{\epsfig{file=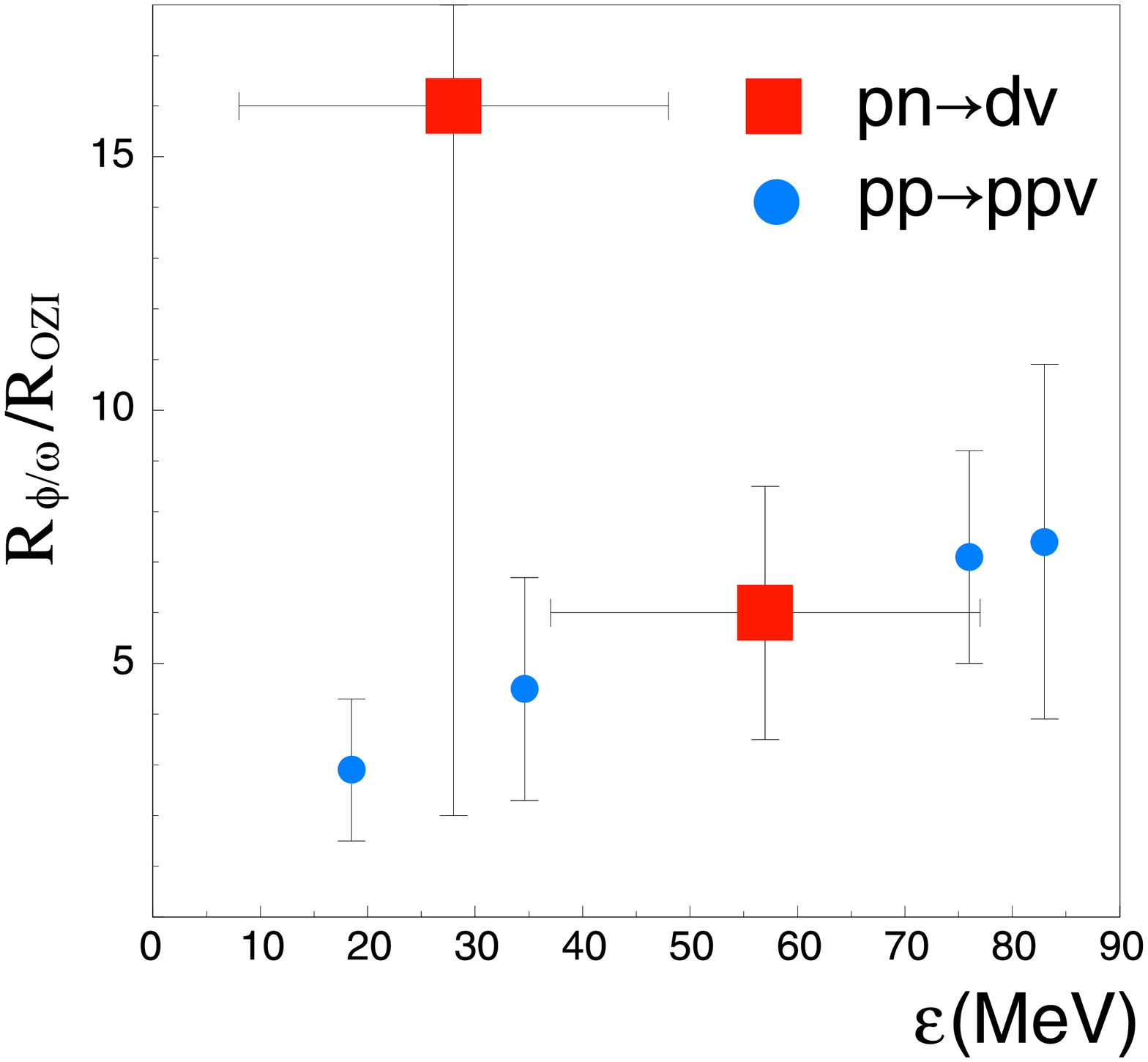,height=6cm}}%
\caption{Cross sections for vector meson production in
nucleon-nucleon collisions. \label{Yoshitot}}
\end{center}
\end{figure}

Preliminary values of the $pn\to d\phi$ and $pp\to pp\phi$ total
cross sections measured at ANKE are shown in Fig.~\ref{Yoshitot}
and compared to the energy dependence expected from phase space,
\emph{viz}.\ $\sqrt{Q}$ and $Q^2$ respectively. The ratio of these
cross sections is much less than that observed for $\eta$
production~\cite{Stina}, indicating that the isospin dependence is
much weaker for $\phi$ than $\eta$ production.

If the $\phi$ meson were an ideal mixture containing only strange
quarks then its production by hadrons containing no strangeness
would be forbidden by the Okubo--Zweig--Iizuka (OZI) rule, which
does not allow diagrams with disconnected quark lines~\cite{OZI}.
Deviations from ideal mixing are small and these suggest that,
under similar kinematic conditions, the ratio of single $\phi$ to
$\omega$ production should be about $4.2\times 10^{-3}$. Using
$\omega$ data from $pp\to pp\,\omega$~\cite{Hibou,TOFomega} and
$pn\to d\,\omega$~\cite{SBIL}, it is possible to quantify
deviations from the OZI rule in both the $pp$ and $pn$ channels.
As seen in Fig.~\ref{Yoshitot}b, there is some evidence that the
deviation from the OZI rule prediction increases slightly with
excess energy.

\subsubsection{Polarisation measurements}

\begin{figure}[ht!]
\begin{center}
\epsfig{file=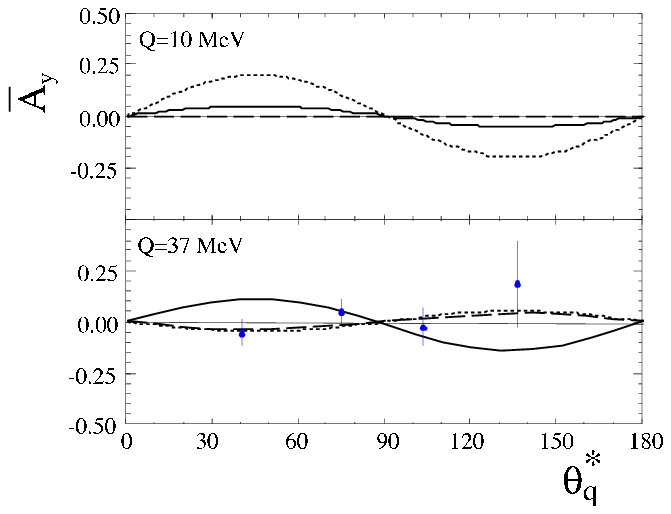, height=8cm} \caption{Proton analysing
power in the reaction $\vec{p}p\to pp\eta$ as a function of the
$\eta$ cm angle. The data~\cite{Winter} are compared to the
theoretical predictions of Ref.~\cite{FW5} (dotted curves) and
\cite{Nakayama} (solid and broken curves).} \label{Hiver}
\end{center}
\end{figure}

Very little is known about the spin dependence in the production
of mesons heavier than the pion. The only analysing power
measurement in $\vec{p}p\to pp\eta$ was carried out using the
COSY11 spectrometer~\cite{Winter}. The signal shown in
Fig.~\ref{Hiver} is not very strong and does not constrain
seriously the theoretical calculations~\cite{FW5,Nakayama}.
Clearly, far more detailed polarisation measurements are needed to
provide clues on the underlying dynamics.

At threshold only a single spin amplitude survives in each of the
$pp\to pp X$ ($X=\eta,\,\omega,\,\eta',\,\phi$) reactions and a
similar uniqueness follows for $pn\to d X$ and the isoscalar
channel in $pn\to pnX$. Under these conditions, measurements of
the values of the spin correlation parameter $C_{nn}$ would serve
little purpose because they are already fixed by spin--parity
constraints~\cite{Wilkinphi,Rekalophi}.
One has therefore to measure away from threshold in
order to provide extra tests on the reaction dynamics. Two obvious
candidates for such experiments in missing--mass mode are
$\vec{p}\,\vec{d}\to p_sd\,\eta/\omega$, with the spectator proton
measured in the silicon telescope providing the vertex
determination.

There are connections between $C_{nn}$ and the spin alignment
(tensor polarisation) of the produced vector meson that can be
measured through the angular dependence of one of the decay
products in $\omega\to \pi^+\pi^-\pi^0$ or
$\phi\to K^+K^-$~\cite{Wilkinphi,Rekalophi}. For
$\phi$ production through $pp\to pp\phi$ in the forward direction
at small $pp$ excitation energies the alignment determines
$C_{nn}$ completely, but this is an extreme case.

\newpage
%
%
\subsection{Meson production in $\vec{p}\,\vec{d}\to\,
^3\textrm{He}\,X$ reactions} \label{He3}

Because of the double charge on the helium nucleus, the $pd\to\,
^3\textrm{He}\,X$ reaction can often be cleanly isolated on a
large background. We here describe what one might learn from
measurements of $\pi$ and $\eta$ production with polarised beam
and target. However, this is far from being a complete list, and
it would also be interesting to study the excitation of the
\emph{ABC} enhancement of the two--pion
spectrum~\cite{ABC,Banaigs}, with the spin observables being
pinned down.
%
%

\subsubsection{$\vec{p}\,\vec{d}\to\,^3\textrm{He}\,\pi^0$}
\label{He3pi}

The data on the $\vec{d}\,p\to\,^3\textrm{He}\,\pi^0$ reaction
near threshold are remarkable in that for a pion cm energy of only
2.7{\ww}MeV ($p_{\pi} = 27${\ww}MeV/c) the ratio of the forward to
backward pion production is about a factor of
SIX~\cite{Pickar,Kerboul}. Since at threshold the cross section
must be isotropic, this indicates the influence of enormous
$p$--waves (as compared to the $s$--waves) coming in very quickly.
This was quantified in a more extensive Saclay experiment that
measured the deuteron analysing powers as well as the differential
cross section~\cite{Nikulin}. They found that near threshold
\begin{equation}
\frac{\dd\sigma}{\dd\Omega}\propto (1+\alpha\eta\cos\theta)\:,
\end{equation}
where $\theta$ is the cm angle between the incident proton and
final $\pi^0$, $\eta=p_{\pi}/m_{\pi}$ is the pion momentum in pion
mass units, and the parameter had the value $\alpha\approx 4.1$.

\begin{figure}[hbt]
\input epsf
\begin{center}
\epsfig{file=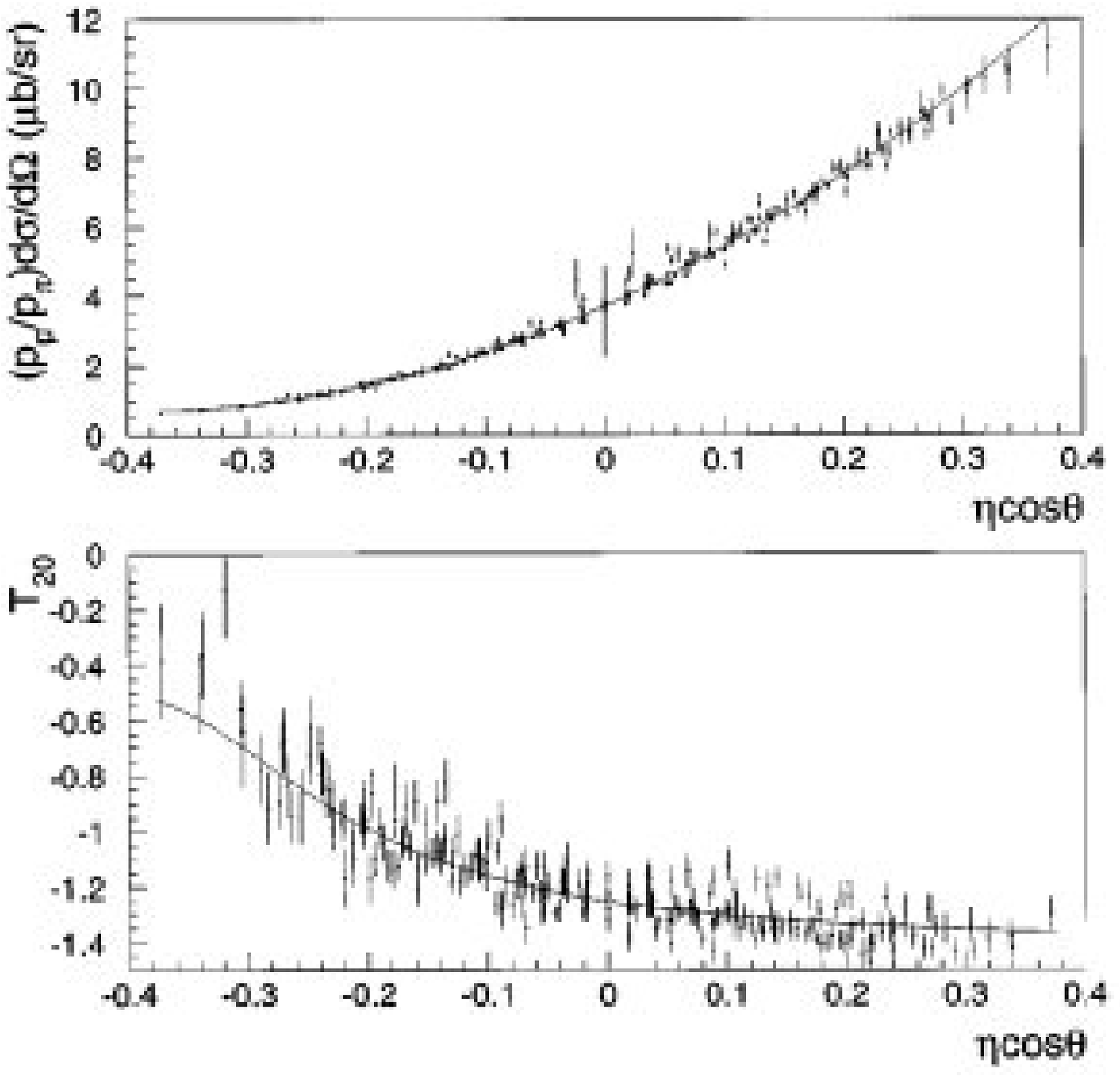, height=10cm} \caption{Averaged amplitude
squared and deuteron tensor analysing power for the reaction
$\vec{d}\,p\to\,^3\textrm{He}\,\pi^0$~\protect\cite{Nikulin}. The
data at different near--threshold beam energies and angles seem to
be universal functions of the parameter $\eta\cos\theta$, where
$\eta$ is the pion cm momentum in pion mass units.}\label{volodia}
\end{center}
\end{figure}

Even more unexpected was the observation that, after including the
phase space factor, the differential cross section and deuteron
tensor analysing power $t_{20}$ were functions of a single
parameter $x=\eta\cos\theta$, whereas $t_{22}$ and $it_{11}$ were
consistent with zero over their whole energy and angular range.
These \emph{universal} plots are illustrated in
Fig.~\ref{volodia}.

Now there are only two independent $pd\to\,^3\textrm{He}\,\pi^0$
amplitudes near threshold (or in the forward direction), and these
may be written as~\cite{collinear}

\begin{equation}
F_{dp\rightarrow\,^3\! He\,\pi^{0}} = \shalf
\bar{u}_{\tau}\bmath{p}_{d}\cdot\left(A\bmath{\epsilon} + iB
\bmath{\epsilon}\times \bmath{\sigma}\right)u_{p}.
\end{equation}
Here $\bmath{\epsilon}$ is the deuteron polarisation vector,
$\bmath{p}_{d}$ and $\bmath{p}_{\pi}$ the deuteron and pion
centre--of--mass momenta and $u_{p}$ and $u_{\tau}$ the initial
and final fermion spinors.

If only the two amplitudes $A$ and $B$ are kept, the unpolarised
cm differential cross section, deuteron tensor analysing power,
and vector transverse spin correlation become
\begin{eqnarray}
\label{sig3} \left(\frac{d\sigma}{d\Omega}\right)
&=& \frac{p_{\pi}p_{d}}{3} \,(\mid A\mid^2 + 2\mid B\mid^2)\:,\\
\label{t203} t_{20} &=&\sqrt{2}\, \frac{\left(\mid B\mid^2 - \mid
A\mid^2\right)} {\left(\mid A\mid ^2 + 2\mid B\mid ^2\right)}\:,\\
\label{cyy3} C_{yy}&=&-\frac{2Re\left(A^*B\right)}{\left(\mid
A\mid^2 + 2\mid B\mid^2\right)}\:,
\end{eqnarray}
whereas $it_{11}$ and $t_{20}$ should both vanish.

The essence of Saclay results is that $|A|\gg |B|$, with $A$ being
a steep but linear function of $\eta\cos\theta$ near threshold,
the other spin amplitudes being negligible. Now, since $B$ is
small, it would be better to investigate it through its
interference with the $A$ amplitude and, as shown in
Eq.~(\ref{cyy3}), this is possible by the measurement of the
transverse spin correlation, which should be possible with
hydrogen in the PIT. The \emph{universal} fit~\cite{Nikulin}
suggests that $C_{yy}\,\dd\sigma/\dd\Omega$ should be a linear
function of $\eta\cos\theta$.

To see why a measurement of $C_{yy}$ might be interesting,
consider the description of the process within an impulse
approximation picture~\cite{GW7,collinear}, which gives a
simplistic but effective zeroth--order description of the
experimental data~\cite{Pickar,Kerboul,Nikulin}. Here pion
production takes place on one nucleon in the target, with the
other being a spectator, and this is reasonable near threshold (or
in the forward direction) where the minimum spectator momentum is
not excessive. If one neglects $D$--state effects in the deuteron
and $^3$He, then $B$ gets a contribution only from $pp\to
pp\pi^0$, whereas $A$ gets its largest contribution from $pn\to
pn(d)\pi^0$~\cite{GW7,collinear}. Because the $\Delta$ contributes
only to the latter, it is dominant over the former, and it also
leads to the strong $p$--wave effects observed
here~\cite{Nikulin}. Though there are potentially significant
corrections to this description, including multiple scatterings
and initial state distortions, it does offer the tantalising
prospect of determining the relative phases of the $pp\to pp\pi^0$
and $pn\to pn\pi^0$ amplitudes near threshold, which is hard to
determine in
other ways.\\
%
%
\subsubsection{$\vec{p}\,\vec{d}\to\,^3\textrm{He}\,\eta$}
\label{He3eta}

The $pd\to\,^3\textrm{He}\,\eta$ reaction near threshold also
shows a very striking energy dependence but of a completely
different nature~\cite{Berger,Mayer}. Despite the angular
distribution remaining essentially isotropic, the square of the
amplitude decreases by a factor of three over a few MeV in excess
energy. In contrast, at higher energies, structure is seen in the
angular distribution~\cite{Bilger2}. The general feeling is that
the threshold behaviour is due to a very strong final state
interaction between the $\eta$ and the $^3$He, suggesting that
this system has a nearby pole in the complex momentum plane. It is
not at all clear whether this corresponds to an $\eta$--nucleus
quasi--bound state~\cite{Liu,Wilkin2,Sibirtsev:2003db} or not,
depending largely upon the sign of the imaginary part of the pole
position. If the \emph{fsi} interpretation is correct, the effect
should depend only weakly upon the characteristics of the entrance
channel and it is very important to verify this.

Data on $\gamma\, ^3\textrm{He} \to\eta\, ^3\textrm{He}$ from
Mainz~\cite{Pfeiffer} show an even stronger energy dependence but
these data are not as precise as the Saclay results and cannot be
used to constrain the pole position. However, their back--to--back
$\pi^0pX$ results indicate an anomalous behaviour just below the
$\eta$ threshold, consistent with a possible decay channel of the
$_{\eta}^{3}\textrm{He}$ nucleus, but the interpretation is not
unambiguous~\cite{Hanhart:2004qs}.\vspace{-3mm}

\begin{figure}[hbt]
\input epsf
\begin{center}
\epsfig{file=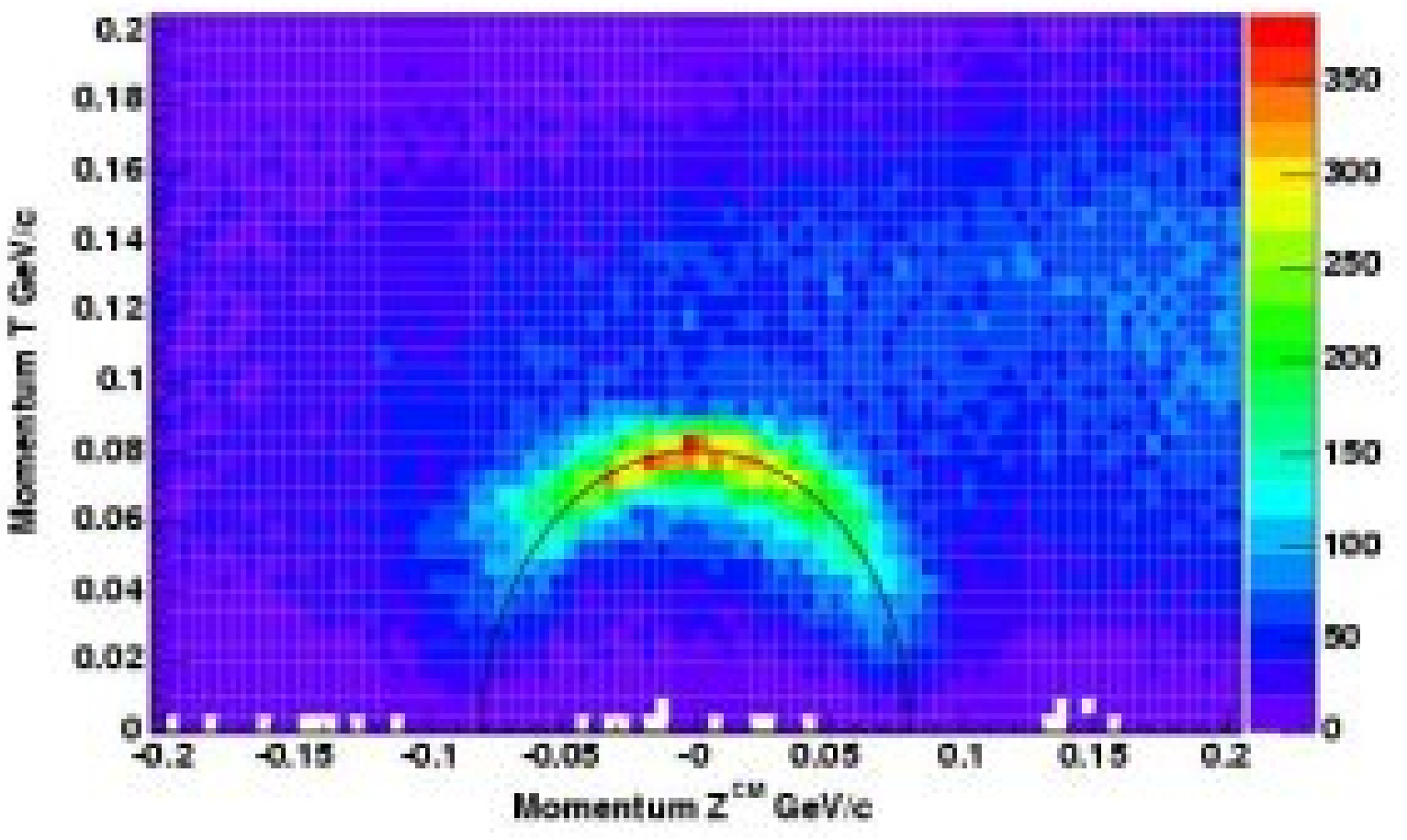, height=7cm} \caption{Transverse
\emph{vs} longitudinal momentum components of charged particles
produced in $\vec{d}p$ collisions at $T_d=1811{\ww}$MeV showing
the momentum ellipse of events corresponding to the
$\vec{d}p\to\,^3\textrm{He}\,\eta$
reaction.}\label{eta_He3_ellipse}
\end{center}
\end{figure}
\vspace{-3mm}

As discussed in \S\ref{He3pi}, close to threshold there are two
independent $dp\to\,^3\textrm{He}\,\eta$ amplitudes $A$ and $B$,
both corresponding the same $J^p=\half^-$ $s$--wave in the final
$\eta\,^3$He system. If the \emph{fsi} interpretation is correct,
one should see the same enhancement in both amplitudes. Unlike the
pion case, the first measurement of $t_{20}$ showed that the
near--threshold values of $A$ and $B$ were of similar magnitude,
though the data were clearly insufficient to determine the energy
dependence of the amplitudes separately~\cite{Berger}. This
separation will be studied in detail with a polarised deuteron
beam at ANKE~\cite{Alfons} and should indicate whether the whole
of the violent energy dependence is due to the final state
interaction or whether the \emph{bare} reaction mechanism
contributes.

An even more refined test would come from the measurement of the
spin correlation in the reaction since, by Eq.~(\ref{cyy3}), this
measures the interference between the two amplitudes and is
sensitive to differences between the two \emph{fsi}. $C_{yy}$
will, of course, be influenced by the different (though coupled)
initial state interactions in the two cases, but this will be
slowly varying over the few MeV over which the \emph{fsi} is
significant.

\begin{figure}[hbt]
\input epsf
\begin{center}
\epsfig{file=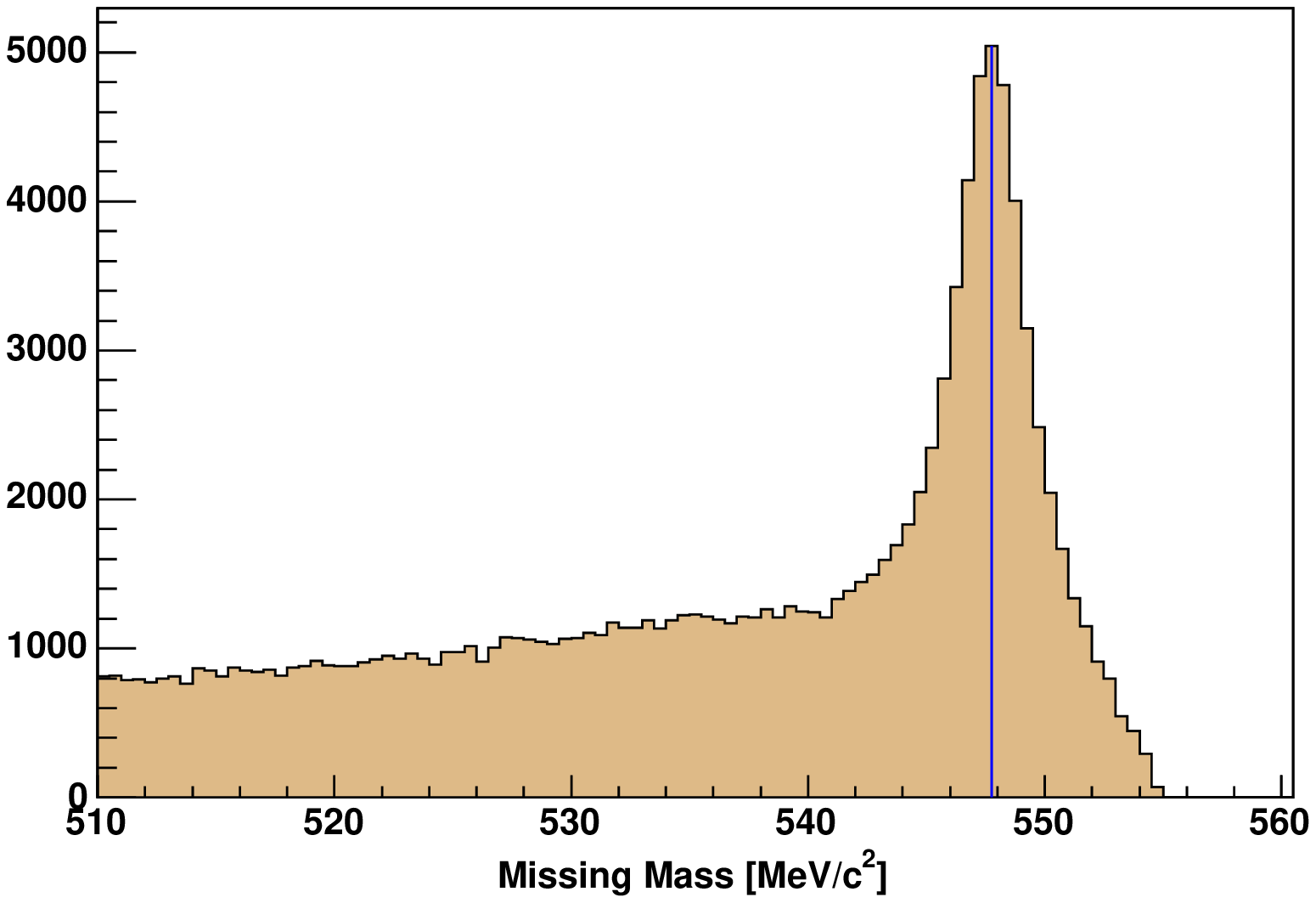, height=6cm} \caption{Missing mass
spectrum for the events shown in Fig.~\ref{eta_He3_ellipse}. The
$\eta$ peak is clearly seen sitting on a multipion
background.}\label{eta_He3_mm}
\end{center}
\end{figure}

Data on the polarised $\vec{d}p\to\,^3\textrm{He}\,\eta$ reaction
were taken in parallel with the deuteron charge--exchange
experiment during the February 2005 beam time at an incident
momentum of 3174{\ww}MeV/c ($T_d=1811{\ww}$MeV), which is
equivalent to an excess energy of $Q=8{\ww}$MeV. In a preliminary
analysis of these data the momentum ellipse corresponding to the
reaction is seen clearly in Fig.~\ref{eta_He3_ellipse}. This is
confirmed by the missing--mass distribution presented in
Fig.~\ref{eta_He3_mm}, where a clean $\eta$ peak shows up above a
physical background due to multipion production.

%
%
\subsubsection{$\vec{p}\,\vec{d}\to\,^3\textrm{He}\,ABC$}
\label{He3ABC}

The only measurement of the spin dependence of \emph{ABC}
production came as a by--product of the study of the deuteron
tensor analysing power $t_{20}$ in single $\pi^0$ production near
the forward and backward directions~\cite{Kerboul}. It should,
however, be noted that the spectrometer was not optimised for such
a study. The resulting data~\cite{AB}
show that $t_{20}$ is consistent with being
constant with a value around $0.1$ for both $\theta=0^{\circ}$ and
$180^{\circ}$.

In collinear kinematics, there are only two independent amplitudes
\begin{equation}
F= \bar{u}_{\tau}\left(A\bmath{\epsilon}_{\parallel}
\bmath{\sigma}_{\parallel}+B\bmath{\epsilon}_{\perp}
\cdot\bmath{\sigma}_{\perp}\right)u_{p}.
\end{equation}
The differential cross section and the deuteron analysing power
together fix the magnitudes of $A$ and $B$ and the smallness of
$t_{20}$ shows that $|A|\approx |B|$. However, a measurement of
the spin correlation would fix also the relative phase of $A$ and
$B$ and it would be interesting to see if this is also independent
of the beam energy.

%
%
\newpage
\section{Production of Strange Mesons and Baryons}
\label{PSMB}

It is well--known that the light mesons and baryons can be
arranged according to the irreducible representations of the group
$SU(3)$. The mass splittings within a multiplet can be well
accounted for by the number of strange quarks in the baryon or
meson. However, not much is known about the dynamics of systems
that contain strangeness. Many phenomenological models for,
\emph{e.g.}\ hyperon--nucleon
scattering~\cite{nimYN,rijken,juel1,juel2} use flavour $SU(3)$ to
fix the meson--baryon couplings.  The remaining unknowns, such as
the cut--off parameters, are then fit to the data. As we discuss
below, so far the existing data base for hyperon--nucleon
scattering is insufficient to judge if this procedure is
appropriate.

As was stressed in \S\ref{tbf}, effective field theories provide
the bridge between the hadronic world and QCD. For systems with
strangeness, there are still many open questions and up to now it
is not clear if the kaon is more appropriately treated as heavy or
light particle. In addition, in order to establish the counting
rules it is important to know the value of the $SU(3)$ chiral
condensate. For a review of this very active field of research, as
well as a list of relevant references, we refer the reader to
Ref.~\cite{strangeulf}.

To improve further our understanding of the dynamics of systems
containing strangeness, better data are needed. The insights to be
gained are relevant, not only for few--body physics, but also for
the formation of hypernuclei~\cite{nogga}, and might even be of
significance for the structure of neutron
stars~\cite{neutronstars}). Naturally, the hyperon--nucleon
scattering lengths are the quantities of interest in this context.

\begin{figure}[t]
\begin{center}
\epsfig{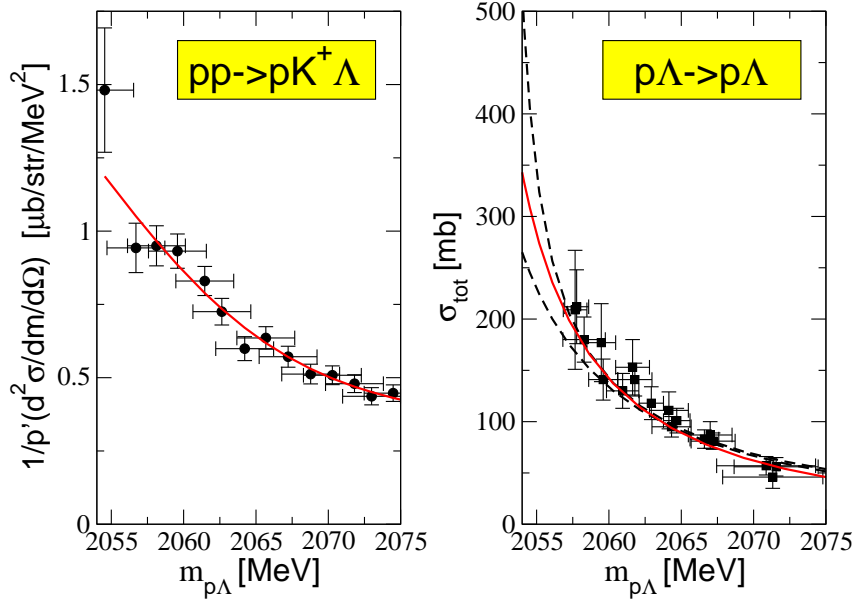} \caption{Comparison of
the quality of available data for the reactions $pp\to K^+\Lambda
p$ at $T_{Lab}$=2.3{\ww}GeV~\protect\cite{saclay} and $\Lambda p$
elastic scattering~\protect\cite{sechi,eisele,alexander68}). In
both panels the solid curve corresponds to a best fit to the data.
In the right
  panel the dashed lines represent the spread in the energy behaviour allowed
  by the data, according to the analysis of Ref.~\cite{alexander68}; analogous
  curves in the left panel would lie almost on top of the solid line and are
  therefore not shown explicitly.}
\label{elastprodcomp}
\end{center}
\end{figure}

\begin{figure}[t!]
\begin{center}
\epsfig{file=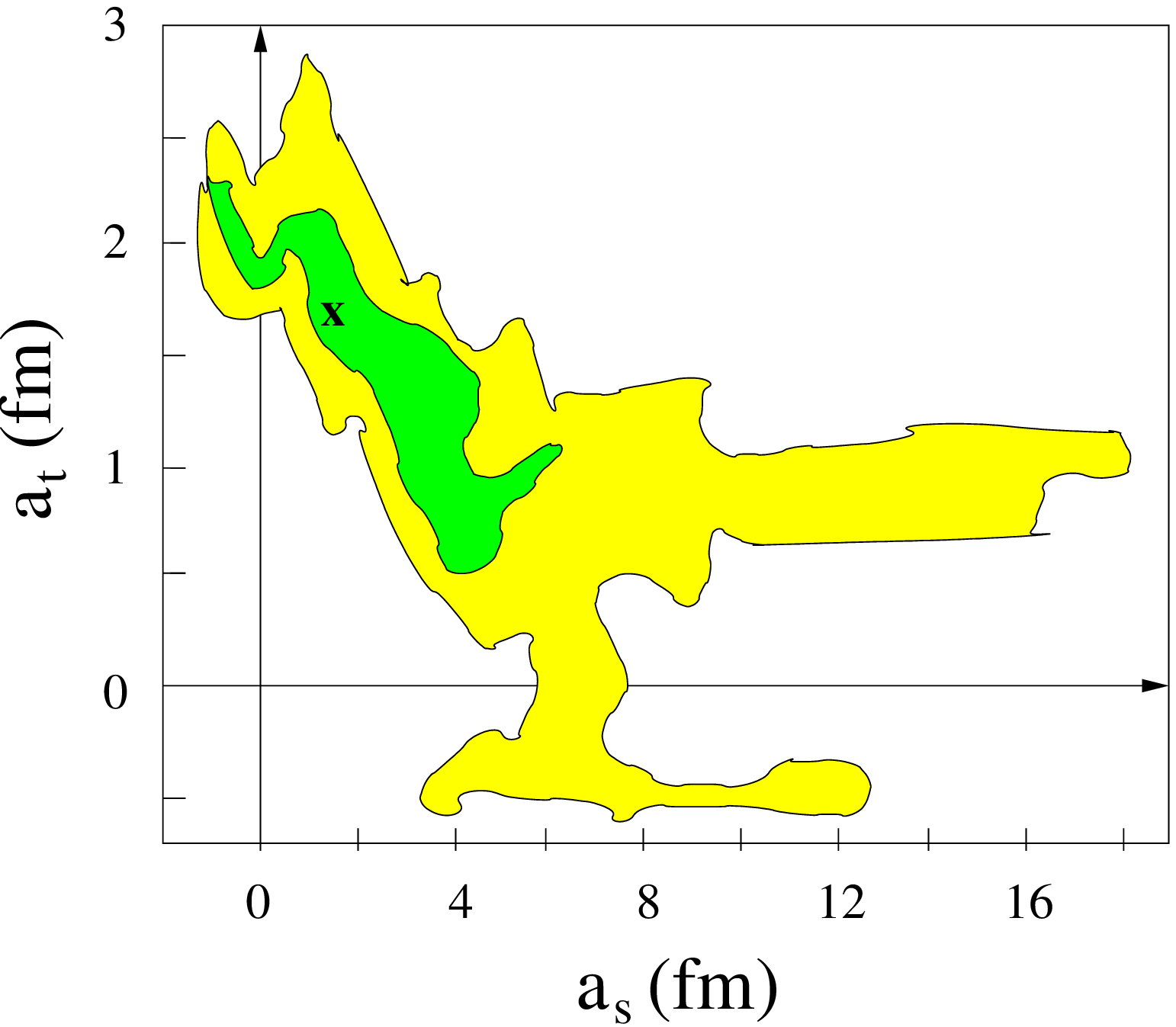, height=10cm} \caption{Values of the
spin--singlet and spin--triplet scattering lengths allowed by the
$\Lambda N$ elastic scattering data according to
Ref.~\protect\cite{alexander68}. The dark shaded area denotes the
1$\sigma$ range for the parameters and the light shaded area the
2$\sigma$ range. The cross shows the best fit value ($a_s=1.8$ fm
and $a_t=1.6$ fm).} \label{lamnelast}
\end{center}
\end{figure}

In the right panel of Fig.~\ref{elastprodcomp} we show the World
data set for elastic $\Lambda p$ scattering. In
Ref.~\cite{alexander68} a likelihood analysis, based on the
elastic scattering data, was performed in order to extract the low
energy $\Lambda p$ scattering parameters. The resulting contour
levels, shown in Fig.~\ref{lamnelast}, clearly demonstrate that
the available elastic hyperon--nucleon scattering data do not
significantly constrain the scattering lengths: the data allows
for values of singlet and triplet scattering lengths of
$(a_s,a_t)=(-1, 2.3)${\ww}fm as well as $(6,1)${\ww}fm.\footnote{Note that
we work within a different sign convention for the scattering
length from Ref.~\cite{alexander68}.}

In subsequent work, potential models were used to extrapolate the
data but even then the scattering lengths could not be pinned down
accurately. For example, in Ref.~\cite{rijken}, six different
models were found that describe equally well the available data
but with $S$--wave scattering lengths that range from 0.7 to
2.6{\ww}fm in the singlet channel and from 1.7 to 2.15{\ww}fm for the
triplet.

There is evidence from hypernuclei that the binding energy of the
spin--singlet state in $_{\Lambda}^{4}$He~\cite{Bedjidian} is
about 1{\ww}MeV stronger than for the triplet and a similar feature
has been observed for $_{\Lambda}^{7}$Li~\cite{Tamura}. Though the
relation of this to the $\Lambda N$ interaction is not
straightforward, it does suggest that the singlet force should be
more attractive.

Production reactions offer a promising alternative approach. In
the literature the reactions
 $K^- d\to \gamma \Lambda n$~\cite{gibbsln},
 $\gamma  d\to K^+\Lambda n$ (Ref.~\cite{gammad5} and references therein) and
 $pp\to pK^+\Lambda$~\cite{COSY11} have all been suggested.

ANKE can contribute to the study of the production of a variety of
strange baryons in nucleon--nucleon collisions and it is likely
that spin--selection will clarify some of the signals. However, we
shall only discuss here two aspects of polarised $\Lambda$
production, where the Physics cases can be argued particularly
clearly.
%
%
\subsection{Determination of the spin--triplet $\Lambda-N$ scattering length}
\label{LNsl}

In Ref.~\cite{ynfsi} a method was developed that allows one to
extract a scattering length directly from data from a production
reaction, in terms of an integral over the invariant mass
distribution, with proper kinematical factors included. A natural
question that then arises is over the quality of data needed
\emph{e.g.} for the reaction $pp\to pK^+\Lambda$ in order to
significantly improve our knowledge of the hyperon--nucleon
scattering lengths. In Ref.~\cite{ynfsi} it was demonstrated that
data of the quality of the Saclay experiment for $pp\to
K^+X$~\cite{saclay}, shown in the left panel of
Fig.~\ref{elastprodcomp} that had a mass resolution of 4{\ww}MeV,
allow the extraction of a scattering length with an experimental
uncertainty of only 0.2{\ww}fm. However, the actual value of the
scattering length extracted from these data is not meaningful,
since they  represent the incoherent sum of the $^3\!S_1$ and the
$^1\!S_0$ hyperon--nucleon final state with unknown relative
weights.

It was shown in the previous section that the two $\Lambda N$
$S$--wave spin states can be separated in the near--threshold
through by using double--polarisation measurements. However, it is
not clear whether such low energy will allow one to explore the
$\Lambda N$ system over a sufficiently wide range of excess
energies to use the techniques of Ref.~\cite{ynfsi} to their full
potential.

Now it is seen from Eq.~(\ref{spin_ob}) that the combination
\begin{equation}
\left(1-C_{NN}\right)\,\frac{\dd\sigma}{\dd\Omega}(pN\to
K^+\Lambda N)
\end{equation}
leads only to spin--triplet final states near threshold. It can be
shown~\cite{report} that this is also true at higher energies
provided that the $K^+$ is detected in the forward direction.
These conditions, which are very suitable for ANKE, would allow
one to investigate higher $\Lambda N$ excitation energies. As a
test on the methodology, one should obtain in this way consistent
results for the scattering lengths for the proton and neutron
targets.

To isolate uniquely the production singlet final states through
forward $K^+$ production far from threshold would require
longitudinally polarised beam and target, neither of which is yet
planned for ANKE. However, as we see below, there are further
constraints in the near--threshold region.

%
%
\subsection{The $NN\rightarrow NK\Lambda$ reaction near threshold}
\label{LNthreshold}

The reaction mechanism for kaon production in nucleon--nucleon
collisions is still far from understood. The
COSY11~\cite{COSY11,Kowina} and
COSY--TOF~\cite{COSY-TOF,COSY-TOF2} collaborations have both made
detailed measurements of $pp\to pK^+\Lambda$ and $pp\to
pK^+\Sigma^0$ production near their respective thresholds, but
these are insufficient to determine what the principal forces are
driving the reactions. To make any progress in this respect, one
needs high quality data on the spin and isospin dependence of the
reactions, and ANKE is well equipped to provide these, especially
in the $\Lambda$ case on which we now concentrate.

At threshold there is a total of only three amplitudes $W_{i,s/t}$
which describe the $I=1$ and $I=0$ $NN\rightarrow NK\Lambda$
reactions~\cite{FW-Lambda}:

\[
\mathcal{M}_1= \left[W_{1,s}\, {\eta}_f^{\,\dagger}\,
\bmath{\hat{p}} \cdot \bmath{\epsilon}_i +
iW_{1,t}\,\bmath{\hat{p}}\cdot ( \bmath{\epsilon}_i \times
             \bmath{\epsilon}_f^{\,\dagger}) \right]\,
\bmath{\chi}_f^{\,\,\dagger}\cdot\bmath{\chi}_i\:,
\]\vspace{-10mm}

\begin{equation}
\mathcal{M}_0= W_{0,t}\,
   \bmath{\hat{p}} \cdot \bmath{\epsilon}_f^{\,\,\dagger}\,{\eta}_i
\ \phi_f^{\,\dagger}\,\phi_i \:,
\label{L_1}
\end{equation}
where \bmath{p} is the incident cm beam momentum.

At least five observables are required in order to isolate these
amplitudes fully (up to an overall phase). Two of these are
provided by the unpolarised cross sections on proton and neutron
targets, which are proportional to the intensities:
\begin{equation}\label{Ipp}
I(pp\!\to\! pK^+\Lambda)= \fourth \left(\mid W_{1,s}\mid^2+ 2\mid
W_{1,t}\mid^2\right),
\end{equation}
\begin{equation}
I(pn\!\to\! nK^+\Lambda)=
\fmn{1}{16}\left(\mid W_{1,s}\mid^2+ 2\mid W_{1,t}\mid^2+ \mid
W_{0,t}\mid^2\right), \label{Ipn}
\end{equation}

In the vicinity of the threshold, both the proton analysing power
and the $\Lambda$ polarisation must vanish and only tensor
combinations can be non--zero. Of these, the most \emph{easily}
accessible at ANKE are the transverse spin--correlation
($C_{NN}=A_{yy}$) and the spin--transfer parameters to the
$\Lambda$ ($D_{NN}$), which are given by~\cite{FW-Lambda}
\begin{eqnarray}
\nonumber
I(pp\!\to\! pK^+\Lambda)\,C_{NN}(\vec{p}\vec{p}\!\to\! pK^+\Lambda)
&=&\fourth\mid\! W_{1,s}\!\mid^2\:,\\
\nonumber
I(pn\!\to\! nK^+\Lambda)\,C_{NN}(\vec{p}\vec{n}\!\to\! nK^+\Lambda)&=&
\fmn{1}{16}\!\left(\mid\! W_{1,s}\!\mid^2\!-\!\mid\! W_{0,t}\!\mid^2\right)\!,\\
\nonumber
I(pp\!\to\! pK^+\Lambda)\,D_{NN}(\vec{p}p\!\to\! pK^+\vec{\Lambda})
&=&-\half\textrm{Re}(W_{1,s}W_{1,t}^*),\\
\nonumber
I(pn\!\to\! nK^+\Lambda)\,D_{NN}(\vec{p}n\!\to\! nK^+\vec{\Lambda})
&= &-\fmn{1}{8}\textrm{Re}\left\{(W_{1,s}+W_{0,t})W_{1,t}^*\right\},\\
I(pn\!\to\! nK^+\Lambda)\,D_{NN}(p\vec{n}\!\to\! nK^+\vec{\Lambda})
&= &-\fmn{1}{8}\textrm{Re}\left\{(W_{1,s}-W_{0,t})W_{1,t}^*\right\}\:.
\label{spin_ob}
\end{eqnarray}

Measurements of the unpolarised cross section on the proton and
neutron, plus the spin--correlation on the neutron, and the
spin--transfer parameters in $\vec{p}p$ and $\vec{p}n$ collisions
would allow one to extract the magnitudes of the three amplitudes
and determine (up to two discrete ambiguities) the relative phases
of $W_{1,s}$, $W_{1,t}$, and $W_{0,t}$ in a model--independent
way.

The above amplitude analysis cannot, of itself, tell us whether
the singlet amplitudes will be big enough to convey sufficient
information to allow the study of final state interaction effects.
To obtain some indication of whether this might be possible, we
turn to a simple one--boson--exchange model. Neglecting
distortions arising from the phase shifts in the initial $NN$
system, the amplitudes can be expressed in terms of the
$\pi,\,\rho,\,\eta,\,\omega$ and the two kaon--exchange terms
as~\cite{FW-Lambda}
\begin{eqnarray}
\nonumber W_{1,s}&=& 2\mathcal{B}_{\rho} + 2\mathcal{B}_{\omega}-
\mathcal{D}_{\pi}-\mathcal{D}_{\eta}+\mathcal{D}_K^1\:,\\
\nonumber W_{1,t}&=&
\mathcal{D}_{\pi}+\mathcal{D}_{\eta}+\mathcal{D}_K^1\:,\\
\label{exch_amps}%
W_{0,t}&=& 6 \mathcal{B}_{\rho}-2\mathcal{B}_{\omega}+
3\mathcal{D}_{\pi}-\mathcal{D}_{\eta}+\mathcal{D}_K^0\,.
\end{eqnarray}
From this it is clear that one has no automatic right to assume
that the population of $\Lambda N$ final spin states is governed
purely by statistical factors.

F\"aldt and Wilkin~\cite{FW-Lambda} speculated that $\rho$ and to
a lesser extent $\pi$--exchange should provide the dominant terms.
If this proves to be the case then it would mean that a large part
of the $pp\to pK^+\Lambda$ cross section leads to spin--singlet
$\Lambda p$ final states, allowing this system to be studied.
However, the initial $NN$ phase shifts must be included in
Eq.~(\ref{exch_amps}) before these are compared with the
spin--transfer observables, which depend sensitively upon
interference effects.

Such experiments cannot, of course, be carried out strictly at
threshold but, from experience with $\eta$ production, the
$S$--wave formulae are typically valid up to excess energies of
say 20--30{\ww}MeV. Over this range in energies, the only strong
variation in the $W$ amplitudes comes from the $\Lambda N$
final--state interactions, which are discussed in a more refined
approach in \S\ref{LNsl}. Though the energy limitation due to the
$S$--wave selection reduces the available phase space, the
different dependence of the extracted values of $\mid\!
W_{1,s}\!\mid^2$ and $\mid\! W_{0,t}\!\mid^2$ on the $\Lambda N$
excitation energy will allow us to compare the effects of the
spin--singlet and triplet $\Lambda-N$ scattering lengths.

It might be possible to check the spin dependence of the
scattering lengths using the spin--transfer information. If, for
simplicity of presentation, we neglect phases arising from the
different $NN$ initial state interactions, and na\"ively
parameterise the $\Lambda N$ \emph{fsi} purely in terms of a
scattering length,
\begin{equation}
W_{s/t} =\frac{A_{s/t}}{(1+ika_{s/t})}
\end{equation}
with real and slowly varying coefficients $A_{s/t}$, then
\begin{equation}
\label{naive}
\frac{\left[\textrm{Re}\{W_{1,s}W_{0,t}^{*}\}\right]^2} {\mid\!
W_{1,s}\!\mid^2\mid\! W_{0,t}\!\mid^2} = 1+
\frac{k^2(a_s-a_t)^2}{(1+k^2a_s^2)(1+k^2a_t^2)}\:.
\end{equation}
Here $k$ is the $\Lambda N$ relative momentum and the $k^2$
variation in Eq.~(\ref{naive}) is directly a measure of the
difference in the scattering lengths. When there is, in addition,
a smoothly varying overall phase coming from the initial--state
interactions, the above combination contains also a term linear in
the scattering length difference. Though the above
phenomenological description is undoubtedly crude, it does suggest
strongly that there is a sensitivity to $a_s-a_t$.
%
%
\subsection{Experimental considerations}
\label{YuV}%

The ANKE magnetic spectrometer provides a unique opportunity to
investigate kaon production on a neutron target. The realisation
of the ambitious programme described in the previous sections will
rely on two of the advantages of the ANKE system, \emph{viz}
spectator particle detection, which allows one to access double
polarised measurements on the neutron, and $K^{+}$ identification
using time-of-flight techniques or the range--telescope structure.
The $K^{+}$ production cross section is low, so that experiments
will have to use the storage cell target. This then requires the
full set of spectator detectors for the reconstruction of the
interaction point and determination of the spectator proton
momentum.

In order to be able to perform such a challenging experiment with
the polarised storage cell target, the natural first step would be
to undertake a sequence of the experiments with polarised beam and
unpolarised deuteron target. Such a study would lead to
measurements of the cross section and single polarised observables
in the $pn\to n K^{+}\Lambda$ and $pp\to p K^{+}\Lambda$ channels.

\begin{figure}[ht]
\centering
\includegraphics[height=13.0cm,angle=-90]{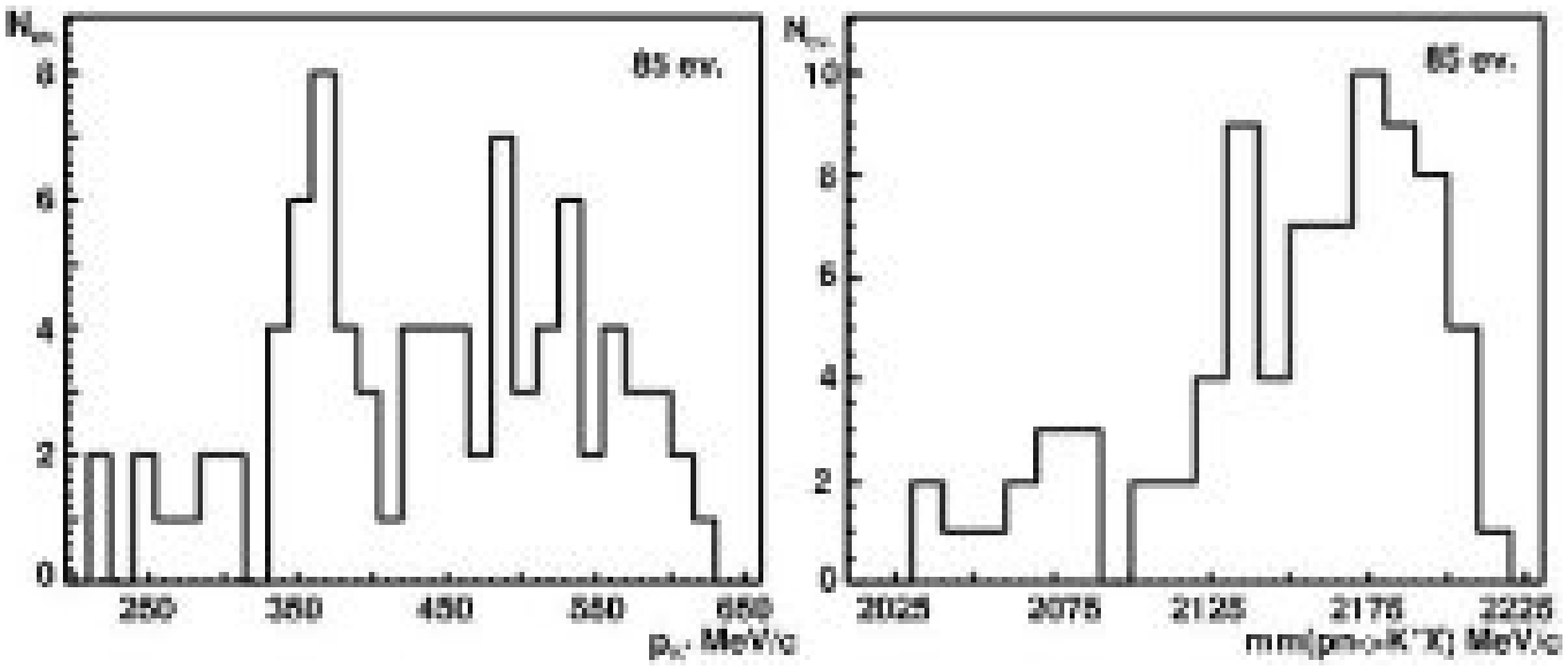}
\caption{Experimental data for the $pd\to p_{sp}K^{+}X$ reaction
collected at 2.01{\ww}GeV. }\label{fig:exp_data_pnK}
\end{figure}

In Fig.~\ref{fig:exp_data_pnK} are presented experimental data for
the $pd\to p_{sp}K^{+}X$ reaction collected at 2.01{\ww}GeV during the
August 2001 beam time. It shows that it is possible to observe
coincidences between the spectator and side--detection system
without imposing any limitation on the detected $K^{+}$ momentum.
The spectator detector employed was a prototype, which had an
acceptance only of the order of $10^{-3}$. The new spectator
detection system, described in \S\ref{SST}, will have much bigger
acceptance and should not introduce any restrictions on the
detected $K^{+}$ momentum and angle.\\

\begin{figure}[ht]
\begin{center}
\epsfig{file=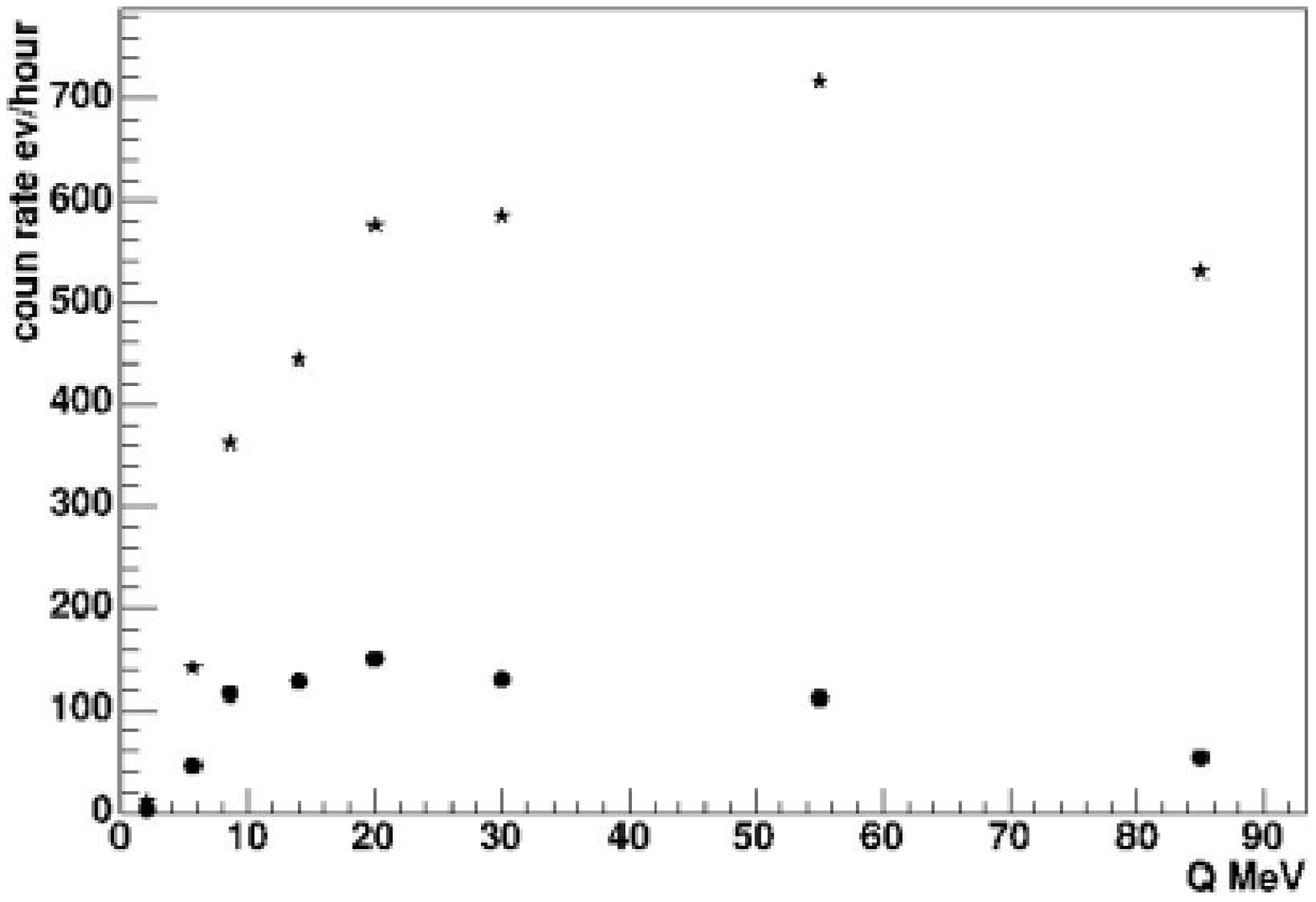, height=6cm} \caption{
   Expected count rate for the $ pp\to p K^{+}\Lambda$ reaction as a
   function of excess energy $Q$. Stars represent conditions where
   only the $K^{+}$ is detected; circles correspond to the $K^{+}$
   and the proton from $\Lambda$ decay being detected in coincidence.
} \label{fig:counts_pKL}
\end{center}
\end{figure}

The count rates to be expected for $K^{+}$ singles and $K^{+}
p_{\Lambda}$ coincidences from the $pp\to p K^{+}\Lambda$ reaction
are shown in Fig.~\ref{fig:counts_pKL}. The estimations, which
were made at proton beam momenta where the total cross section for
the $\Lambda$ production is known~\cite{COSY11,COSY-TOF}, have
been made using the PLUTO event generator and AnkeRoot simulation
packages~\cite{Simulations}.

The highest possible field was taken for the D2 magnet, with the
point--like target being placed in the planned ABS position. A
luminosity of $10^{30}${\ww}cm$^{-2}$s$^{-1}$ was assumed in the count
rate estimations. The time--of--flight start--stop combinations
was excluded from the simulations because it has yet to be chosen
for the new target position. However, the probability for the
$K^{+}$ to decay in flight was included. Simulations which take
into account the new spectator telescopes acceptance as well as
the storage--cell target density distribution will be carried out
in the near future.

It is important to emphasise that, even at the highest excess
energy considered ($Q=85${\ww}MeV), the momentum of the $\Lambda$ in
the centre--of--mass system is at most 20\% of that of the proton
coming from the $\Lambda$ decay. It is therefore not necessary to
measure the $\pi^-$ from the decay in coincidence in order to use
the decay distribution to measure the $\Lambda$ polarisation in
the $pn\to n K^{+}\Lambda$ reaction channel.

Up to the highest excess energy that we have considered, ANKE
covers the full $K^{+}$ angular range in the CM system. The mass
resolution in $n\Lambda$ system, which is also crucial for the
\emph{fsi} analysis described in \S\ref{LNsl} is expected to be
better than 4--5{\ww}MeV ($\sigma$). It can be improved by operating
the side chambers in the drift--chamber mode with new electronics
and the installing a thinner window.

ANKE was built for the particle identification around $0^{o}$. If
a longitudinally polarised beam and target would become available
it is perfectly suited for the spin-singlet $\Lambda - N$
scattering length determination, since this measurement should
preferably be done in the forward direction.

\clearpage
%
%
\section{Test of Time--Reversal Invariance in Proton\---Deuteron Scattering}
\label{TRIC}

\subsection{Overview} At COSY--J\"ulich a novel ($P$--even, $T$--odd) null
test of time--reversal invariance can be performed to an accuracy
of $10^{-4}$ (Phase 1) or $10^{-6}$ (Phase 2). The
parity--conserving, time--reversal--violating observable is the
total cross section asymmetry $A_{y,xz}$. The measurement is
planned as an internal target transmission experiment at the
cooler synchrotron COSY. $A_{y,xz}$ is measured using a polarised
beam and a tensor polarised deuteron target. In this experiment
the COSY ring serves as an accelerator, ideal forward
spectrometer, and detector.

\subsection{Introduction}

So far, the only link to a violation of time--reversal symmetry is
given \emph{via} the $CPT$--theorem and the observation of $CP$ violation
in the neutral Kaon-- and B--system. Although the $CP$ violation
could be accommodated by a complex phase in the Kobayashi--Maskawa
matrix~\cite{r1} or the $\theta$ $T$--term~\cite{r2} allowed by
QCD, other explanations go beyond the standard model. These are,
for instance, the extension of the Higgs sector~\cite{r3}, the
superweak interaction~\cite{r4}, or the left--right symmetric
models~\cite{r5}. Such extensions lead to interactions that are
not related to the observed $CP$-- or $T$--violation. Since the
origin of the $CP$-- or $T$--violation is not clear, further
experimental tests are necessary to probe the manifestation of the
interaction responsible for the observed or possible new $CP$-- or
$T$--violating effects. In this context, more direct information
is to be expected from tests involving elementary particles as
compared to ones involving complex nuclei. In addition, we intend
to probe time--reversal invariance with parity being conserved in
contrast to experiments which test parity and time reversal
invariance (TRI) simultaneously (\emph{cf}.\ tests of the
electric--dipole moments of elementary particles).

Usually $P$--even TRI tests compare two observables (\emph{cf}.\
tests of detailed balance or $P$--$A$ tests) and this limits the
experimental accuracy to about $10^{-3}-10^{-2}$~\cite{r6}. The
accuracy would be increased by orders of magnitude if a true null
experiment could be performed \emph{i.e.}\ one where a
non--vanishing value of a single observable proves that the
symmetry involved is violated. An example of this kind is the
measurement of the parity violating quantity $A_z$ in
proton--proton scattering~\cite{r7}, which has been measured to
some $10^{-8}$. In this context the term \emph{true} stresses the
concept that the intended test has to be completely independent
from dynamical assumptions. Therefore, the interpretation of the
result is neither restricted nor subject to final state
interactions, special tensorial interactions or, Hamiltonians of a
certain form. True null tests are based on the structure of the
scattering matrix only, as determined by general conservation
laws~\cite{r8}.

It has been proven~\cite{r8} that there exists no true null test
of TRI in a nuclear reaction with two particles in and two
particles out, except for forward scattering, from which a total
cross section can be measured. Based on this exception,
Conzett~\cite{r9} showed that a transmission experiment can be
devised, which constitutes a true TRI null test. He suggested
measuring the total cross--section asymmetry $A_{y,xz}$ of vector
polarised spin--half particles interacting with tensor polarised
spin--one particles.

\begin{figure}[ht!]
\begin{center}
\epsfig{file=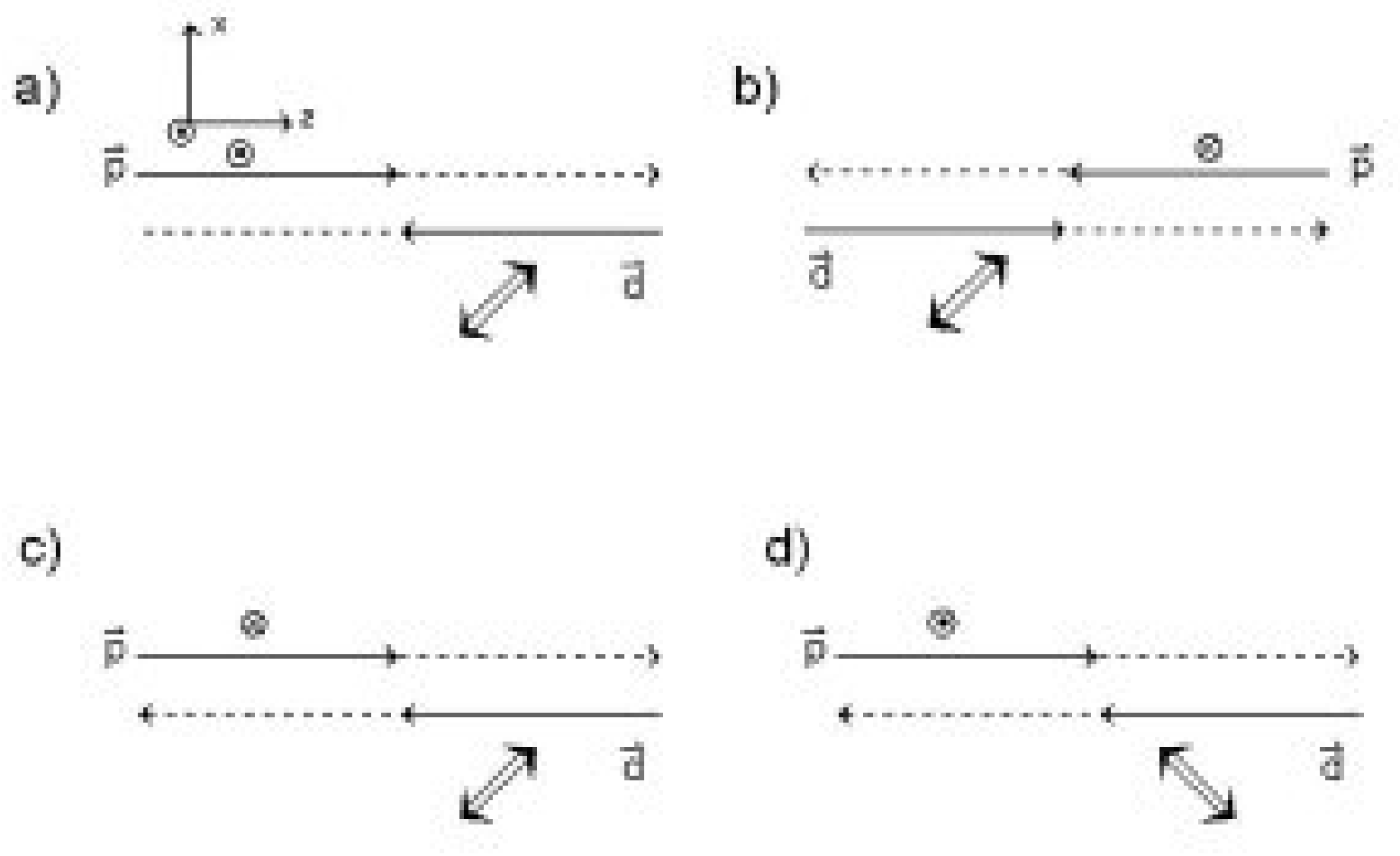, height=8cm} \caption{Pictorial
demonstration that a time--reversed situation is prepared by
either a proton or a deuteron spin--flip. a) The basic system is
shown. b) The time--reversal operation is applied (momenta and
spins are reversed and the particles are exchanged). In order to
have a direct comparison between situation a) and b), two
rotations $R_y(p)$ or $R_x(p)$ by $180^{\circ}$ about the $y$-- or
$x$--axes are applied, leading to the situations c) and d),
respectively. This is allowed, since the scattering process is
invariant under rotations. In the figures we have used notations
whereby $\odot = $ proton spin up ($y$--direction), $\otimes = $
proton spin down, and $\Longleftrightarrow\ = $ deuteron tensor
polarisation. } \label{tricfig1}
\end{center}
\end{figure}

We intend to study the time reversal violating (TRV) quantity
$A_{y,xz}$ in a transmission experiment using an internal deuteron
target in the cooler synchrotron COSY. The tensor polarised
deuteron target is prepared using the polarised atomic beam target
facility of the COSY experiment \#5 (phase 1). Phase 2 is
characterised by an improved control of systematic error
contributions and the addition of a target cell to the polarised
atomic beam target. The transmission losses of the circulating
polarised proton beam change its lifetime. The lifetime as a
function of the vector-- and tensor--polarisation $P_y$ and
$P_{xz}$ is measured by COSY's high precision beam current
transformers. Thus, for this experiment, the COSY facility is not
only used as an accelerator but also as an ideal forward
spectrometer and detector. The preparation of a time--reversed
situation is depicted in Fig.~\ref{tricfig1}.

\subsection{Experimental set--up}

Since the TRV processes are of short--range nature, the long range
contributions for these processes may be parameterised by the
$\rho$ vector meson or $f_1$ axial--vector meson
exchanges~\cite{r9}. Using this, Beyer has shown that the most
favorable momentum, where the experiment has its highest
sensitivity to TRV effects, is at about 500{\ww}MeV/c~\cite{r9}.
Therefore, the COSY ring is presently prepared with: i) injection
of polarised protons (polarisation $> 0.8$) at 40 MeV, ii)
stacking injection to increase the circulating proton beam, and
iii) electron cooling, not only for the stacking injection, but
also during the acceleration and flat--top. At the flat--top, the
beam coasts for about 1{\ww}h, during which time the alignment of the
target--polarisation is flipped. The beam is then decelerated and
dumped and the next injection prepared, with the proton
polarisation being reversed eventually from $+Y\to -Y$. Consistent
results are expected for two out of the four possible combinations
of the target/beam alignments/polarisations, respectively.

The resulting slopes of the decreasing circulating proton beam are
measured with the high precision beam current transformers of the
COSY ring. In the case that there is a TRV effect, the total cross
section, and hence the slopes, would be different.

\subsection{First results}

A new front--end electronics and data acquisition system has been
built and set up, so that the precision of the beam current
transformers can be fully employed. Cooling of the beam turned out
to be essential since, otherwise, the inherent emittance growth of
the proton beam would cause the beam--halo to hit some aperture in
the beam pipe after some minutes. This in turn would lead to a
fast decrease of the beam intensity, which is not compatible with
the usual exponential reduction of the current.

\begin{figure}[ht!]
\begin{center}
\epsfig{file=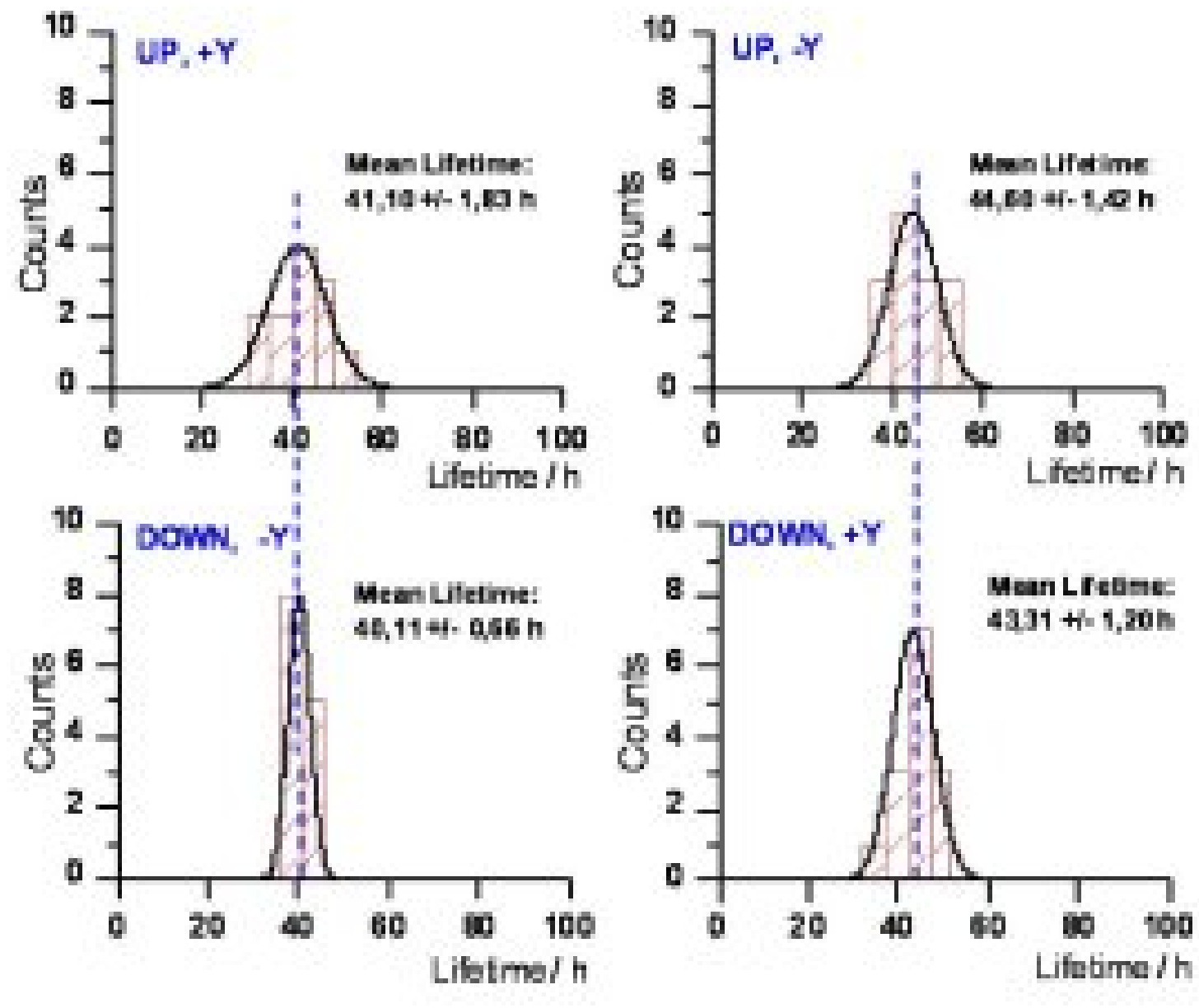, height=10cm} \caption{Results of the total
asymmetry $A_{y,y}$ in $\vec{p}\,\vec{p}$ scattering at
1690{\ww}MeV/c. The agreement of the spectra for spins parallel (left)
and antiparallel (right), as well as their mutual difference in
lifetime, demonstrates that total cross section differences of
polarised particles show up in life--time differences and that
these can be measured in an internal experiment. The effect
corresponds to about 3{\ww}mb~\cite{r10}. UP/DOWN refers to the target
polarisation, $+Y/-Y$ refers to the beam polarisation. }
\label{tricfig2}
\end{center}
\end{figure}

As a first test, the total cross--section asymmetry $A_{y,y}$ in
$\vec{p}\,\vec{p}$ scattering has been measured at 1690{\ww}MeV/c
(with stochastic cooling, polarisation of beam and target $>
0.8$). In $\vec{p}\,\vec{d}$ scattering this quantity is the only
one that through slight misalignments can fake a TRV effect.
However, in contrast to the total cross--section asymmetry
$A_{y,y}$ in $\vec{p}\,\vec{p}$ scattering, its value is not
known. Fig.~\ref{tricfig2} shows that the novel method to measure
total cross sections of polarised particles works in practice. The
cross section difference in the literature~\cite{r10} corresponds
to a fractional lifetime difference of 6.7\% compared to the COSY
measurement of $8.6\pm 4.5\%$. The accuracy achieved so far is
already sufficient to determine $A_{y,y}$ in $\vec{p}\,\vec{d}$
scattering, which can be accomplished by exchanging the gas supply
for the target from hydrogen to deuterium.
%
%
\clearpage
\section{Spin Rotation and Birefringence in Storage Rings}

It is well known in nuclear and particle physics how to measure a
total spin-dependent cross-section in, e.g. proton-proton ($pp$),
proton-deuteron ($pd$), proton-nucleus ($pA$), and deuteron-nucleus
($dA$) interactions. However, measurements of the real part of the
coherent elastic forward scattering amplitude $\Re[f(0)]$ poses some
difficulties.

It has been shown in Refs~\cite{birefringence,VG+S} (and references
therein) that there is an unambiguous method which makes possible the
direct measurement of the real part of the spin-dependent forward
scattering amplitude by observation of proton (deuteron, or
antiproton) beam spin rotation in a polarised nuclear target and by
deuteron spin rotation and oscillation in an unpolarised target.

Spin rotation and oscillation experiments also allow one to carry out
new experiments to study $P$-- and $T$--odd interactions~\cite{birefringence,VG+S}.

Considering the evolution of the spin of a particle in a storage ring
one should take into account several interactions.  The equation for
the particle spin wavefunction considering all these interactions can
be written as
\begin{equation}
i\hbar\frac{\partial\Psi(t)}{\partial
t}=\left(\hat{H}_{0}+\hat{V}_{EDM}+\hat{V}_{\vec{E}}+\hat{V}_{\vec{B}}+\hat{V}_E^{nucl}+
\hat{V}_{{B}}^{nucl}\right)\Psi(t) \label{1}
\end{equation}
where $\Psi(t)$ is the particle spin wavefunction, {$\hat{H}_{0}$ is
the Hamiltonian describing the spin behavior caused by the interaction
of the magnetic moment with the electromagnetic
field\footnote{Equation~(\ref{1}) with $\hat{H}_{0}$ alone, is
equivalent to the well-known Bargmann--Michel--Telegdi (BMT) equation.},
$\hat{V}_{EDM}=-d\left(\vec{\beta}\times\vec{B}+\vec{E}\right)\vec{S}$
describes the interaction of the particles electric dipole moment
(EDM) with the electric field, $\hat{V}_{\vec{E}} = -\frac{1}{2}
\hat{\alpha}_{ik} (E_{eff})_{i}(E_{eff})_{k}$ describes the
interaction of the particle with the electric field due to the tensor
electric polarisability, $\hat{\alpha}_{ik}$ denotes the electric
tensor polarisability of the particle, $\vec{E}_{eff} =
(\vec{E}+\vec{\beta} \times \vec{B})$ is the effective electric field.
The latter expression can be rewritten as $\hat{V}_{\vec{E}} =
\alpha_{S}E^{2}_{eff}-\alpha_{T}E^{2}_{eff}\left(\vec{S}\vec{n}_{E}\right)^{2}$,
where $\vec{n}_{E} = \left.\left(\vec{E} + \vec{\beta}
\times\vec{B}\right)\right/|\vec{E} + \vec{\beta}\times\vec{B}|$, and
$\alpha_{S}$ and $\alpha_{T}$ are the electric scalar and tensor
polarisabilities of the particle, respectively.  A particle with spin
$S \ge 1$ also possesses a magnetic polarisability, which is described
by the magnetic polarisability tensor $\hat{\beta}_{ik}$, and the
interaction of the particle with the magnetic field due to the tensor
magnetic polarisability is $\hat{V}_{\vec{B}} =
-\frac{1}{2}\hat{\beta}_{ik}(B_{eff})_{i}(B_{eff})_{k} $, where
$(B_{eff})_{i}$ are the components of the effective magnetic field
$\vec{B}_{eff}=(\vec{B}-\vec{\beta} \times \vec{E})$;
$\hat{V}_{\vec{B}}$ can be expressed \emph{via} $ \hat{V}_{\vec{B}} =
\beta_{S}B_{eff}^{2} - \beta_{T}B_{eff}^{2}
\left(\vec{S}\cdot\vec{n}_{B}\right)^{2}$, where $\vec{n}_B =
\left.\left(\vec{B}-\vec{\beta} \times \vec{E}\right)\right/\left(|\vec{B}-\vec{\beta} \times
\vec{E}|\right)$, and $\beta_{S}$ and $\beta_{T}$ are the scalar and tensor
magnetic polarisabilities of the particle, respectively.
$\hat{V}_{{B}}^{nucl}$ describes the effective potential energy of the
interaction with the nuclear pseudomagnetic field of the
target~\cite{birefringence,VG+S}, and $\hat{V}_E^{nucl}$ denotes the
effective potential energy of the interaction with the nuclear
pseudoelectric field of the target~\cite{birefringence,VG+S}. It
should be emphasised that $\hat{V}_{{B}}^{nucl}$ and
$\hat{V}_E^{nucl}$ include contributions from strong interactions as
well as those caused by weak interactions, violating $P$ and $T$
invariance.

Let us consider particles moving in a storage ring with low residual
gas pressure ($10^{-10}$ Torr) and without other targets inside the
storage ring.  In this case we can omit the effects caused by the
interactions $\hat{V}_{{B}}^{nucl}$ and $\hat{V}_E^{nucl}$ altogether.
Let us further consider a particle with $S=1$ (for example, deuteron)
moving in such a ring.  According to the above analysis, the spin of
such a particle can not be described by the BMT--equation.  The
equations for the particle spin motion, including contributions from
the tensor electric polarisability, were obtained in
Refs.~\cite{birefringence,VG+S}. Considering that the deuteron
possesses also a tensor magnetic polarisability, after adding the
terms caused by this interaction to the equations obtained in
Refs.~\cite{birefringence,VG+S}, we finally obtain
\[
\frac{d\vec{P}}{dt}=
\frac{e}{mc}\left[\vec{P}\times\left\{\left(a+\frac{1}{\gamma}\right)\vec{B}
-a\frac{\gamma}{\gamma+1}\left(\vec{\beta}\cdot\vec{B}\right)\vec{\beta}-
\left(\frac{g}{2}-\frac{\gamma}{\gamma+1}\right)\vec{\beta}\times\vec{E}\right\}\right]\]
\begin{equation} +
\frac{d}{\hbar}\left[\vec{P}\times\left({\vec{E}}+\vec{\beta}\times\vec{B}\right)\right]
-\frac{2}{3}\frac{\alpha_{T}E^{2}_{eff}}{\hbar}[\vec{n}_{E}\times\vec{n}_{E}^{\prime}]
-\frac{2}{3}\frac{\beta_{T}B^{2}_{eff}}{\hbar}[\vec{n}_{B}\times\vec{n}_{B}^{\prime}],
\end{equation}
\[
\frac{dP_{ik}}{dt}  =
-\left(\varepsilon_{jkr}P_{ij}\Omega_{r}+\varepsilon_{jir}P_{kj}\Omega_{r}\right)
-
\frac{3}{2}\frac{\alpha_{T}E^{2}_{eff}}{\hbar}\left([\vec{n}_{E}\times\vec{P}]_{i}n_{E,\,k}
+n_{E,\,i}[\vec{n}_{E}\times\vec{P}]_{k}\right)\]
\begin{equation} -
\frac{3}{2}\frac{\beta_{T}B^{2}_{eff}}{\hbar}\left([\vec{n}_{B}\times\vec{P}]_{i}n_{B,\,k}
+n_{B,\,i}[\vec{n}_{B}\times\vec{P}]_{k}\right),
\label{50}\end{equation}
where $m$ is the particle mass, $e$ is the charge, $\vec{P}$ is the
spin polarisation vector, $P_{xx}+P_{yy}+P_{zz}=0$, $\gamma$ is the
Lorentz-factor, $\vec{\beta}=\vec{v}/c$, $\vec{v}$ is the particle
velocity, $a=(g-2)/2$, $g$ is the gyromagnetic ratio, $ \vec{E}$ and
$\vec{B}$ are the electric and magnetic fields at the position of the
particle, $\vec{E}_{eff}=(\vec{E}+\vec{\beta} \times \vec{B})$,
$\vec{B}_{eff}=(\vec{B}-\vec{\beta} \times \vec{E})$,
$\vec{n}=\vec{k}/k$, $\vec{n}_{E} = (\vec{E} +
\vec{\beta}\times\vec{B})/|\vec{E} + \vec{\beta}\times\vec{B}|$,
$\vec{n}_B=(\vec{B}-\vec{\beta} \times \vec{E})/|\vec{B} -
\vec{\beta} \times \vec{E}|$, $n_{i}^{\prime} = P_{ik}n_{k}$,
$n_{E,\,i}^{\prime} = P_{ik}n_{E,\,k}$, $n_{Bi}^{\prime} =
P_{il}n_{Bl} = P_{i3}$.  The components of the vector
$\vec{\Omega}(d)$ are given by $\Omega_{r}(d)$ (the indices $r=1,2,3$
correspond to $x,y,z$, respectively),
\begin{eqnarray}
\vec{\Omega}(d) & = &
\frac{e}{mc}\left\{\left(a+\frac{1}{\gamma}\right)\vec{B}
-a\frac{\gamma}{\gamma+1}\left(\vec{\beta}\cdot\vec{B}\right)\vec{\beta}-
\left(\frac{g}{2}-\frac{\gamma}{\gamma+1}\right)\vec{\beta}\times\vec{E}\right\} \nonumber \\
& &
+\frac{d}{\hbar}\left({\vec{E}}+\vec{\beta}\times\vec{B}\right).
\label{2.24}
\end{eqnarray}
From Eq.~(\ref{50}) it follows that the magnetic polarisability leads
to spin rotation with two frequencies $\omega_1$ and $\omega_2$
instead of $\Omega$ and, therefore, experiments beating with the
frequency $\Delta \omega=\omega_1-\omega_2=2 \sqrt{\Omega_T^{\mu}
\Omega_T^{\prime \mu}}=\beta_T B_{eff}^2/\hbar$.  According
to the evaluation in Refs.~\cite{birefringence,VG+S}, the tensor
magnetic polarisability $\beta_T \sim 2 \times 10^{-40}$, therefore for
the beating frequency one obtains $\Delta \omega \sim 10^{-5}$ in a
field of $B \sim 10^4$ gauss.  The measurement of the frequency of
this beating makes possible the determination of the tensor magnetic
polarisability of the deuteron (nuclei).  Thus, due to the presence of
the tensor magnetic polarisability, the horizontal component of the
spin rotates around $\vec{B}$ with two frequencies
$\omega_1,~\omega_2$ instead of the expected rotation with the
frequency $\Omega$. The resulting motion of the spin is beating:
$P_1(t) \sim \cos \Omega t \sin \Delta \omega t$.  This is the reason
for the component $P_3$, caused by the EDM, to exhibit a similar
beating.

Another class of experiments deals with the use of polarised
targets. One should recall that the density of a polarised gas
target is substantially lower than the one of an unpolarised
target, the anticipated density of the polarised target of ANKE,
described in \S\ref{PIT}, is about $d_t = 10^{14}$ cm$^{-2}$. When
a particle (proton, antiproton, deuteron ...) propagates through a
nuclear polarised target, precession of the particle spin occurs
In a polarised target the particles can be characterised by two
refraction indices, $N_{\uparrow \uparrow }$ for particles with
spin parallel to the target polarisation vector, and $N_{\uparrow
\downarrow }$ for particles with the opposite spin orientation
$N_{\uparrow \uparrow }\neq N_{\uparrow \downarrow }$. According
to \cite{birefringence,VG+S}, in the target with polarised nuclei,
there is a nuclear pseudomagnetic field and the interaction of an
incident particle with this field leads to spin rotation. The
experiments with slow neutrons proved the existence of this effect
\cite{birefringence,VG+S}.

The effective potential energy of a particle in a pseudomagnetic
nuclear field $\vec{G}$ of matter can be written as:
\begin{equation}
\hat{V}_{B}^{nucl}=-\vec{\mu} \vec{G},
 \label{Vnucl}
\end{equation}
where $\vec{\mu}$ is the magnetic moment of the particle and $\vec{G}$
can be expressed as~\cite{birefringence,VG+S} $\vec{G} = \vec{G}_s +
\vec{G}_w$, where
\begin{equation}
\vec{G}_s = \frac{2 \pi \hbar^2}{\mu m}\rho [A_1
\langle \vec{J} \rangle + A_2 \vec{n} (\vec{n}\cdot\langle \vec{J}
\rangle)+...],\end{equation}
\begin{equation}
\vec{G}_w = \frac{2 \pi \hbar^2}{\mu m}\rho [b
\vec{n} + b_1 [\langle \vec{J} \rangle \times \vec{n}]+b_2 \vec{n}_1
+b_3\vec{n}(\vec{n}\cdot\vec{n}_1)+b_5 [\vec{n} \times
  \vec{n}_1]+...].\end{equation}
Here $\vec{n}=\vec{v}/v$, $\vec{J}$ is the nuclear spin, $\langle
\vec{J} \rangle=\textrm{Sp}\left\{\rho_{nucl} \vec{J}\right\}$ is the average value
of the nuclear spin, $\vec{n}_1$ are the components $n_{1j}=\langle
Q_{ij} \rangle n_j$, where $\langle Q_{ij} \rangle=\textrm{Sp}\left\{
\rho_{nucl} Q_{ij}\right\}$ is the polarisation tensor.

It is easy to see that the interaction, described by Eq.~\ref{Vnucl})
looks like the interaction of a magnetic moment with a magnetic field,
thus the field $\vec{G}$ contributes to the change of the particle
polarisation similar to the one of a magnetic field. It should be
mentioned that $\hat{V}_{B}^{nucl}$ contains both a real part, which
is responsible for spin rotation, and an imaginary part, which
contributes to the spin dichroism (beam absorption depending on the
relative spin orientation). A detailed analysis of the effects caused
by the nuclear pseudoelectric field is described in
Ref.~\cite{birefringence}.

The interaction with the field $ \vec{G}=\vec{G}_s+\vec{G}_w$ contains
two terms: $\vec{G}_s$ corresponds to the strong interaction, which is
$T$, $P$--even, while $\vec{G}_w$ describes spin rotation by the weak
interaction, which has contributions from both $T$, $P$--odd (the term
containing the constant $b_1$) and $T$--odd, $P$--even (the term containing
the constant $b_5$) interactions.  If either the vector or the tensor
polarisation of the target rotates, the effects provided by
$\vec{G}_s$ and $\vec{G}_w$ vary periodically as a function of time,
i.e. Eq.~(\ref{1}) can be written as
\begin{equation}
i\hbar\frac{\partial\Psi(t)}{\partial
t}=\left(\hat{H}_{0}+\hat{V}_{EDM}+\hat{V}_{\vec{E}}+\hat{V}_{\vec{B}}+\hat{V}_E^{nucl}(t)+
\hat{V}_{{B}}^{nucl}(t)\right)\Psi(t). \label{1(t)}
\end{equation}
This equation coincides with the well-known equation for a
paramagnetic resonance. Thus, in a strong magnetic field, orthogonal
to the weak nuclear pseudomagnetic field $\vec{B} \perp \vec{G}$,
$\vec{G}$ either rotates or oscillates with the frequency
corresponding to the splitting caused by $\vec{B}$, a resonance occurs
and this leads to the conversion of horizontal spin components to
vertical ones with the frequency determined by the spin precession
frequency in the field $\vec{G}$~\cite{VG+S}. Thus, one can measure
all the constants contained in $\vec{G}_s$ and $\vec{G}_w$: $A_i$
yields the spin-dependent part of the elastic coherent forward
scattering amplitude of proton (deuteron, or antiproton); the
amplitudes $b_i$ provide a measurement of the constants of $T$-- and
$P$--odd interactions.

The $T$--odd nucleon-nucleon interaction of protons (or antiprotons) and
deuterons with polarised nuclei is described by $V_{P,T} \sim \vec{S}\cdot
\left( \vec{p}_N \times \vec{n} \right)$, where $\vec{P}_N(t)$ is the
polarisation vector of target.  The interaction $V_{P,T}$ leads to the
spin rotation around an axis given by the unit vector $\vec{n}_T$,
parallel to $\left[ \vec{P}_N(t) \times \vec{n} \right]$.  Spin
dichroism also appears with respect to this vector $\vec{n}_T$, i.e. a
proton (deuteron) beam with spin parallel to $\vec{n}_T$ exhibits
different absorption cross-sections depending on the spin direction.

$P$--even, $T$--odd spin rotation, oscillation and dichroism of deuterons
(nuclei with $S \ge 1$) caused by the interactions of $V_T \sim
\vec{S}\cdot(\vec{P}_N(t) \times \vec{n})(\vec{S}\cdot\vec{n})$ could be
observed~\cite{birefringence}; $P$--even, $T$--odd spin rotation and
dichroism for a proton, deuteron (or more generally for a nucleus with
spin $S \ge 1/2$) $V^{\prime}_T \sim b_5 [\vec{n} \times
\vec{n}_1(t)]$ could be observed~\cite{birefringence,VG+S} in
paramagnetic resonance conditions as well.

The observation of spin rotation and birefringence of particles stored
in a ring provides the opportunity to measure the real part of the
coherent elastic zero-angle scattering amplitude, as well as tensor
electric and magnetic polarisabilities.  The same information could be
obtained also from particles and nuclei in a storage ring by the
paramagnetic resonance method, induced by time varying nuclear
pseudoelectric and pseudomagnetic fields.
%
%
\newpage
\section{Preparatory Work for the FAIR Project}
\label{PWFP}
\subsection{The PAX proposal at FAIR}

The PAX Collaboration is proposing a programme of spin experiments
with intense beams of polarised antiprotons at FAIR. A practical
and viable scheme to reach polarisations of the stored antiprotons
at HESR--FAIR of about 30\% has been presented~\cite{ap}. The
above performance is expected based on the electron--to--proton
spin transfer interpretation~\cite{meyer,horowitz-meyer} of the
FILTEX experiment~\cite{rathmann}; such an approach is routinely
used at Jefferson Laboratory for the separation of electromagnetic
form factors~\cite{JlabFF}.

The PAX Technical Proposal was submitted in January 2004. The
physics programme was reviewed by the QCD Programme Advisory
Committee (QCD--PAC) in June 2004~\cite{paxweb}. Following the
QCD--PAC report and the recommendation of the Chair of the
Committee on Scientific and Technological Issues (STI) and the
FAIR project coordinator~\cite{paxweb}, the PAX collaboration has
optimised the technique to achieve a sizable antiproton
polarisation. The studies are contained in the recently submitted
Technical Proposal~\cite{paxweb}. The goal of achieving the
highest possible figure of merit requires that the antiprotons be
polarised in a dedicated low--energy ring. The transfer of
polarised low--energy antiprotons into the HESR ring would then
require pre--acceleration to about 1.5{\ww}GeV/c in a dedicated
booster ring (CSR).  The incorporation of this booster ring into
the HESR complex would, quite naturally, open up the possibility
of building an asymmetric antiproton--proton collider.

The main features of the accelerator setup, shown in
Fig.~\ref{fig:CSRring}, are:
\begin{enumerate}
\item An Antiproton Polariser Ring (APR), built inside the HESR
area, having the crucial goal of polarising antiprotons at kinetic
energies around $50${\ww}MeV ($p_{\bar{p}}\approx 300
${\ww}MeV/c). These would
subsequently be accelerated and injected into the other rings.%
\item A second Cooler Synchrotron Ring (CSR) in which protons or
antiprotons with momenta up to 3.5{\ww}GeV/c could be stored. This
ring, whose parameters would be rather similar to those of COSY,
should have a straight section running parallel to the
experimental
straight section of HESR, where a PAX detector could be installed.%
\item By deflecting the HESR beam into the straight section of the
CSR, both collider and fixed--target modes become feasible.%
\end{enumerate}

It is worthwhile stressing that, through the employment of the
CSR, a second interaction point is effectively formed that has a
minimum interference with PANDA. The proposed solution opens the
possibility of running simultaneously two different experiments.

\begin{figure}[h]
\begin{center}
  \includegraphics[width=0.9\linewidth]{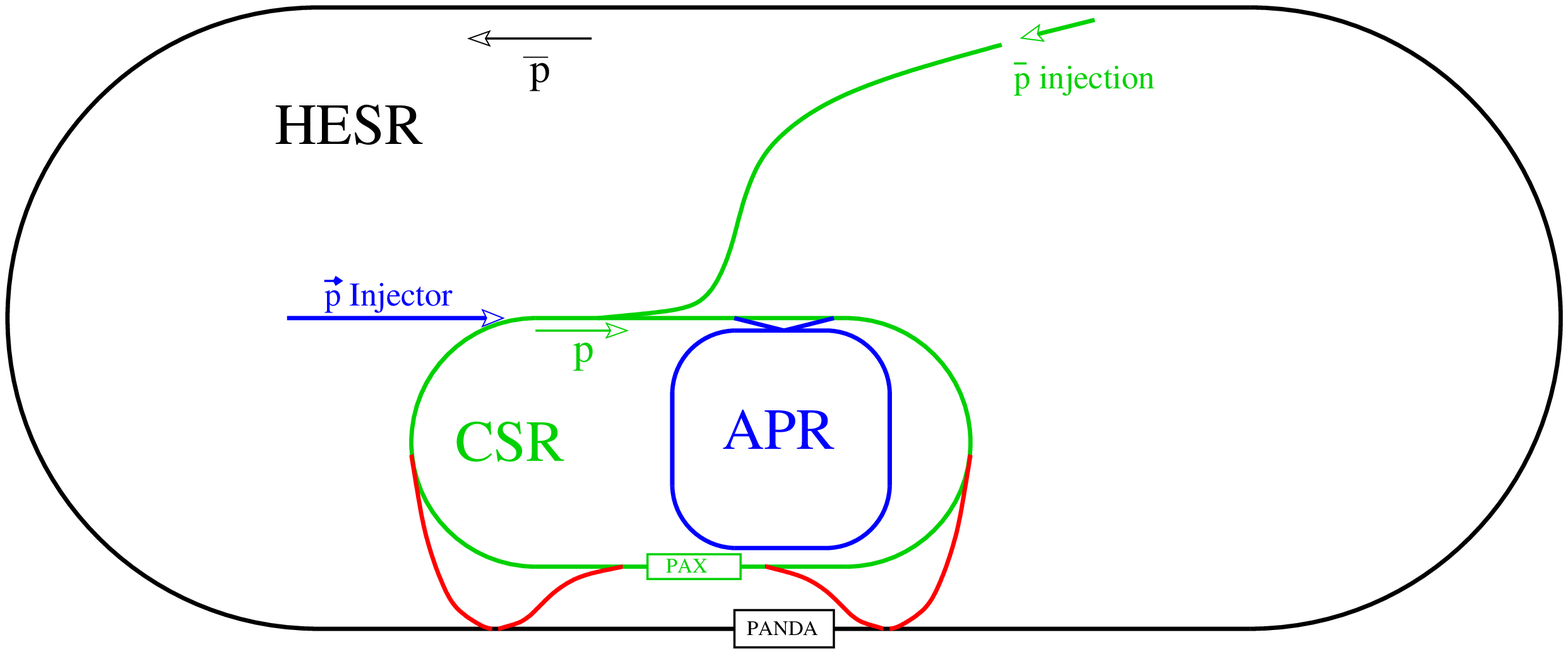}
  \parbox{14cm}{\caption{\label{fig:CSRring}\small The proposed
      accelerator setup at the HESR (black), with the equipment to be
      used by the PAX collaboration in Phase--I: CSR (green), APR,
      beam transfer lines and polarised proton injector (all blue). In
      Phase--II, by adding two transfer lines (red), an asymmetric
      collider is set up. It should be noted that, in this phase,
      fixed target operation at PAX is also possible. (The figure is
      drawn to scale.)}}
\end{center}
\end{figure}

\subsection{Preparatory phase}
The objectives arising from the PAX proposal for the immediate
future (3--4 years) concern mainly the polarisation buildup
process through which polarised antiprotons will be produced. The
PAX collaboration suggests carrying out spin--filtering
experiments in storage rings to achieve a better understanding of
these processes in general. Furthermore, for antiprotons, the
experimental basis for polarisation buildup in a stored beam is
practically non--existent. Therefore, it is of highest priority
for PAX to perform spin--filtering experiments using stored
antiprotons, \emph{e.g.}\ at the AD of CERN. Once the experimental
data base is available, the final design of the APR can be
targeted. The physics of the buildup of polarisation in a storage
ring involves a number of features absent in conventional
scattering experiments, especially the into--the--beam scattering
of target particles, which have to be well understood before one
embarks on polarising antiprotons. A few simple spin--filtering
experiment carried out at COSY, would certainly enhance
substantially our understanding of these processes.  The PAX
proposal calls eventually for an asymmetric antiproton--proton
collider at the HESR. To this end, machine studies using
state--of--the--art simulation tools (\emph{e.g.}\ BETACOOL) and
in depth analyses of the phase space cooling processes, intrabeam
scattering effects and beam--beam interactions have to be carried
out. The goal of these studies is to optimise the collider and
fixed target scenarios of the PAX experiment with respect to
luminosity. This naturally involves also machine studies for the
dedicated Antiproton Polariser Ring (APR), which is at the core of
the PAX proposal.

\subsection{Spin Filtering Experiments with protons at COSY
and with antiprotons at the CERN AD}%

Central to the PAX proposal is spin filtering of stored
antiprotons by multiple passage through an internal polarised gas
target. The feasibility of the technique has been convincingly
demonstrated in the FILTEX experiment at TSR~\cite{rathmann}: for
23~MeV stored protons, a transverse polarisation rate of $\dd
P/\dd t = 0.0124 \pm 0.0006$ per hour has been reached with an
internal polarised atomic hydrogen target of areal density $6
\times 10^{13}$ atoms/cm$^2$. In view of the fundamental
importance of spin filtering for the PAX experiment, experiments
at COSY would provide the necessary data to test our present
understanding of spin--filtering processes in storage rings.
However, since for PAX the design of a dedicated APR is foreseen,
it is crucial that such experiments be performed eventually with
antiprotons at a suitable antiproton storage ring, \emph{e.g.}\ at
the CERN AD. For a proton impinging on a polarised hydrogen gas
target, the spin--dependent interaction leading to the buildup of
polarisation in the beam is well known; recent
investigations~\cite{milstein} have, however, shown that an
unambiguous quantitative understanding and interpretation of the
FILTEX result~\cite{rathmann} may not yet be available. In
addition, there are no experimental data available on the spin
correlations, or more generally, on any of the two--spin
observables in antiproton--proton scattering. The final goal is to
provide the experimental data base of antiproton--proton
interactions that is necessary to define the optimum working
parameters of the dedicated APR, suggested for FAIR. Ultimately,
these measurements are needed in order to arrive at the final
design parameters for the APR. The new data will, moreover,
provide new insight into the $\bar{p}p$ physics more generally.

\subsection{Design and construction of the Antiproton Polariser Ring}

Tuning and commissioning of the APR will require a beam of
polarised protons.  Such a beam and a hall, including the
necessary infrastructure, are available at COSY. This makes the
Institut f\"ur Kernphysik of the Forschungszentrum J\"ulich the
ideally suited site for the design, construction and testing of
the APR. A floor plan of the polariser ring is shown in
Fig.~\ref{fig:APR-floor}. For more details, see Ref.~\cite{paxweb}
\begin{figure}[htb]
\begin{center}
  \includegraphics[width=0.70\linewidth]{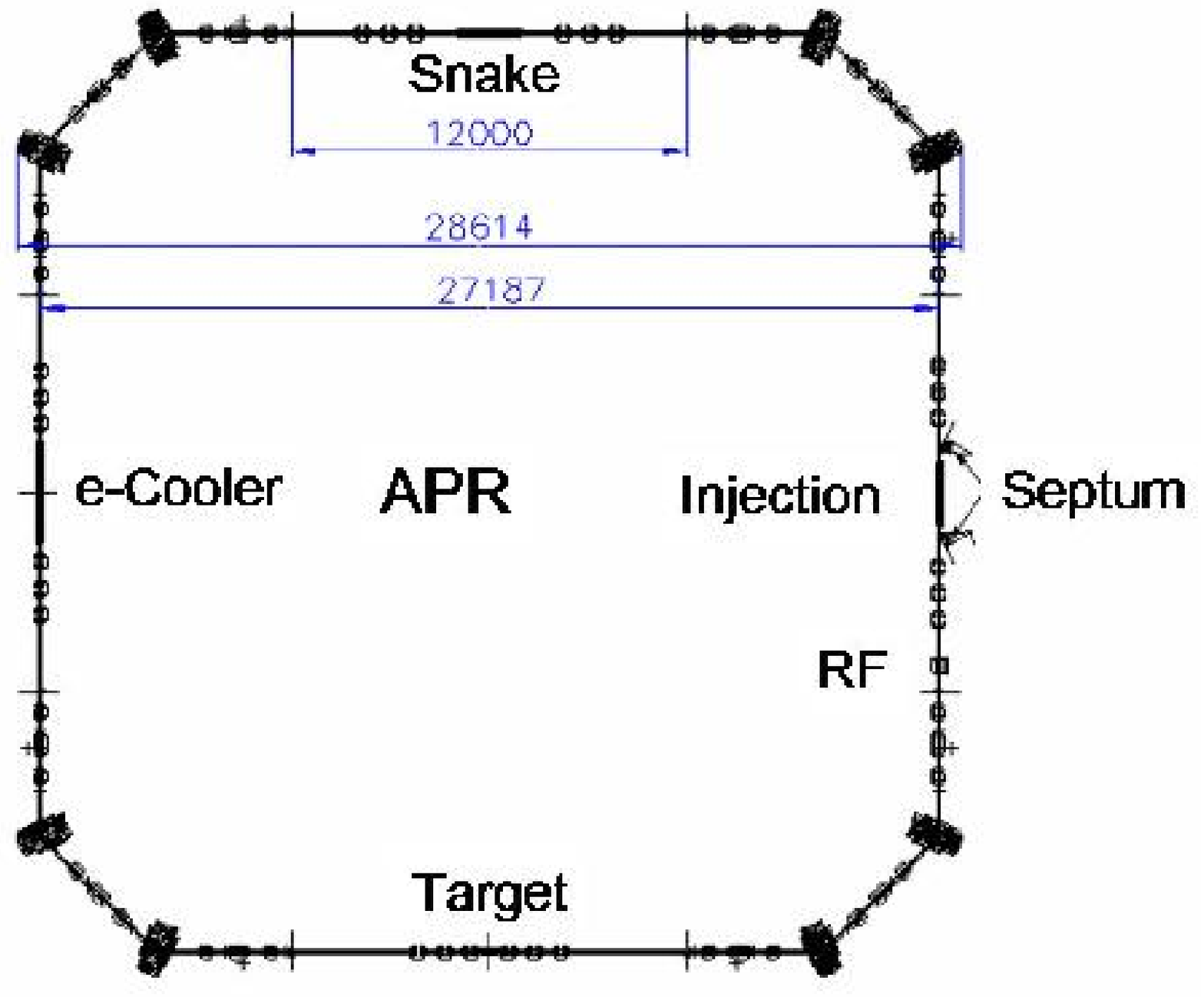}
  \parbox{14cm}{\caption{\label{fig:APR-floor} \small Floor plan of
      the APR lattice.}}
\end{center}
\end{figure}

The technical feasibility of the polarisation transfer method was
verified in the FILTEX experiment at TSR--Heidelberg in 1992 with
a 23{\ww}MeV proton beam~\cite{rathmann}. As mentioned above, a
pre--APR test can be performed at COSY by verifying
$\sigma_{EM_\perp}$ at 40, 70 and 100{\ww}MeV, using the polarised
internal target at the ANKE interaction point. The measurement can
be performed by injecting pure states $\ket{1}$ or $\ket{3}$ in a
weak transverse target guide field (10{\ww}G). Under these
conditions the electron target polarisation $Q_e$ is equal to the
proton target polarisation $Q_p$. The latter can be measured
through $pp$ elastic scattering with the help of the ANKE silicon
telescope system described in \S\ref{SST}~\cite{Schleichert:2003}.

\subsection{Development of polarised sources}

The polarisation mechanism relies on an having an efficient,
high--intensity, source of polarised hydrogen atoms. Most of the
World--expertise on polarised sources is already present within
the PAX collaboration and a programme for the development of a new
generation of high--intensity atomic beam sources has to be
started during the preparatory phase of the PAX experiment. The
COSY experience in polarised sources and polarised beams will be
crucial in this development.

\subsection{Staging of the PAX experiment and the role of COSY}

The final goal of the PAX collaboration is a polarised
proton--antiproton asymmetric collider, in which 3.5{\ww}GeV/c
polarised protons will collide head--on with polarised antiprotons
with momenta up to 15{\ww}GeV/c. However, the collaboration
proposes an approach composed of different stages, during which
the major milestones of the project can be tested and optimised.

\subsection{Phase--I: transfer of APR and CSR to FAIR}

APR and CSR will be placed inside the HESR.  The straight sections
of CSR and HESR will be parallel, allowing an additional
interaction point, independent of that of PANDA, to be formed.

A beam of unpolarised or polarised antiprotons, with momenta up to
3.5{\ww}GeV/c, will be available in the CSR ring. Collisions with
a polarised hydrogen target in the PAX detector can then be
studied. This first phase, which is independent of the HESR
performance, will allow for the first time the measurement of the
time--like proton form factors in single-- and double--polarised
reactions from close to threshold up to 3.5{\ww}GeV/c. It will be
possible to determine several (single and double) spin asymmetries
in the elastic $p\bar{p}\to p\bar{p}$ scattering. By detecting
back--scattered antiprotons, one can also explore the hard
scattering regions of large $t$. (Reaching the same region of $t$
in proton--proton scattering requires twice as high an energy.)

\subsection{Phase--II: HESR modifications to the collider mode or to the polarised
internal target}
A chicane for CSR and HESR would have to be built to bring the
proton beam of the CSR and the antiproton beam of the HESR to a
collision point at the PAX detector. This phase will allow the
first ever direct measurement of the quark transversity
distribution $h_1$, by studying the double transverse spin
asymmetry $\Att$ in the Drell--Yan processes $p^{\uparrow}
\bar{p}^{\uparrow} \rightarrow e^+ e^- X$ as a function of Bjorken
$x$ and $Q^2$ (= $M^2$).  Two possible scenarios might be foreseen
to perform this measurement.

\begin{figure}[hbt]
 \begin{center}
 \includegraphics[width=0.49\linewidth]{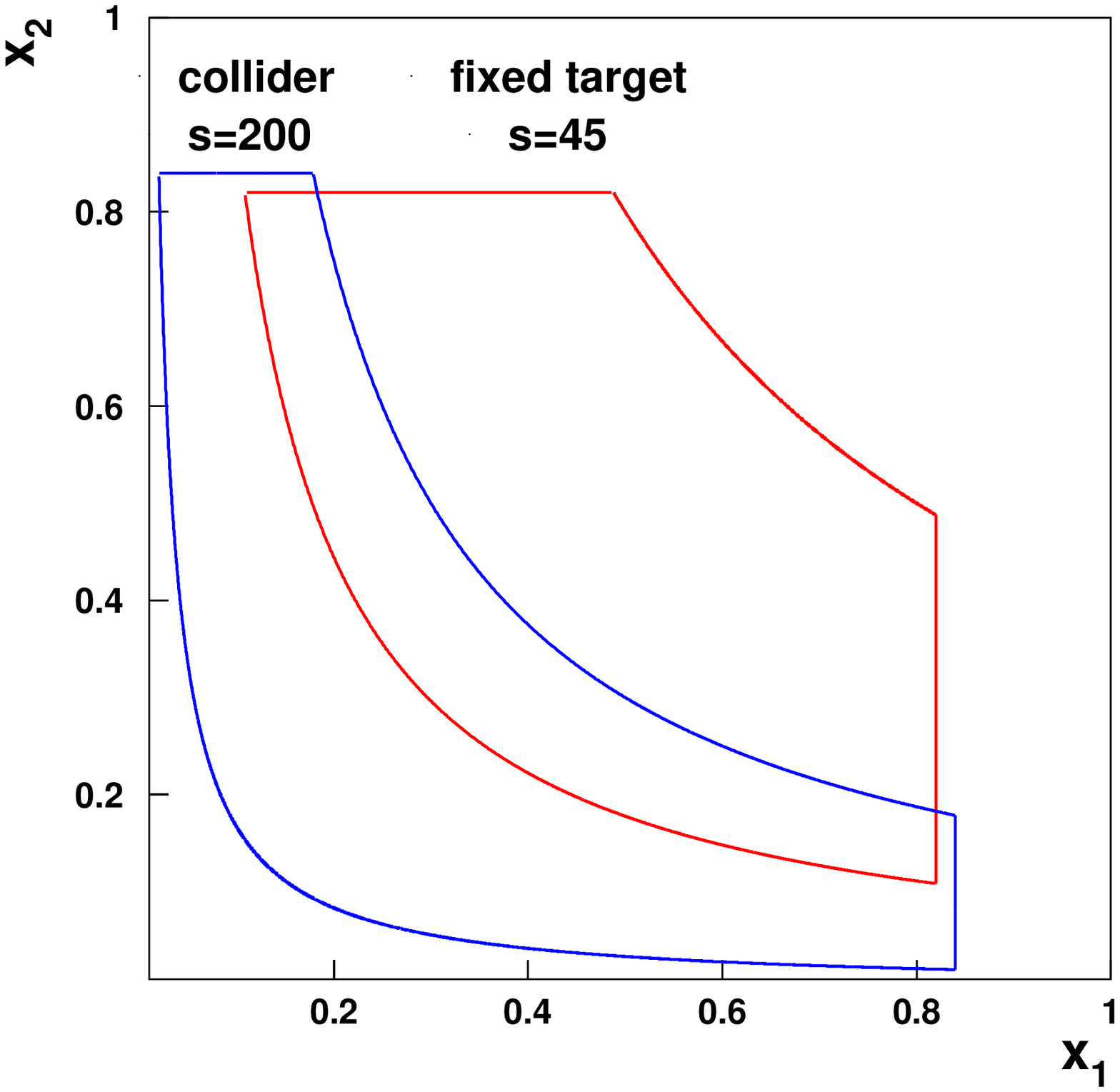}
  \includegraphics[width=0.49\linewidth]{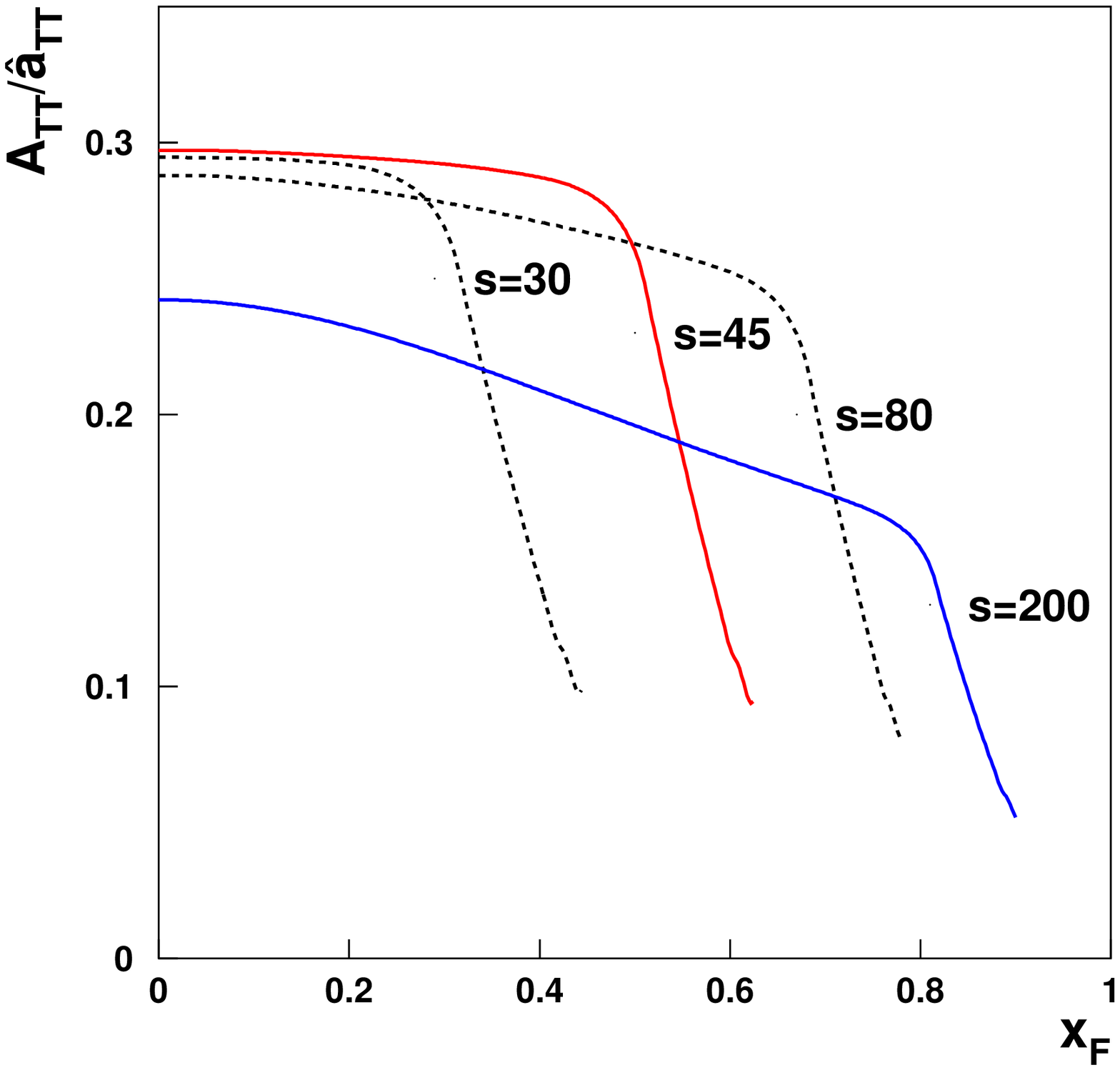}
  \parbox{14cm}{\caption{\label{Figphysics}\small Left: The kinematic
      region covered by the $h_1$ measurement at PAX in phase II. In
      the asymmetric collider scenario (blue), antiprotons of
      15{\ww}GeV/c impinge on protons of 3.5{\ww}GeV/c at cm energies
      of $\sqrt{s}\sim \sqrt{200}${\ww}GeV and $Q^2>4${\ww}GeV$^2$.
      The fixed target case (red) represents antiprotons of
      22{\ww}GeV/c colliding with a fixed polarised target
      ($\sqrt{s}\sim\sqrt{45}${\ww}GeV).  Right: The expected
      asymmetry as a function of Feynman $x_F$ for different values of
      $s$ and $Q^2=16${\ww}GeV$^2$.}}
\end{center}
\end{figure}

\begin{itemize}
\item[(a)] A beam of polarised antiprotons from 1.5{\ww}GeV/c up
to 15{\ww}GeV/c, circulating in the HESR, collides with a beam of
3.5{\ww}GeV/c polarised protons circulating in the CSR. This
scenario requires one to demonstrate that a suitable luminosity is
reachable. Deflection of the HESR beam to the PAX detector in the
CSR is necessary (see Fig.~\ref{fig:CSRring}).  By properly
varying the energy of the two colliding beams, this setup would
allow a measurement of the transversity distribution $h_1$ in the
valence region of $0.1<x<0.8$ with the corresponding $Q^2$ in the
range $4<Q^2<100${\ww}GeV$^2$ (see Fig.~\ref{Figphysics}). As
shown in Fig.~\ref{Figphysics}, the double transverse spin
asymmetry ($A_{TT}$) in Drell--Yan processes with \emph{both
transversely polarised beam and target} is predicted to be greater
than 0.2 in the full kinematic range~\cite{abdn,gms}.
\begin{equation}
A_{TT} \equiv \frac{d\sigma^{\uparrow\uparrow} -
d\sigma^{\uparrow\downarrow}} {d\sigma^{\uparrow\uparrow} +
d\sigma^{\uparrow\downarrow}} = \hat{a}_{TT} \> \frac{\sum_q e_q^2
\, h_1^q(x_1, M^2) \, h_1^{\bar q}(x_2, M^2)} {\sum_q e_q^2 \,
q(x_1, M^2) \, \bar q(x_2, M^2)}\>, \label{att}
\end{equation}
where $q = u, \bar u, d, \bar d, ...$, $M$ is the invariant mass
of the lepton pair, and $\hat{a}_{TT}$ is the double spin
asymmetry of the QED elementary process, $q \bar q \to \ell^+
\ell^-$,
\begin{equation}
\hat a_{TT}= \frac{\sin^2\theta}{1 + \cos^2\theta}\cos 2\phi \,,
\end{equation}
with $\theta$ the polar angle of the lepton in the $l^+l^-$ rest
frame and $\phi$ the azimuthal angle with respect to the proton
polarisation.

Such an experiment for $h_1$ can be considered as the analogue of
polarised deep inelastic scattering for the helicity distribution
$\Delta q$. The kinematical coverage in $(x,Q^2)$ will in fact be
similar to that of the HERMES experiment~\cite{HERMES}.

\item[(b)] Should the desired luminosity for the collider not be
reachable, a fixed target experiment can be performed.  A beam of
22{\ww}GeV/c (or 15{\ww}GeV/c) polarised antiprotons circulating
in the HESR, can be used to collide on a polarised internal
hydrogen target.  This scenario also requires the deflection of
the HESR beam to the PAX detector in the CSR (see
Fig.~\ref{fig:CSRring}).
\end{itemize}

\newpage
%
\section{Time Schedules}
\label{TS}
\subsection{General remarks}

At present, we consider that the measurements within our spin
programme should take place during the four years 2006 -- 2009. We
assume that during this period COSY is fully operational,
providing beams from its polarised ion source, and that the ANKE
polarised internal target, together with the ANKE spectrometer,
are also in operation. However, the time schedule for the
spin--physics programme ANKE@COSY will, of course, depend to some
extent on the planning and progress for the realisation of the
FAIR project.

There are currently eight PhD students (D.~Chiladze, K.~Grigoriev,
D.~Gussev, A.~Mussgiller, V.~Leontiev, T.~Mersmann, T.~Rausmann,
and Yu.~Valdau) working on spin--physics subjects within the
framework of the ANKE collaboration.  Their projects involve
Monte--Carlo simulations for the different experiments as well as
data analysis of experiments with polarised deuteron and proton
beams.  Recent data taken with the (unpolarised) storage cell
target, which will provide a test for the upcoming
double--polarised experiments, are also in the course of analysis.
The strong involvement of members of the theory group in J\"ulich,
as well as of external experts, provides an excellent theoretical
support of the spin--physics programme at ANKE.

\subsection{Submissions and requests}

\begin{itemize}
\item  April 2005  -- elements of the spin document included in
the submission to the POF evaluation committee were favourably
received~\cite{POF2005}.%
\item  October 2005  -- presentation of the full document to
the COSY PAC.%
\item Spring 2006  -- beam--time request for $np$
double--polarised experiment.
\end{itemize}

\subsection{Timelines}

As described in this document, the spin--physics programme can be
broadly grouped into four categories, addressing the following
subjects:\\[1ex]
1.\ Proton--neutron spin physics,\\%
2.\ Non--strange meson production,\\%
3.\ Production of strange mesons and baryons,\\%
4.\ Preparatory work for the FAIR project.\\

\newpage
\noindent The following time schedule is anticipated for the
delivery of this programme:\\

\begin{tabular}{rcp{10cm}}
6/2005        & --- & Installation of the PIT at the ANKE target position\\
until 6/2006  & --- & Commissioning of the PIT\\
7/2006        & --- & First double--polarised $np$ experiments \\
2007          & --- & Data taking with ANKE for item 1 ($\approx 3$ months)\\
2008          & --- & Data taking with ANKE for item 2 ($\approx 3$ months)\\
2009          & --- & Data taking with ANKE for item 3 ($\approx 3$ months)\\
2007--11      & --- & Data analysis \& publication of results
\end{tabular}
\vspace{5mm}

\addtocounter{table}{1} \noindent A list of some of the possible
experiments that will be carried out in parallel is presented in
Table~\arabic{table}. The last item (4), preparations for FAIR,
mainly addresses the design, building and commissioning of the
Antiproton Polariser Ring, for which the time estimates are:\\

\begin{tabular}{rcp{10cm}}
 2006--2007 & --- & APR machine design\\
           &     & Preparation of ``ready--to--be--built'' documents\\
           &     & Development of a new generation of high--intensity  ABS for the APR\\

 2008--9    & --- & Construction of APR at IKP\\
 2010--2011 & --- & Commissioning  of APR\\
 $>2012$   & --- & Transfer of APR and CSR to HESR
\end{tabular}
\newpage
\baselineskip 3ex \vspace*{-1.3cm}

\centerline{\textbf{Table~\arabic{table}:} Table of experiments
that will be done simultaneously.}\vspace{5mm}

\noindent {\bf Phase I:} Polarised deuteron beam \& polarised
hydrogen storage-cell target $(dp)$;\\
Energy range: $1.2<T_d<2.3${\ww}GeV;\\
Instrumentation:    ANKE, silicon telescopes, polarised
storage-cell target.\\
Methodical advantage: High intensity polarised deuteron beam, high
count rate, clean identification of many two-body reactions,
polarimetry of target can be done without Lamb-shift polarimetry,
polarimetry standards are already achieved including polarisation
export technique, preparation of the polarised hydrogen target
relatively easy.
\\[1ex]
\textbf{Reactions:} \\[1ex]
$dp\rightarrow(2p)n$ small momentum transfer
charge-exchange\\
$dp\rightarrow(2p)\Delta^0$ at minimum momentum transfer\\
$dp\rightarrow dp$ small angle scattering\\
$dp\rightarrow dp$ large angle scattering\\
$dp\rightarrow (2p)n$ large momentum transfer charge-exchange\\
$dp\rightarrow\,^3\textrm{He}\,\pi^0$\\
$dp\rightarrow\,^3\textrm{He}\,\eta$\\
$np\rightarrow d\pi^0$ quasi-free\\
$np\rightarrow d X (\pi\pi)$\\
$pp\rightarrow pp$ quasi-free  \\[2ex]
\textbf{Phase II:} Polarised proton beam \& Polarised deuteron
storage-cell target $(pd)$;\\[1ex]
Energy range $T_p$ = 1.0{\ww}GeV up to 3.0{\ww}GeV;\\
Instrumentation: ANKE, silicon telescopes, polarised storage-cell
target, Lamb-shift polarimeter.\\
Methodical aspects: need good adjustment of the trigger
conditions.
\\[1ex]
\textbf{Reactions:} \\
$pd\rightarrow(2p)n$ small momentum transfer charge-exchange\\
$pd\rightarrow p_{sp} p(n)$     quasi-elastic scattering (pn)\\
$pd\rightarrow p_{sp} p(\Delta) (pn\rightarrow \Delta p)$\\
$pd\rightarrow pd$ small angle scattering\\
$pd\rightarrow pd$ large angle scattering\\
$pd\rightarrow (2p)n$ large momentum transfer charge-exchange\\
$pd\rightarrow p_{sp} pp \pi^- (pn\rightarrow pp \pi^-)$\\
$pd\rightarrow n_{sp} pp \pi^0 (pp\rightarrow pp \pi^0)$\\[2ex]
$pd\rightarrow p_{sp} K^+ X   (pn\rightarrow n K^+ \Lambda)$\\
$pd\rightarrow n_{sp} K^+ X   (pp\rightarrow p K^+
\Lambda)$\\[2ex]
$pp\rightarrow pp \eta, \omega, \cdots$\\
$pn\rightarrow d \phi$\\
$pn\rightarrow d X$

%
%

%
%
\end{document}